\documentclass[
 journal=pasa,
 manuscript=research-paper, 
 year=2020,
 volume=37,
]{cup-journal}
\usepackage{microtype,siunitx,booktabs}
\usepackage[nolist]{acronym} 
\usepackage{graphicx}
\usepackage{amsmath}	
\usepackage{amssymb}	
\usepackage{multicol} 
\usepackage{bm}		
\usepackage{pdflscape}	
\usepackage{aas_macros}
\usepackage{hyperref} 
\usepackage[T1]{fontenc}
\usepackage{ae,aecompl}
\usepackage{pdflscape}
\usepackage{longtable}
\usepackage{subfigure}
\usepackage{xcolor}

\definecolor{byzantine}{rgb}{0.74, 0.2, 0.64}

\newcommand{\lsun}{\ifmmode{{\rm ~L}_\odot}\else{~L$_\odot$}\fi}

\newcommand{\Msun}{\ifmmode{{\rm ~M}_\odot}\else{~M$_\odot$}\fi}
\newcommand{\degr}{$^{\circ}$}
\newcommand{\sqdeg}{\,deg$^2$}
\newcommand{\arcsec}{\,arcsec}
\newcommand{\arcmin}{\,arcmin}
\newcommand{\ujybm}{\,$\mu$Jy/beam}

\newcommand{\rfr}{$a/b$}

\def\arcmin{\hbox{$^\prime$}}
\def\arcsec{\hbox{$^{\prime\prime}$}}

\newcommand{\ytag}{$_{\rm n}$}
\newcommand{\mtag}{$_{\rm m}$}
\begin{acronym}
\acro{3C}{Third Cambridge catalogue of Radio Sources}
\acro{AAT}{Anglo-Australian Telescope}
\acro{ACIS}{Advanced CCD Imaging Spectrometer}
\acro{AGN}{Active Galactic Nuclei}
\acro{ASKAP}{Australian Square Kilometre Array Pathfinder}
\acro{ATCA}{Australia Telescope Compact Array} 
\acro{ATNF}{Australia Telescope National Facility}
\acro{ATOA}{Australia Telescope Online Archive}
\acro{BCG} {Brightest Cluster Galaxy}
\acro{CABB}{Compact Array Broad-band Backend}
\acro{CASDA}{CSIRO Australian Square Kilometre Array Pathfinder Science Data Archive}
\acro{CNN}{Convolutional Neural Network}
\acrodefplural{CNN}{Convolutional Neural Networks}
\acro{CSIRO}{Commonwealth Scientific and Industrial Research Organisation} 
\acro{CSM}{Circumstella Medium}
\acro{DEC}{Declination}
\acro{DES}{Dark Energy Survey}
\acro{DESI}{Dark Energy Spectroscopic Instrument}
\acro{DRAGN}{Double Radio source associated with Active Galactic Nucleus}
\acrodefplural{DRAGN}{Double Radio sources associated with Active Galactic Nuclei}
\acro{EM}{Electromagnetic}
\acro{EMU}{Evolutionary Map of the Universe}
\acro{eROSITA}{extended ROentgen Survey with an Imaging Telescope Array}
\acro{FAST}{Five-hundred-meter Aperture Spherical Telescope}
\acro{FIR}{Far Infrared}
\acro{FIRST}{Faint Images of the Radio Sky at Twenty-centimeters \citep{Becker1995}}
\acro{FR}{Fanaroff-Riley}
\acro{FR0}{Fanaroff-Riley Class 0}
\acrodefplural{FR0}{Fanaroff-Riley Class 0's}
\acro{FR1}{Fanaroff-Riley Class I}
\acrodefplural{FR1}{Fanaroff-Riley Class I's}
\acro{FR2}{Fanaroff-Riley Class II}
\acrodefplural{FR2}{Fanaroff-Riley Class II's}
\acro{FoV}{Field of View}
\acro{GAMA}{Galaxy And Mass Assembly}
\acro{GLEAM}{GaLactic and Extragalactic All-sky Murchison Widefield Array}
\acro{GMRT}{Giant Metrewave Radio Telescope}
\acro{GRG}{Giant Radio Galaxy}
\acrodefplural{GRG}{Giant Radio Galaxies}
\acro{HERG}{High Excitation Radio Galaxy}
\acrodefplural{HERG}{High Excitation Radio Galaxies}
\acro{HI}[H\textsc{I}]{Atomic Hydrogen}
\acro{HII}[H\small{II}]{Ionised hydrogen}
\acro{HST}{Hubble Space Telescope}
\acro{HyMoRS}{HYbrid MOrphology Radio Sources}
\acro{ICM}{Intracluster Medium}
\acro{ID}{Identification}
\acro{IFRS}{Infrared-Faint Radio Source}
\acro{IGM}{Intergalactic Medium}
\acro{ISM}{Interstellar Medium}
\acro{IR}{infrared}
\acro{KDE}{Kernel Density Estimation}
\acro{KiDS}{Kilo-Degree Survey}
\acro{LERG}{Low Excitation Radio Galaxy}
\acrodefplural{LERG}{Low Excitation Radio Galaxies}
\acro{LMC}{Large Magellanic Cloud}
\acro{LINER}{Low Ionisation Nuclear Emission-line Region}
\acrodefplural{LINER}{Low Ionisation Nuclear Emission-line Regions}
\acro{LOFAR}{Low-Frequency Array \citep{lofar}}
\acro{LoTSS}{LOFAR Two-Metre Sky Survey}
\acro{MIR}{Mid Infrared}
\acro{MIRIAD}[\textsc{miriad}]{Multichannel Image Reconstruction, Image Analysis and Display}
\acro{Mpc}{Megaparsec}
\acrodefplural{Mpc}{Megaparsecs}
\acro{MWA}{Murchison Widefield Array}
\acro{NIR}{Near Infrared}
\acro{NRAO}{National Radio Astronomy Observatory}
\acro{NVSS}{the NRAO VLA Sky Survey, \citep{condon98}}
\acro{PA}{Position Angle}
\acrodefplural{PA}{Position Angles}
\acro{PAF}{Phased Array Feed}
\acrodefplural{PAF}{Phased Array Feeds}
\acro{PCA}{Principal Component Analysis}
\acro{PINK}{Parallelized rotation and flipping Invariant Kohonen}
\acro{PSPC}{Position Sensitive Proportional Counter}
\acro{QSO}{Quasi-stellar Object}
\acrodefplural{QSO}{Quasi-stellar objects}
\acro{RA}{Right Ascension}
\acro{RG}{Radio Galaxy}
\acrodefplural{RG}{Radio Galaxies}
\acro{RMS}{Root Mean Square}
\acro{ROSAT}{ROentgen SATellite}
\acro{SDSS}{Sloan Digital Sky Survey}
\acro{SED}{Spectral Energy Distribution}
\acro{SFG}{Star forming galaxy}
\acrodefplural{SFG}{Star forming galaxies}
\acro{SKA}{Square Kilometre Array}
\acro{SMBH}{Supermassive Black Hole}
\acrodefplural{SMBH}{Supermassive Black Holes}
\acro{SN}{Supernova}
\acro{SNR}{Supernova Remnant}
\acrodefplural{SNR}{Supernova Remnants}
\acro{solar mass}[M$_\odot$]{Solar Mass}
\acro{TGSS}{TIFR GMRT Sky Survey}
\acro{UV}{UltraViolet}
\acro{VLA}{Very Large Array}
\acro{VLASS}{Very Large Array Sky Survey \citep{lacy20}}
\acro{VLBI}{Very Long Baseline Array}
\acro{VLTS}{Very Large Telescope}
\acro{WISE}{Wide-field Infrared Survey Explorer}
\acro{XMM-Newton}[\emph{XMM-Newton}]{\emph{X-ray Multi-Mirror Mission}}
\end{acronym}

\newcommand{\total}{3557} 
\newcommand{\dragn}{3557} 
\newcommand{\catwise}{3182} 
\newcommand{\fri}{238}
\newcommand{\frii}{1410}
\newcommand{\frx}{1055}
\newcommand{\LTS}{696}
\newcommand{\bt}{243}
\newcommand{\hht}{17}
\newcommand{\HyMoRS}{42}
\newcommand{\oss}{34}
\newcommand{\dd}{20}
\newcommand{\ssh}{37}
\newcommand{\xsh}{12}
\newcommand{\tsh}{33}

\newcommand{\wtf}{15}
\newcommand{\complex}{59}
\newcommand{\nondragn}{36}

\title{EMU and the DRAGNs I: A Catalogue of DRAGNs}

\author{Ray P. Norris}
\affiliation{ATNF, CSIRO Space \& Astronomy, P.O. Box 76, Epping, NSW 1710, Australia}
\alsoaffiliation{Western Sydney University, Locked Bag 1797, Penrith, NSW 2751, Australia}
\email[Ray P. Norris]{ray.norris@csiro.au}

\author{Miranda Yew}
\affiliation{Western Sydney University, Locked Bag 1797, Penrith, NSW 2751, Australia}

\author{Evan Crawford}
\affiliation{Western Sydney University, Locked Bag 1797, Penrith, NSW 2751, Australia}

\author{Nikhel Gupta}
\affiliation{ATNF, CSIRO Space \& Astronomy, P.O. Box 76, Epping, NSW 1710, Australia}

\author{ Lawrence Rudnick}
\affiliation{Minnesota Institute for Astrophysics, University of Minnesota, 116 Church St. SE, Minneapolis, MN 55455, USA}

\author{H. Andernach}
\affiliation{Th\"uringer Landessternwarte, Sternwarte 5,
D-07778 Tautenburg, Germany}
\alsoaffiliation{Departamento de Astronom\'ia, Universidad de Guanajuato,
DCNE, Callej\'on de Jalisco s/n, Guanajuato, C.P. 36023, GTO, Mexico} 

\author{Miroslav D. Filipovi\'c}
\affiliation{Western Sydney University, Locked Bag 1797, Penrith, NSW 2751, Australia}

\author{Yjan A. Gordon}
\affiliation{Department of Physics, University of Wisconsin-Madison, 1150 University Avenue, Madison, WI 53706, USA}

\author{Andrew M. Hopkins}
\affiliation{School of Mathematical and Physical Sciences, 12 Wally's Walk, Macquarie University, NSW 2109, Australia}

\author{Laurence Park}
\affiliation{Western Sydney University, Locked Bag 1797, Penrith, NSW 2751, Australia}

\author{Michael J. I. Brown}
\affiliation{School of Physics \& Astronomy, Monash University, Clayton, VIC, Australia}

\author{Ana Jimenez-Gallardo}
\affiliation{European Southern Observatory, Alonso de C\'ordova 3107, Vitacura, Regi\'on Metropolitana, Chile}

\author{S. S. Shabala}
\affiliation{School of Natural Sciences, University of Tasmania, Hobart, Australia}
\alsoaffiliation{ARC Centre of Excellence for All Sky Astrophysics in 3 Dimensions (ASTRO 3D)}

\received {dd Mmm YYYY}
\revised  {dd Mmm YYYY}
\accepted {dd Mmm YYYY}
\published{dd Mmm YYYY}


\begin{document}

\begin{abstract}

We present a catalogue of \total\
\acp{DRAGN} 
from the First Pilot Survey of the \ac{EMU}, observed at 944 MHz with the \ac{ASKAP} telescope, covering 270~deg$^{2}$. We have extracted and identified each source by eye, tagged it with a morphological type and measured its parameters.
The resulting catalogue
will be used in subsequent papers to explore the properties of these sources, to train machine-learning algorithms for the detection of these sources in larger fields, and to compare with the results of Citizen Science projects, with the ultimate goal of understanding the physical processes that drive DRAGNs. Compared with earlier, lower sensitivity, catalogues, we find more diffuse structure and a plethora of more complex structures, ranging from wings of radio emission on the side of the jets, to types of object which have not been seen in earlier observations. As well as the well-known FR1 and FR2 sources, we find significant numbers of rare types of radio source such as Hybrid Morphology Radio Sources and one-sided jets, as well as a wide range of bent-tail and head-tail sources.

\end{abstract}

\section{INTRODUCTION }
\label{sec:intro}

The Evolutionary Map of the Universe \citep[EMU,][]{emu, pilot, hopkins25} is a large radio continuum survey being conducted using the Australian Square Kilometre Array Pathfinder \citep[ASKAP,][]{johnston07, johnston08, hotan21}. ASKAP consists of 36 12-metre antennas with an innovative phased-array feed at the focus of each, which provides an instantaneous field of view up to 30 \sqdeg, producing a much higher survey speed than that of previous synthesis arrays.
The EMU survey is expected to generate a catalogue of tens of millions of radio sources, although extracting, identifying, and classifying them will be a major challenge.
Here, we use data from the EMU Pilot Survey 1 \citep[hereafter EMU-PS1, ][]{pilot} to extract and classify a sample of Double Radio sources associated with Active Galactic Nuclei (DRAGNs)\citep{leahy93}.

Double-lobed radio sources dominate the population of extragalactic radio sources at high flux densities, and were classified by \citet{FR} as being Fanaroff-Riley Type II (FR2) sources, which are edge-brightened, and Fanaroff-Riley Type I (FR1) sources, where the maximum of the lobes occurs closer to the host galaxy. At lower flux densities, many more morphological types of \ac{AGN} appear that do not fit the simple FR1/FR2 dichotomy, especially low-surface brightness structures that were not recognised in the earlier AGN catalogues
{  \citep[e.g.][]{pilot, mingo19, mightee}}. In addition, at low flux densities the population becomes dominated by star-forming galaxies \citep{pilot}. In the EMU survey, most star-forming galaxies are unresolved, except for those at low redshift, where they are often observed as diffuse blobs of emission, roughly coincident with the optical image of the star-forming galaxy or sometimes with prominent spiral arms or other morphological features.

In this paper we focus on \acp{DRAGN}, following the definition of a \ac{DRAGN} given by \citet{leahy93}, which includes all radio activity that has evidence of \ac{AGN} outflows, including jets, hot-spots, lobes, double structure, and cores.

We therefore focus on sources with double structure, and \citep[like][]{leahy93} we do not include unresolved sources, or sources whose morphology is a single resolved Gaussian. Many such sources may be star-forming galaxies, but many will be AGN. Similarly, we also include one-sided jets, head-tail (HT) sources, and other complex morphologies. In these cases, we recognise that these are often not strictly ``double'' radio sources, although they may appear two-sided if observed with sufficient sensitivity and resolution.

Throughout this paper, we use the term ``jet'' to include any elongated structure between the core and the lobe, or radiating from the core, including curved elongated structures. We do not necessarily mean a collimated or relativistic jet. 

\subsection{Scientific Motivation}
This paper is primarily motivated by the need to understand the physical processes that drive the formation of radio jets and other structures around supermassive black holes, which we observe as DRAGNs.

Most of our existing knowledge of DRAGNs is based either on radio surveys covering large areas of sky with relatively low sensitivity, or on surveys with high sensitivity but covering only a small area. As a result, we know little about unusual sources at low flux densities or low surface brightness. In addition, most of these earlier survey have relatively poor sensitivity to diffuse, low surface brightness objects. Modern survey telescopes such as \ac{ASKAP}, \ac{LOFAR}, and MeerKAT \citep{jonas16} overcome both these limitations of earlier surveys by having a high survey speed, enabling the surveys to cover a wide area, and a spatial distribution of antennas that results in a high sensitivity to low surface brightness objects.

The EMU survey therefore samples a region of parameter space {  (e.g. sensitivity to low surface brightness, observing frequency, area surveyed)} that has not hitherto been observed. In this series of papers, we examine the properties of DRAGNs, and compare them to the properties of those found in earlier observations. Our motivation for doing so is threefold:
\begin{itemize}
    \item To find previously unknown classes of source -- the 
    ``unknown unknowns''. \citet{norris17b} described their WTF (Widefield ouTlier Finder) project which aimed {  to} make  such discoveries, and WTF has since been widely adopted as the acronym for such objects. \citet{norris17b} showed that WTFs are likely to be widespread in the next generation of radio surveys.  The first such discovery from the EMU survey were the Odd Radio Circles \citep{orc}. We present additional WTFs in this paper.
    \item To find features and unusual morphologies of DRAGNs that were not apparent in earlier observations. We present several examples in this paper, such as diffuse emission and unusual structures associated with DRAGNs, that may challenge existing AGN models and perhaps trigger new insights.
    \item To examine how the properties of \acp{DRAGN} change as we probe to fainter populations of sources, and to explore whether existing models can accommodate the changes that we find. 
\end{itemize}

\subsection{Classifying Sources in Large Radio Surveys}
Classifying radio sources and cross-identifying them with the host galaxies in radio/IR surveys is extremely difficult because of the diverse morphologies and because a radio source may consist of two components separated by arcminutes on either side of the host galaxy. The problem is exacerbated by the large numbers (up to tens of millions) of sources found in modern radio surveys \citep{norris17a}. Classifying and cross-identifying such a large number of sources by eye is impossible, and yet the classifications and identifications are essential to extract the astrophysics. 

There are two main approaches to this challenge of classifying millions of radio sources.

One is the use of citizen science (CS) programs such as RGZ \citep{banfield15, wong25}, RGZ-LOFAR \citep{hardcastle23}, and RGZ-EMU \citep{bowles23}. For example, in RGZ, thousands of citizen scientists classified about a hundred thousand radio sources and identified them with IR sources, and also identified peculiar classes of radio sources such as the spiral DRAGNs \citep[][and Keel et al., in preparation]{keel22}.

The second approach is the use of machine learning (ML) algorithms. Several different ML approaches have been published \citep[e.g.][]{wu19,galvin20,gupta22,segal23,gupta23,gupta24a}, but most require representative samples of ``ground truth'' identifications  for training, {  and even unsupervised algorithms \citep[e.g.][]{galvin20,mostert21,vantyghem24} require a ground truth dataset for} labelling, which can be provided either by expert manual identification or by CS programs. Similarly, CS approaches also need representative samples against which they can be validated to test their accuracy. It should be noted that these representative samples need to be chosen from a survey with very similar observing parameters as the survey to be targeted by the ML or CS projects, as the properties of radio sources vary significantly with sensitivity, resolution, and observing frequency. For example, an algorithm trained for the \ac{LOFAR} cannot necessarily be used directly for ASKAP data, and vice versa.

A third, less-widely adopted, approach, has been the development of non-ML algorithms to identify particular classes of source \citep[e.g.][]{proctor11, gordon23}. However, these algorithms still need a ``ground truth'' validation set to develop and test them.

A key goal of the present paper is to provide such a training/validation set that will facilitate CS and ML identifications for the main EMU project \citep{emu, hopkins25}. A preliminary version of the catalogue presented here has already been used to train an ML algorithm \citep{gupta23b,gupta23,gupta24a,gupta24b,gupta25submit,gupta25emuse}.

Specifically, the catalogue presented in this paper has been obtained by scanning images by eye from EMU-PS1, without the use of any source finders. We hope, by doing so, to have compiled an unbiased catalogue of nearly all DRAGNs in EMU-PS1  larger than about 1~arcmin and limited only by the observing parameters of the telescope. However, we also acknowledge that such a human process inevitably introduces human error, and we cannot guarantee that some sources have not been missed, mis-classified, or mis-identified. The results from this paper are available as an online version which will be periodically updated to correct any errors reported to the authors.

\subsection{ Host Identification and Classification (Tagging)}
\label{sec:tags}
In the discussion of our methods here it may  appear that the host identification and classification processes are independent, but in reality they are inter-dependent. For example, the host identification process might reveal that a source that superficially looks like an FR2 double-lobed source has no host galaxy between the two lobes, but there is a galaxy coincident with each of the lobes. In this case, it is likely that the source is not an FR2 at all, but is more likely to consist of the radio emission from two separate galaxies, and so is not a DRAGN. Conversely, a bent-tail source is likely to have the host galaxy at one end of the radio source, whereas an FR2 radio source is likely to have the host near the centre. Host identification for such double radio sources is particularly easy when the source is core-dominated. In these cases the host is nearly always coincident with the core. 

For this paper, we started by making a tentative classification based on radio morphology alone, followed by cross-identification with infrared data. However, in many cases the host identification process would then trigger a joint re-evaluation of the classification and host identification.

Rather than classifying a source as either ``class~A'' or ``class~B'', we adopt the use of tags, as advocated by \citet{rudnick21}, and discussed further in Section \ref{sec:morphology}. The advantage of tags is that a source may be tagged as belonging to several different intersecting classes, as opposed to a binary classification which assumes that classes do not intersect. For example, the tags \#FR1 and \#BT (bent-tail) are explicitly non-exclusive. Furthermore, users may select a combination of {  tags} to create a sample of interest.

In addition, the use of a tag signals that this is a descriptor that is based on a specific survey, and set of definitions, not an absolute box into which the source falls. It could be that a source with \#FR1 here may be classified as a \#HyMoRS in another survey using different frequency and spatial resolution. To circumvent this problem, we recommend that tags be subscripted to indicate which definition is being used. In 
this paper, all tags that we assign to sources are given the subscript ``n'', indicating the first author (Norris) of this paper. In FITS and other machine-readable files, we write the subscript as (e.g.) FR1\_n.

In the extended discussion of each tag in Section \ref{sec:morphology}, we define the meaning of each of our tags, to avoid ambiguity resulting from confusion with previous definitions.

 \subsection{Source classification in context}
The above discussion about the use of tags, which can differ between surveys at different frequencies, resolutions, and sensitivities, leads to the question of their fundamental purpose. What meaning do they have, for example, if they do not reflect source properties alone, but also the particulars of a specific survey and analysis technique? To answer that, we must examine their intended use.

At the simplest level, classifications and tags are useful in defining samples for further study. Whether those are follow-ups at other wavelengths, or higher-resolution radio observations, finite resources require us to frame science questions, and use {  tags} to define samples that will help us address those questions. Ultimately, the usefulness of classifications and different versions of {  tags}, depends on their scientific utility.

For DRAGNs, the underlying physical questions fall into four main categories: (a)~the origins of the jets, (b)~the physics of jet propagation including both the magnetised thermal plasma and coupled radiating cosmic ray electrons, (c)~the interactions of the jets with the surrounding medium, and (d)~the effects that the radio emitting plasma has on its environment. Whether jets create a termination shock and backflow, or are deflected, or go unstable, or entrain significant external plasma, or can re-energise the relativistic plasma, or heat the surrounding medium, or create cavities, are all examples of issues to be probed. In that context, specific tags can be chosen and evaluated, based on their ability to answer such questions. New tags can be developed as our insights change, and old ones dropped if they are no longer useful. It thus becomes critical to put this work into context: the purpose of the classifications is \emph{not}, e.g., so we can explain some idealised picture of an FR1, a trap into which papers often fall, but to ask instead, how do the sources classified here as \#FR1 help us clarify the underlying science questions? In the former case, we remain stuck in a clearly imperfect specific characterisation of a source; in the latter case, we can modify, replace, or supplement the classifications as our science issues become better refined.

\subsection{Structure of this series of papers}
This paper is the first of a series of four papers exploring the DRAGNs detected by ASKAP in the EMU-PS1 field. 

In this paper (Paper I) we describe our techniques for extracting and classifying the sources, and present a catalogue of the DRAGNs in the EMU-PS1 field, with morphological tags, measured parameters, and tentative cross-identifications to the \ac{WISE} catalogue. We discuss the morphological classes of the radio sources we find, and briefly compare our catalogue with previous catalogues.

In Paper II we compare the orientations of the major axis of neighbouring DRAGNs, to explore the possible alignment of nearby sources as previously suggested by
\citet{taylor16, contigiani17, osinga20} but not confirmed by \citet{simonte23}.

Paper III will explore the relationship of bent-tail and head-tail galaxies to known clusters and groups.

Paper IV will add spectral index information to the data, extend the host identifications using a wider range of multiwavelength data, including redshift estimates, and attempt to understand the diverse morphologies in terms of the physical processes that drive them, including the relationship between FR1 and FR2 sources.

In this paper we first describe the observations and source extraction, followed by a detailed discussion of the classification process and the resulting classes (tags), followed by a discussion of the results.

\section{Observations and Source Extraction}
\subsection{ASKAP and the EMU survey}
\ac{ASKAP} is operated by the \ac{CSIRO} at the Inyarrimanha Ilgari Bundara, the CSIRO Murchison Radio-astronomy Observatory, in Western Australia. It consists of 36 $\times$ 12~m diameter radio antennas which are separated by baselines ranging from 37~m to 6~km resulting in angular scales of 10\arcsec -- 3\arcmin\,at frequencies from 700 to 1800~MHz. All but six of the antennas are within a region of 2.3~km diameter, with the outer 6 extending the baselines up to 6~km. In the observations described here, as many of the 36 antennas were used as possible. However, in some cases a few antennas were omitted because of maintenance or hardware issues. 

At the prime focus of each antenna is a \ac{PAF}, which subtends a solid angle of about 30 square degrees of the sky. The \ac{PAF} consists of 192 single-polarisation dipole receivers. A weighted sum of the outputs of groups of these receivers is used to form 36 dual-polarisation `beams', which together cover an area of about 30 square degrees. 
A full description of \ac{ASKAP} may be found in \citet{hotan21}.

The two largest projects for which ASKAP was designed were EMU \citep{emu, hopkins25} and WALLABY \citep{wallaby}. Here we use data from the EMU-PS1  survey of EMU.
The EMU-PS1  field was observed with \ac{ASKAP} at a central frequency of 944 MHz, in the period 15 July to 24 November 2019. The observations and data processing are fully described by \citet{pilot} so here we give only a brief overview. The data were taken in 10 tiles, covering 270~deg$^{2}$, and taking a total of about 100 hours of observing time. Here we use the ``native'' resolution product, giving a synthesised beam of about $11 \times 13$ arcsec with an rms sensitivity of about 25~$\mu$Jy/beam.

\subsection{Source Extraction and Characterisation}
\label{extraction}

\begin{figure*}
  \includegraphics[width=18cm]{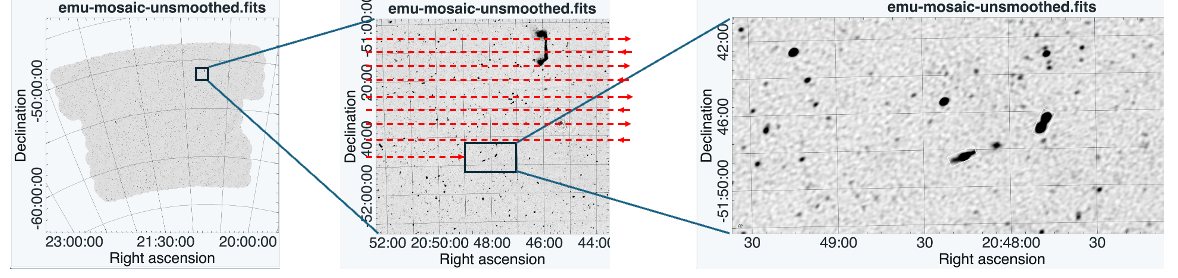}
\caption{The process of manually scanning the image. The left-hand panel shows the area of the EMU-PS1 survey. The middle panel shows the raster pattern employed. The lines are separated by about 5 arcmin, and the raster box is about $10 \times 20$ arcmin, and is moved by about 10 arcmin in \ac{RA} for each inspection, so that each point on the image is inspected at least 4 times. Thus, to cover the whole field requires about 500,000 manual inspections. As the whole field was scanned more than once, over a million inspections were performed in total.}
    \label{fig:raster}
\end{figure*}

{  We have found \citep[e.g.][]{boyce23} that traditional automated source extractors such as {\it Selavy} \citep{whiting12} and {\it PyBDSF} \citep{mohan15} are relatively insensitive to extended structures, so } 
no automated source extraction was used in this work. Instead, the sources were identified by visual inspection. We acknowledge that this may introduce its own bias and subjectivity. 
Our primary goal is to compile a catalogue of DRAGNs, so we ignore unresolved sources, and include other complex sources only if we think they may contain a DRAGN.

We first displayed the radio images on a large screen using CARTA \citep{carta}, using a linear transfer function, and setting the greyscale level so the 3-$\sigma$ noise peaks were just visible (see right-hand panel of Figure \ref{fig:raster}), and setting the spatial scale so an area of 10 $\times$ 20 arcmin was displayed on the screen. We then searched the image by eye, in three stages, for sources that appeared to be DRAGNs, as follows:

\begin{itemize}
\item We first performed a careful visual raster scan through the entire radio image using CARTA, starting from the top left corner, moving across and down in a zig-zag pattern to the bottom right, identifying candidate sources, as shown in Figure \ref{fig:raster}. 
\item For each source, we drew a rectangular ``region'' around the source, whose position angle was matched, where possible, to the position angle of the jets.
 The major axis of the rectangle was adjusted visually to bound the lowest visible emission of the source, which  corresponds roughly to the $5\,\sigma$ rms noise level of the image.  Because the Gaussian image of a  point source increases in size with source strength, this will  result 
in a relatively larger box for strong sources, particularly FR2 sources in which the peak is close to the edge of the bounding box.
\item The third stage is to repeat the scan, focusing on finding any sources that might have been missed (i.e., are not marked by a rectangular region in the CARTA display) or whose total extent may have been under-estimated due to extended low-level emission. This last step was repeated several times by three different members of the author team (RPN, MY, and NG) until we were satisfied that no major sources had been missed. Our goal in doing so was to maximise the number of \ac{DRAGN}-like sources larger than one arcmin, rather than prioritising any particular morphology.

\item For each source consisting of two lobes, we also measured the distance $a$ between the maxima (identified visually) either side, using CARTA's ``Distance Measurement'' tool, for use in the FR1/FR2 classification, as described in Section \ref{sec:frx}.

\item For each source identified, we recorded the parameters of the region (centroid RA and Declination, major and minor axes $b$ and $c$, and position angle). This RA and Declination is used to form the \ac{ID} of the source in Table \ref{catalogue}. The parameter $b$ is often called the ``Largest Angular Size'' in the literature.

\item For each source, we recorded an initial classification based on the morphology (e.g. FR1, bent-tail, X-shaped) and any further comments on the appearance (e.g. ``overlaps with nearby source'').

\end{itemize}

These three stages resulted in the identification of a total of about 4000 candidate DRAGNs.
During subsequent steps of this project (host identification, etc.), a few more sources were also found and added to the catalogue, while some sources were removed. We also identified and corrected any duplications, errors, and mis-classifications made in earlier stages. Our final list (hereafter referred to as ``our catalogue'') of DRAGN candidates contains \dragn\ sources.

\subsection{Completeness}
As the sources are identified by eye, this process is necessarily subjective and our catalogue of sources should not be considered a complete sample. Instead, the sources listed in this paper should be regarded as a representative, but substantially complete, sample of the sources in the image. 
We discuss the actual distribution of sizes and flux densities  in Section \ref{sec:results}.

\begin{figure}
  \includegraphics[width=4cm]{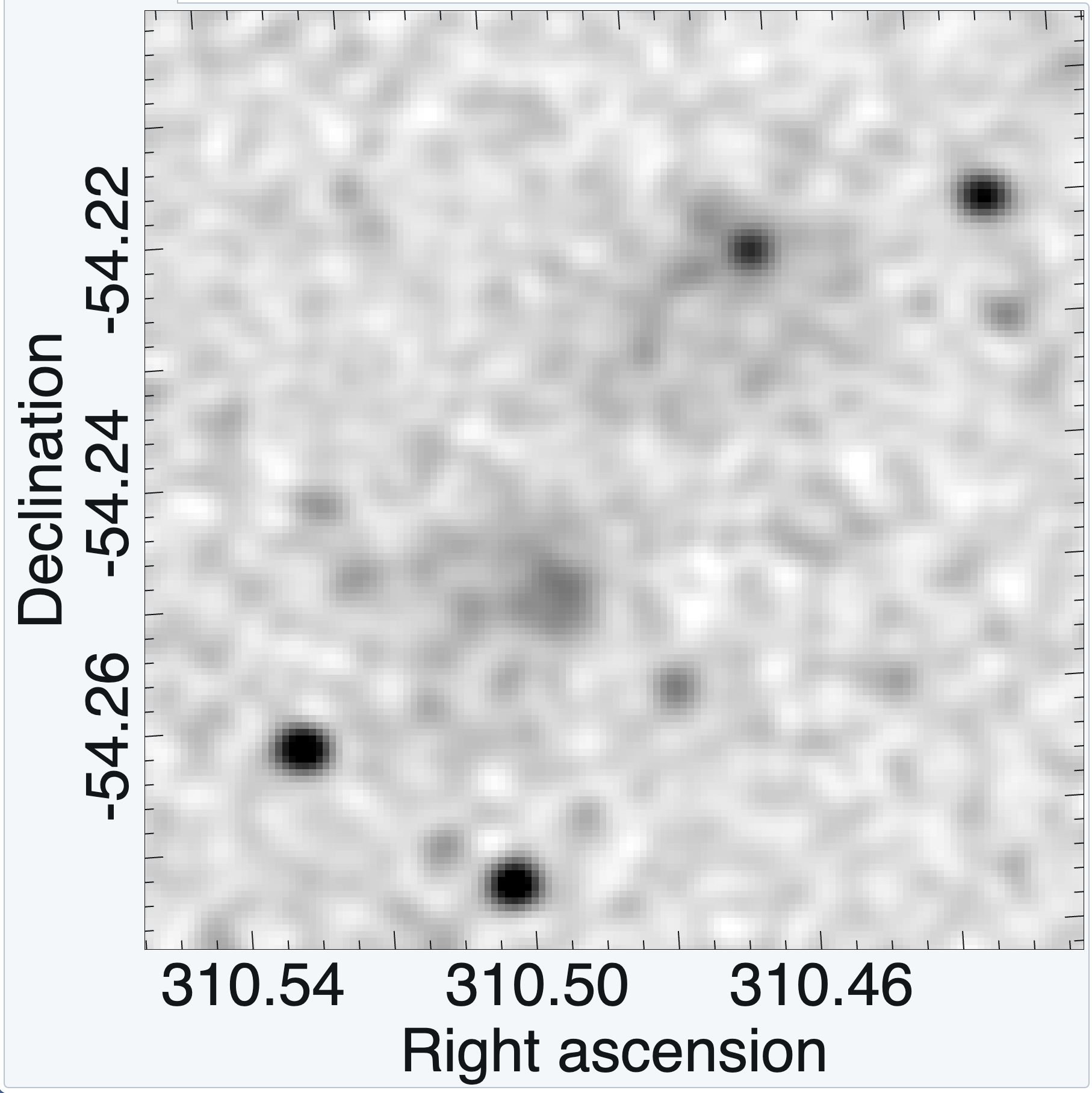}
  \includegraphics[width=4cm]{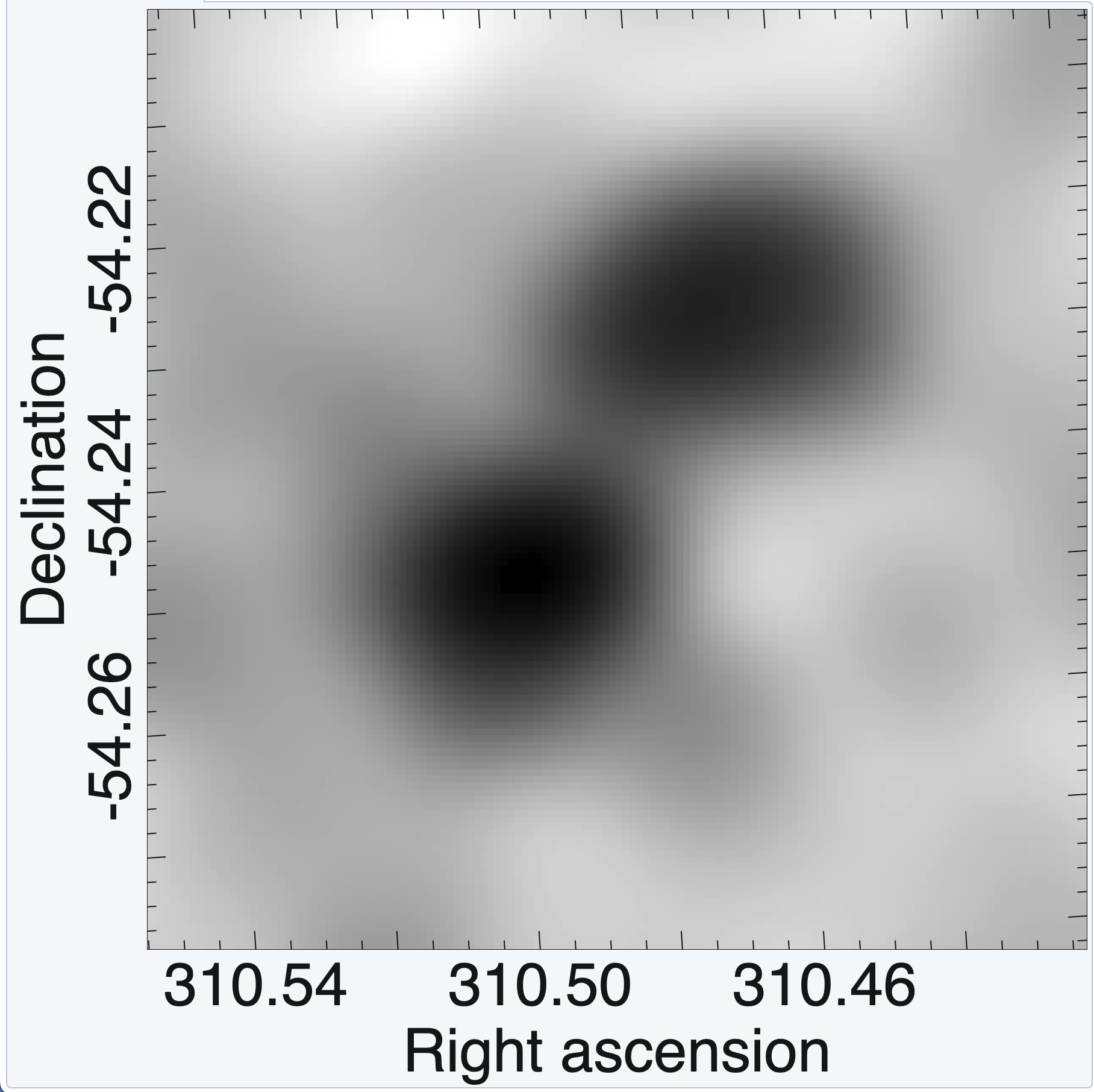}
  \caption{An example of a faint low-surface-brightness source that is not easily visible in the raw data but becomes obvious after spatial filtering, as described in Section 2.3. Left: raw unfiltered data. Right: filtered data. Although both lobes have some compact emission which is visible in the unfiltered EMU image, they appear as two disconnected compact sources and so were not included in our catalogue. Filtering reveals   a  radio galaxy with two connected low surface brightness lobes characteristic of aging electrons. Such sources that are visible only after spatial filtering are not included in our catalogue. }
  \label{fig:fuzzy}
\end{figure}

While we clearly cannot detect sources that are below our sensitivity limit, we note that spatial filtering of the sources, as described by \citet{rudnick02}, can effectively increase the sensitivity of the data to low-surface brightness sources. A further exploration of this filtering, and the availability of python scripts to execute it, are found in \citet{duchesne24}. Using this technique, we find several low-surface brightness sources in each of the 10 pointings that were not detected in our initial manual search. One of these is shown in Fig \ref{fig:fuzzy}.  Another  radio galaxy (J204837.7-491145) with low-surface-brightness lobes is bright enough to appear in our catalogue, and is shown in Figure 23 of \citet{pilot}. That example is a remnant radio galaxy \citep[e.g.][]{brienza17}, which is a galaxy in which the AGN has turned off, leaving two remnant radio lobes of aging electrons. Fainter examples may be identified using filtering, but are not sufficiently obvious in the full resolution images to be included in our catalogue. This is a conscious decision, implemented to maintain the consistency of our selection criteria.

\section {Host Galaxy Identification}
\label{ir-xid}

To identify the host galaxies of the DRAGNs, we need to cross-match their centres with potential host galaxies, as listed in optical and infrared catalogues. \citet{norris06} showed that this host identification was more reliable if the cross-match was first done with an infrared source catalogue, and so here we use data from the Wide-field Infrared Survey Explorer telescope \citep[WISE;][]{wright10} for this purpose.
The cross-matching took place in three stages, each of which is described in a subsection below.

\subsection{Identify the likely position of the host in the radio image}
For DRAGNs that have a radio core we expect the host should coincide, or be very close to, the radio core \citep[e.g.][]{gordon23}. In sources with a jet connected to the lobes, the host should lie on the line of the jet \citep[e.g.][]{barkus22}. In double sources, the host often (but not always) lies roughly on the major axis of the source (or emission ridge), and is roughly equidistant from the two lobes. In sources with lobes of unequal brightness, the core tends to be closer to the brighter lobe \citep[e.g.][]{mackay71, rosa19}.

\subsection{Identify the host in the WISE images}
We use the unWISE data release 6 images \citep{meisner17, meisner19} which are derived from coadds of the WISE images over 5 years, covering the entire sky at 3.4 and 4.6 $\mu$m, giving an angular resolution of $6.1\arcsec$ and $6.4\arcsec$ respectively. They include the images from the NEOWISE phase of the WISE mission, resulting in five times the exposure time of the coadds from the initial WISE survey.

Having identified likely positions of the host in the radio image, we cross-matched the radio images with the unWISE images by visual inspection, by displaying two linked images in CARTA. In each case we then searched for an \ac{IR} source close to the expected host position. Our priority was reliability rather than completeness. In cases where no likely host was found in the IR image, we then re-examined the radio image to find likely positions of the host in the radio image that coincided with likely IR candidates.

If no plausible host was found, or our ability to find one was impacted by confusion or a nearby bright star, we do not associate the radio source with a host.

\subsection{Identify the host in the CATWISE2020 catalogue}
We matched the position of the IR source, measured from the unWISE \citep{schlafly19} image, with a source in the CATWISE2020 catalogue \citep{catwise} which is deeper and more reliable than the earlier unWISE catalogues. The cross-matching was done using the match function in TopCat \citep{topcat}.
To determine an appropriate search radius, we measured the number of cross-matches between the manually determined WISE \ac{IR} hosts and the CATWISE2020 catalogue as a function of separation. We also estimated the false-ID rate by shifting the radio declination positions by 1~arcmin and then repeating the cross-match. 

 As a result, we adopted a maximum search radius of 2.0 arcsec, at which we get a 3.2\% false-ID rate and a 90.0\% total cross-match rate. 

\subsection{Summary of host identifications}

The resulting numbers of sources are listed in Table~\ref{samplesize}, and the identifications themselves are listed in Table \ref{catalogue}. 

\begin{table}
\small
  \begin{center}
    \caption{Numbers of sources remaining after each stage of the host identification of EMU-PS1 data.} 
    \label{samplesize}
    \begin{tabular}{lcc}
      \toprule
      Criterion	&	\# Sources & \% of \\
                & & sample	\\
      \midrule
      Initial candidate source extraction & $\sim$4000 & n/a\\
      Sources classified as DRAGNs & \dragn & 100 \\
      CATWISE2020 {  Cross-identification} & \catwise & 89.4\\
      \bottomrule
    \end{tabular}
    \medskip\\
  \end{center}
\end{table}

\subsection{Caveats}
Because our source extraction and host identification processes are largely subjective, we stress that these results should be regarded as tentative, and subject to revision.

For example, to cross-match the initial manual identification of the IR host in the unWISE image to the CATWISE Survey has a quantifiable false-ID rate discussed above. However, we have no way of accurately estimating a false-ID rate of the initial match of the radio source to the IR sources visible in the unWISE image. Even when two different authors repeat the host identification, they are likely to be subject to the same biases and preconceptions. Our best guess, based on the number of sources which had an obvious host ID, compared to those where it was uncertain, suggest that our false-ID rate is likely to be about 10\%. 

It should also be emphasised that the above description may suggest that the source extraction, classification, and host identification processes occurred in a linear sequence. However, at each stage, earlier decisions were revisited, new sources were added to the catalogue, and some sources were dropped (e.g. when an apparent FR2 turned out during the host identification process to be two nearby interacting galaxies).

We plan to revisit the host identification in Paper IV of this series, where we will use all available multi-wavelength data to aid in the initial host identification, including higher-resolution radio data where available, and assign a reliability score to each host identification.

\section{Results}
\label{sec:results}
We present a complete catalogue of the extracted sources in Table \ref{catalogue}, and a table showing the number assigned to each tag in Table \ref{tags}. We also show a thumbnail image of each source in the Supplementary Material.

In Figure \ref{fig:sldists}, we show the distributions of the major axis of the bounding box (Panel a) and total flux density (Panel b) as grey solid histograms.
The major axis of the bounding box approximates the largest angular size (LAS) for DRAGNs.  
As discussed in Section \ref{extraction}, this may be slightly over-estimated for strong FR2 sources.

Panel (a) of Figure \ref{fig:sldists} shows the size distribution for our DRAGNs peaks around 1 arc minute.
Below 1 arc minute, the count sharply decreases as our selection is less sensitive to smaller sources.
At box sizes larger than 1 arc minute, the count decreases approximately proportionally to $\theta^{-2}$ (where $\theta$ is the major axis of the box) as radio galaxies spanning larger angles than subtended by the bounding box become rarer. 
The total flux densities of our sample (Panel b) range from $\sim100\,\mu$Jy to $\gtrsim 1\,$Jy, with most on the order of a few tens of mJy.

\begin{figure}
    \centering
    \subfigure[]{\includegraphics[width=\columnwidth]{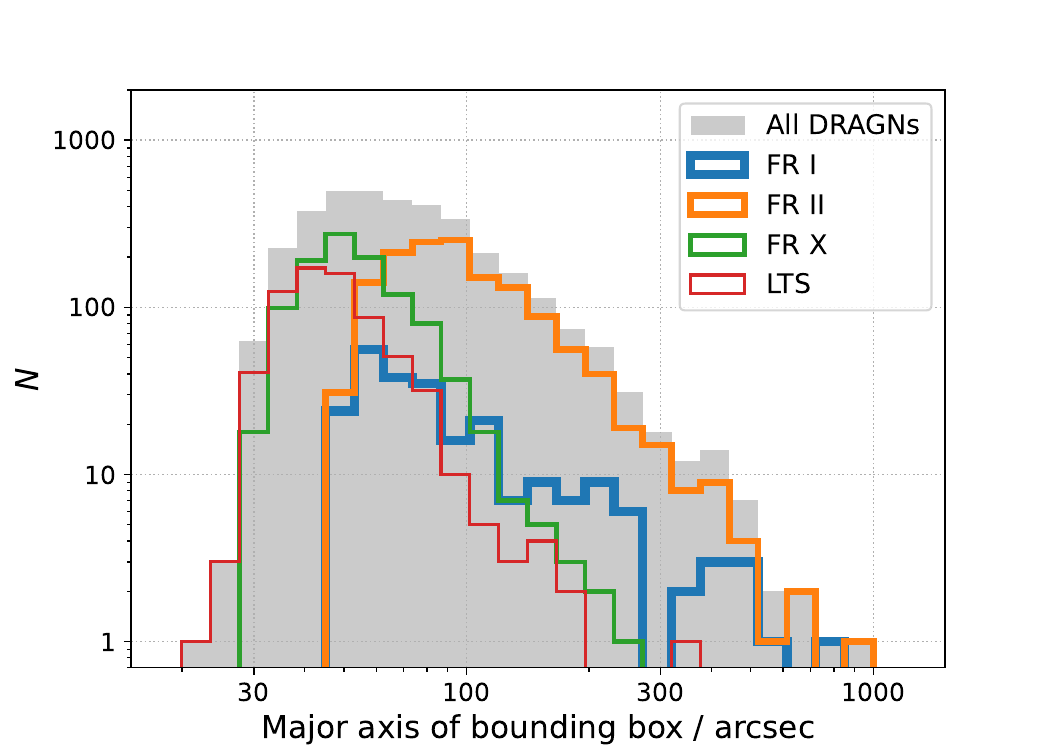}}
    \subfigure[]{\includegraphics[width=\columnwidth]{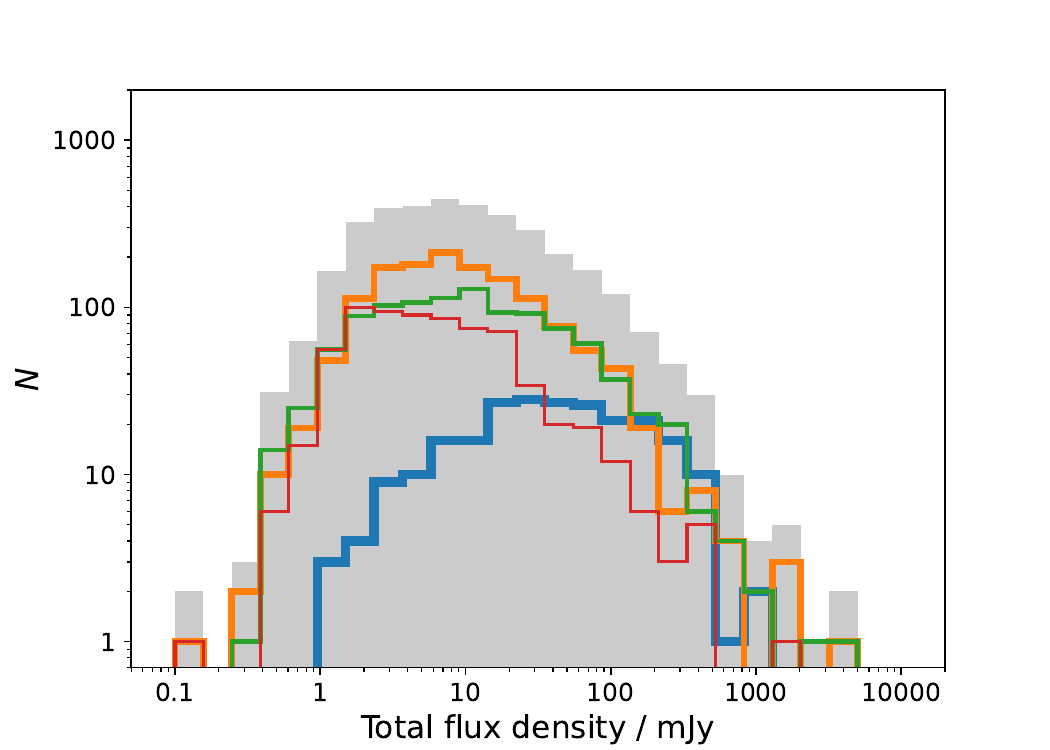}}
    \caption{Distributions of (a)~the major axis size of the bounding box, and (b)~total flux density of our DRAGNs.
    In both panels, the grey solid histograms show the distribution for the entire catalogue, while the coloured lines show the distributions for sources with some of the more common tags in our catalogue (see Section \ref{sec:morphology}).}
    \label{fig:sldists}
\end{figure}

\subsection{DRAGNs in the context of other radio sources}
In the early days of radio astronomy, the vast majority of detected radio sources were DRAGNs \citep{norris17a}. But, partly because the fraction of radio sources that are star-forming galaxies increases rapidly with decreasing flux destiny,  the  sources in our catalogue are only a small subset of all sources in EMU-PS1. 
There is currently no easy way to measure the total number of sources in the EMU-PS1 field, but we can make a rough estimate as follows.

\citet{pilot} describe how sources were initially extracted from the EMU-PS1  data using the Selavy source finder \citep{whiting12}. Selavy identifies individual Gaussian ``components'', and also ``islands'' which consist of one or more components linked by low-level emission. Both of these are distinct from ``sources'' which describe all the emission from one astrophysical object, such as a DRAGN. 

\citet{pilot} found a total of 220,102 components with the Selavy source finder. Of these, 178,921 were ``simple'' sources meaning that an island contained only one component. We note that this fraction (81\%) of sources that were compact in EMU-PS1   is similar to the fraction (89\%) found in \ac{LoTSS} \citep{shimwell19, williams19}. The comparison with LoTSS will be discussed further in Section \ref{LoTSS}. 

Selavy also identified 198,217 islands in the EMU-PS1  data, of which 178,921 were the same simple objects as above, leaving 19,286 islands which contain more than one component. If we 
assume that each island corresponds to one physical source, then that suggests there are 19,286 sources consisting of more than one component. In this paper, we identify \total\ DRAGNs, 
leaving 15,755 extended sources which are not in our catalogue of DRAGNs. These will presumably include (a)~DRAGNS smaller than 1 arcmin, (b)~extended star-forming galaxies, (c)~diffuse emission from clusters (e.g. halos, relics, etc.), (d)~imaging artefacts, (e)~DRAGNs larger than 1 arcmin that we have missed (including DRAGNs which are detected as two or three components but which show no evidence of being connected), (f)~WTFs, i.e., objects that we do not recognise as being in a known category of source. This number will not be accurate because (a)~not all sources consist of only one island (e.g., several sources in our DRAGN catalogue consist of widely separated components which Selavy would identify as distinct islands, (b)~not all islands will consist of only one source (e.g., several sources in our DRAGN catalogue consist of two or more adjacent or overlapping sources), (c)~like all point source finders, Selavy is poor at identifying extended sources, and so will miss some sources that are in our catalogue of DRAGNs.

Nevertheless we can use this calculation to make a rough estimate that EMU-PS1  contains a total of about 200,000 sources, consisting of 178,921 simple (i.e.compact) sources, \dragn\ DRAGNs larger than $\sim$ 1 arcmin, and $\sim$ 16000 other extended sources. Thus, we estimate that DRAGNs larger than $\sim 1$\,arcmin constitute about 1.8\% of all radio sources at this sensitivity and resolution.  
{  We note that this is similar to the fraction (1.6\%) of sources larger than 1 arcmin in the LoTSS DR2
catalogue \citep{hardcastle23}.}

If we assume that about one quarter of all detected sources are AGN and the rest are star-forming galaxies \citep{hopkins25}, this implies that about 7.2\% of radio-loud AGN in this survey are DRAGNs larger than $\sim 1$\,arcmin. 
{  Again, this is very similar to the fraction (6\%) of AGN larger than 1 arcmin in the \citet{hardcastle25} sample.}

We may compare this with theoretical predictions. \citet{bonaldi19} produced the T-RECS  simulated distribution of radio galaxies as a function of redshift. If we select AGN from that simulation with an integrated 1\,GHz flux density larger than 1\,mJy,  then about 6\% have a projected size > 1 arcmin, or  5\% for S > 0.1\,mJy. That number is comparable with the number (7.2\%) for our sample, suggesting good agreement between that simulation and our data.

\section {Morphological Classification}
\label{sec:morphology}
\subsection{Classification process}
Each source was inspected by eye and initially assigned one or more of the tags shown in Table~\ref{tags}, based primarily on radio morphology. This classification was subsequently revised (a)~by using an algorithm, described below, to determine the FR1/FR2 classification, and (b)~by taking into account the position of the host in the host identification step, such as a candidate FR2 source which turns out to be two (presumably interacting) galaxies with a connecting bridge of emission.

In the initial classification, there were four further tags that are not shown in Table \ref{tags}:
\begin{itemize}
    \item Out of field: Part of the source lies beyond the outer boundary of the EMU-PS1  field
    \item Artefact: The source appeared to be dominated by an imaging artefact
    \item Single: The source appears to be a single component with no significant extension
    \item Error: All other errors including human errors
\end{itemize}
Sources tagged with any of these four tags have not been included in the results shown in this paper. Every source in Table \ref{catalogue} has at least one of the tags shown in Table \ref{tags}, and the mean number of tags/source is 1.15.

We note that these morphological tags are, of course, based on the two-dimensional projection of one or more three-dimensional objects, and it is possible that some unusual morphological shapes may be due to a  projection of (e.g.) a bent-tail galaxy, or the superposition of two or more unrelated objects.

\begin{table*}
\small
  \begin{center}
    \caption{Tags used to classify sources. A source may have more than one tag, in which case it will appear in more than one row in this table. Tags are appended with the subscript \ytag\ as discussed in Section \ref{sec:tags}. The total number of DRAGNs includes all tagged sources except Non-DRAGNs.
    }
    \label{tags}
    \begin{tabular}{lll}
      \toprule
      Tag 		 & \# sources & Description\\
      \midrule
FR1\ytag\ &\fri &  Fanaroff-Riley type I {  (an edge-dimmed source} - see section \ref{sec:fr12})\\      
FR2\ytag\ &\frii& Fanaroff-Riley type II {  (an edge-brightened source} - see section \ref{sec:fr12})\\      
FRX\ytag\ &\frx&  FR1 or FR2 but data inadequate to distinguish between them (see section \ref{sec:frx})\\
HyMoRS\ytag\ &\HyMoRS& a radio source which is FR1 on one side of the host and FR2 on the other.(see Section \ref{sec:HyMoRS})\\
LTS\ytag\ &\LTS& Linear triple source - a core-dominated source with two linear features extending from the host, \\
&& but either one or both do not have the maximum that would be required to class them FR1 or FR2 \\
&& (possibly due to inadequate resolution) (see Section \ref{sec:LTS}).\\
OSS\ytag\ &\oss&  One sided source, such as a host with a jet on one side. (see Section \ref{sec:oss})\\
BT\ytag\ &\bt& A bent-tail source, including {  narrow-angled tail (NAT) and wide-angled tail} (WAT) (see Section \ref{sec:bt})\\
HT\ytag\ &\hht& A head-tail source, with the host at one end of a single or double tail (see Section \ref{sec:bt}) \\
DD\ytag\ &\dd&A double-double radio source which is sometimes interpreted as a ``restarted'' radio galaxy.(see Section \ref{sec:dd})\\
ZRG\ytag\ &\ssh& S-shaped or Z-shaped radio galaxy (see Section \ref{sec:ssh})\\
XRG\ytag\ &\xsh& X-shaped radio galaxy (see Section \ref{sec:xsh})\\
TRG\ytag\ &\tsh& T-shaped radio galaxy (see Section \ref{sec:tsh})\\
WTF\ytag\ &\wtf& a radio source with weird or unusual features that cannot easily be explained and merits further \\ & & investigation(see Section \ref{sec:wtf})\\
CPLX\ytag\ &\complex& a radio source, or group of radio sources, that is too complex to classify unambiguously.\\
Non-DRAGN\ytag\ &\nondragn& Two or more nearby blobs of radio emission which resemble a DRAGN but which have no other \\
&& evidence to connect them, or appear on the IR image as multiple galaxies. These sources are listed \\
&&in table \ref{tab:non-AGN} to enable identification but are not included in the catalogue or in the tally of DRAGNs.\\
\midrule   
      \bottomrule
    \end{tabular}
    \medskip\\
  \end{center}
\end{table*}

\subsection{The Fanaroff-Riley classification}
\label{sec:fr12}

\begin{figure}
  \includegraphics[width=6cm]{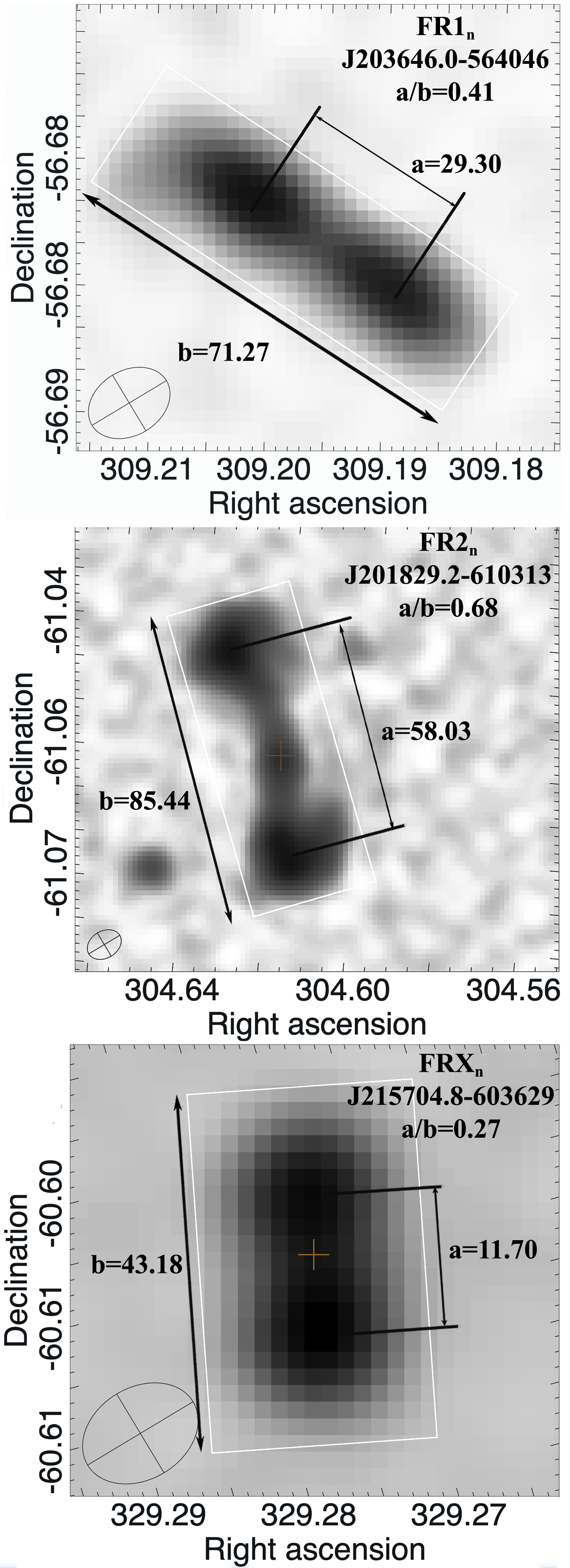}
\caption{Classifying a source as FR1\ytag\ or FR2\ytag\ relies on the measurement of $a$ (the distance between the peaks) and $b$ (the total extent of the source). FR1\ytag\ have $a/b<0.5$ and FR2\ytag\ have $a/b > 0.5$. The top image satisfies the FR1\ytag\ criterion ($a/b < 0.5)$, and the centre image satisfies the FR2\ytag\ criterion ($a/b > 0.5)$. The bottom image satisfies the FR1\ytag\ criterion ($a/b < 0.5)$ but the small size combined with the measurement uncertainty makes the classification uncertain. We label such borderline cases as FRX\ytag, as described in Section \ref{sec:frx}. All distances are in arcsec.}
    \label{fig:frx}
\end{figure}

\citet{FR} showed that a selection of radio-loud \acp{AGN} could be classified into two morphological types: \ac{FR1} and \ac{FR2}, according to the position of the brightness peaks of the radio emission, although the cause of the difference between the edge-brightened FR2 and the edge-dimmed FR1 is still unclear. FR2 jets seem to remain relativistic along their length until they abruptly end in a hotspot. FR1 jets, on the other hand, gradually decelerate along their length \citep[e.g.][]{bicknell95, laing02, tchekhovskoy16}, as shown in Fig.\ref{fig:frx}. 

Factors determining whether a source is FR1 or FR2 may include:

\begin{itemize} 
\item environmental effects, in which the jet loses energy through its interaction with the external medium, which will depend on external density\citep{bicknell86,bicknell95,kaiser07}. These effects include the entrainment or mass loading of a jet \citep{dekel06,bicknell95,gopalkrishna01,krause12, perucho07, perucho14}.

\item the intrinsic properties of the accretion and jet formation processes near the parent \ac{SMBH}. Possible causes include the different content of the jet plasma (electron–positron pairs or normal electron–proton plasma), and the black hole spin \citep{reynolds96,baum95,meier99}. 
\end{itemize}

 \citet{ledlow96} found that the FR1/II luminosity break depends on host-galaxy absolute magnitude, so that \acp{FR1} are found to have higher radio luminosities in more luminous host galaxies (where the density of the \ac{ISM} is assumed to be higher). However, this result is based on highly flux-limited samples, with different redshift distributions and environments for the \acp{FR1} and \acp{FR2}. Therefore, the result does not necessarily apply to the broader population of radio galaxies \citep[e.g.][]{best09,wing11, shabala18}. Furthermore, \citet{krause12} argued that FR1 sources must have previously passed through an FR2 phase, which precludes a fundamental difference based on the host galaxy.

In this paper, we adopt the original \ac{FR} definition, as illustrated by Figure \ref{fig:frx}, which classifies sources based on the ratio \rfr\ of $a$ (the distance between the brightest regions on opposite sides of the central host galaxy) to $b$ (the total extent of the source). The measurement of $a$ and $b$ is described in Section \ref{extraction}.
 Sources with \rfr $<$ 0.5 are classified as the edge-darkened \ac{FR1}\ytag, while those with 
\rfr $>$ 0.5 are classified as the edge-brightened \ac{FR2}\ytag. 
In this project, $a$ and $b$ were measured as part of the Source Extraction and Characterisation process, as described in Section \ref{extraction}.

{  When measuring this ratio, Fanaroff \& Riley say ``The sources are classified using the ratio of the distance between the regions of highest brightness on opposite sides of the central galaxy or quasar; any compact component situated on the central galaxy was not taken into account''.  This definition is open to some interpretation, as it does not, for example, specify how compact a central source must be to be excluded.  We adopt an interpretation that excludes any central maximum (i.e. coincident with the host galaxy) from this classification, and base the classification only on the distance between the strongest maxima either side of the host galaxy. If there are no maxima either side of the core (or a maximum on only one side), then we do not classify it as an FR source, but as a linear triple source (LTS\ytag). 

Other authors may adopt different  interpretations. For example, Figure 2(a) of \citet{mingo19} is classified by them as a ``lobed FR1'',  because of the strong central source, whereas we would classify it as a core-dominated FR2, because the only maxima other than the central compact core are in the outer lobes, which thus  satisfy the \citet{FR} definition of an FR2.

Such differences in classification criteria make it difficult to compare results from different surveys. In this paper we try to mitigate this problem by (a) explicitly defining our classification criteria, and (b) using tags, as discussed in Section \ref{sec:tags}, so that a definition can be traced back to the originating publication. We discuss this further in Section \ref{LoTSS}.
}

\label{sec:frx}
In some cases, a source is clearly double, and yet its small extent, compared to the resolution and measurement accuracy of the data, makes it difficult to determine whether it is FR1\ytag\ or FR2\ytag. We label such sources FRX\ytag.
We introduce two parameters, $b_{min}$ and $\epsilon$, to quantify this uncertainty. We define
 $b_{min}$ as the minimum value of $b$ at which a measurement is meaningful, and
$\epsilon$ as the uncertainty in the measurement of $b$.

We conducted experiments, using both real and simulated sources, to find the optimum value of these parameters. As a result, we adopt the values of $b_{min}$ = 50\arcsec\ and $\epsilon$ = 10\arcsec.

The FR definition then becomes:
\\
FR1\ytag: $a/(b - \epsilon$) < 0.5 AND $b > b_{min}$\\
FR2\ytag: $a/(b + \epsilon$) > 0.5 AND $b > b_{min}$\\
FRX\ytag: $a/(b + \epsilon) < 0.5 < a/(b - \epsilon$) OR $b < b_{min}$\\

Throughout this paper, we use the term ``FR sources'' to mean a source which is one of FR1\ytag, FR2\ytag, or FRX\ytag.

Given that FR1 sources tend to be less luminous than FR2 sources, it might be expected that the high sensitivity of EMU would find a larger number of FR1 sources than the earlier less sensitive surveys, but that is not the case.
In our sample, we find many more FR2\ytag\ than FR1\ytag\ sources. We discuss this later in Section \ref{sec:comparison}. 

Figure \ref{fig:sldists} also  shows that  FR II\ytag\ dominate the larger sources, with their size distribution peaking at $\sim 100^{\prime\prime}$ compared to $\sim 50^{\prime\prime}$ for other morphologies.
However, the flux density distribution of the FR1\ytag\ differs from other morphologies, peaking at $\sim 50\,$mJy, in comparison to $\lesssim 10\,$mJy for the rest of the DRAGNs.
A discussion of the implications of this for the physics of FR1 and FR2 must await Paper IV where we derive source luminosities.

\subsection{HyMoRS\ytag}
\label{sec:HyMoRS}
Hybrid Morphology sources (HyMoRS) are a rare type of radio source that satisfy the FR1 criteria on one side of their host galaxy, and the FR2 criteria on the other side. Here they are tagged as \#HyMoRS\ytag\ and not \#FR1\ytag\ or \#FR2\ytag.

 They were first identified as a class by \citet{gopalkrishna00}. 
\citet{kapinska17} found 25 new candidate \ac{HyMoRS} using Radio Galaxy Zoo \citep{wong25}. Since then several other authors have listed larger samples of HyMoRS \citep[e.g.][]{kumari22}.
We find that \HyMoRS\ of our \dragn\ radio sources are \ac{HyMoRS}\ytag\ 
candidates, or slightly more than 1\% of DRAGNs. Alternatively, HyMoRS\ytag\ represent about 2.6\% of all FR sources (FR1\ytag+FR2\ytag+HyMoRS\ytag), where we exclude FRX\ytag\ because we would not be able to detect HyMoRS\ytag\ morphology in that group. This may be compared with the results of \citet{gawronski06} who estimated that $\lesssim$ 1\% of a sample of strong extended radio sources are HyMoRS. This suggests that HyMoRS may be more common in fainter samples of radio sources. 

HyMoRS provide potentially important clues to understanding the FR1/FR2 dichotomy. For example, intrinsic effects like the black hole spin must be common to both sides, so the very existence of HyMoRS favours an environmental cause for the FR1/FR2 dichotomy \citep{gopalkrishna00, gopalkrishna02}. This support for environmental causes has been reinforced by high-resolution observations \citep{harwood20} and modelling \citep{stroe22}.

\begin{figure}
    \includegraphics[width=4cm]{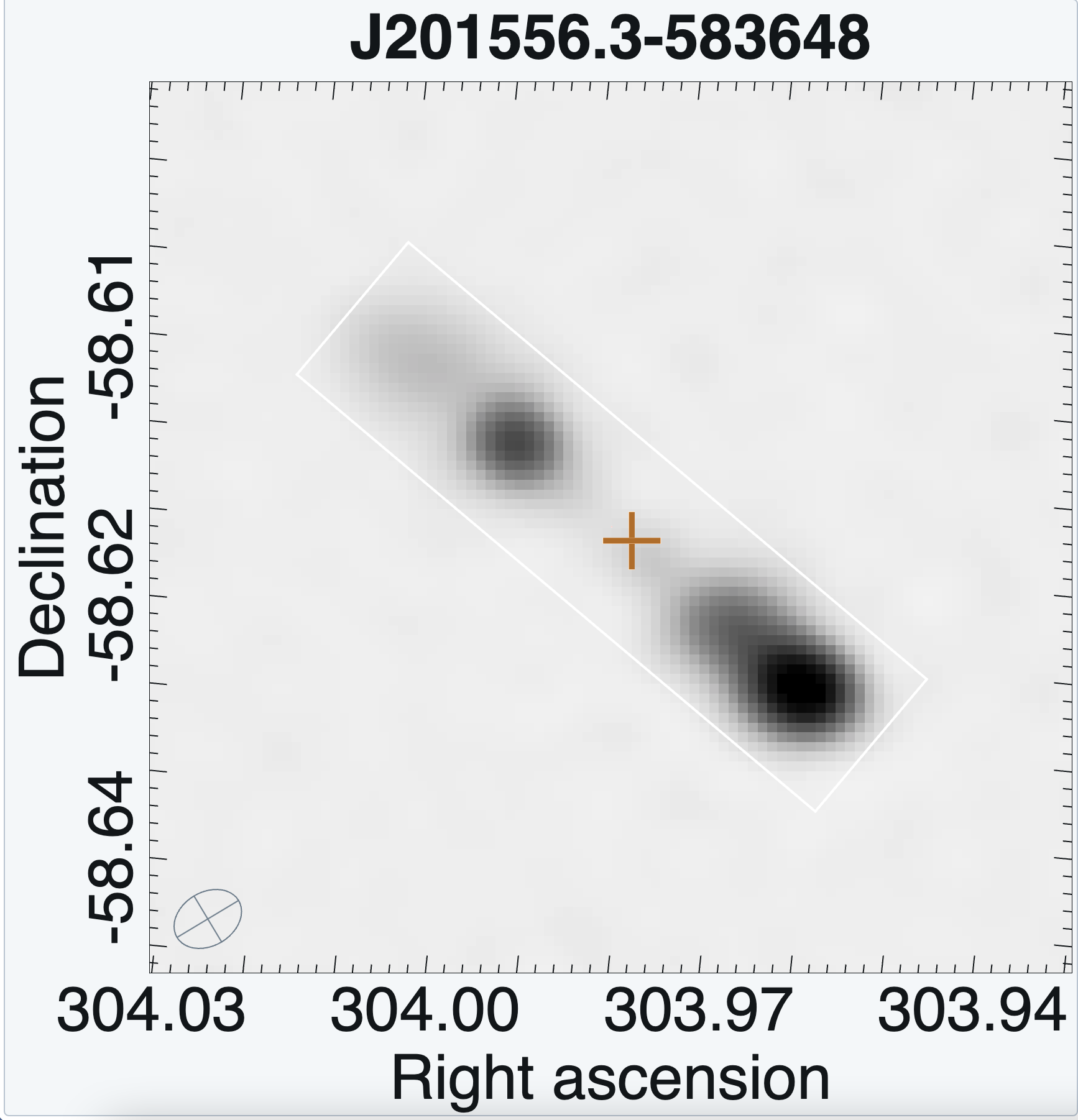}
    \includegraphics[width=4cm]{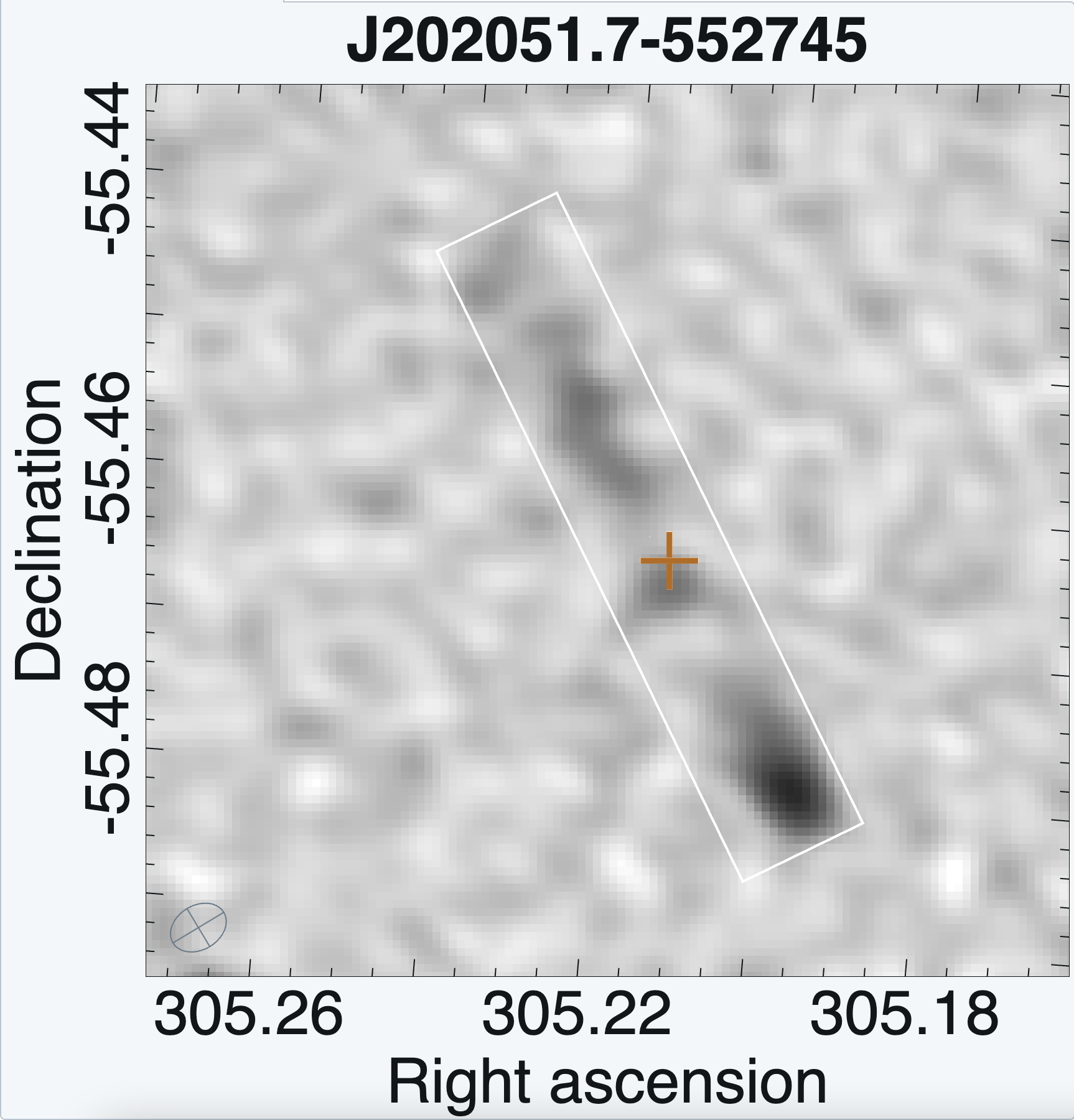}
    \includegraphics[width=4cm]{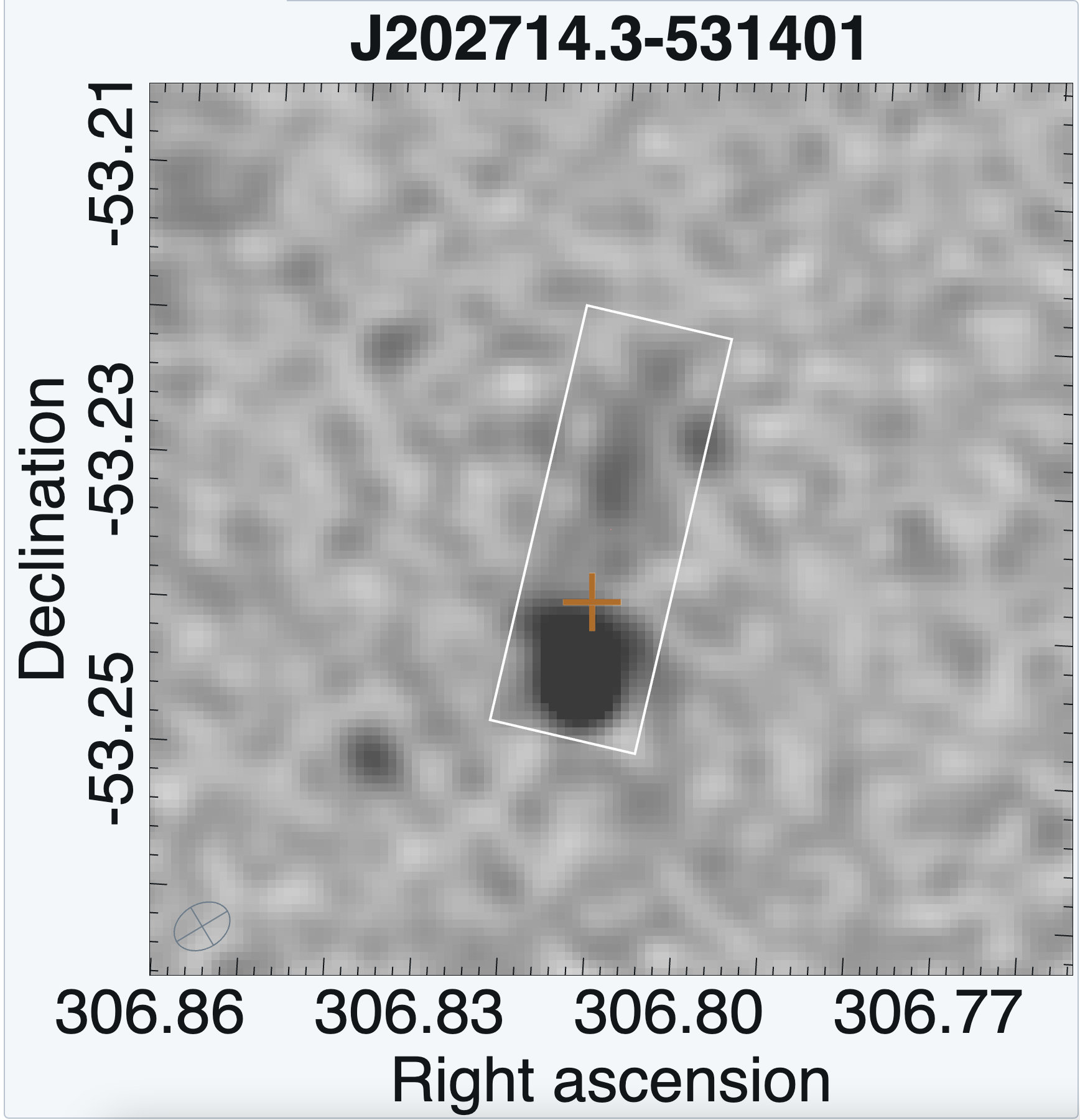}
    \includegraphics[width=4cm]{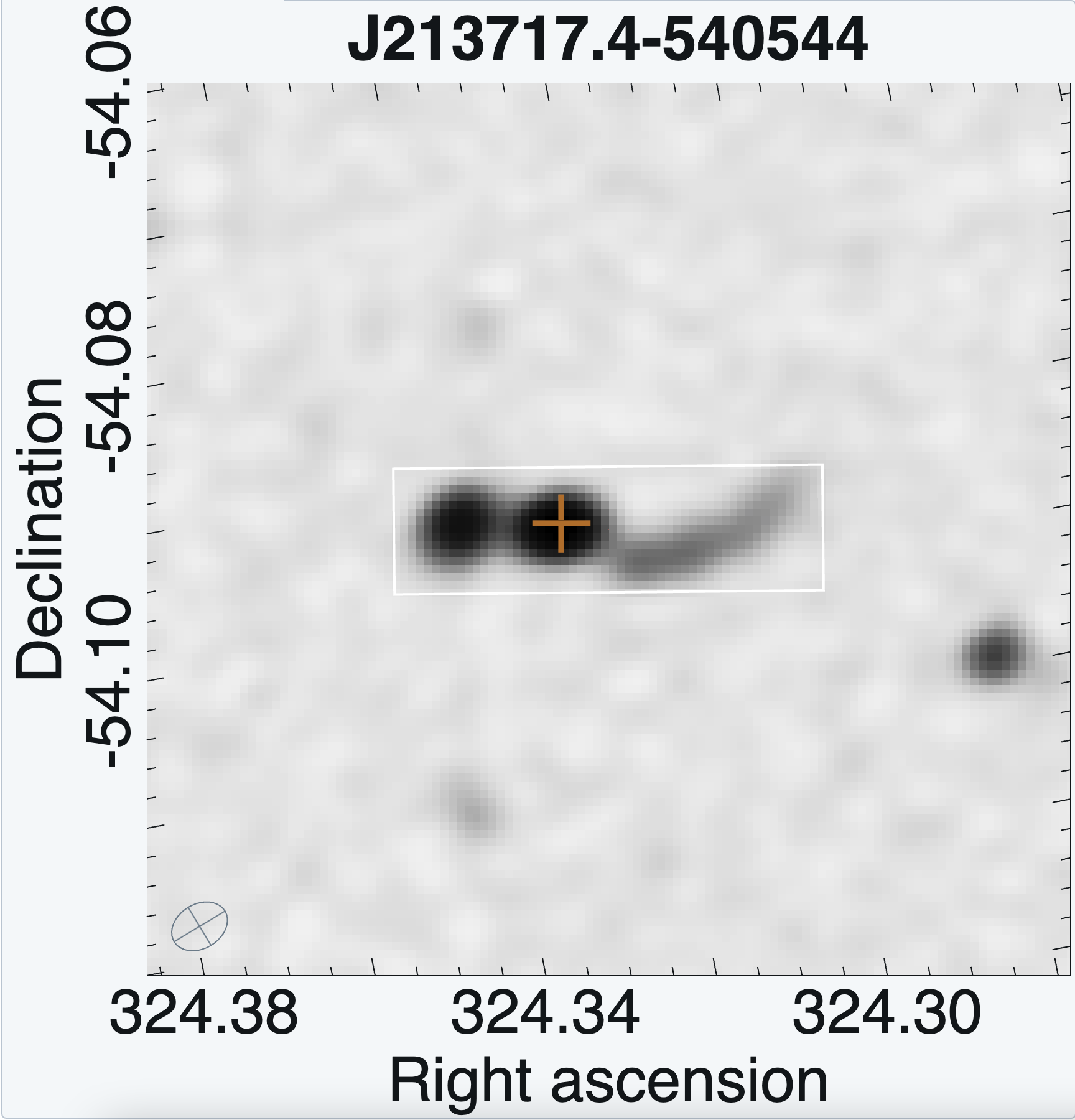}
    \includegraphics[width=4cm]{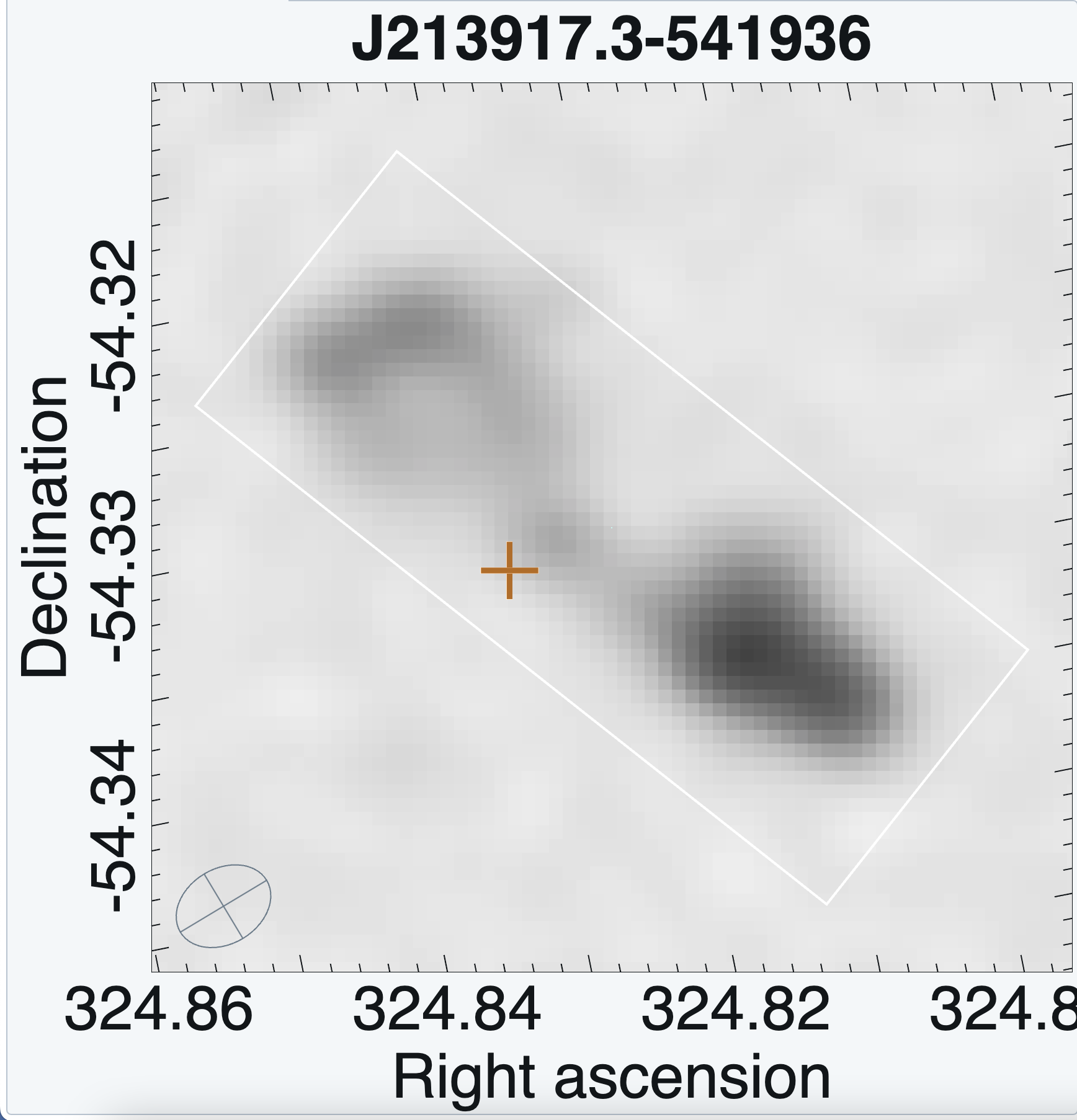}
    \includegraphics[width=4cm]{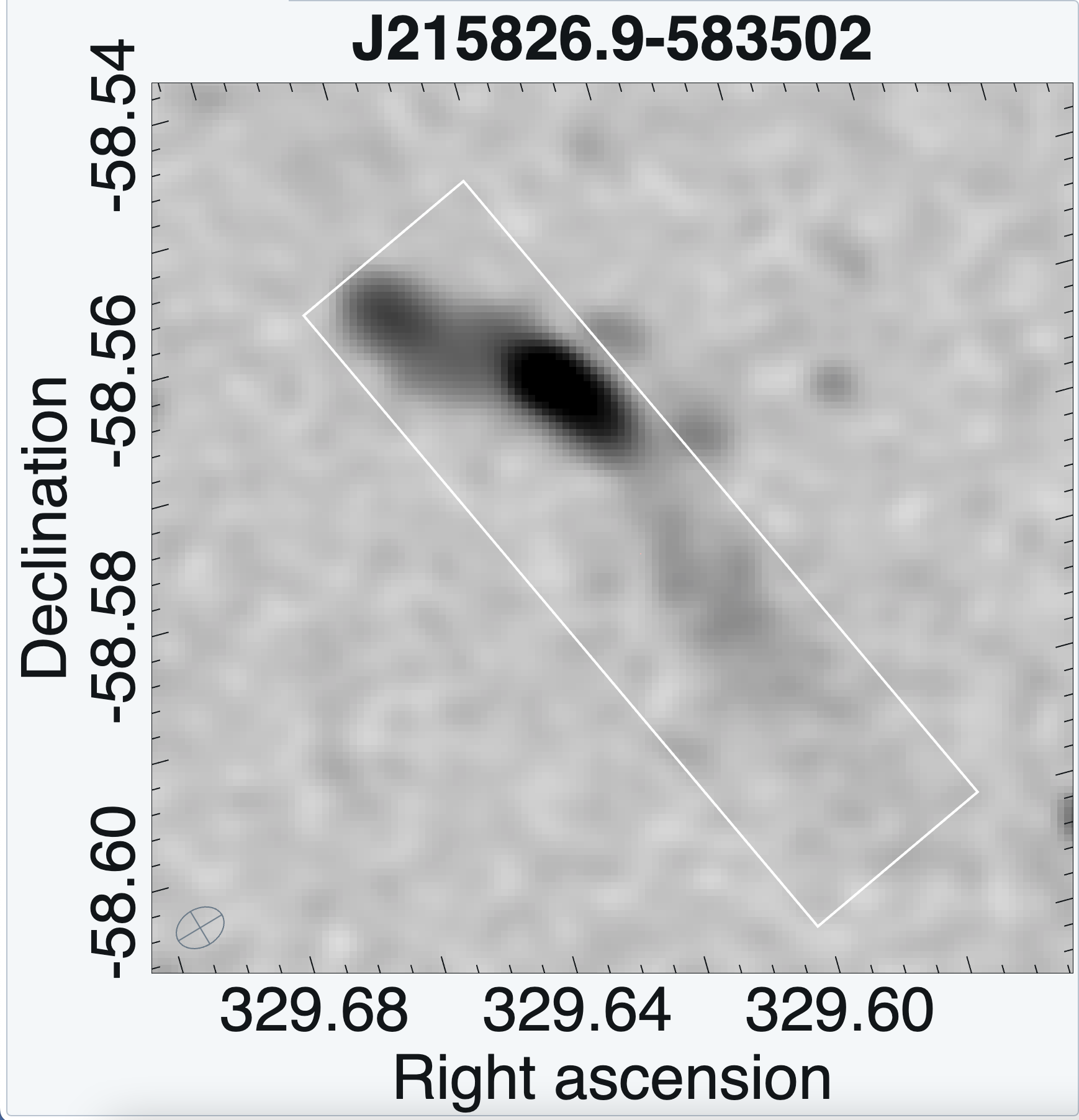}
  \caption{Examples of HyMoRS\ytag\ sources, which appear to be FR1 on one side and FR2 on the other. 
 In these and all subsequent multi-panel Figures, the white box is the rectangular region bounding box described in Section \ref{extraction} and the red cross is the tentative position of the host. If no cross is shown, then we were unable to identify a host.}
  \label{fig:HyMoRS}
\end{figure}

Fig \ref{fig:HyMoRS} shows a few examples of the HyMoRS sources in our sample, as follows.

J201556.3-583648 resembles a double-double source (see Section \ref{sec:dd}, with each side having an inner and outer lobe, but in the northern arm the inner lobe is brighter than the outer, while in the south the outer is brighter than the inner.

J202051.7-552745 is a classic example of a HyMoRS\ytag, with the southern jet being very clearly edge-brightened, while the northern jet gradually fades over its length.

In J202714.3-531401 the southern lobes is a classic edge-brightened FR2 lobe, while the northern lobe peaks a short distance from the core, and then gradually fades.

J213717.4-540544 has a bright hot-spot in the east, while the western jet is a thin bending arc which fades over its length.

J213917.3-541936 resembles a classic FR2, except for the hotspot half-way down its southern jet which is brighter than the hotspot, thus converting it from an FR2\ytag\ to a HyMoRS\ytag. 

J215826.9-583502 has a classic FR2 hot spot terminating its northern jet, while the southern jet reaches a peak a short distance from the core and then fades over its length, although both jets have a similar shape.

While these few examples on their own cannot add significantly to the FR1/FR2 debate, EMU-PS1  contains \HyMoRS\ HyMoRS\ytag\ candidates, so that the whole EMU survey will contain thousands of such objects, which are likely to have a significant influence on our understanding of the FR dichotomy.

\subsection{Linear triple sources (LTS\ytag)}

LTS\ytag\ sources are core-dominated sources that have two linear features extending from the host, but either one or both features do not have the maximum that would enable them to be classified as FR1\ytag\ or FR2\ytag.

\label{sec:LTS}
\begin{figure}
    \includegraphics[width=4cm]{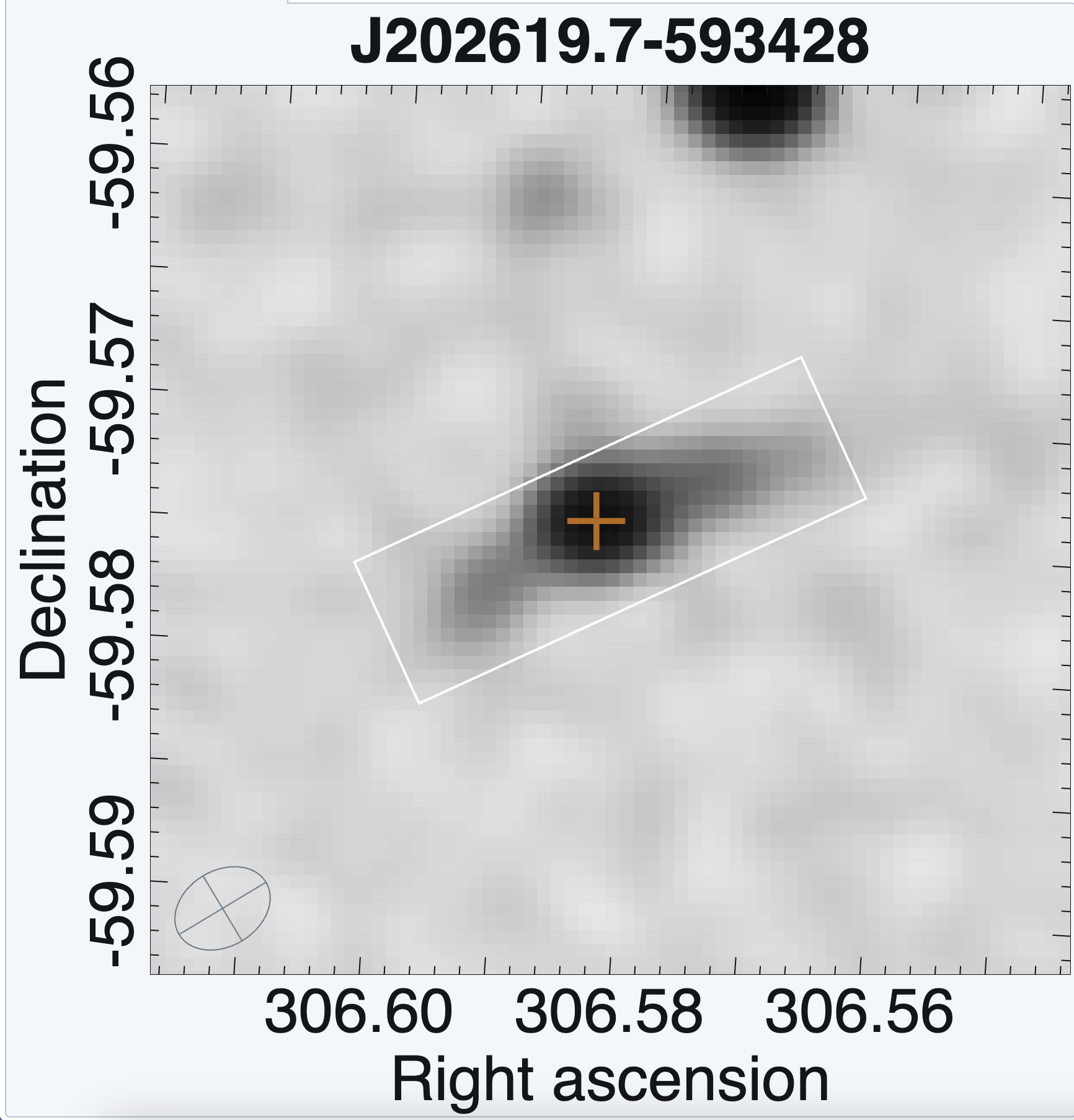}
    \includegraphics[width=4cm]{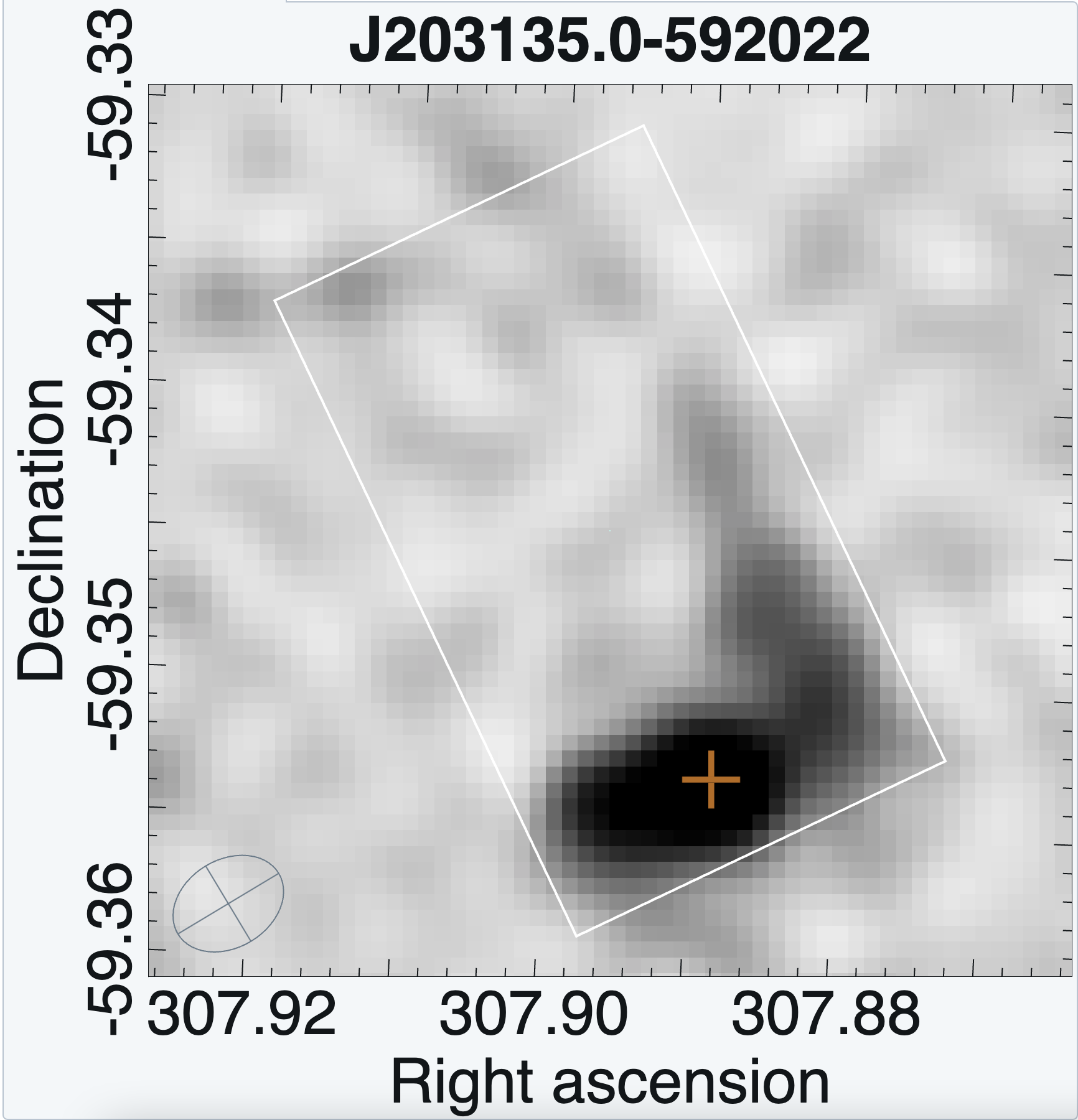}
    \includegraphics[width=4cm]{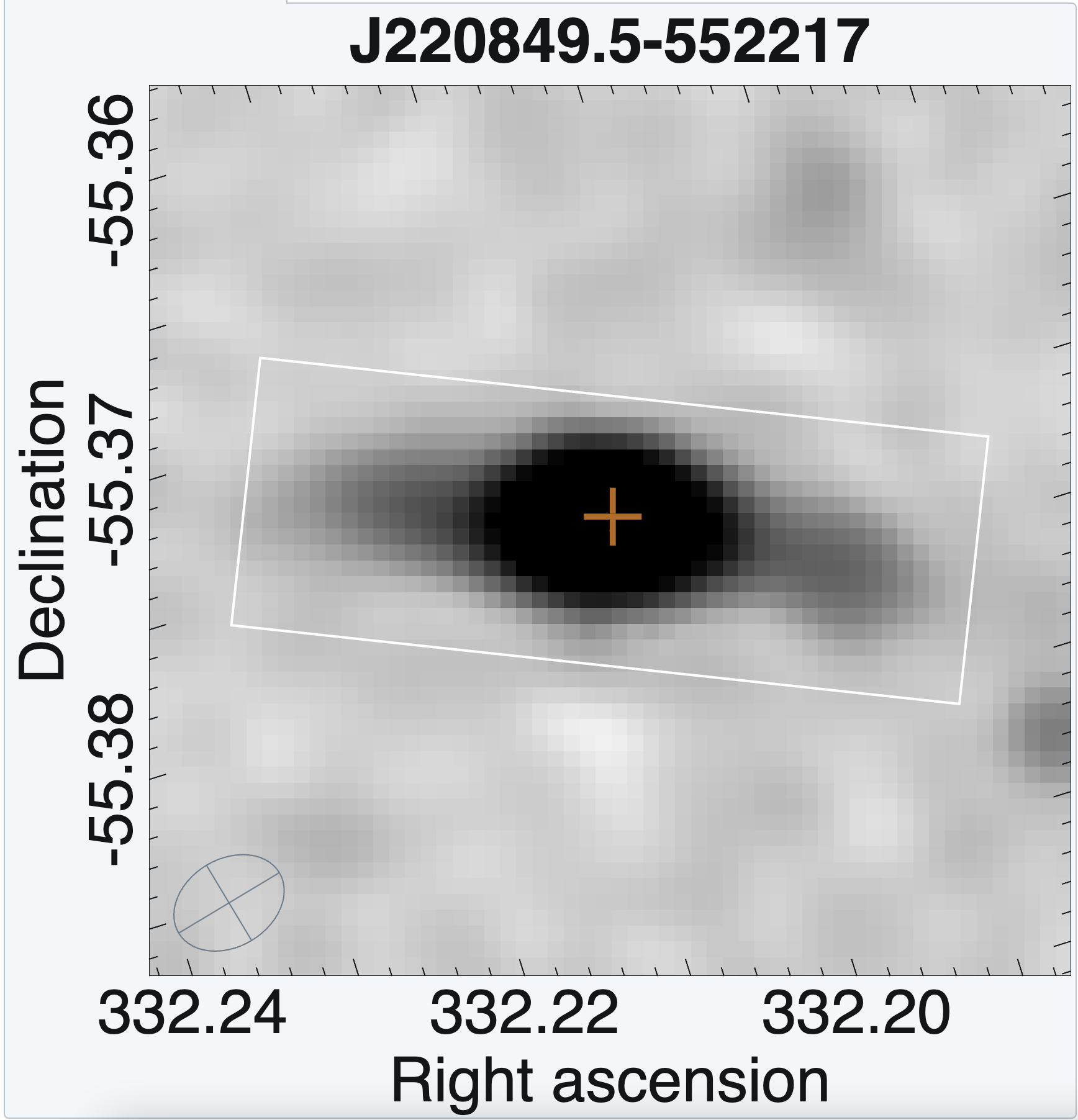}
    \includegraphics[width=4cm]{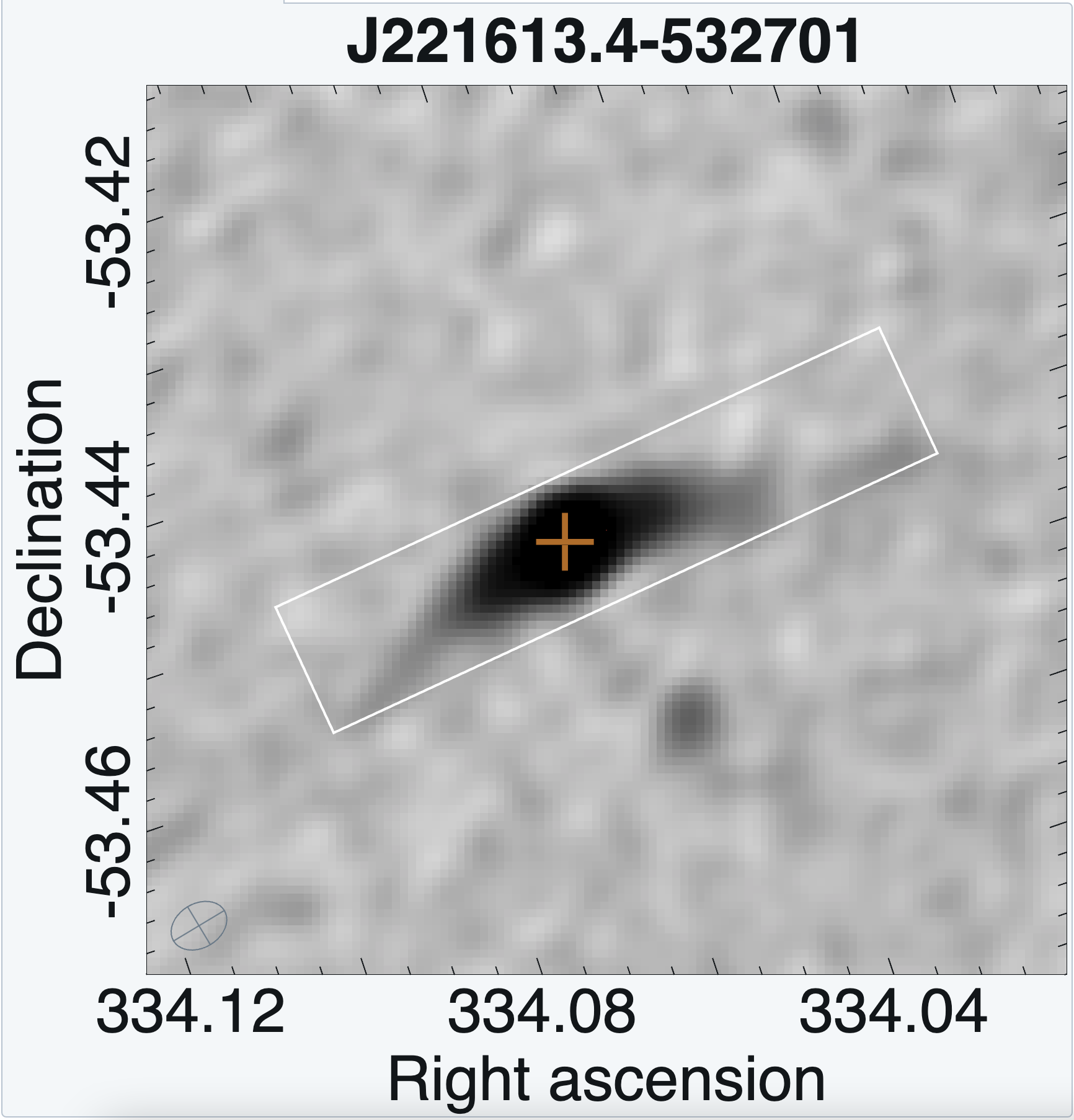}
    \includegraphics[width=4cm]{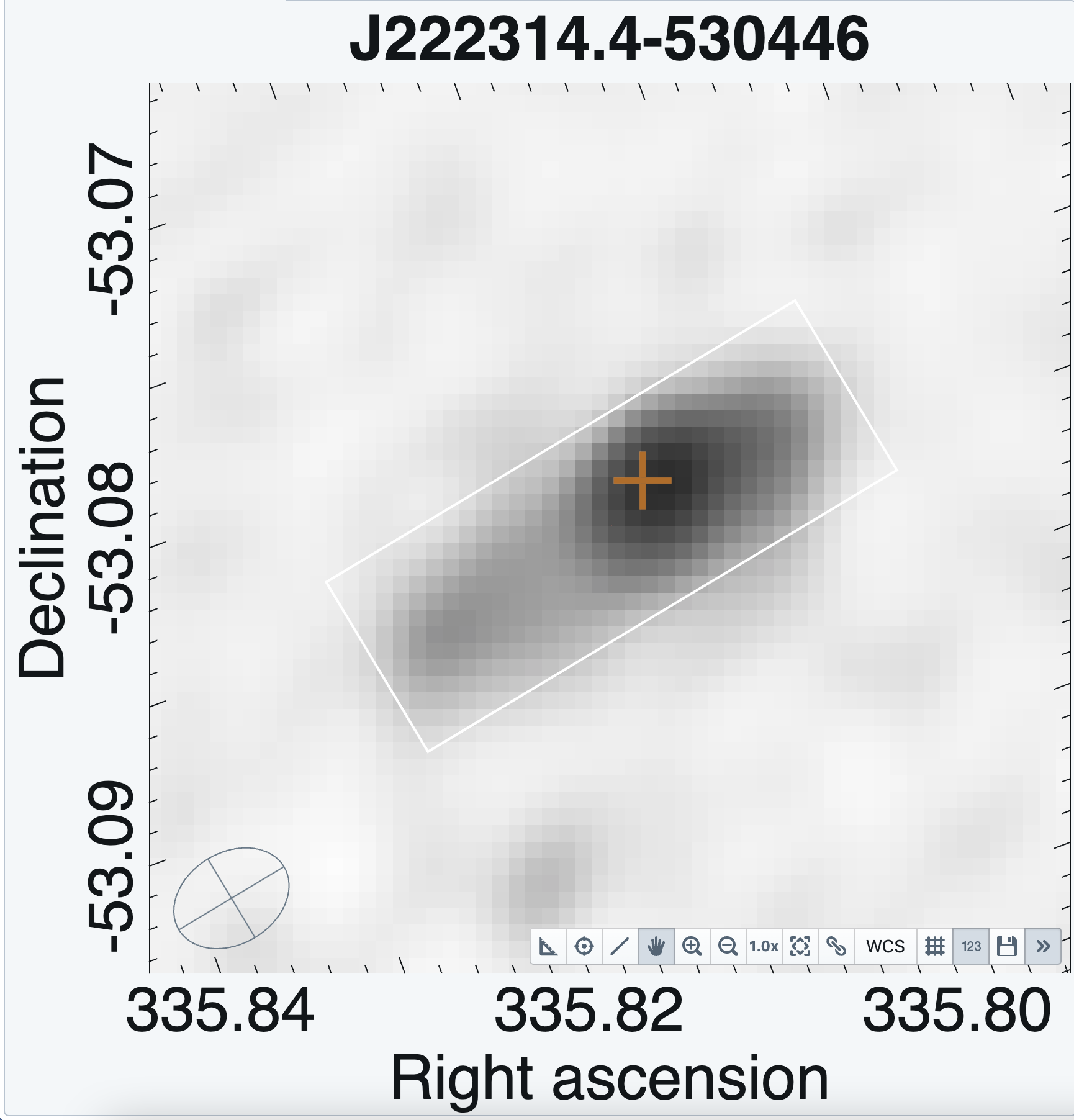}
    \includegraphics[width=4cm]{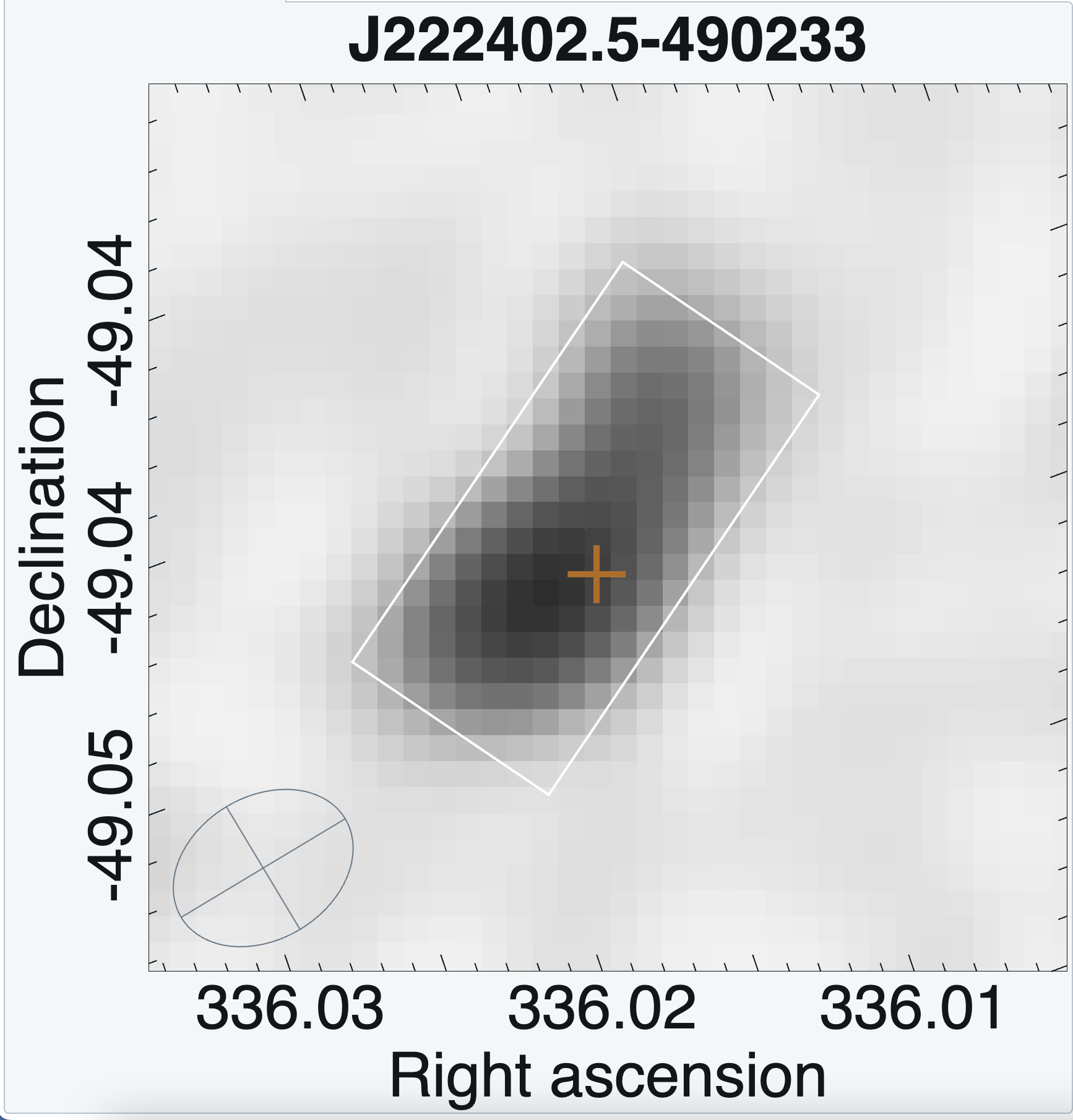}

  \caption{Examples of LTS\ytag\ (Linear triple) sources which have two jets but lack the maxima on both sides that would enable them to be classified as a FR source. The source J203135.0-592022, is very asymmetrical and has a curious 90\degr bend in one jet. 
  }
  \label{fig:LTS}
\end{figure}

We find \LTS\ of these sources, a sample of which are shown in Fig. \ref{fig:LTS}. It is possible that in higher resolution observations, some might show a maximum near the core that would qualify them as FR1\ytag. Alternatively, they could be high-redshift FR2 sources where the lobes have been diminished by inverse Compton losses \citep{turner20}. In either case,   they still represent a distinct morphological class. To the best of our knowledge, this class has not previously been identified in the literature, although they are probably similar to the ``naked jets'' identified by \citet{parma87} and \citet{ruiter90}. The large number found in this survey, and their relative absence in earlier surveys, suggests they they are largely confined to the low-luminosity end of the distribution of AGN.

Several of the sources in Figure \ref{fig:LTS} show slight bending of the jets. In the most extreme case, J203135.0-592022, the {  north-west} jet executes an abrupt 90\degr bend shortly after leaving the core.

\subsection{One-sided sources (OSS\ytag)}
\label{sec:oss}

OSS\ytag\ are sources that, in the EMU images, have a jet on one side of the host but no detectable jet can be seen on the other. 

The vast majority of DRAGNs are two-sided, but a significant minority, including the very first radio-loud quasar to be identified, 3C273, have one-sided jets. The jet in 3C273 is observed at radio, optical, and X-ray wavelengths and no counter-jet has ever been detected \citep{courvoisier98}. One-sided jets are relatively common in radio-loud quasars, and are presumably due to relativistic boosting of the nearside jet \citep{kellermann88}.

\begin{figure}
  \includegraphics[width=4cm]{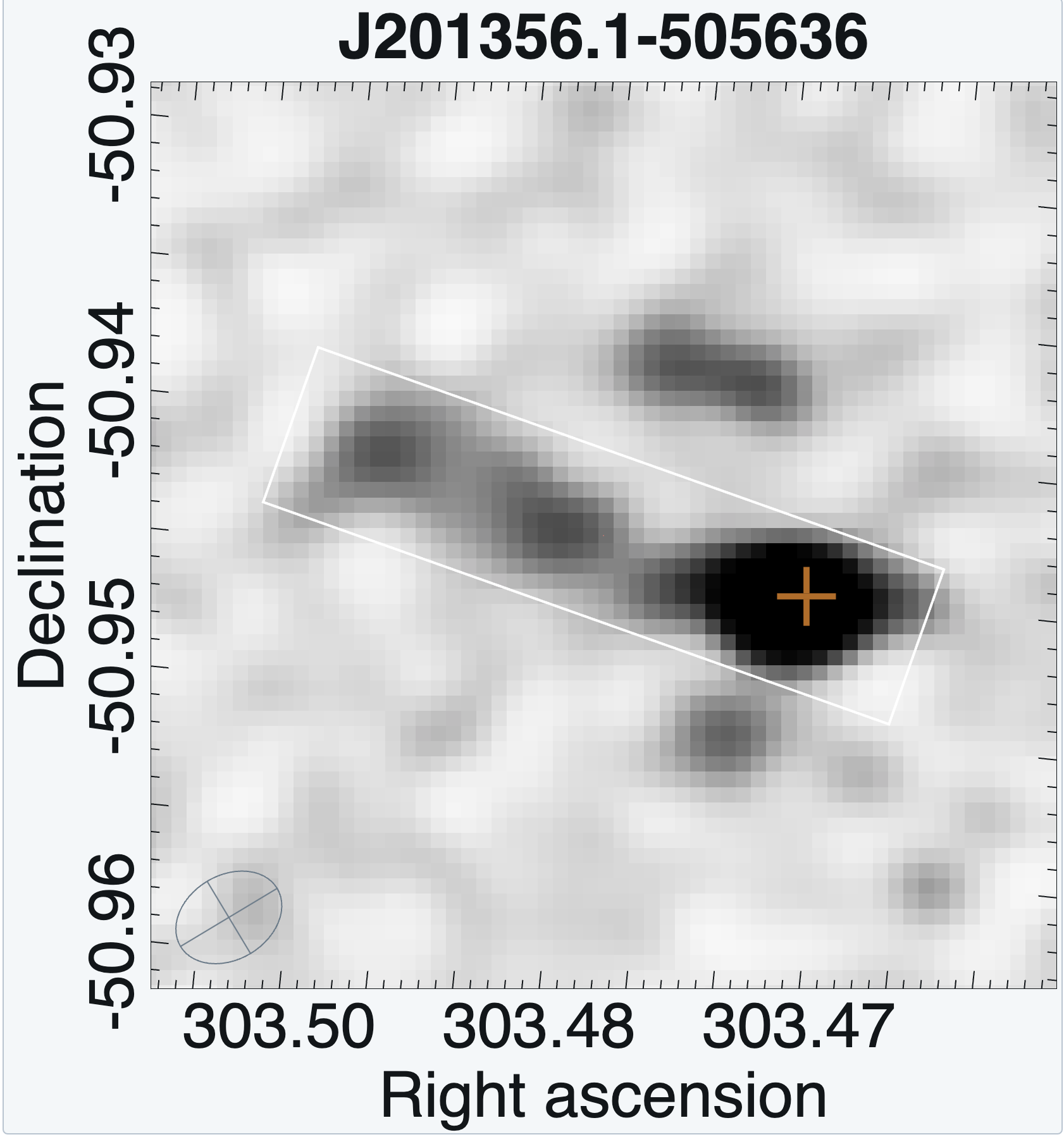} 
  \includegraphics[width=4cm]{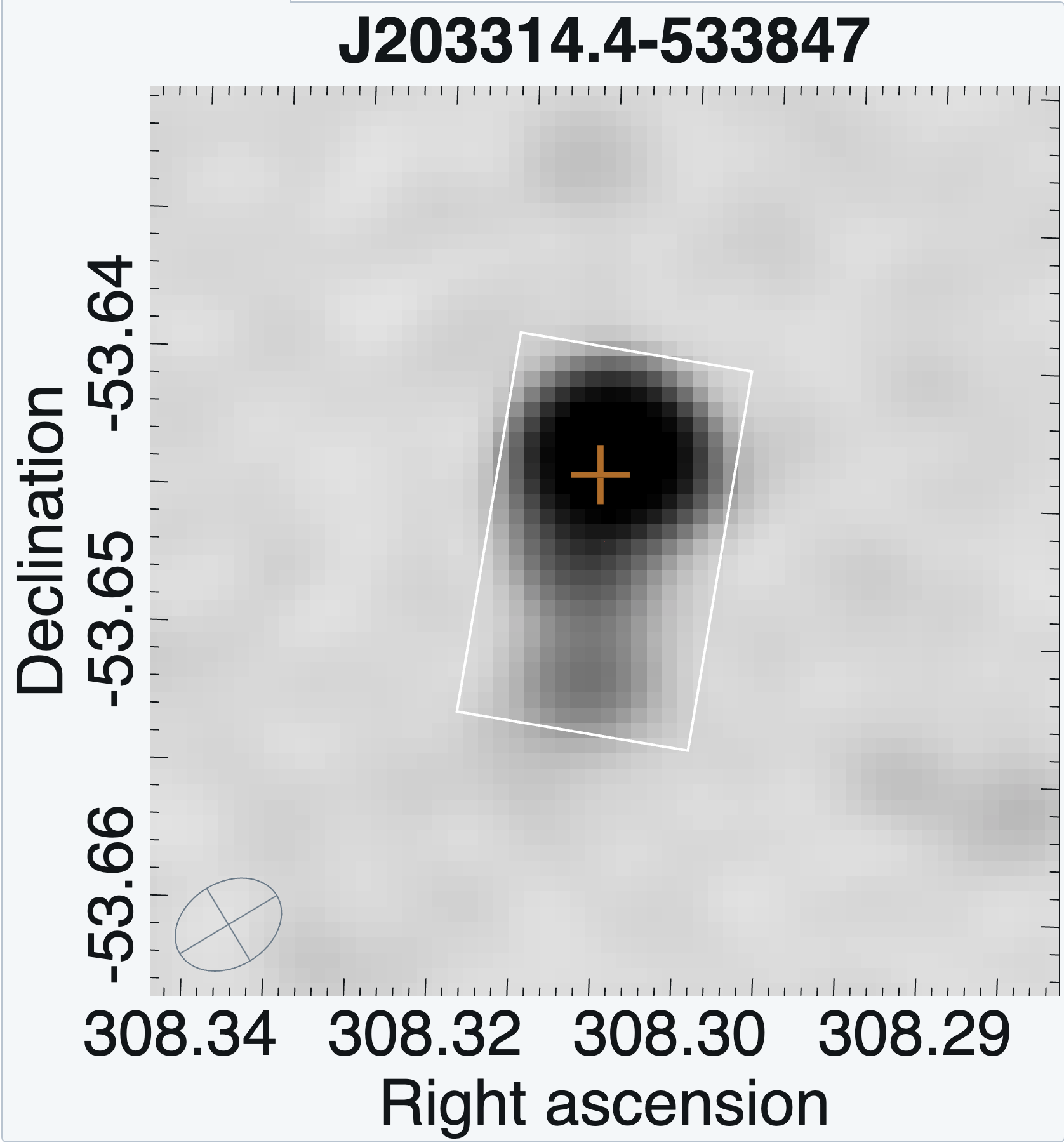} 
  \includegraphics[width=4cm]{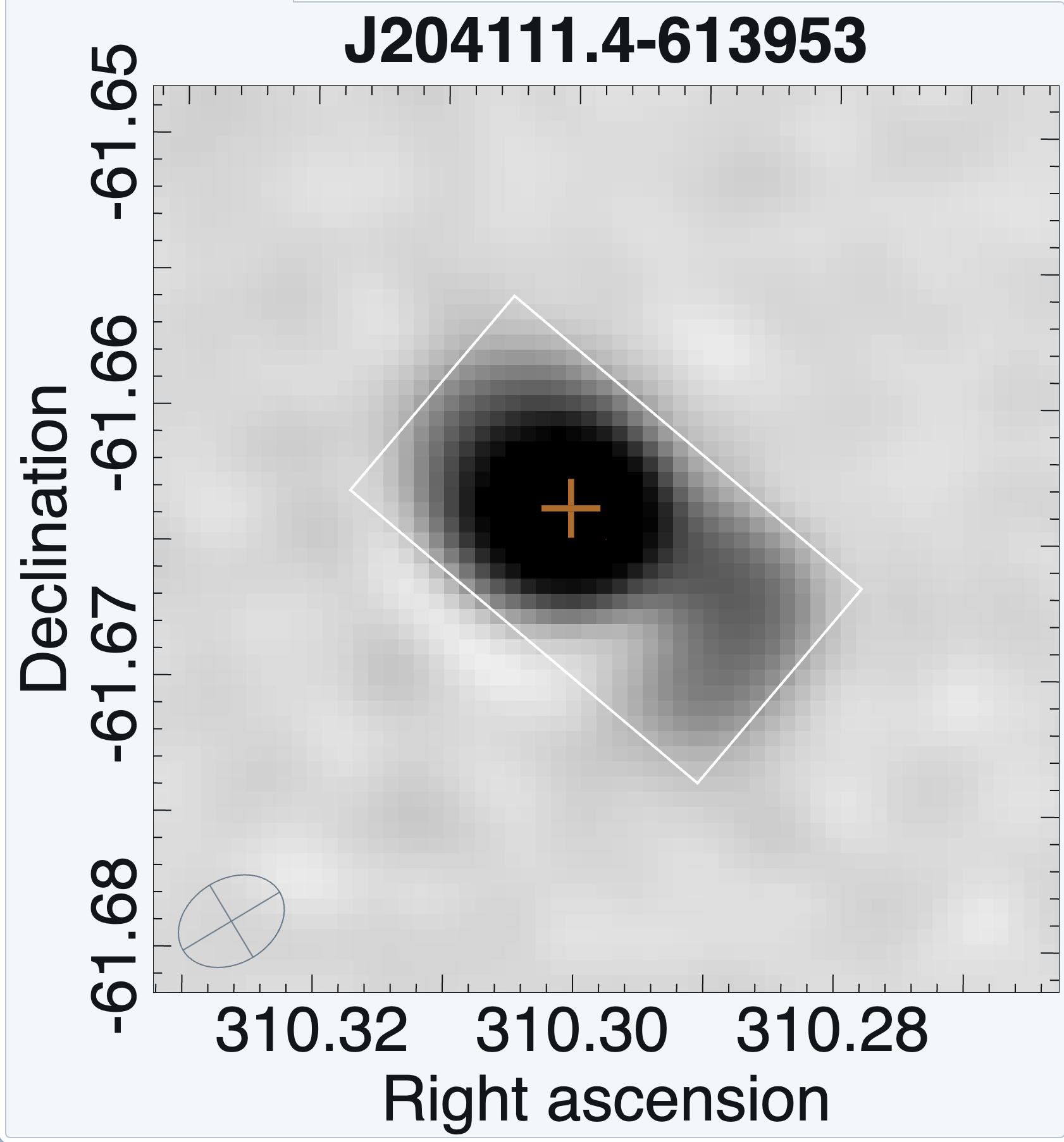}   
  \includegraphics[width=4cm]{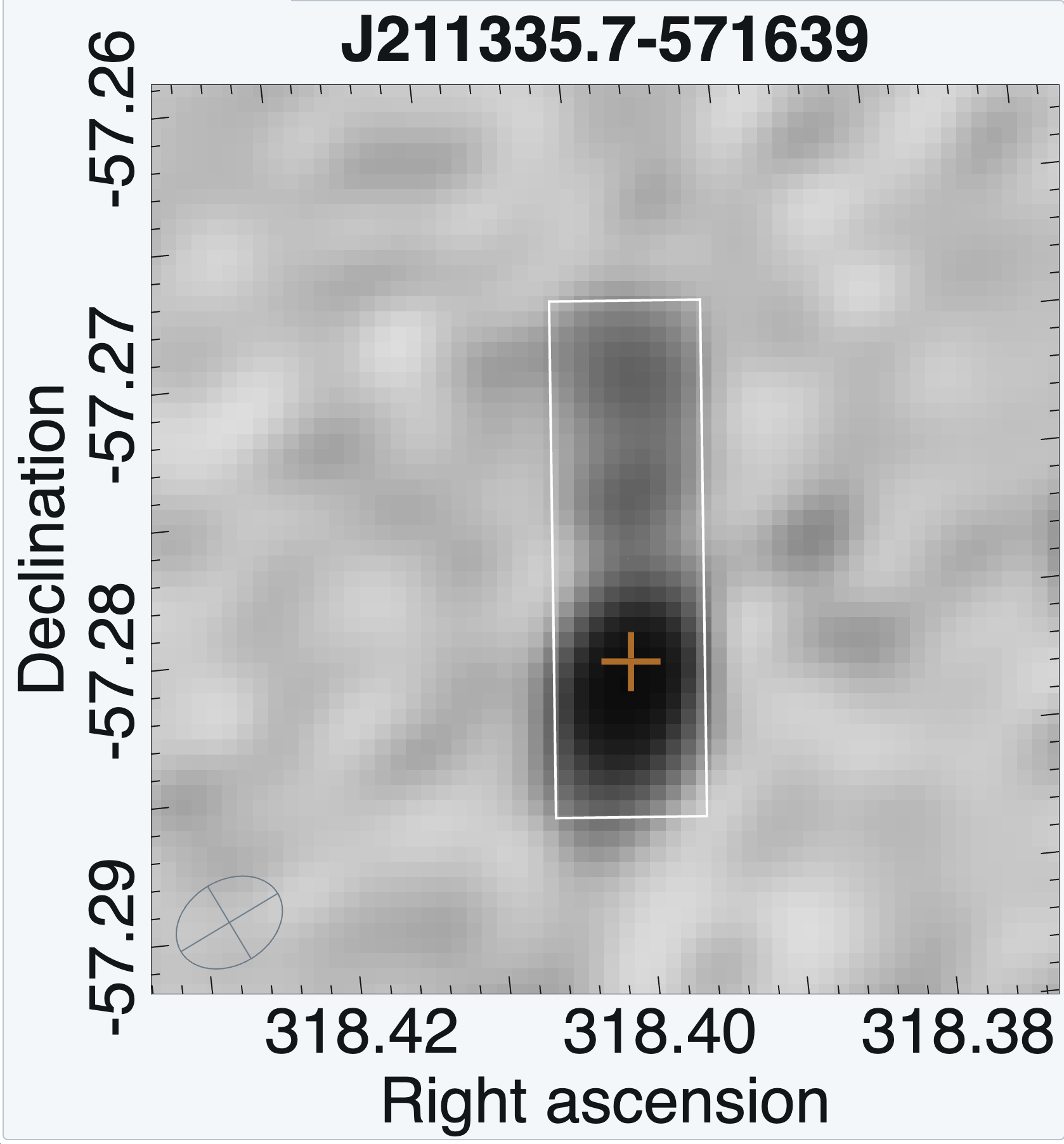}  
  \includegraphics[width=4cm]{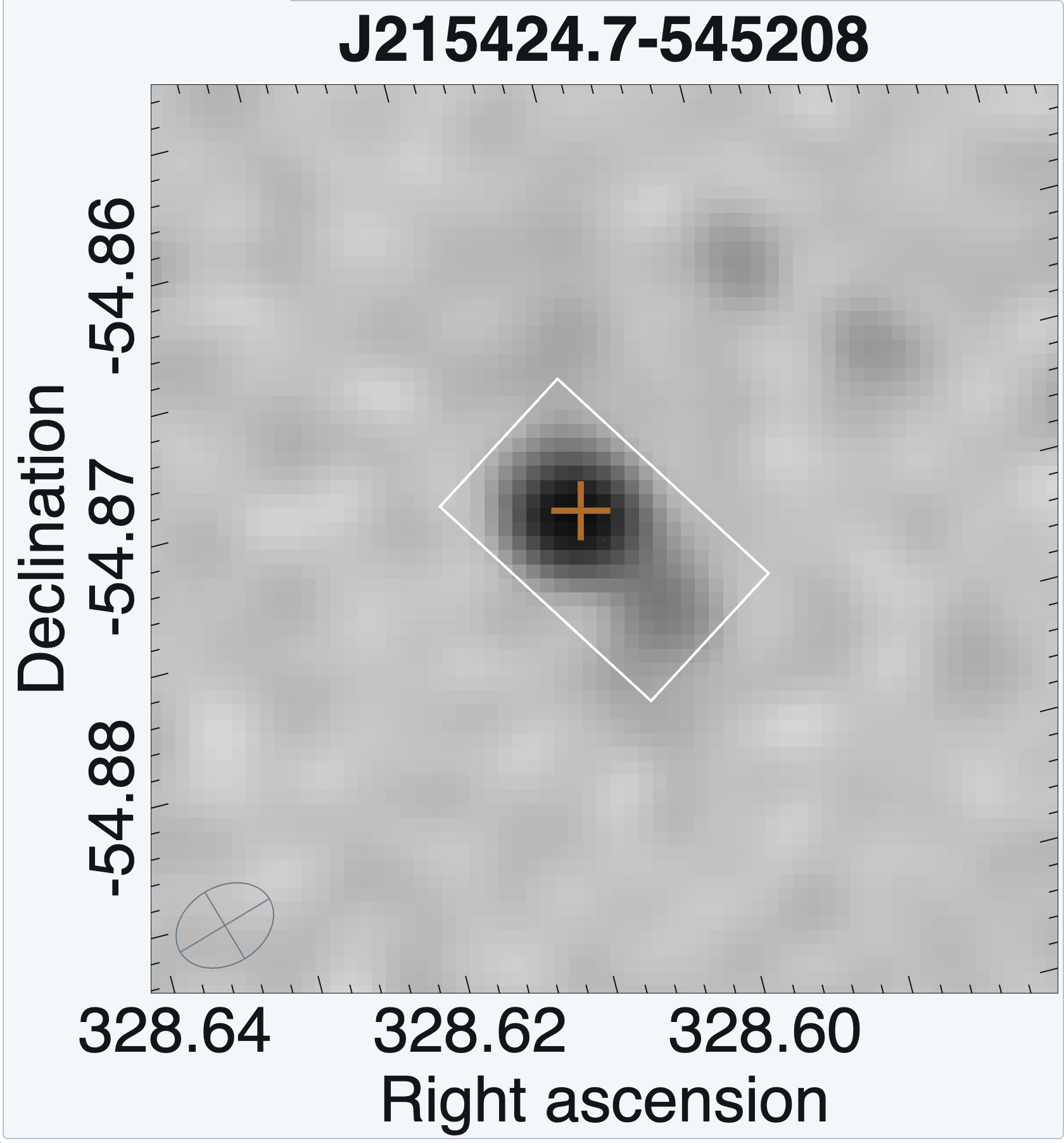}  
  \includegraphics[width=4cm]{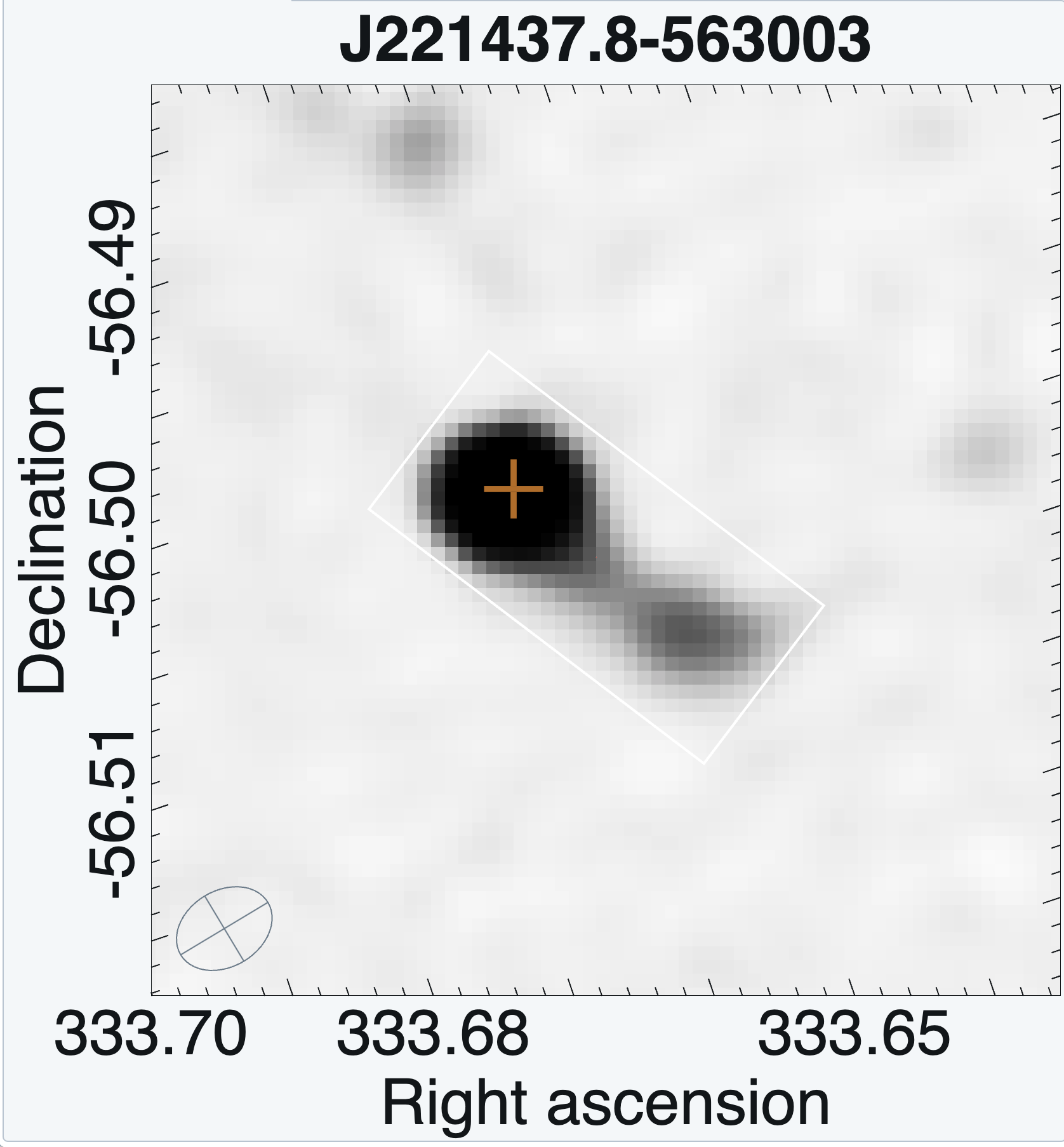}
  
  \caption{Examples of candidate OSS\ytag\ (one-sided) sources. Many show a bent tail, notably J204111.4-613953 and J215424.7-545208 which have a 90\degr bend in the jet close to the host, possibly due to a superposition of sources. }
  \label{fig:oss}
\end{figure}

Also relatively common are double-lobed sources in which the jet is one-sided at small scales, but then becomes two-sided at kpc scales, presumably as the jet loses energy and becomes sub-relativistic, so that relativistic beaming is seen only near the core \citep{kellermann88}. Such one-sided jets at \ac{VLBI} scales are often called core-jet sources \citep[e.g.][]{yang21}.

Much rarer are sources, corresponding to our OSS\ytag\ class, in which the emission at kpc scales is one-sided. Nevertheless, a significant number has been catalogued over the years \citep{kapahi81,saikia86,saikia89, saikia90, saikia96}.

At this scale, in most cases the jet is unlikely to be relativistic, and so the one-sidedness may be attributed to an intrinsically one-sided source, possibly because the source ``flip-flops'' between the two sides \citep{groningen80,laing99, rudnick84, kellermann88}.

In our sample, we found \oss\ candidate one-sided sources. Some have a suggestion of a counter-jet that is much shorter than the main jet, suggesting that these are very asymmetric sources rather than true one-sided sources. Furthermore, to establish that a source is a genuine one-sided source, the position of the host needs to be determined. We show a sample of those that have a host galaxy coincident with the putative core in Fig. \ref{fig:oss}. In two sources, J204111.4-613953 and J215424.7-545208, the jet bends sharply after leaving the core, presumably due to the effect of intergalactic gas. At least two sources (J211335.7-571639 and J204111.4-613953) have a suggestion of a counter-jet.

\subsection{Bent-tail and head-tail sources (BT\ytag, HT\ytag)}
\label{sec:bt}

In the majority of DRAGNs, the two jets/lobes and the host galaxy are roughly in a straight line. In a significant minority, the two jets are bent, presumably by their interaction with the intergalactic medium. 

In this paper we tag any DRAGN where the jet is bent by more than 10\degr (i.e. so the angle between the jets must be less than 170\degr ) as a ``Bent-tail'' (BT\ytag) source.

\begin{figure}
    \includegraphics[width=4cm]{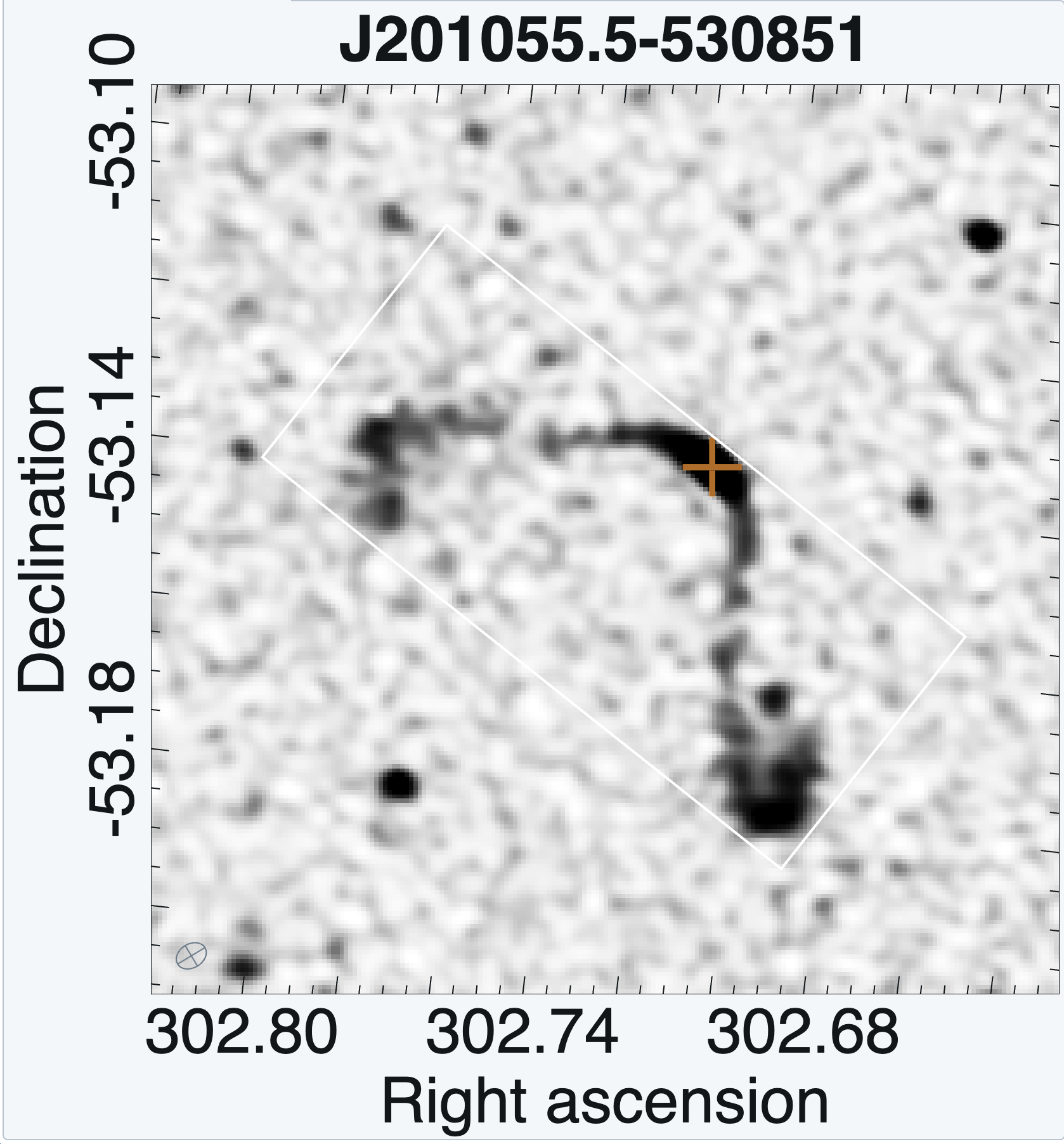}
    \includegraphics[width=4cm]{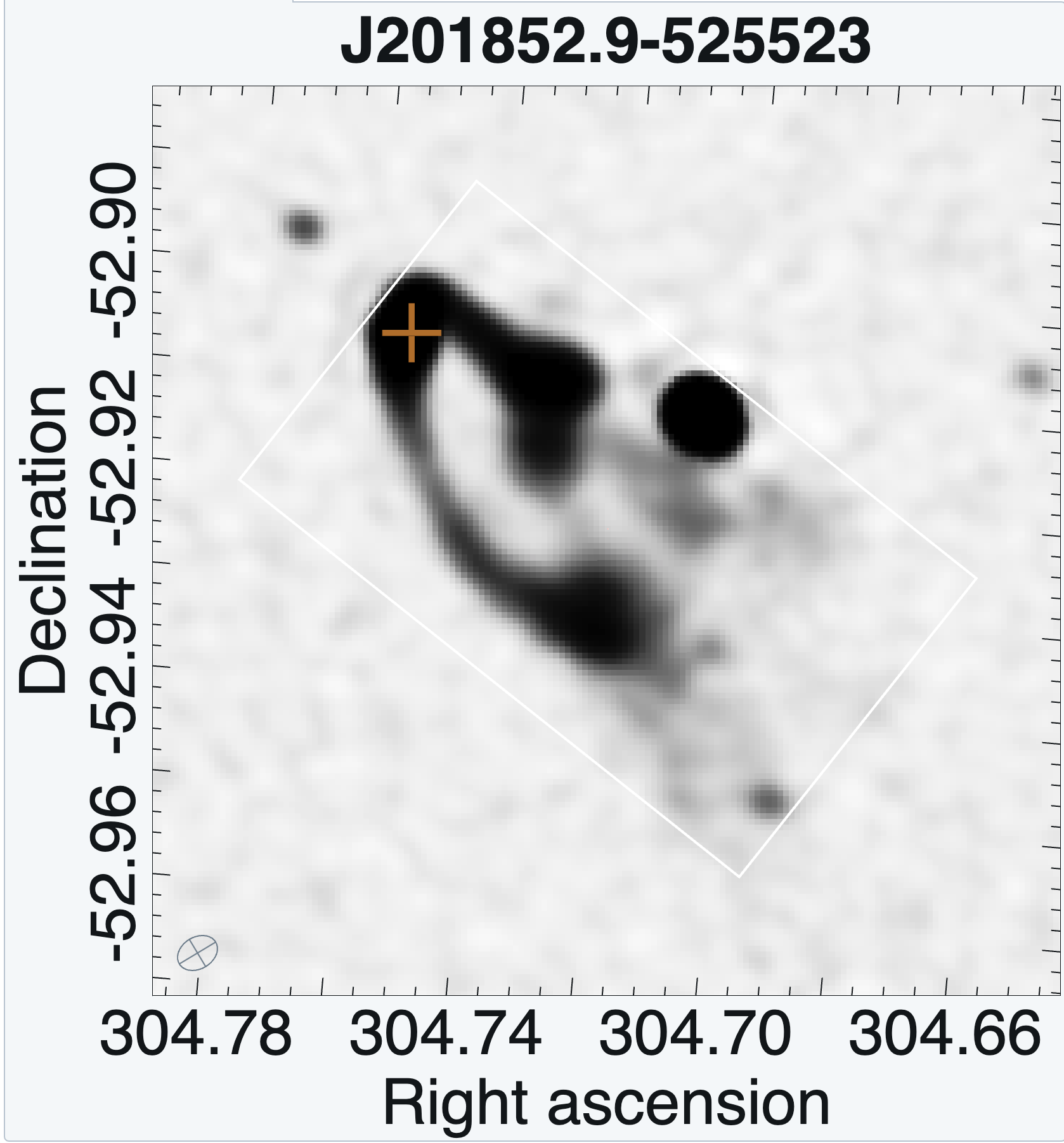}
    \includegraphics[width=4cm]{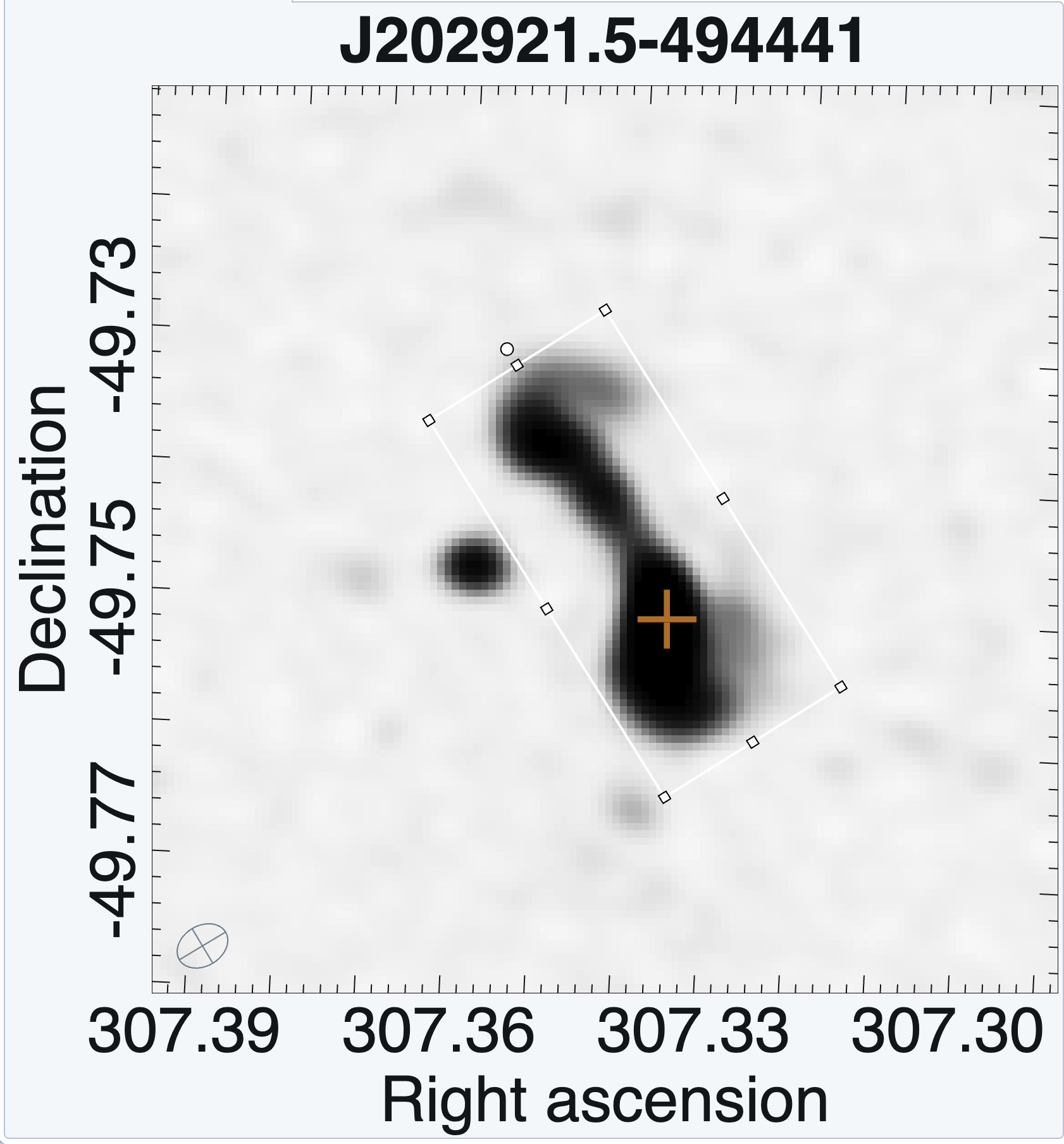}
    \includegraphics[width=4cm]{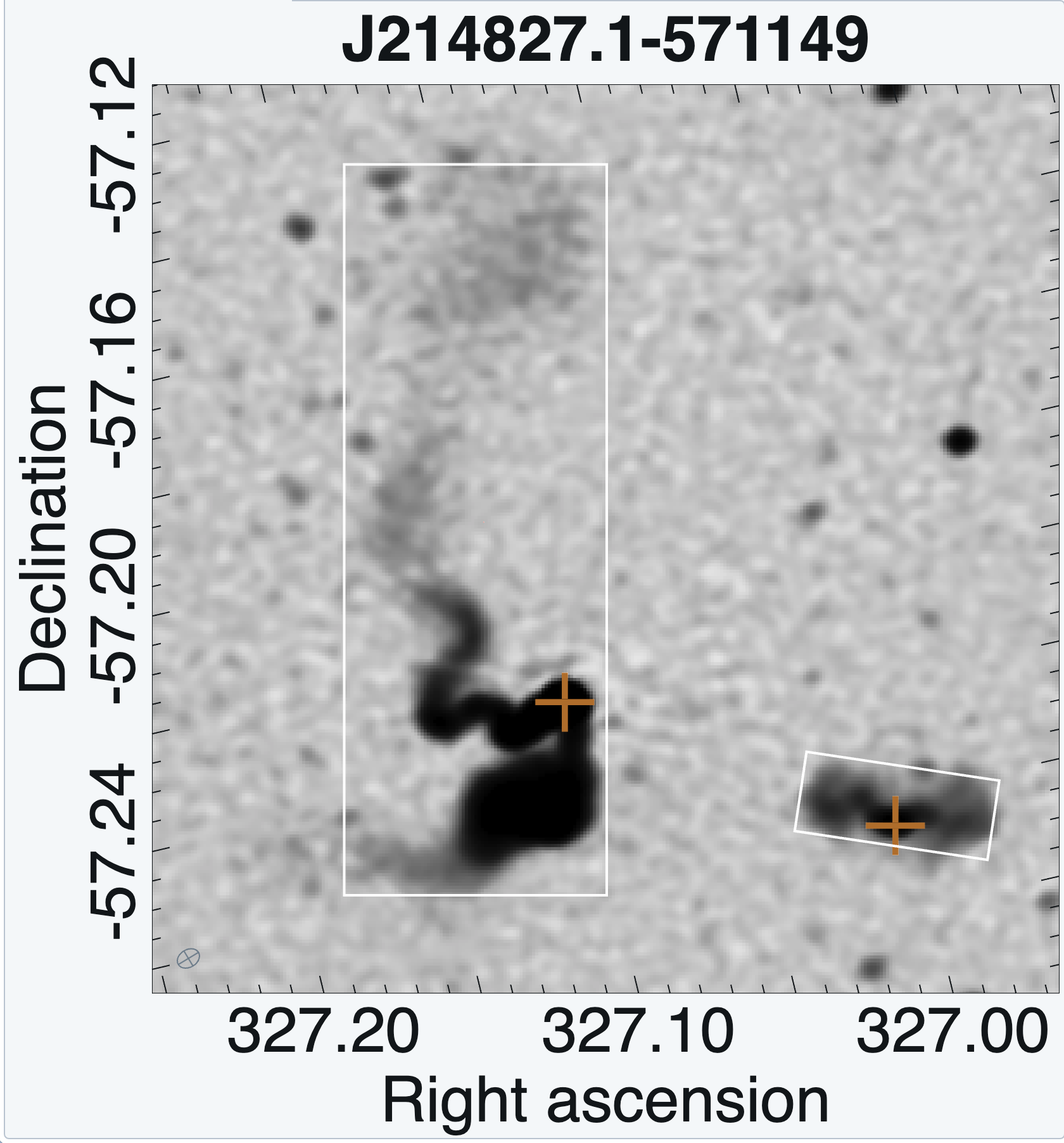}
    \includegraphics[width=4cm]{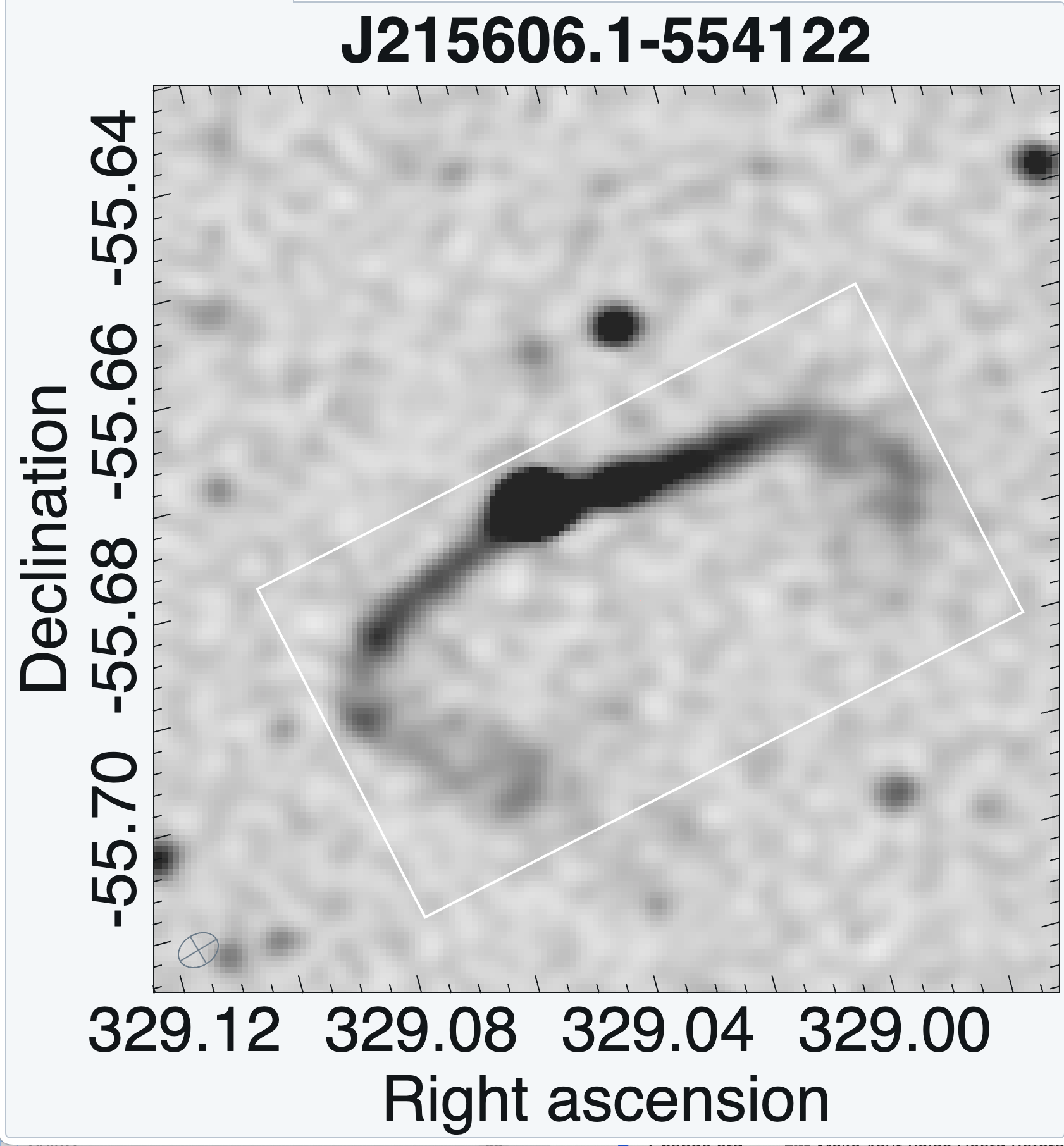}
    \includegraphics[width=4cm]{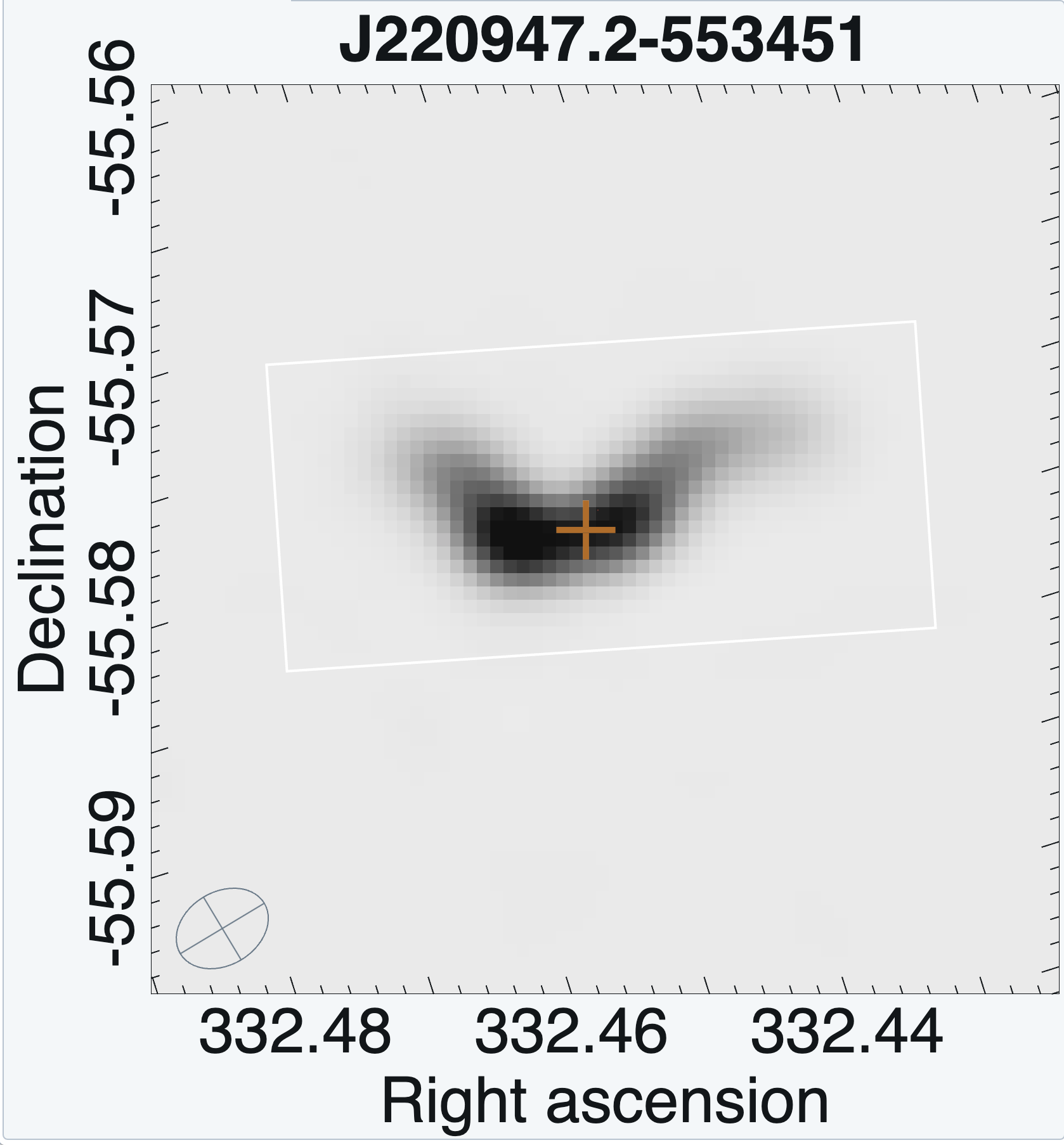}
  \caption{Examples of BT\ytag\ (bent-tail) sources.
  }
  \label{fig:bt}
\end{figure}

BT\ytag\ sources are often divided into three subsets:
\begin{itemize}
    \item Wide-Angle Tail sources (WATs): Sources in which the angle between the two jets is greater than 90\degr. We note that some authors do not include this bending angle in their criterion for a WAT \citep[e.g.][]{missaglia19}.
    \item Narrow-Angle Tail sources (NATs): Sources in which the angle between the two jets is less than 90\degr
    \item Head-tail (HT) sources. Sources in which the tails are bent around so far they they end up parallel, and frequently combine so that only one tail is seen at the largest scales or in low resolution observations.
\end{itemize}

In this paper we do not attempt to tag sources with NAT or WAT (since the classification often changes with distance from the host), but we do identify the HT\ytag\ class.

These definitions are not universally adopted, which leads to much confusion in the literature.
For example, the HT source was first recognised by \citet{ryle68}, although they did not actually use the term. The term ``head-tail'' was first used by \citet{miley72}, and a clear definition was first given by \citet{mao09} as follows:
{\em Head–tail (HT) sources are a subclass of radio galaxy whose radio jets are distorted such that they appear to bend in a common direction. This gives rise to the so-called ‘HT’ morphology where the radio jets are bent back to resemble a tail with the luminous galaxy as the ‘head’}. Here we adopt this definition.

Most authors (including those cited above) effectively define the HT label in the same way \citep[e.g.][]{rudnick76}, but we note that some authors \citep[e.g.][]{sasma22,pal23} adopt a definition of HT to include all bent-tail sources, even WATs. For example, none of the sources in the catalogue of 55 HT sources listed by \citet{pal23} would be recognised as an HT by the Miley or Mao definitions. To some extent this problem can be circumvented by the use of subscripted tags (see Section \ref{sec:tags}) but we would still urge authors to avoid confusion by using widely-adopted definitions where possible.

We show a sample of BT\ytag\ sources in Figure \ref{fig:bt}. They show a variety of morphologies. In most cases (e.g. J201055.5-530851, J202921.5-494441, J220947.2-553451) it is easy to understand the source in terms of the conventional explanation that the source is moving with respect to the intergalactic medium, and that the jet leaves the source with enough energy to withstand the ram pressure, but as it loses energy the jet is blown back by the intergalactic medium. However, two cases (J201852.9-525523 and J214827.1-571149) have such tangled morphologies that a more complex mechanism must be invoked. For example, perhaps the source is immersed in an intergalactic medium with a complex turbulent motion such that the direction of flow varies rapidly along the length of the jet, or perhaps the host galaxy is orbiting a companion \citep[e.g.][]{blandford78}. 

Bent-tail sources were first found when studying the radio emission from clusters \citep{ryle68,miley72}, and they have since traditionally been viewed as tracers of clusters \citep[e.g.][]{owen76, rudnick76, giacintucci09, lao25}. Their relationship to clusters will be explored in Paper III of this series.

\subsubsection{HT\ytag\ candidates}

\begin{figure}
  \includegraphics[width=4cm]{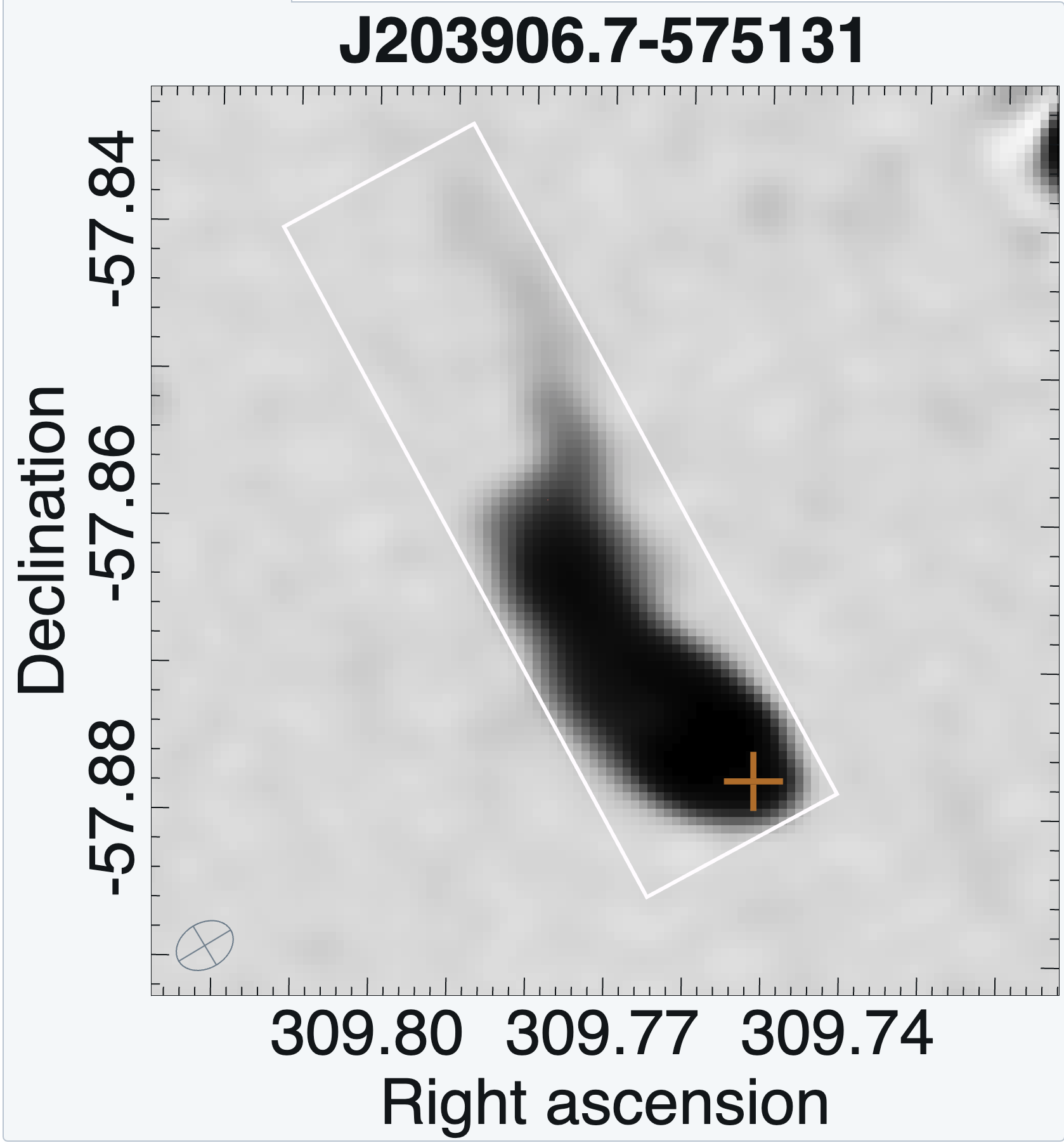}
  \includegraphics[width=4cm]{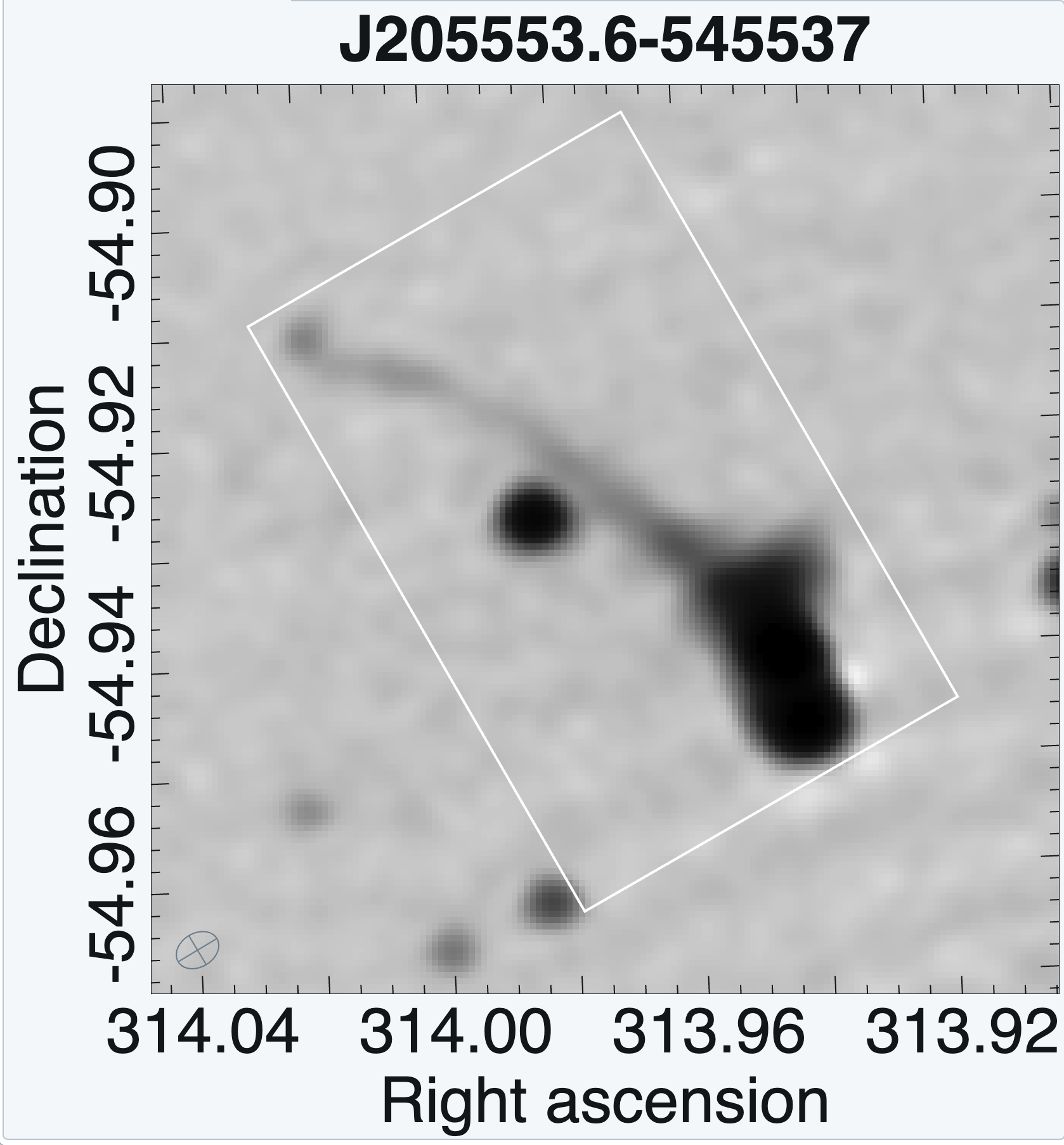}
  \includegraphics[width=4cm]{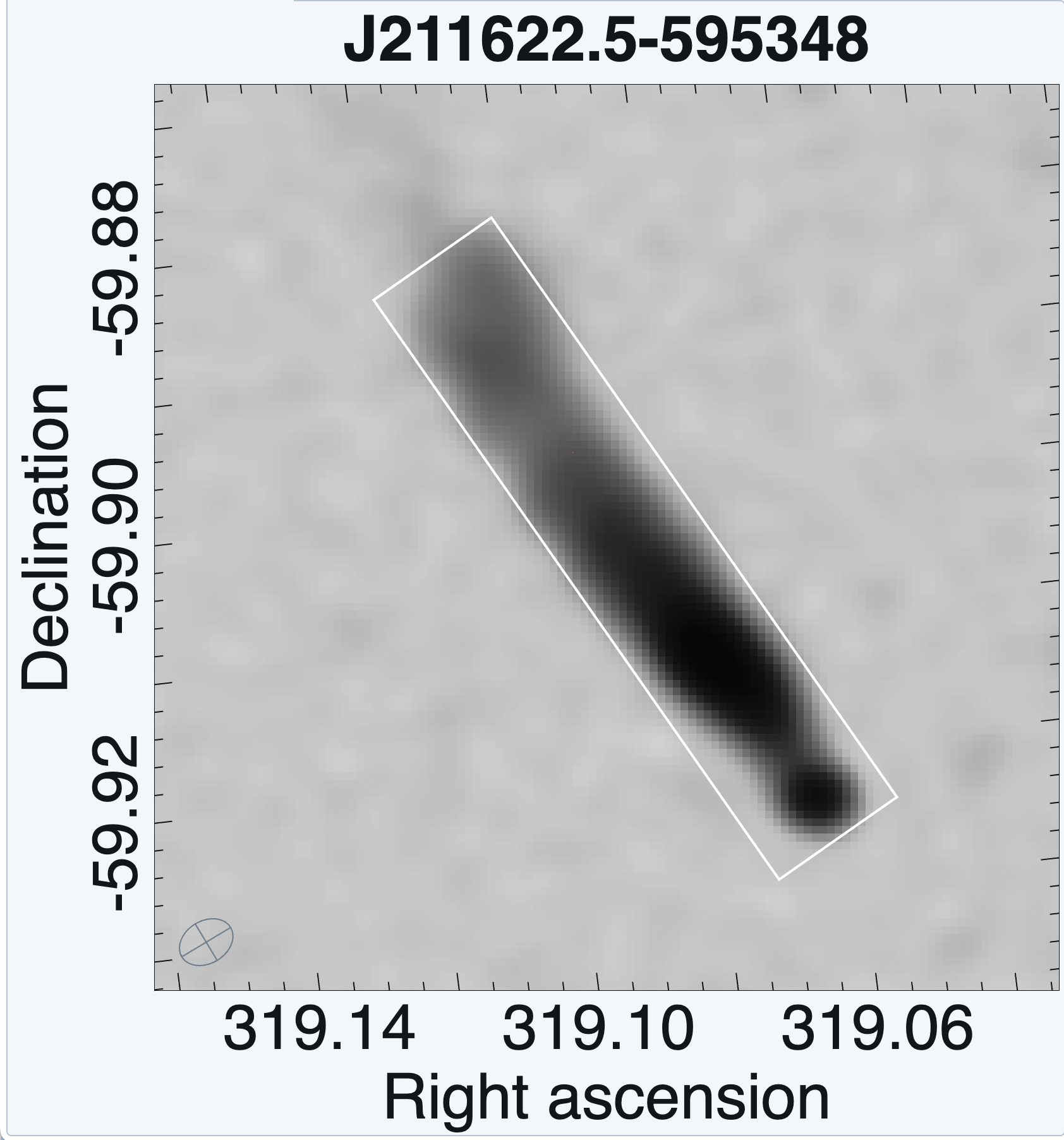}
  \includegraphics[width=4cm]{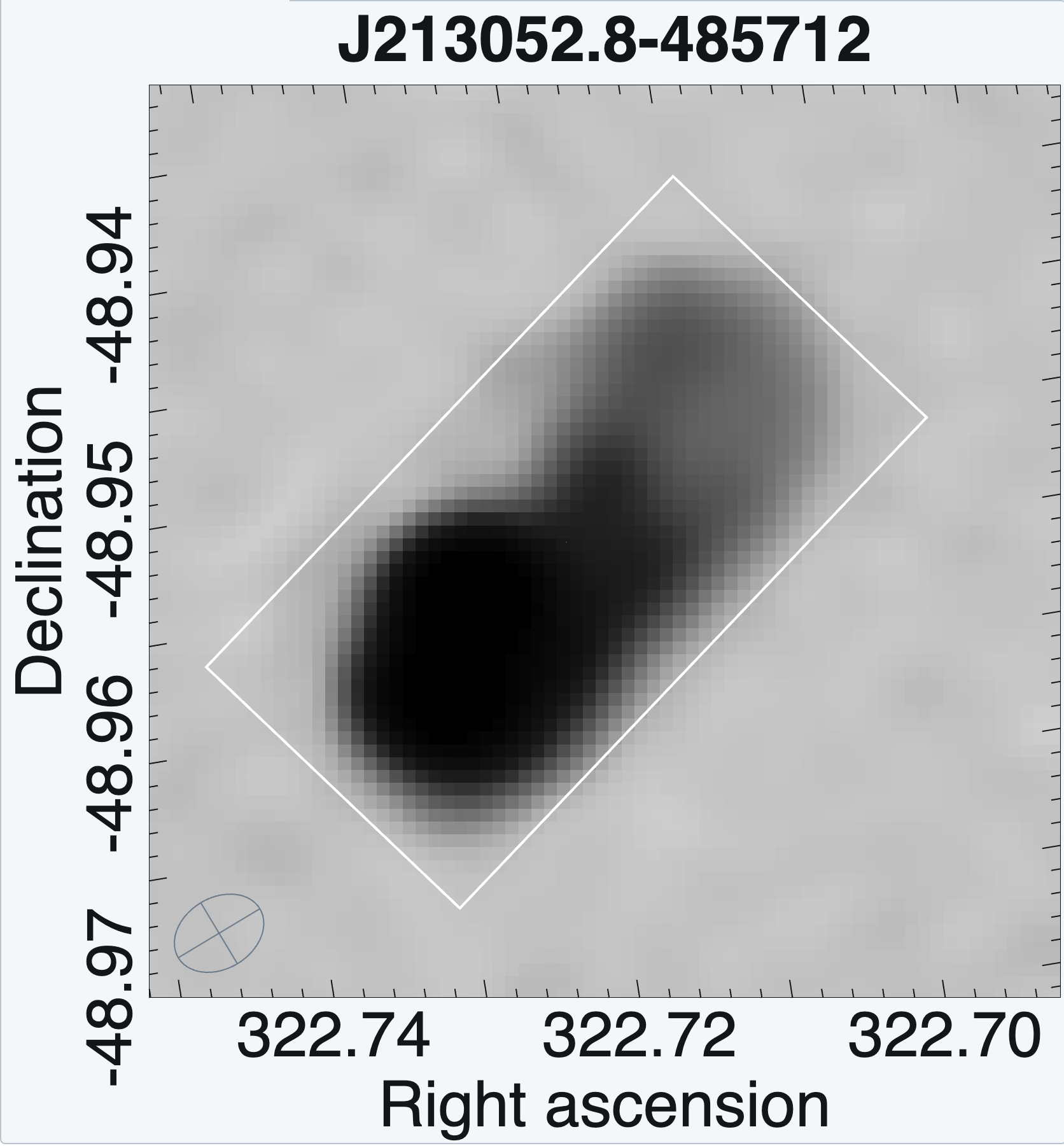} 
  \includegraphics[width=4cm]{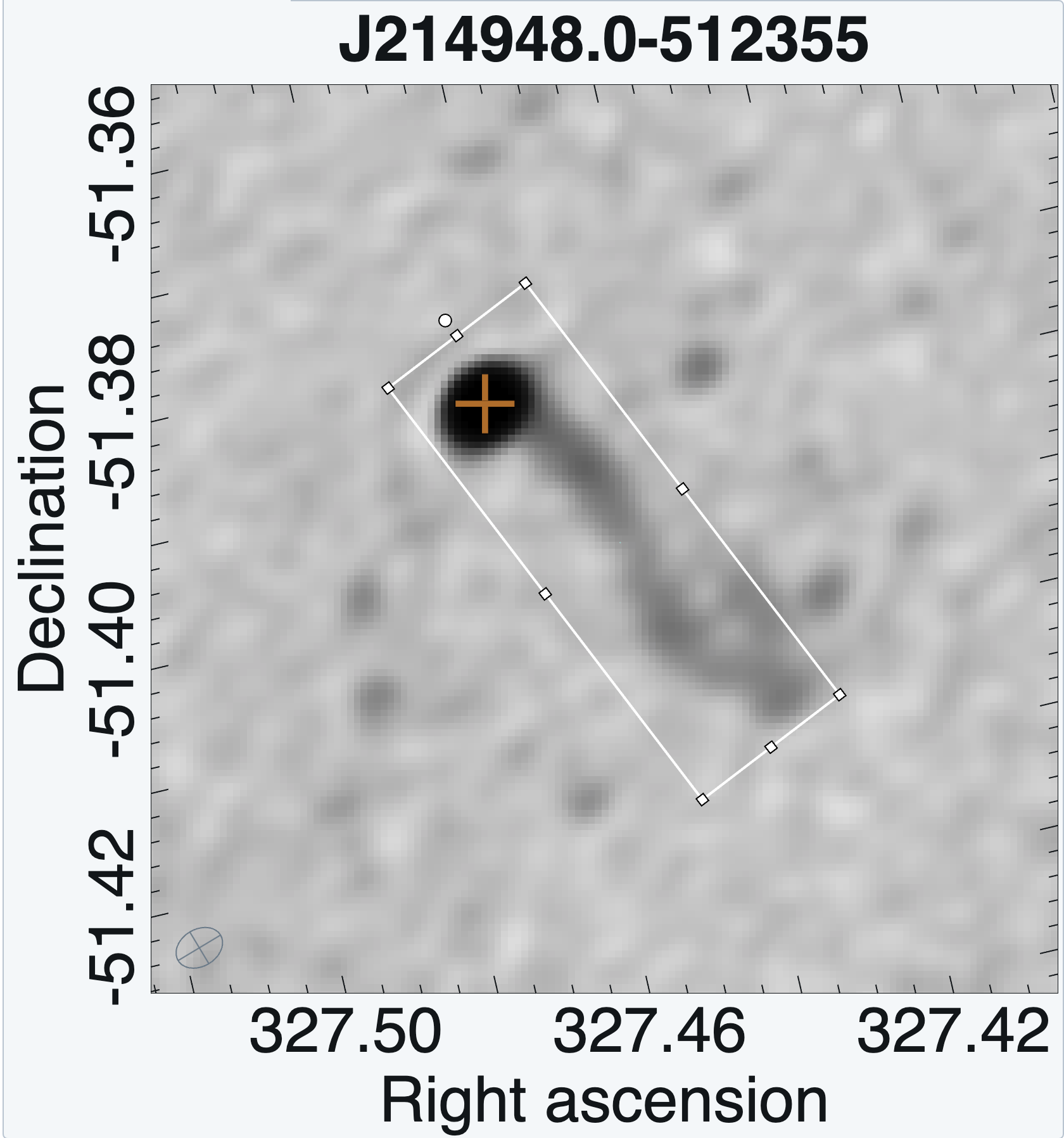}
  \includegraphics[width=4cm]{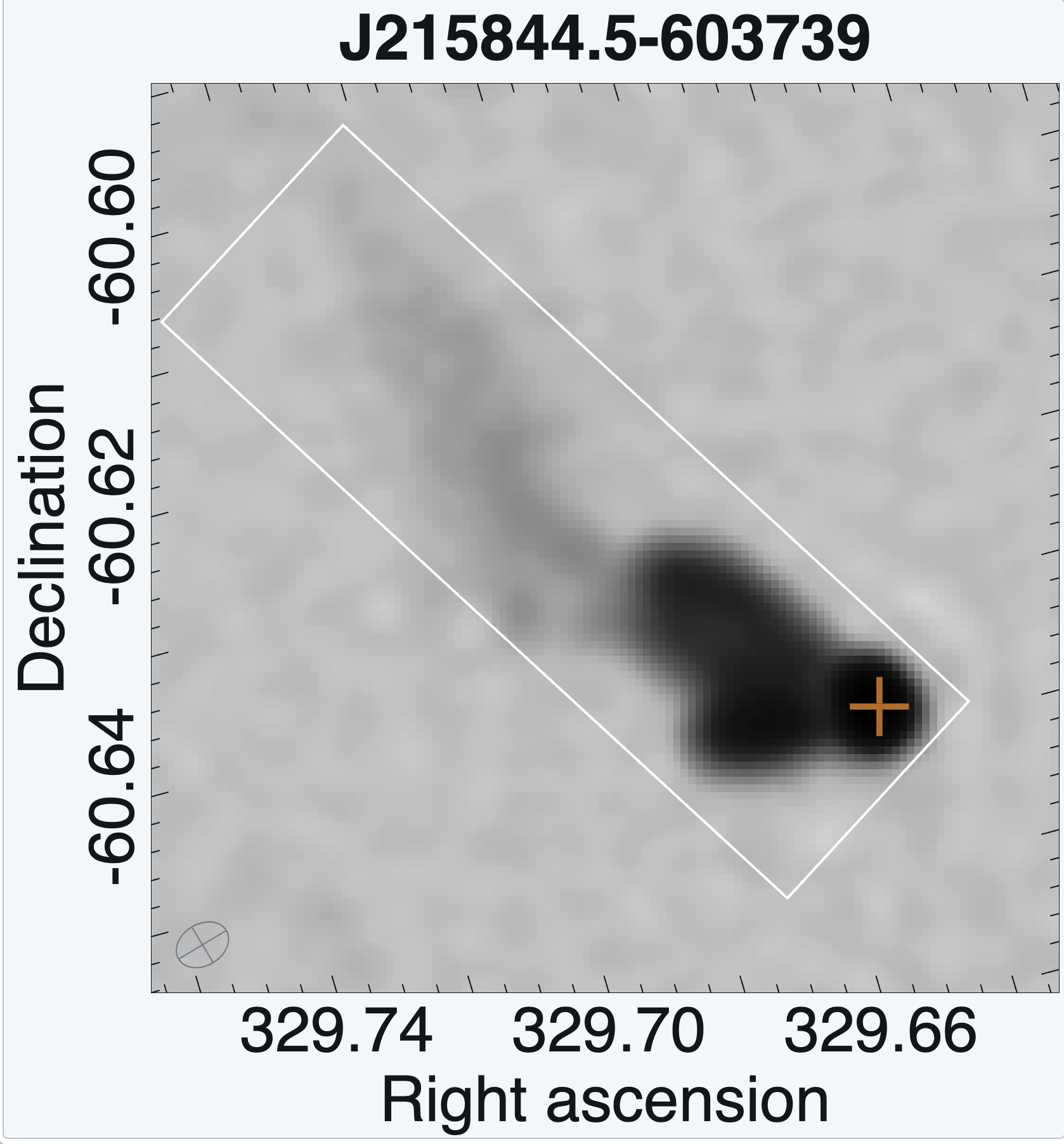}
  \caption{Examples of HT\ytag\ (head-tail) candidate sources, in which the two jets from the SMBH appear to have curved round to form a single tail.}
  \label{fig:ht}
\end{figure}

 HT sources are DRAGNs in which the two jets have been bent round to form a single tail. We show a sample of candidate HT\ytag\ sources in Figure \ref{fig:ht}. We call them ``candidate'' HT\ytag\ sources to acknowledge that the identification of these sources is rarely unambiguous, and some may be alternatively identified as one-sided sources (OSS\ytag). 

To confirm that they really are HT\ytag\ sources would require high-resolution observations to test whether, close to the head, there really are two jets which are then curved around.

\subsubsection{Bending as a function of FR1\ytag\ and FR2\ytag\ class}
If FR1 sources have lower jet powers than FR2 sources, then it might be expected that FR1 sources are more easily bent by the ram pressure of intergalactic gas, and so we might expect BT sources to be more prevalent amongst FR1 sources than FR2 sources. This is largely borne out by observations. For example, \citet{barnes25} find that in a sample of 72 sources {  drawn from the 3CRR survey \citep{laing83}},  23.6\% of 3CRR FR2 sources and 70.6\% of 3CRR FR1 sources are bent by more than 30\degr, so that, in that admittedly small sample, an FR1 source is almost four times more likely to be bent than an FR2.

In Table \ref{tab:bending} we show that, in our sample, while a greater fraction of FR1\ytag\ sources are bent than FR2\ytag, the fraction of bent FR1\ytag\ sources is less than twice that of FR2\ytag\ sources. Furthermore, the fraction of either that are bent is lower than in brighter samples.
This will be explored further in Paper III of this series.

\begin{table}
    \caption{Frequency of BT\ytag\ sources as a function of FR1\ytag/FR2\ytag\ classification.} 
    \label{tab:bending}
    \begin{tabular}{lcc}
      \toprule
      	&	\# FR1\ytag\ & \# FR2\ytag\ \\
      \midrule
      \# Sources classified as BT\ytag\ & 38 & 118 \\
      Total \# Sources  & 239 & 1410\\
      Fraction of sources that are BT\ytag\ & 15.5\% & 8.4\% \\
      \bottomrule
    \end{tabular}
    \medskip\\
\end{table}

\subsection{Double-double Radio Sources (DD\ytag)}
\label{sec:dd}

A ``double-double'' (or DD) radio source is defined by \citet{schoenmakers00} as ``a pair of double radio sources with a common centre''. Implicit in their definition is that the two sources must also have a common axis. They qualify this definition by adding ``Furthermore, the two lobes of the inner radio source must have a clearly extended, edge-brightened radio morphology'' which confines DD sources to FR2 sources. Here we relax this latter criterion to admit the possibility that DD sources may include FR1 sources. However, we must also be careful to distinguish a true DD source from a double source that simply has strong symmetrical knots in its jets, and so add the clause that each of the double radio sources must have the morphology of a normal FR radio source. 

\begin{figure}
 \includegraphics[width=4cm]{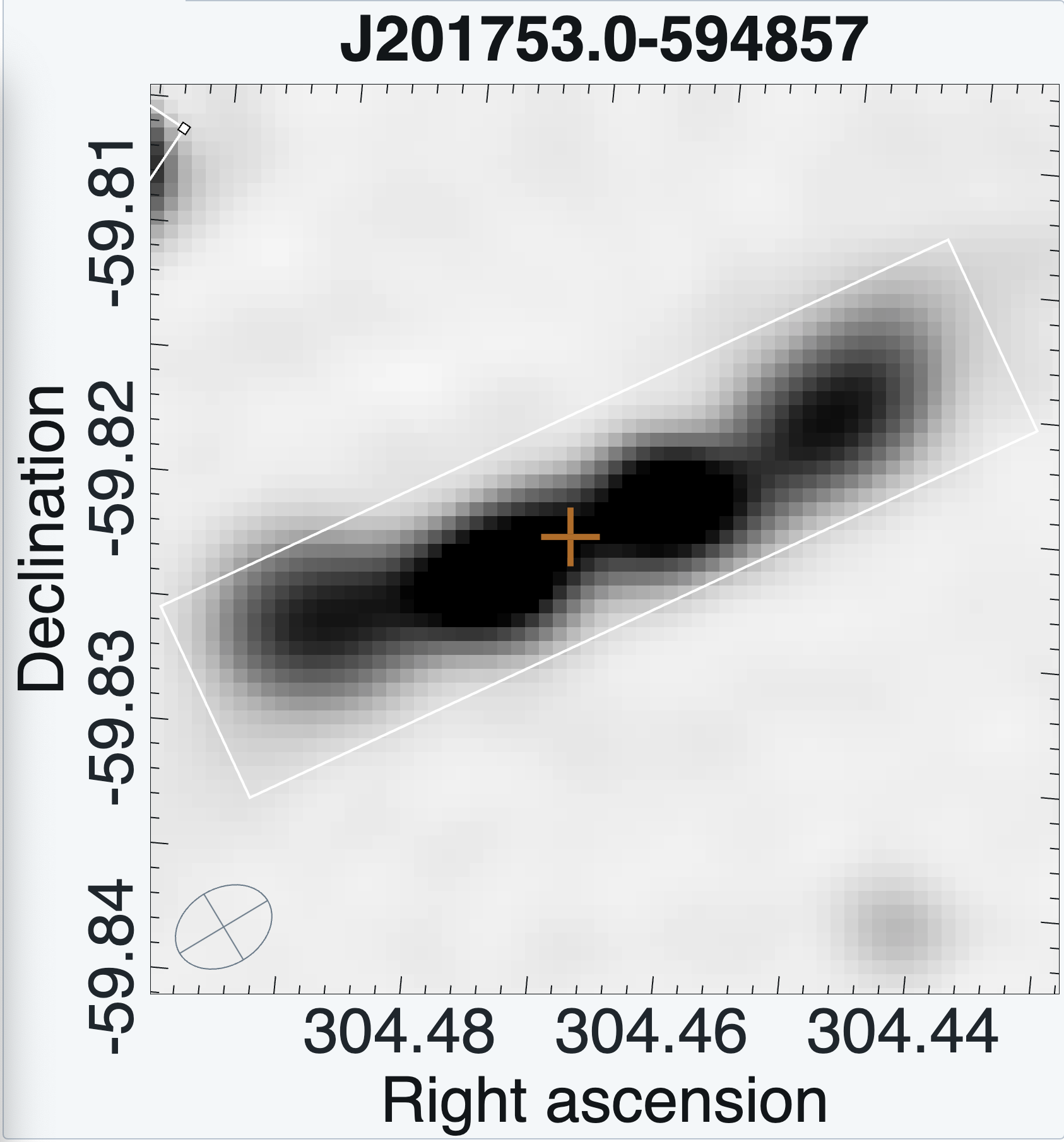}    
 \includegraphics[width=4cm]{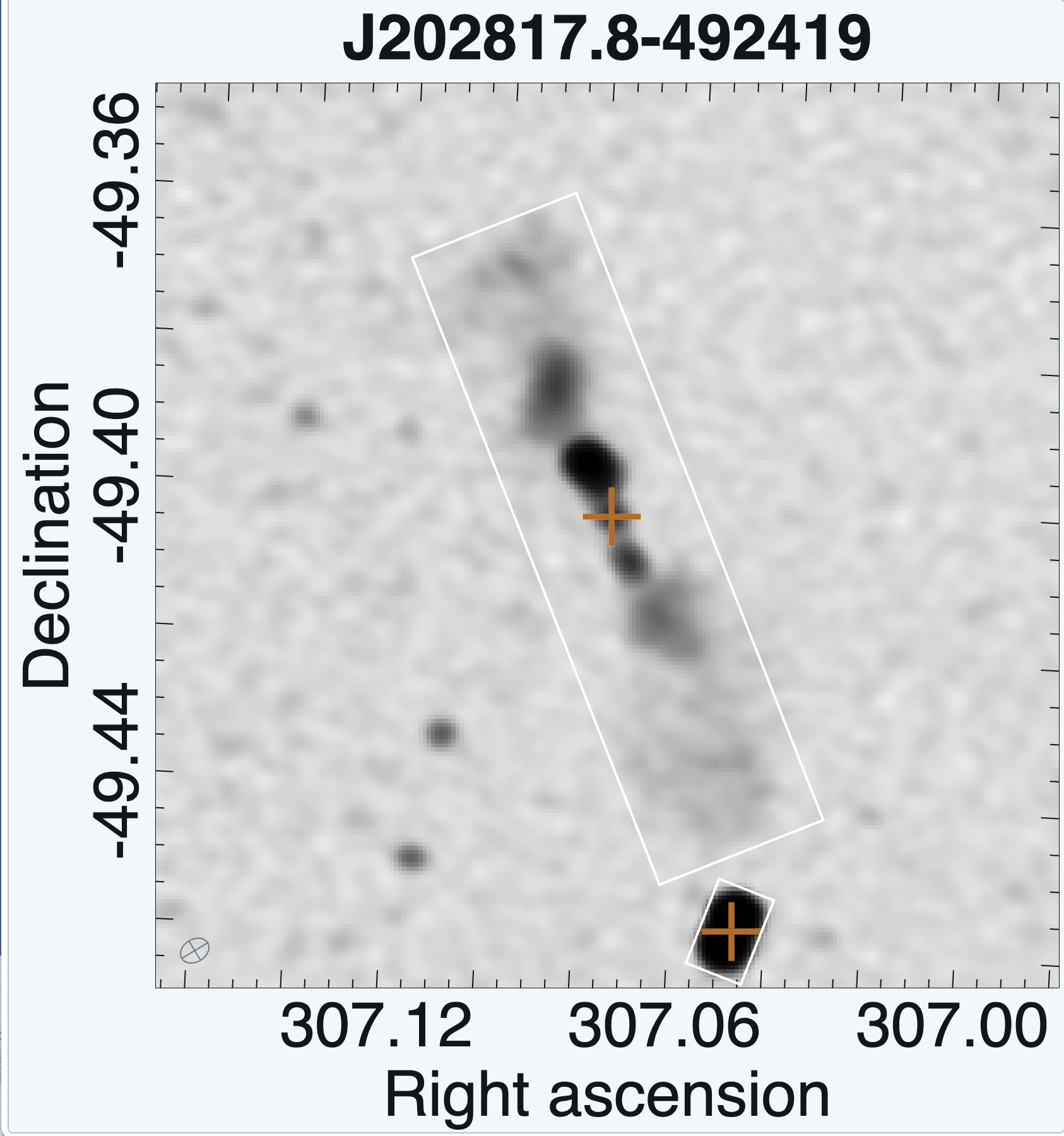}
 \includegraphics[width=4cm]{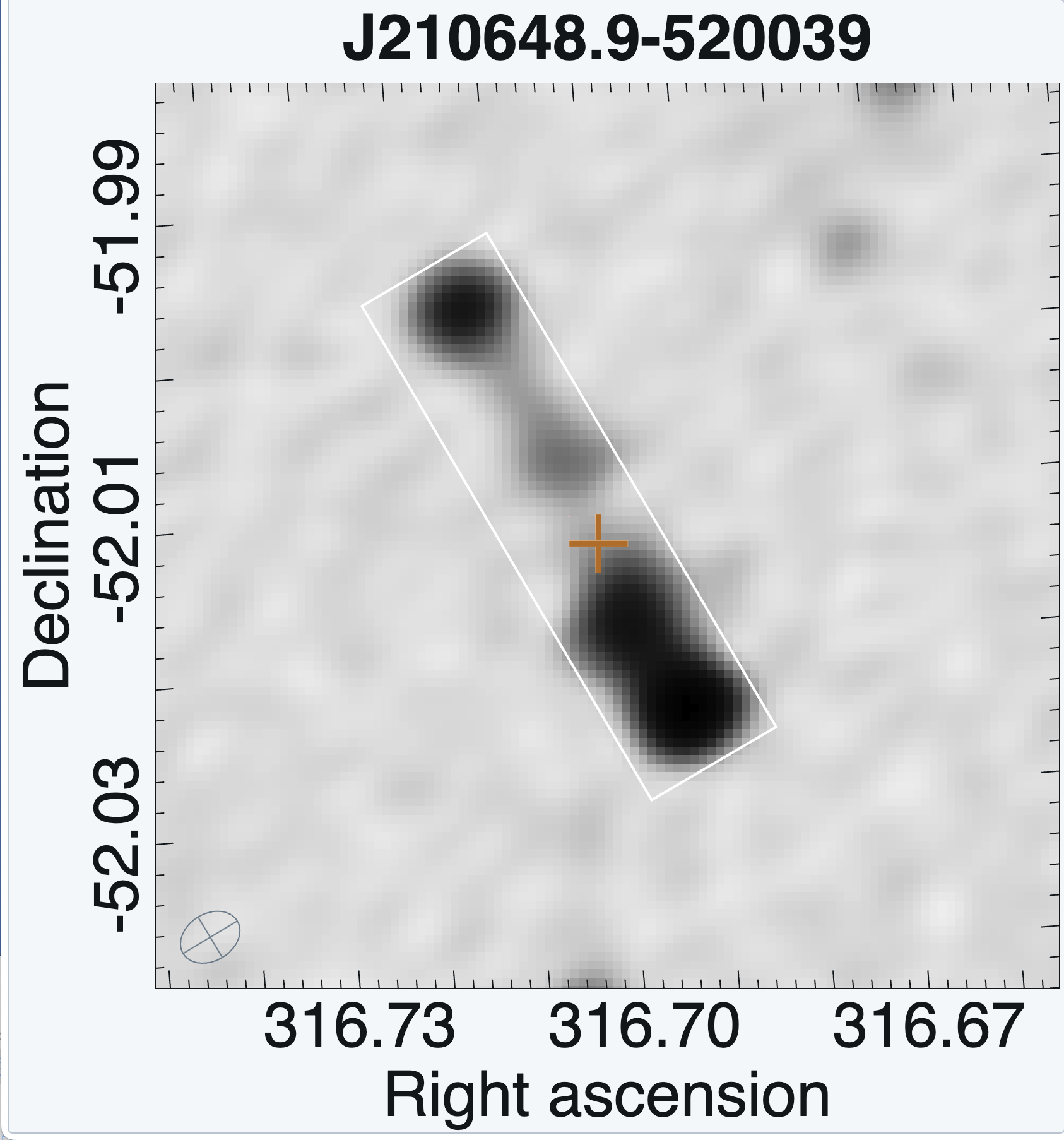}  
 \includegraphics[width=4cm]{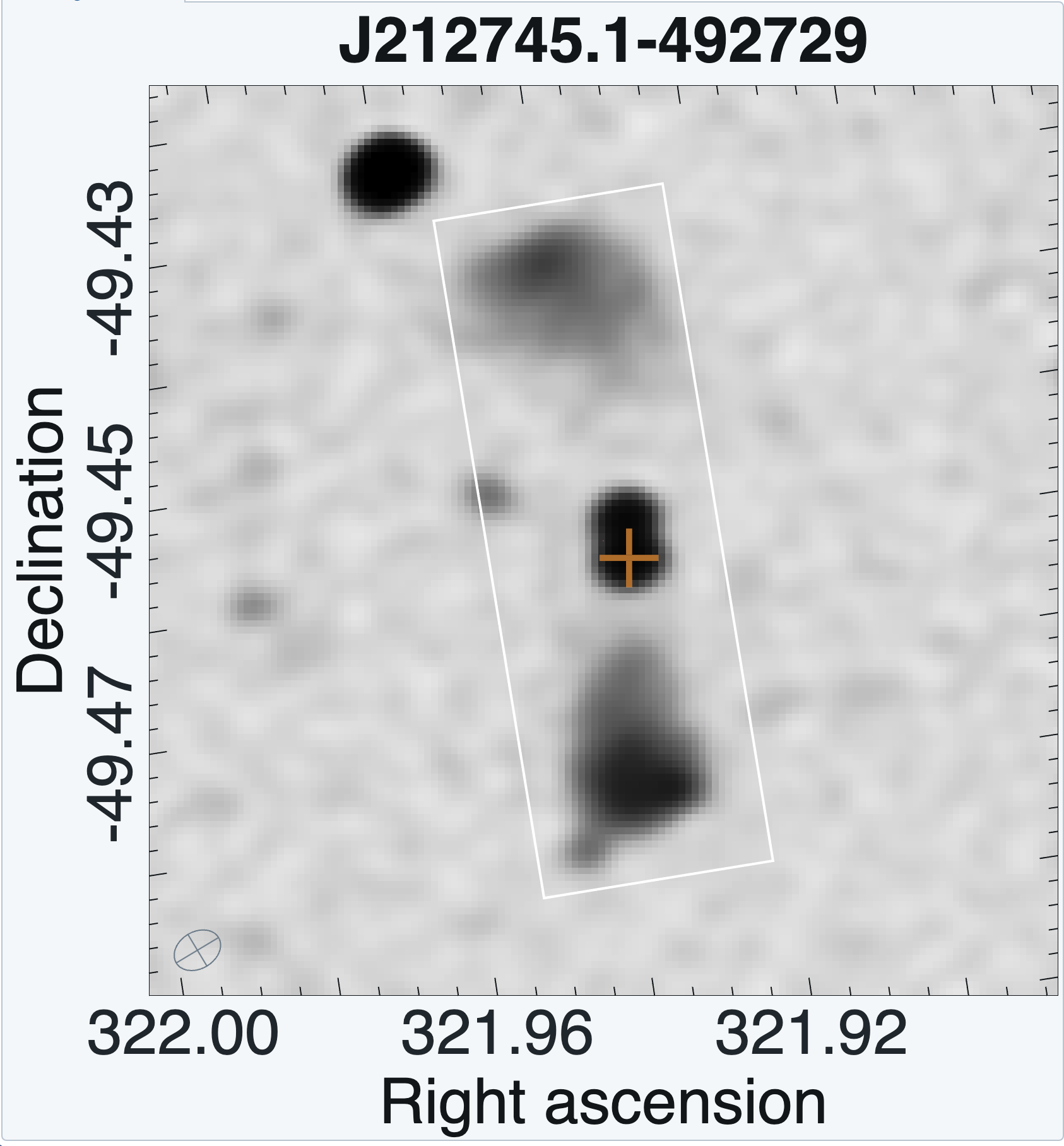}  
 \includegraphics[width=4cm]{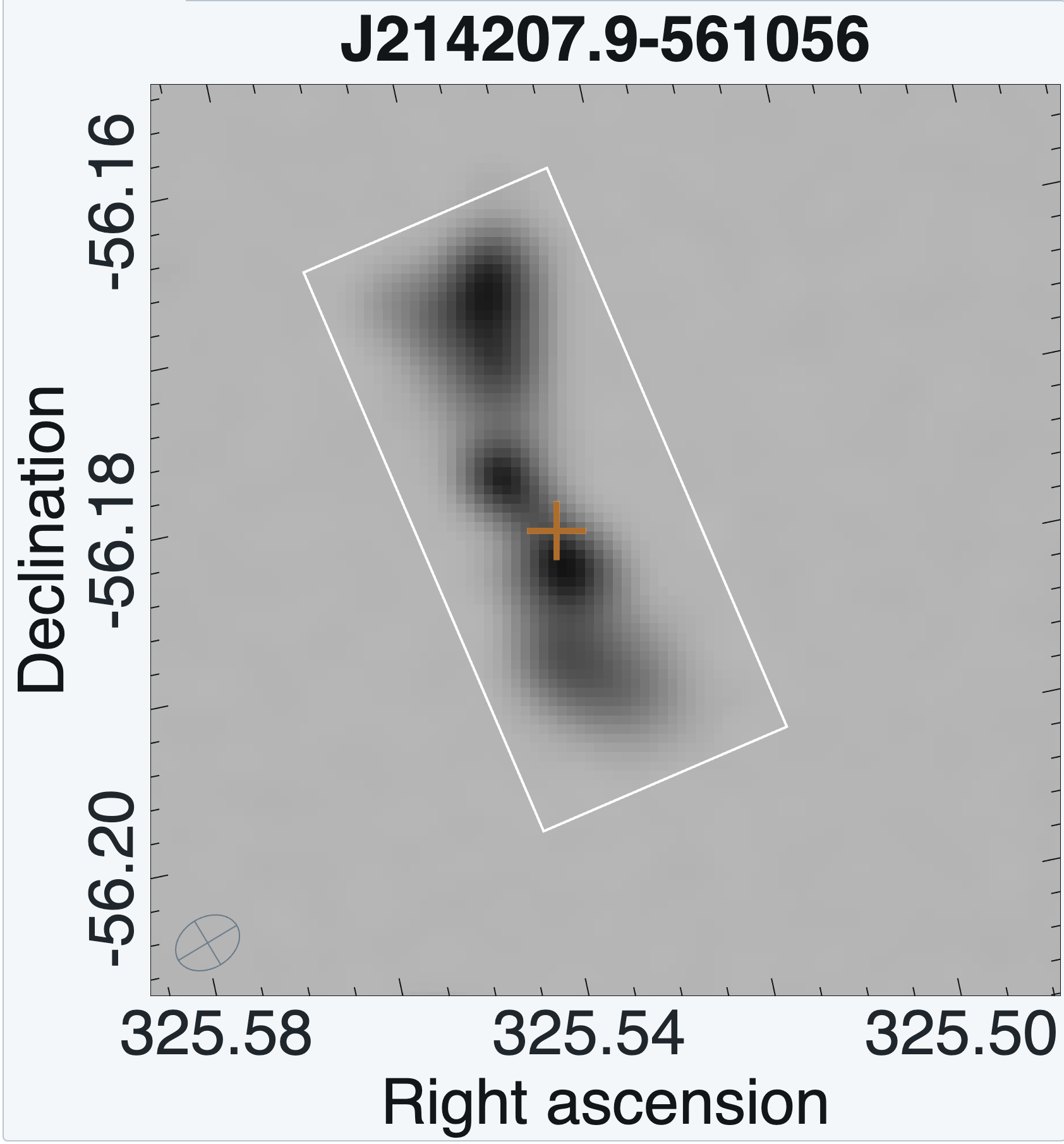}
 \includegraphics[width=4cm]{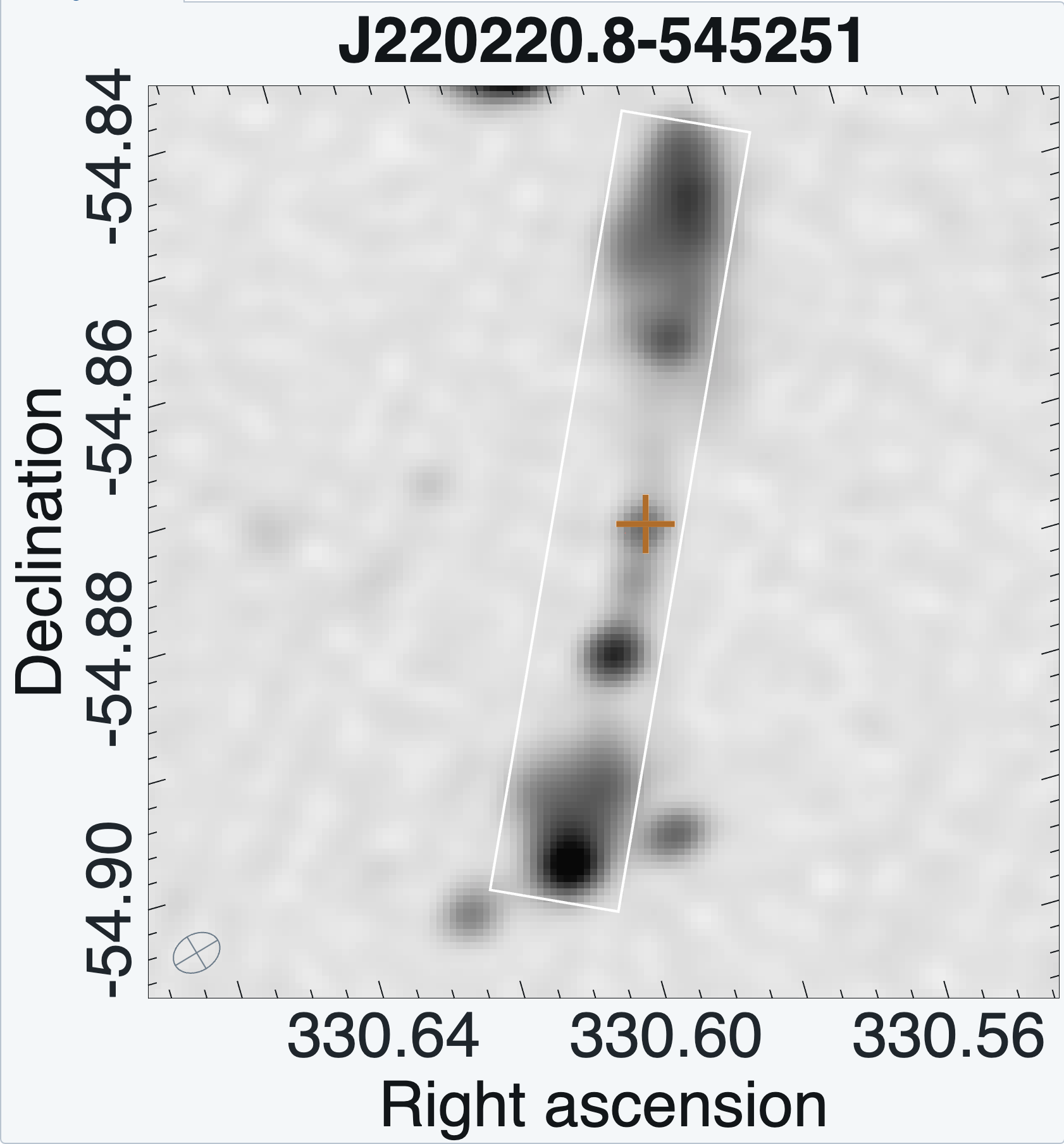} 
\caption{Examples of double-double (DD\ytag) sources, sometimes called ``restarted sources", in which there appears to be a pair of older diffuse lobes surrounding a younger inner double radio source. Sometimes there are more than two ``generations'' of radio lobes. }
  \label{fig:dd}
\end{figure}

We have also found examples, shown in Fig \ref{fig:dd}, in which there are three or more double radio sources. These were first identified by \citet{brocksopp07} and more have since been found \citep[e.g.][]{dabhade24}. Rather than creating more subclasses, we include these in the category of DD\ytag\ sources. We therefore end up with a revised definition of a double-double source, as ``two or more double radio sources with a common centre and a common axis, each of which has the morphology of a normal FR radio source.''

It is now widely accepted that DD sources are caused by an AGN first generating the outer lobes, then switching off and switching on again to generate the inner lobes. They are thus also often called ``restarted'' radio sources \citep{schoenmakers00, mahatma19, saikia06}. In this case we can, in principle, learn much about AGN duty cycles by studying DD sources. An alternative model is that the lobes result from jet-cloud interactions \citep{young25}.

In or catalogue we find \dd\ DD\ytag\ sources, a sample of which are shown in Fig. \ref{fig:dd}.
Of these \dd\ DD\ytag\ sources, 14 have two lobes on each side, one (J210542.7-571912) has three lobes on each side, two (J201404.8-522628, J202817.8-492419) have an unequal number of lobes on each side, and one (J214401.1-563712; see Figure \ref{fig:wtf}) has a ``trident'' morphology which we discuss further in Section \ref{sec:wtf}.

\subsection{Winged radio sources (XRG\ytag, ZRG\ytag, TRG\ytag)}
\label{sec:winged}

\begin{figure}
  \includegraphics[width=4cm]{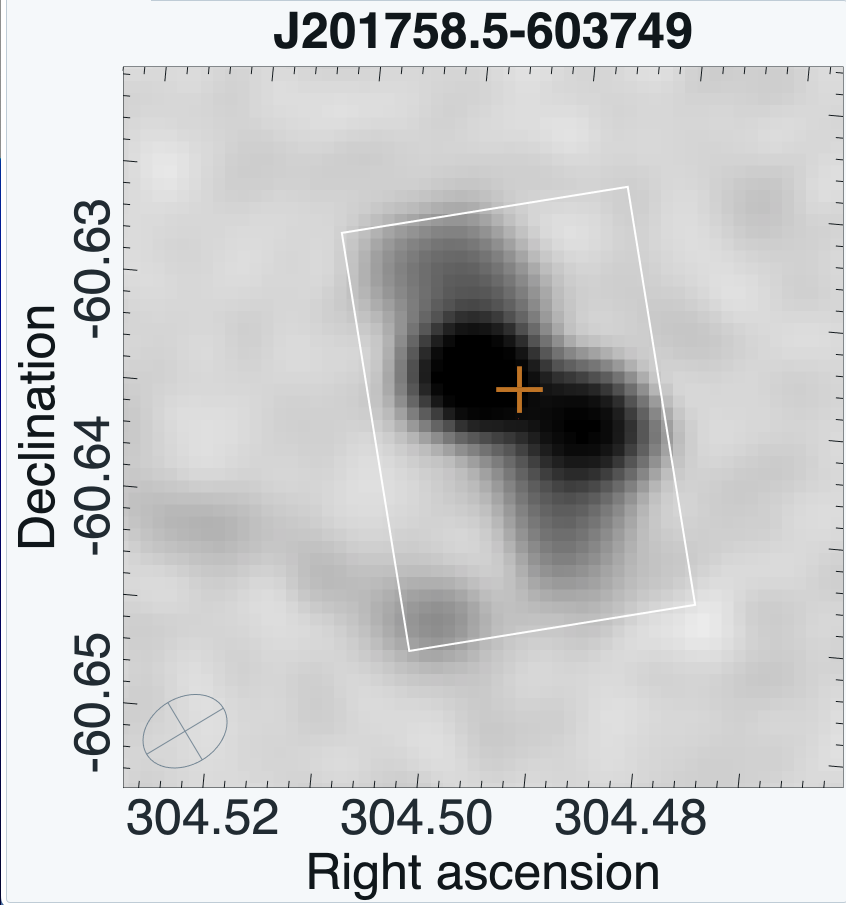} 
  \includegraphics[width=4cm]{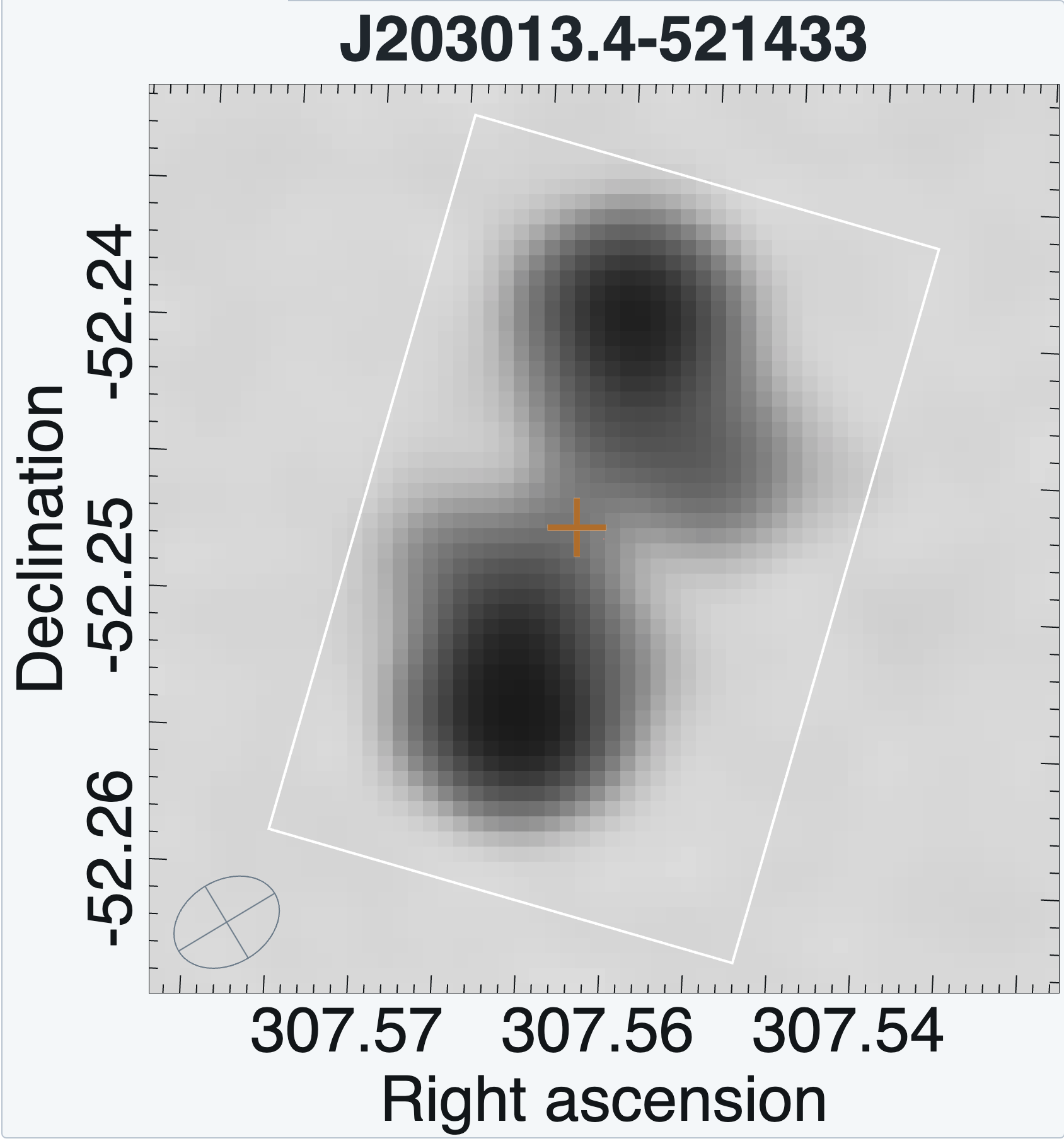} 
  \includegraphics[width=4cm]{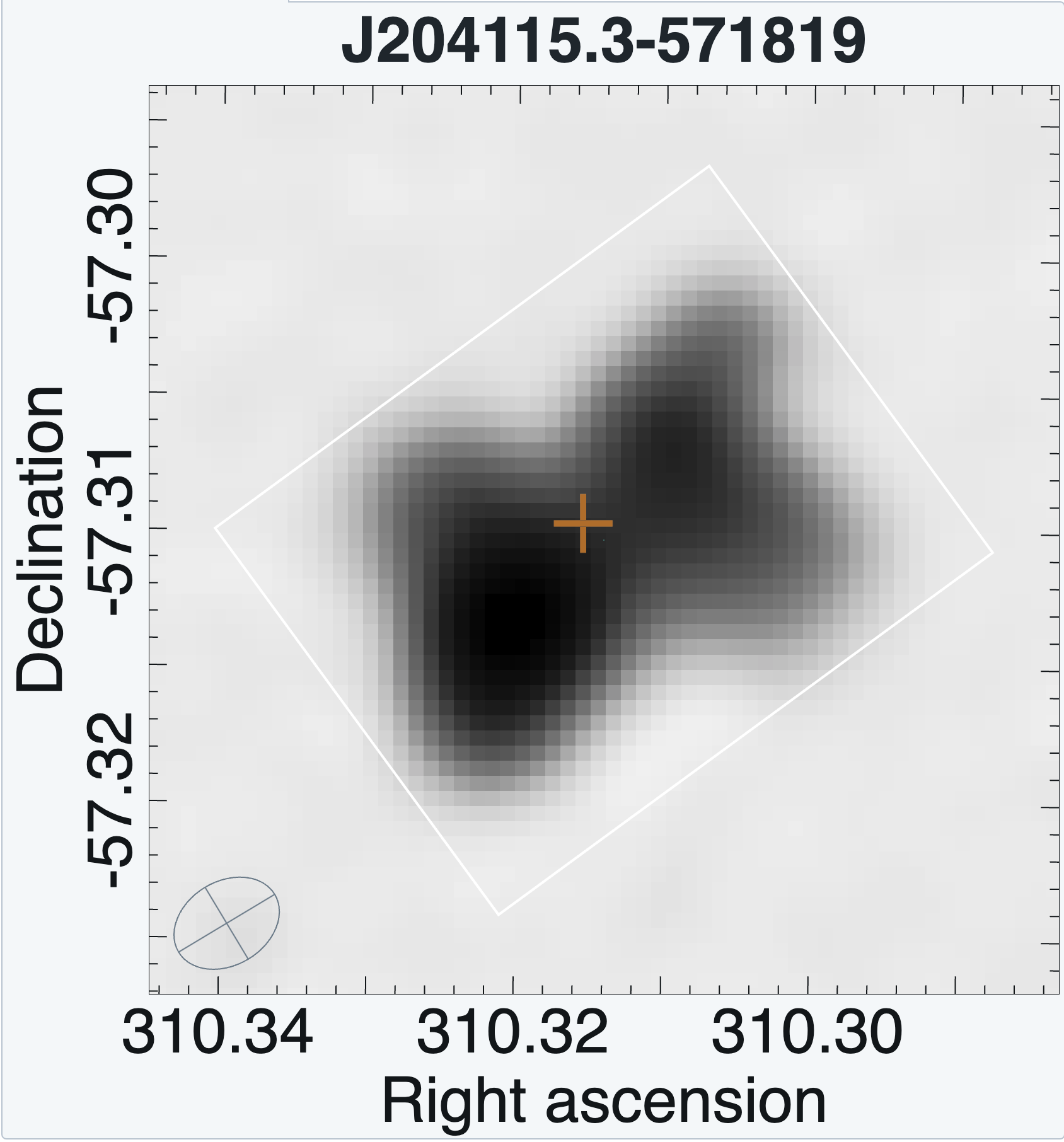} 
  \includegraphics[width=4cm]{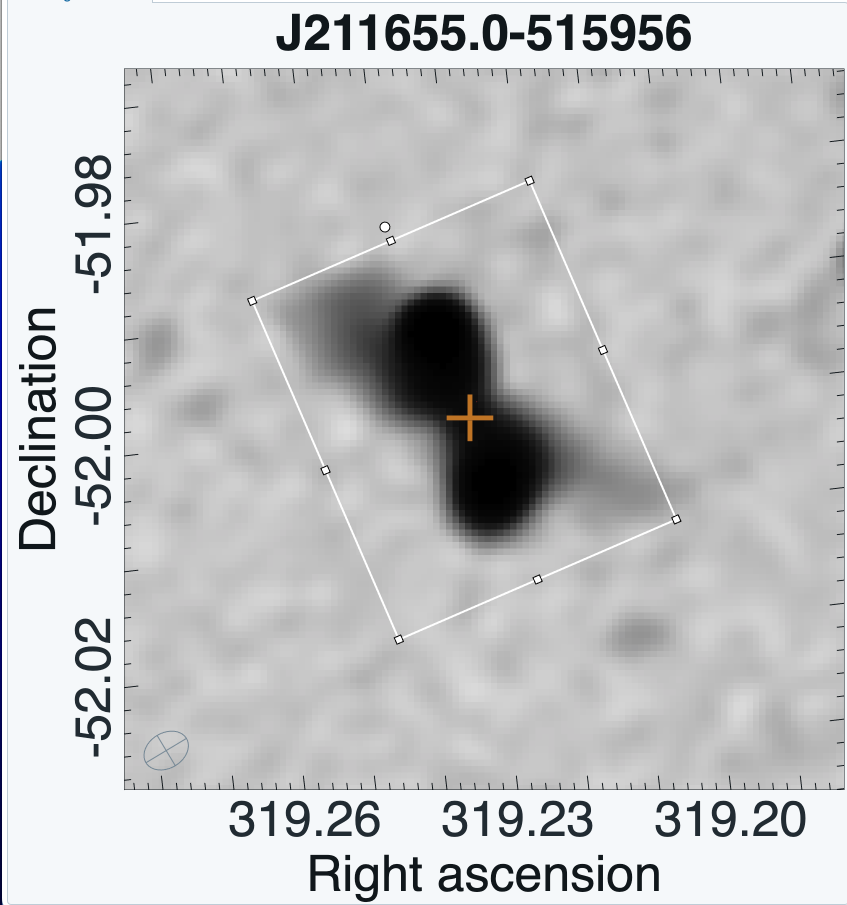} 
  \includegraphics[width=4cm]{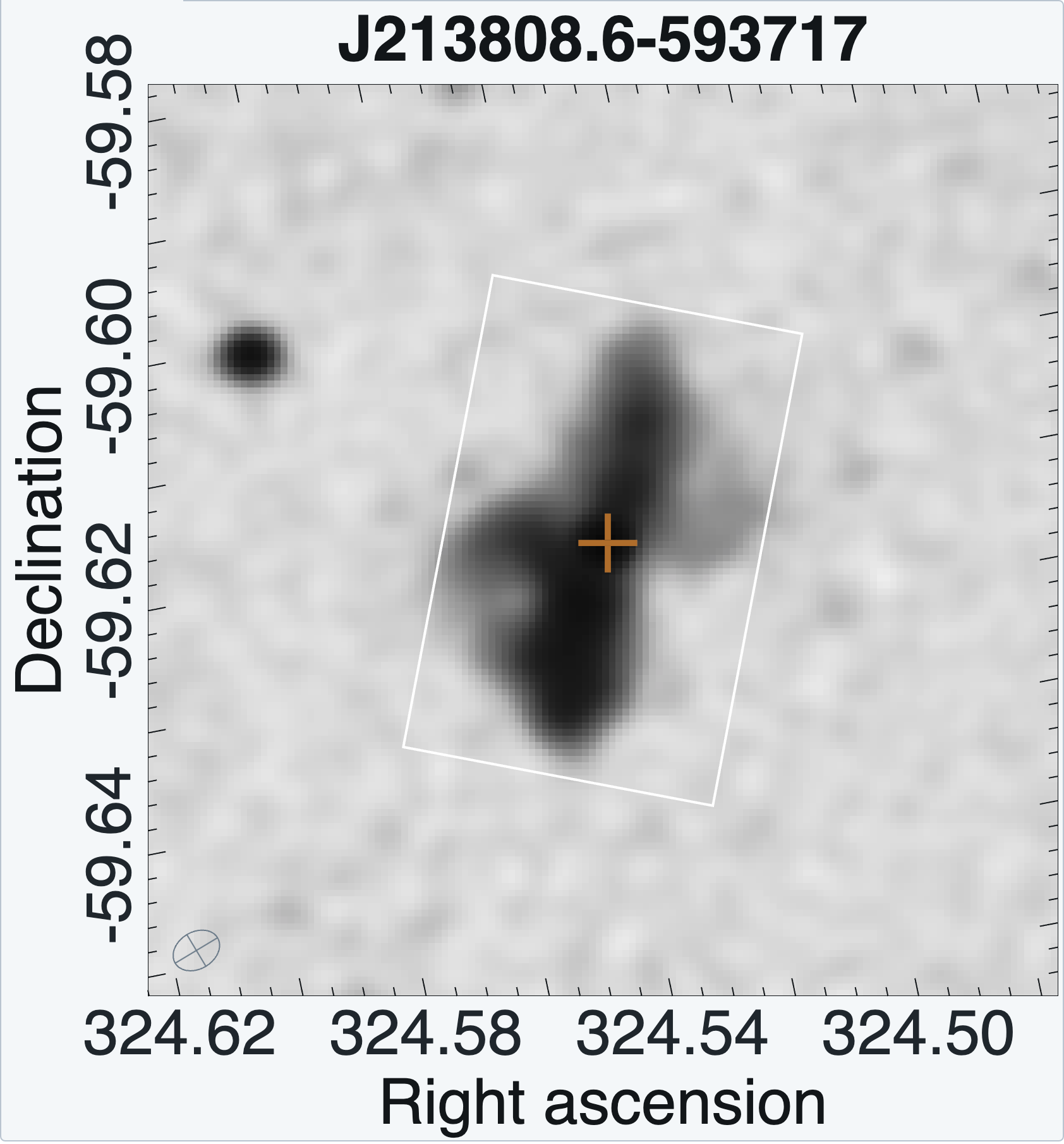}
  \includegraphics[width=4cm]{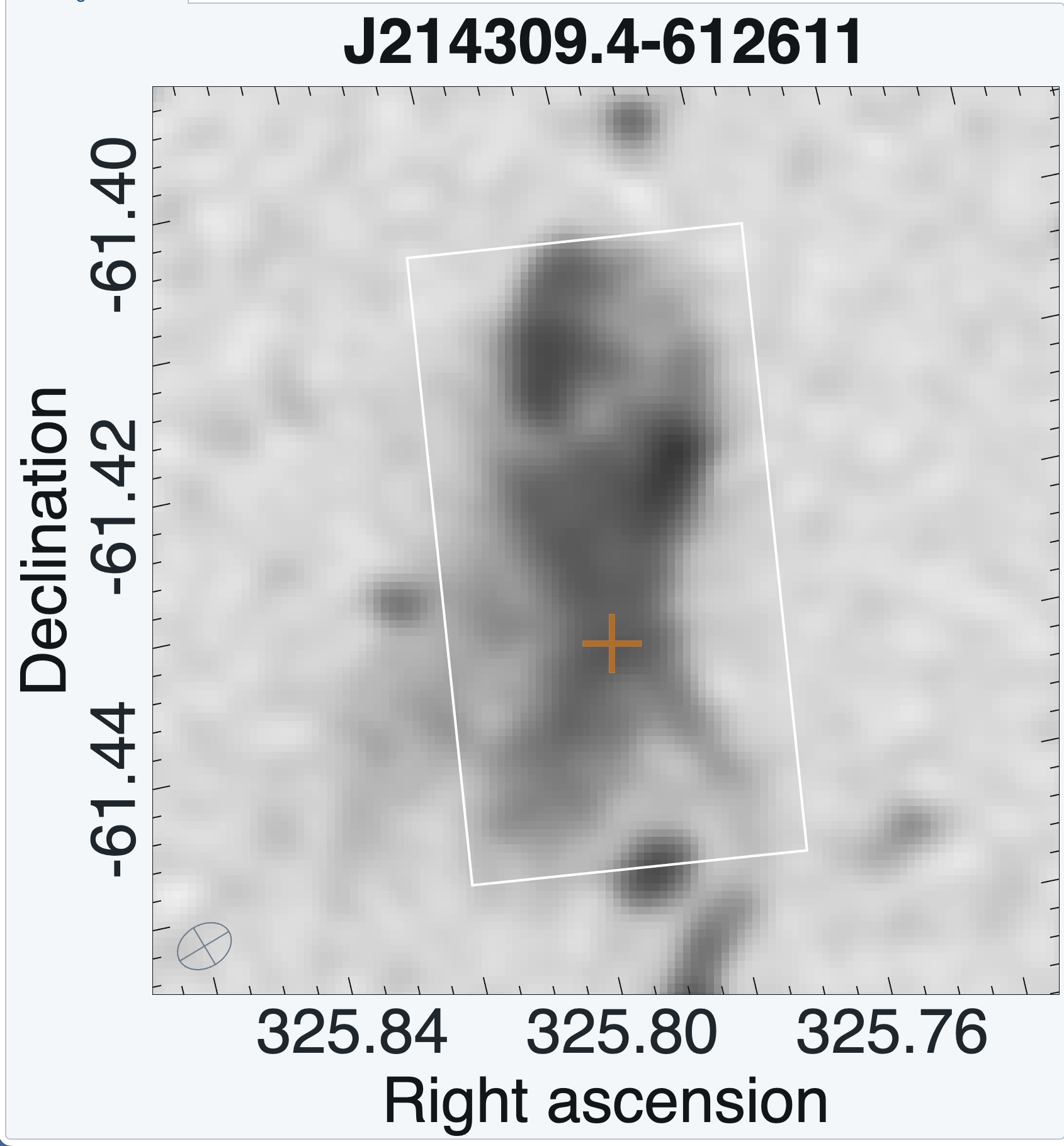}
  \caption{Examples of XRG\ytag\ (X-shaped radio galaxies). 
  }
    \label{fig:xsh}
\end{figure}

Winged radio sources are sources which have one or more linear structures extending from the side of the major axis of the main radio source. Here we identify three types of winged sources:
\begin{itemize}
    \item X-shaped radio galaxies, or XRG\ytag, which have two linear structures emanating from the source to give the radio source the appearance of the letter ``X'', as shown in Figure \ref{fig:xsh}.
    \item S/Z-shaped radio galaxies, or ZRG\ytag,  which have two linear structures emanating from opposite sides of the lobes, giving the radio source the appearance of the letter ``S'' or ``Z'', as shown in Figure \ref{fig:ssh}
    \item T-shaped radio galaxies, or TRG\ytag, which have one or more linear structures emanating from the side of the source, giving the radio source the appearance of the letter ``T'',``L'', or ``M''as shown in Figure \ref{fig:tsh}
\end{itemize}

\subsubsection{X-shaped radio sources (XRG\ytag)}
\label{sec:xsh}
The best-studied of the winged sources are the XRGs \citep{leahy84, cheung07,joshi19,yang19, bera20}, although that term is often broadened to include other types of winged sources. They were often explained in terms of either (a)~a binary SMBH, each of which produced a pair of jets \citep[e.g.][]{lal05}, (b)~precession or a change in the SMBH spin axis \citep[e.g.][]{ekers78}, or (c)~a backflow from the 
main jets, which is redirected by the gas of the host galaxy \citep[e.g.][]{leahy84, saripalli09}. In our sample of DRAGNs, we find \xsh\ XRG\ytag's.

A MeerKAT observation of the XRG PKS 2014-55 by \citet{cotton20} shows clearly that, in this source at least, the X-shape is due to backflows from the main lobe. There is currently no clear evidence for an alternative mechanism, so it is possible that all X-shaped radio sources share this mechanism. It is also possible that this same mechanism explains the other morphologies of winged sources.

\subsubsection{S-and Z-shaped radio sources (ZRG\ytag) }
\label{sec:ssh}

The first known S-shaped source, 3C272.1, was described by \citet{riley72} as having a spiral structure. Many more have been found since, described as having ``rotational symmetry'' or either S-shaped or Z-shaped depending on the viewpoint of the observer \citep[e.g.][]{ekers78, miley80, taylor98, rubinur17, bera20}.
Here we adopt the nomenclature of \citet{bera20} who label them as ZRG.\\

ZRG sources have linear structures of radio emission from the sides of the two lobes, in opposite directions. Thus in some ways they resemble XRG, except that the linear structures emanate from the lobes rather than the core. It is therefore possible that the morphology is driven by a backflow. 

These interesting sources raise the obvious question of why the outflow comes from opposite sides from the two lobes. One possible answer is that the source jet axis has rotated relative to the intergalactic medium, leading to a lower density on the trailing side of the lobe than the leading side. Another possible answer is that it is purely a selection effect, and that emission can come randomly from either side of the lobes. If it comes from the same side, it is classified as a bent-tail source or a T-shaped source (see below), but if the emission comes from opposite sides it is called a S- or Z-shaped source. Further study is needed to determine which if either of these answers is correct.

\begin{figure}
  \includegraphics[width=4cm]{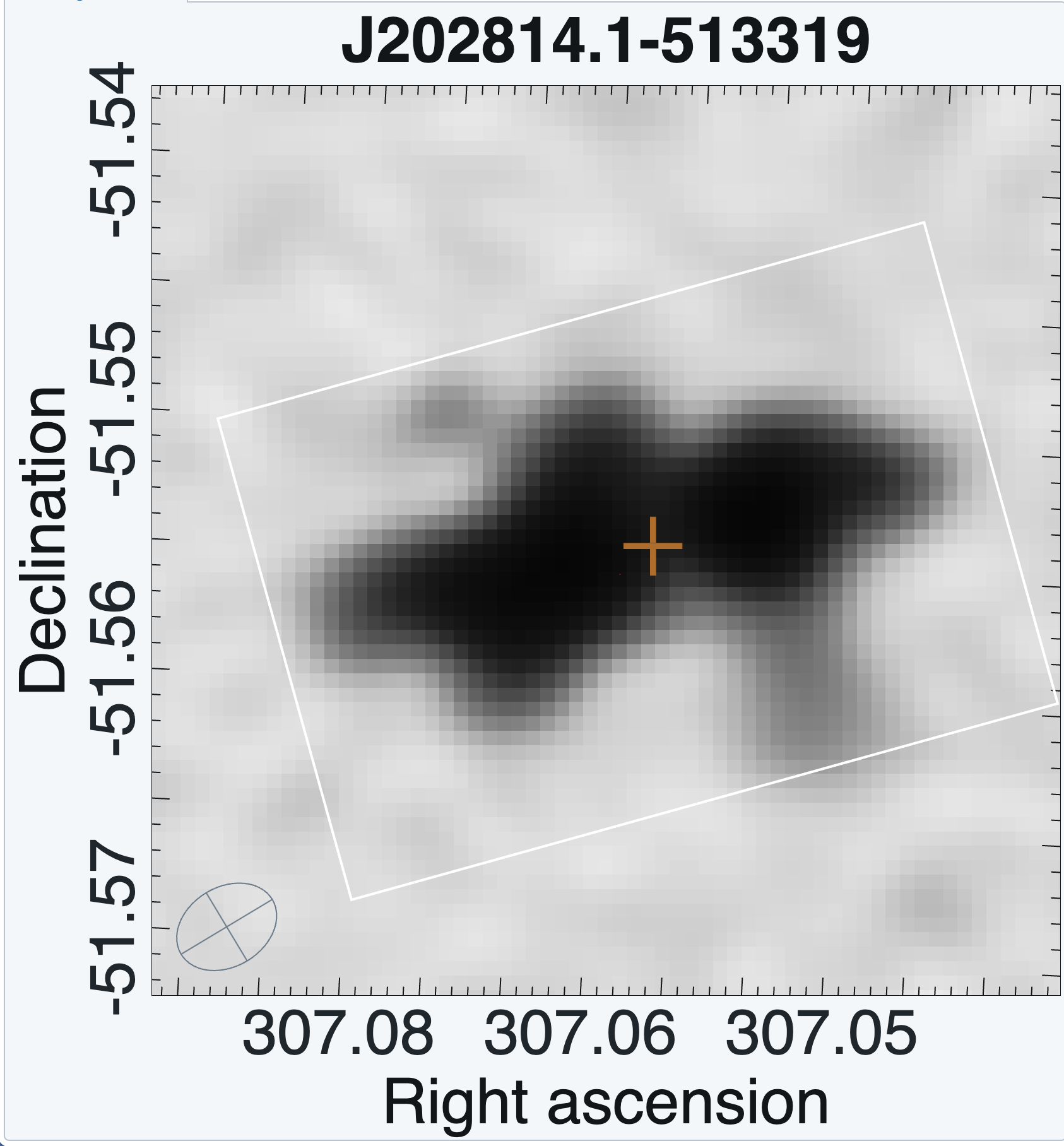}
  \includegraphics[width=4cm]{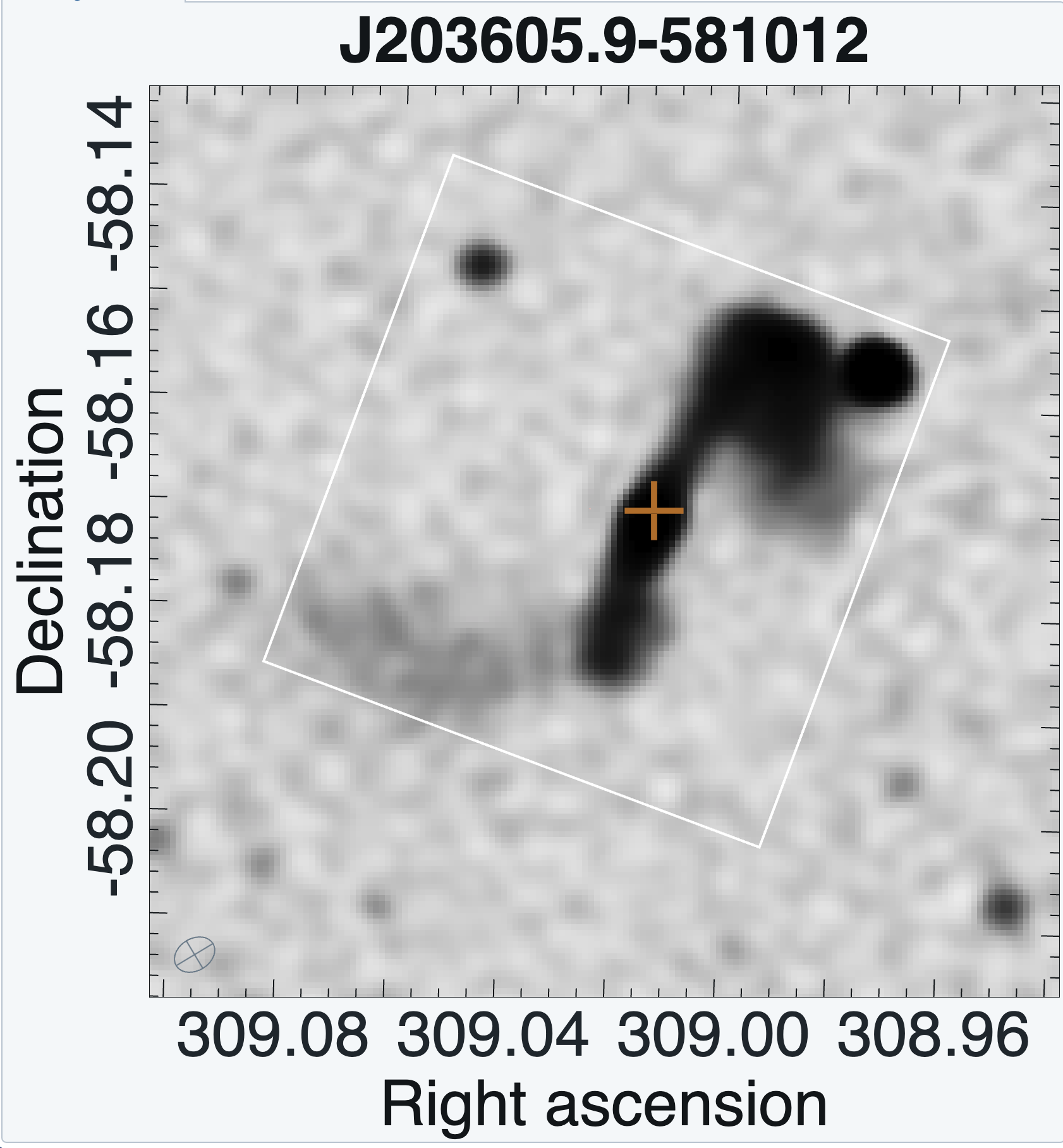}
  \includegraphics[width=4cm]{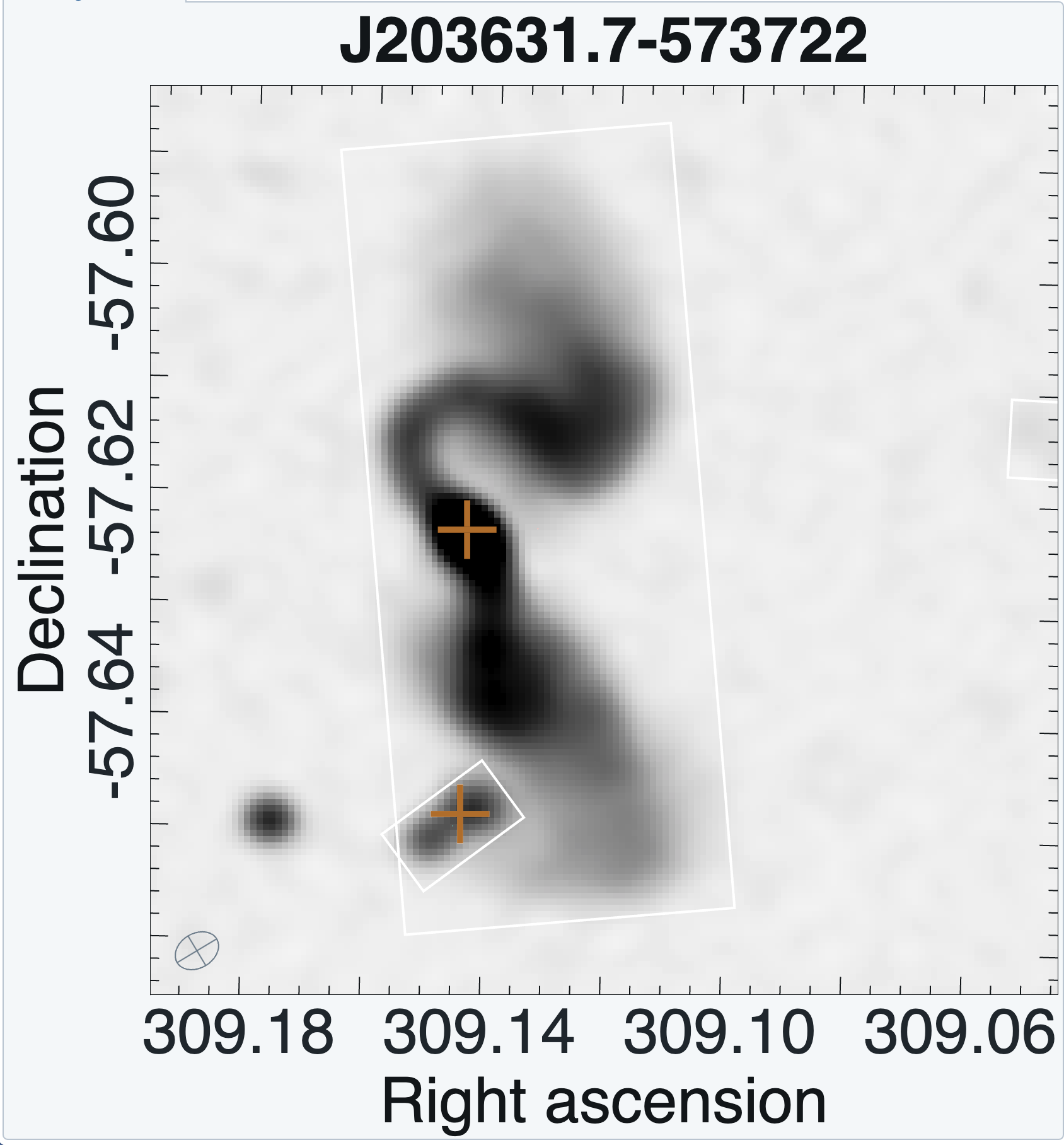} 
  \includegraphics[width=4cm]{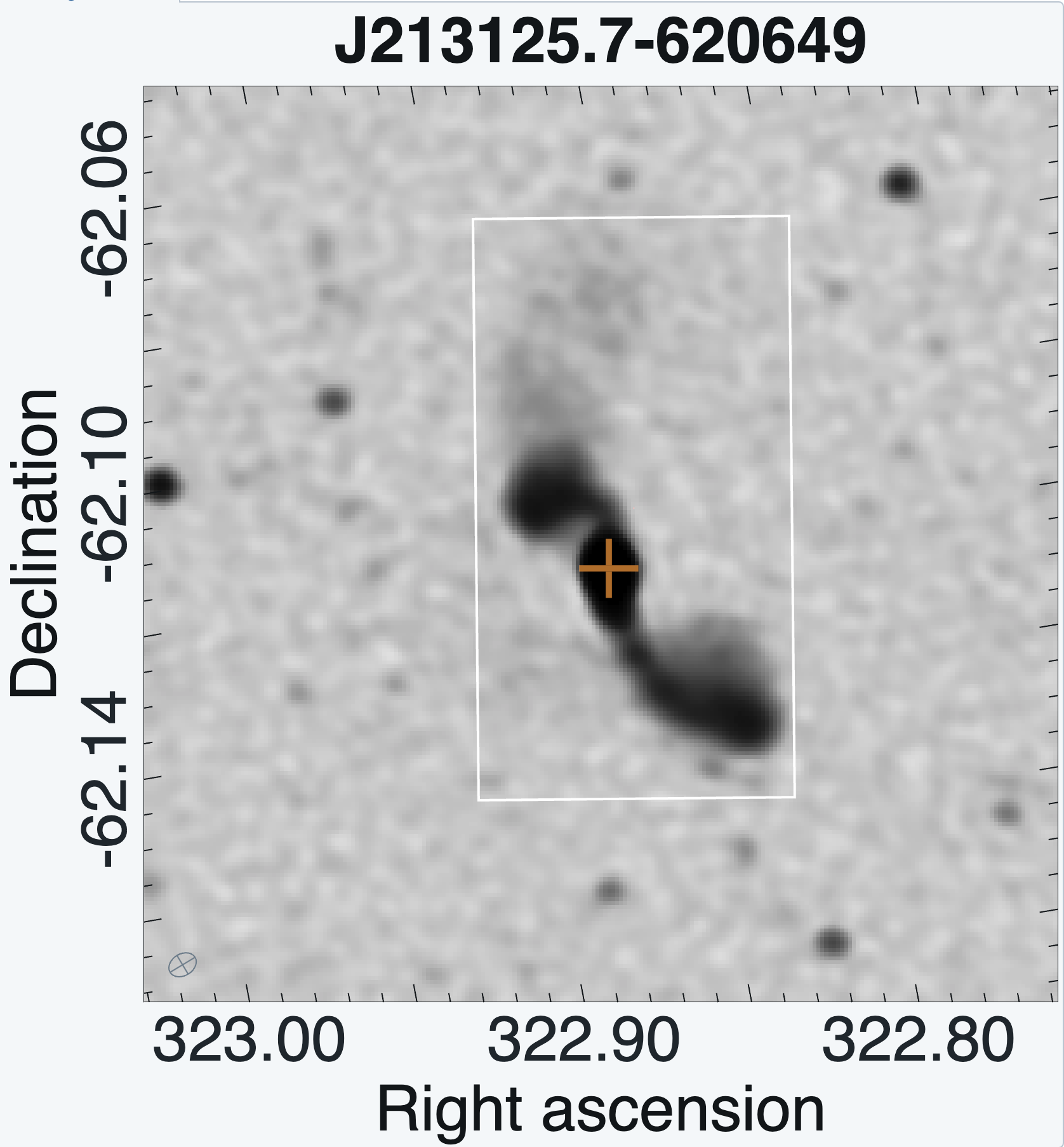}
  \includegraphics[width=4cm]{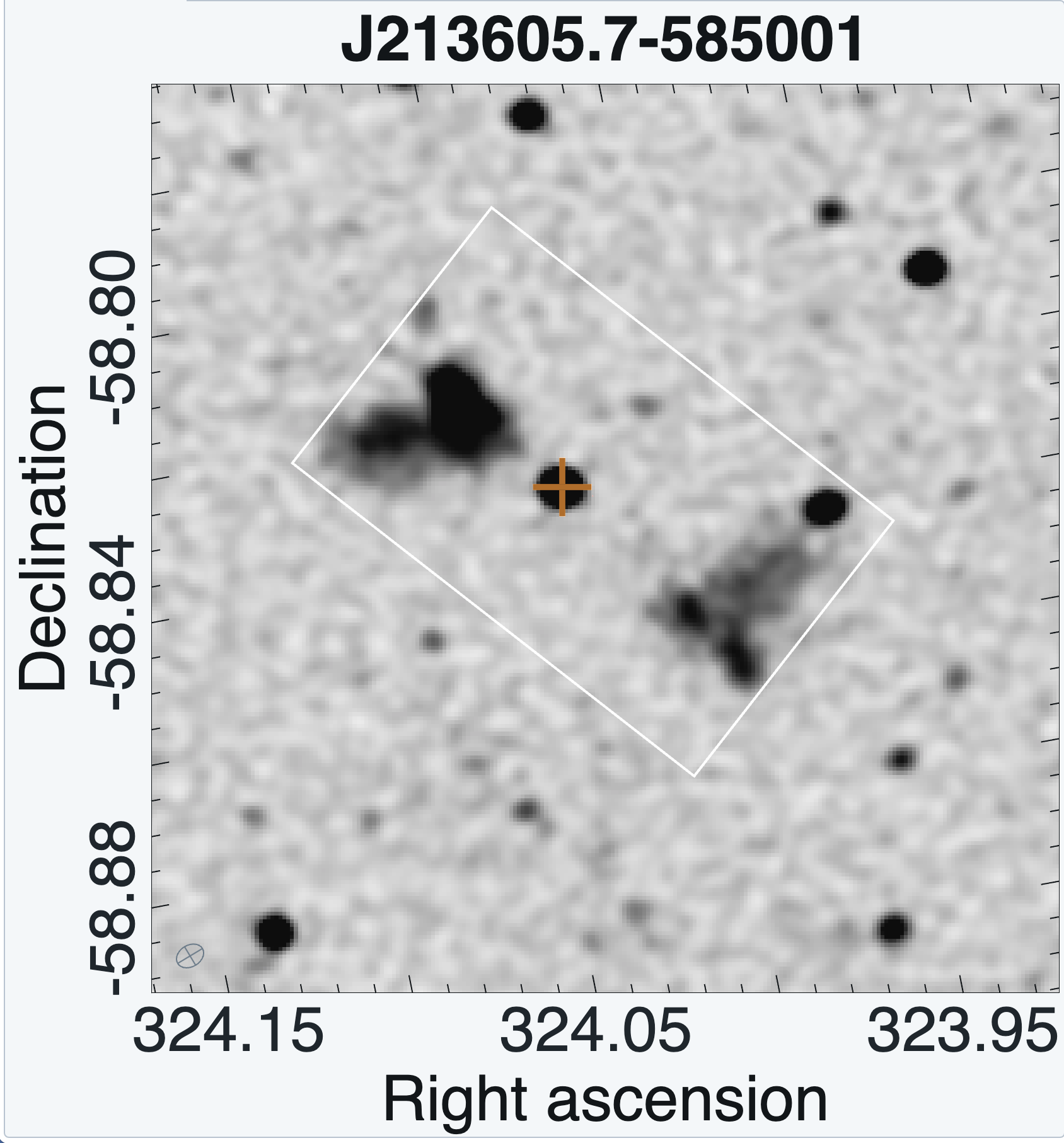}
  \includegraphics[width=4cm]{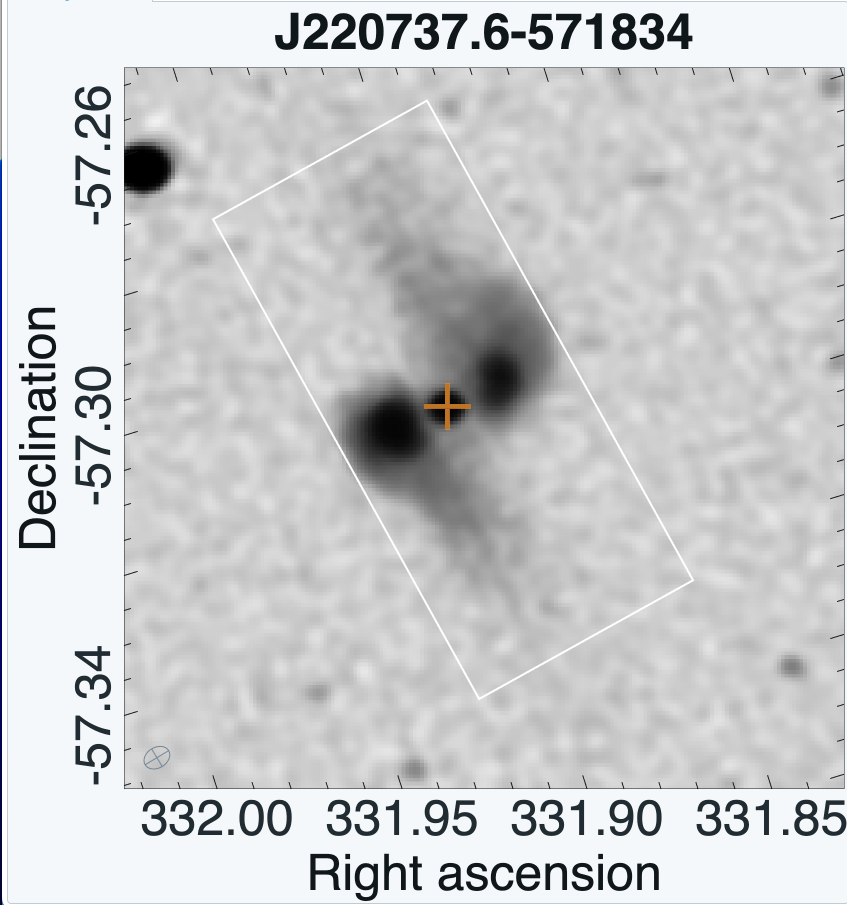}
  \caption{Examples of ZRG\ytag\ (S- and Z- shaped radio galaxies)}
    \label{fig:ssh}
\end{figure}

We show several examples in Figure \ref{fig:ssh} and note that one example (J203631.7-573722) is not only Z-shaped because of the outflow from the two lobes, but is also a sinuous bent-tail source which may illustrate the motion of the source relative to the intergalactic medium, or may be due to a projection effect.

\subsubsection{T-shaped Radio galaxies (TRG\ytag)}
\label{sec:tsh}
\begin{figure}
  \includegraphics[width=4cm]{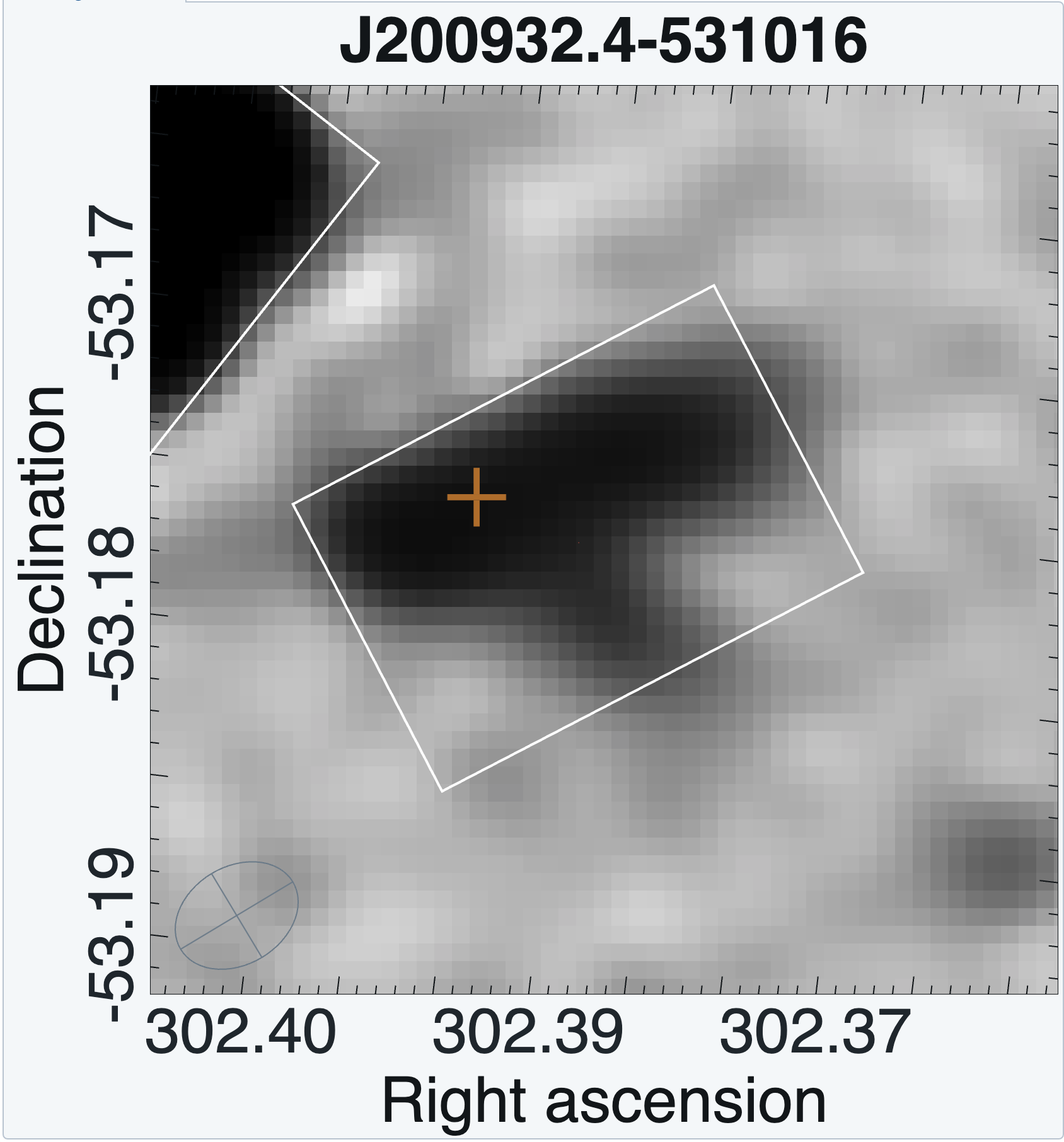} 
  \includegraphics[width=4cm]{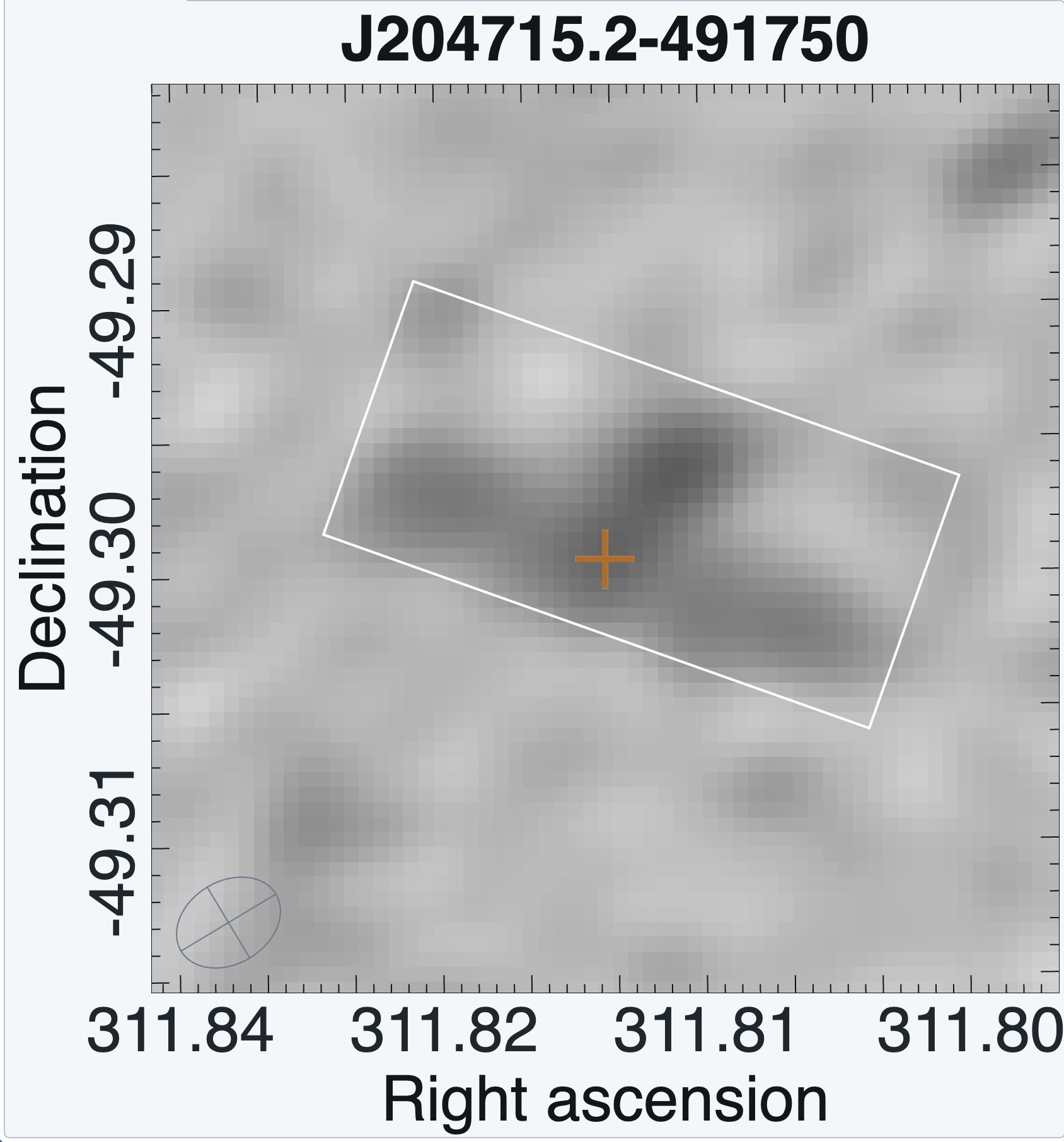}   
  \includegraphics[width=4cm]{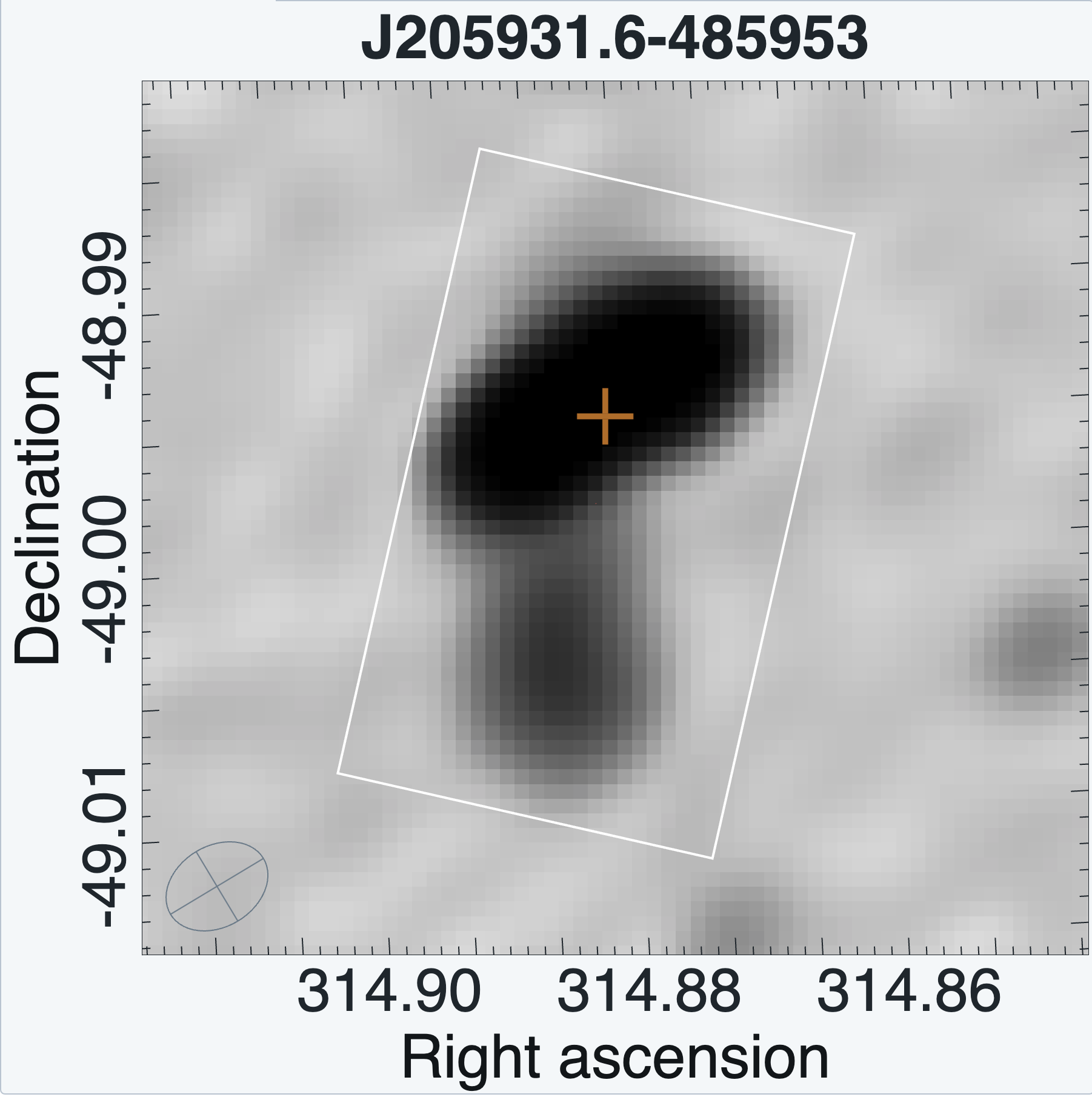} 
  \includegraphics[width=4cm]{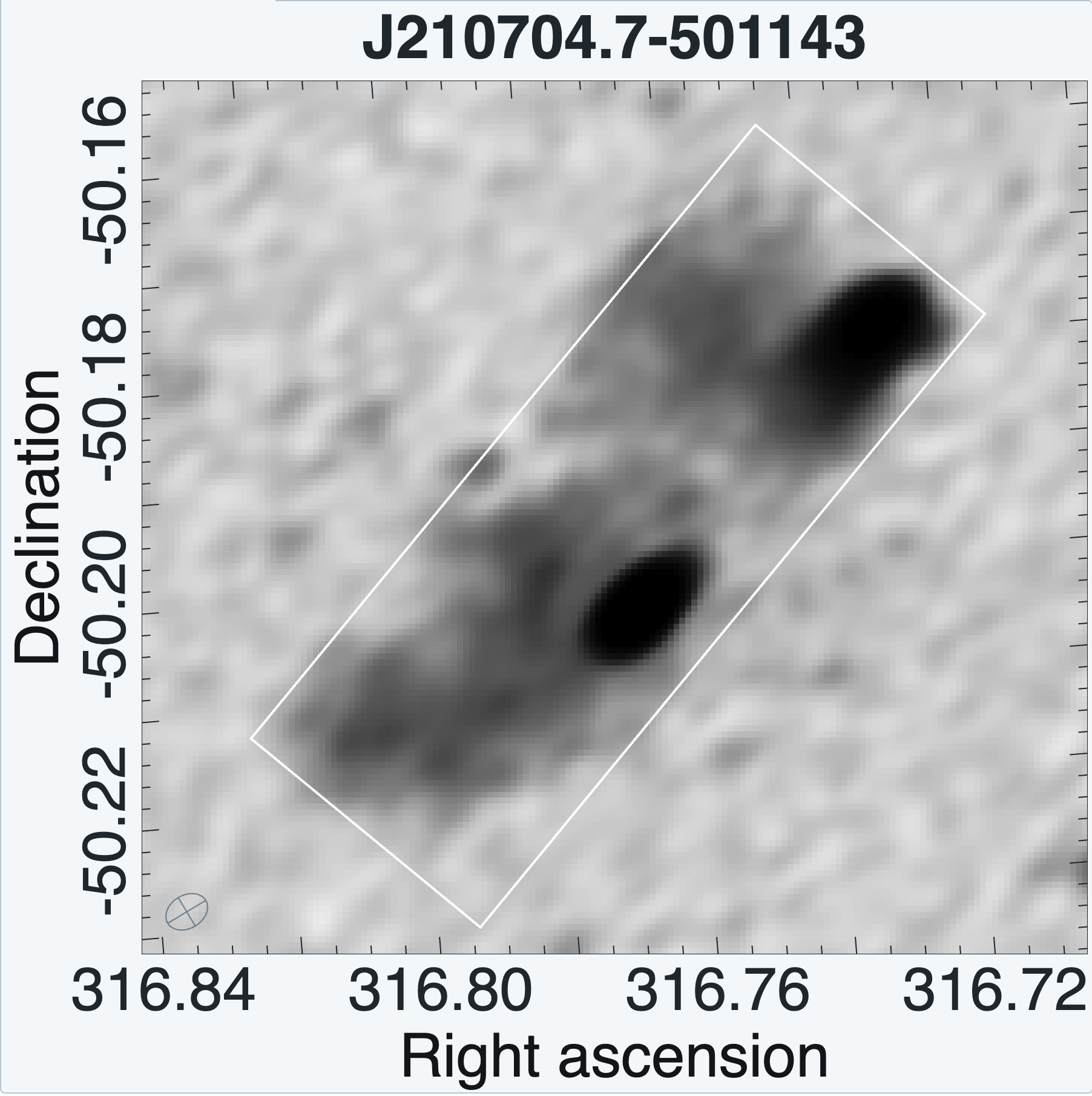}  
  \includegraphics[width=4cm]{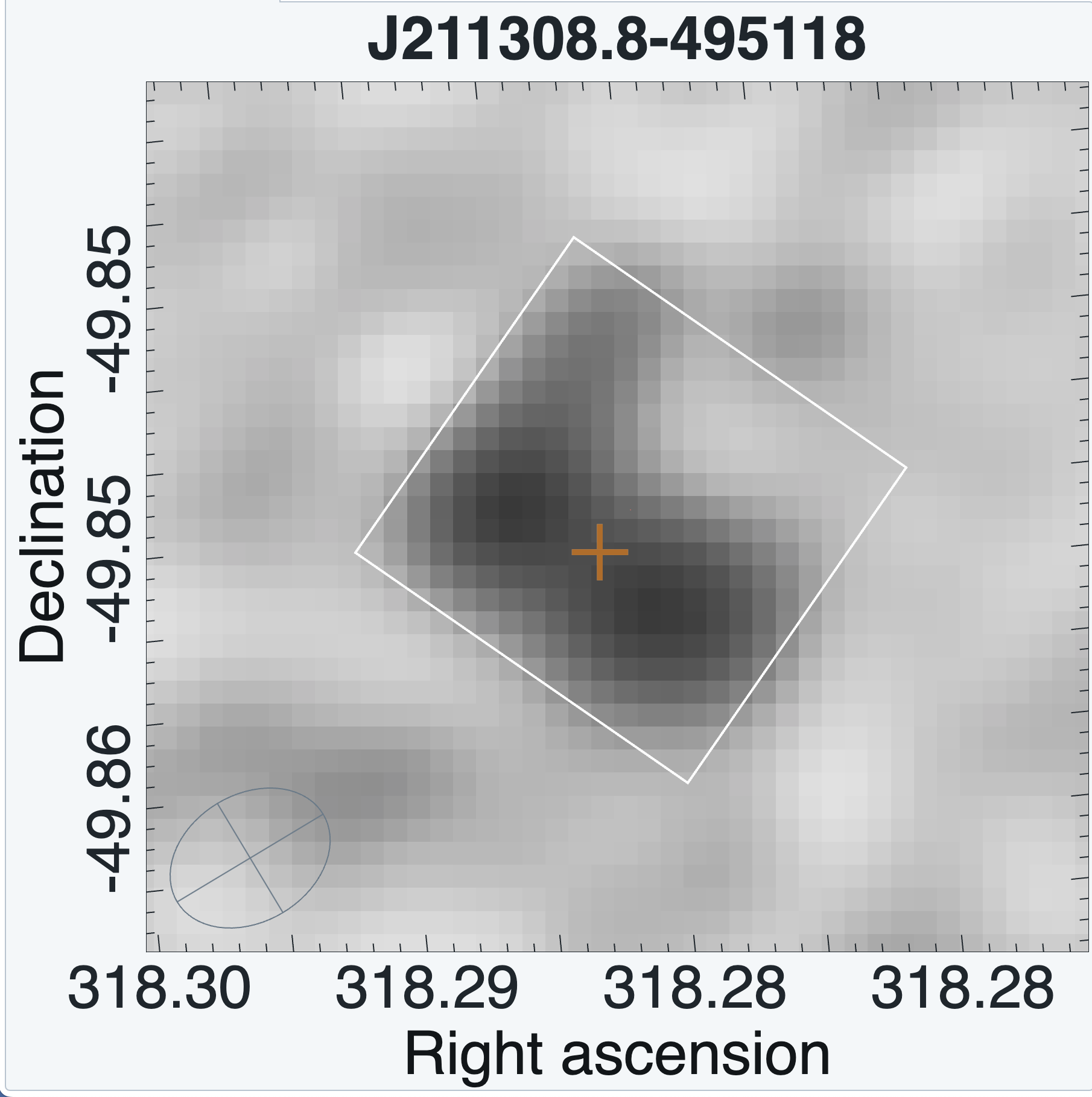} 
  \includegraphics[width=4cm]{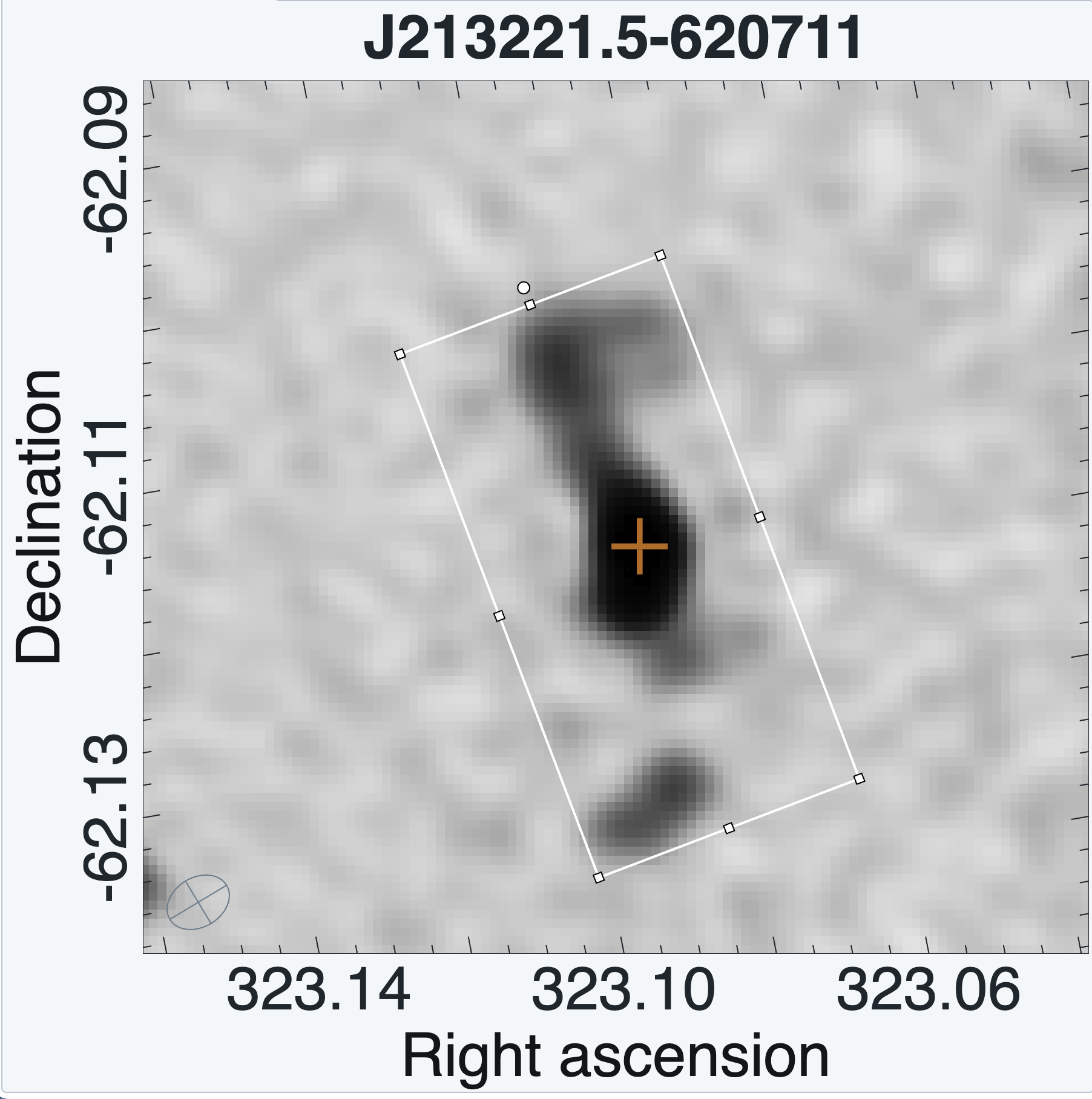}
  \caption{Examples of TRG\ytag\ (T-shaped radio galaxies) with linear extensions emanating from the side of the source. J210704.7-501143 differs from the others in that the protrusions are diffuse, suggesting a different mechanism operating in this case.
  }
    \label{fig:tsh}
\end{figure}

A number of radio sources have a single linear structure protruding from the side of the source, as shown in Figure \ref{fig:tsh}. In some cases they appear to emanate from the core, and in some cases from the lobe. They thus resemble just one side of XRG and ZRG respectively. We call all of these T-shaped radio galaxies (TRG\ytag). 

In a few cases protrusions stick out from both lobes, but on the same side of the galaxy, so are unlike XRG\ytag\ or ZRG\ytag\ sources. They differ from bent-tail sources in that the main axis is straight, and the protrusions from the side are quite separate. We therefore also tag these sources as TRG\ytag.

J210704.7-501143, shown in Figure \ref{fig:tsh}, and also shown in Figure 24 of \citet{pilot}, resembles a TRG\ytag\ source in that it has two relatively compact lobes with outflows from one side, but differs from other TRG\ytag\ sources in that the outflows are very diffuse, possibly as the result of an interaction with the intergalactic medium.

\subsubsection{Mechanisms driving winged radio sources}
We recognise that all these three different morphologies are similar except for the location of the linear protrusion(s). 

We also acknowledge that, with all such peculiar radio morphologies, in some cases the protrusion might be caused by the superposition of two unrelated sources, or be a projection effect, but this is unlikely to be the cause in most cases.

Although we have identified these three different types of morphology, we are not asserting that they are produced by different mechanisms, and it is possible that just one mechanism (e.g. backflow, as suggested by \citet{cotton20} for XRG), is responsible for all these morphologies. That backflow may then burst out of the radio galaxy at different points, depending on the local conditions. Further studies will be required to establish whether this is the cause.

We also note the similarities between the winged morphologies and the ``Trident'' morphologies discussed below in Section \ref{sec:wtf}.

\subsection{Unusual sources (WTF\ytag)}
\label{sec:wtf}

With the advent of large radio surveys with with millions of sources, probing parts of parameter space that have never before been observed, it was recognised several years ago that new types of radio source were almost certain to be found \citep{norris17b}, including ones with surprising morphologies that do not easily fit into the known classes of radio source. \citet{norris17b} coined the name WTF (for ``Widefield ouTlier Finder'') to describe these objects, which we now adopt as a tag to describe such objects.

The first WTF sources to be found in the EMU survey were the Odd Radio Circles (ORCs) \citep{orc}. Although it lies in the EMU-PS1  field covered by our catalogue, ORC 1 does not appear in our catalogue because it does not resemble a DRAGN. However, ORCs 2 and 3 do appear in our catalogue as they do indeed contain a DRAGN, which is now thought to be the causes of ORC2 and 3 \citep{macgregor25}, unlike the other ORCs which are thought to be caused by a shock from the elliptical galaxy at their centre \citep{orc, koribalski25}. 

\begin{figure}
  \includegraphics[width=4cm]{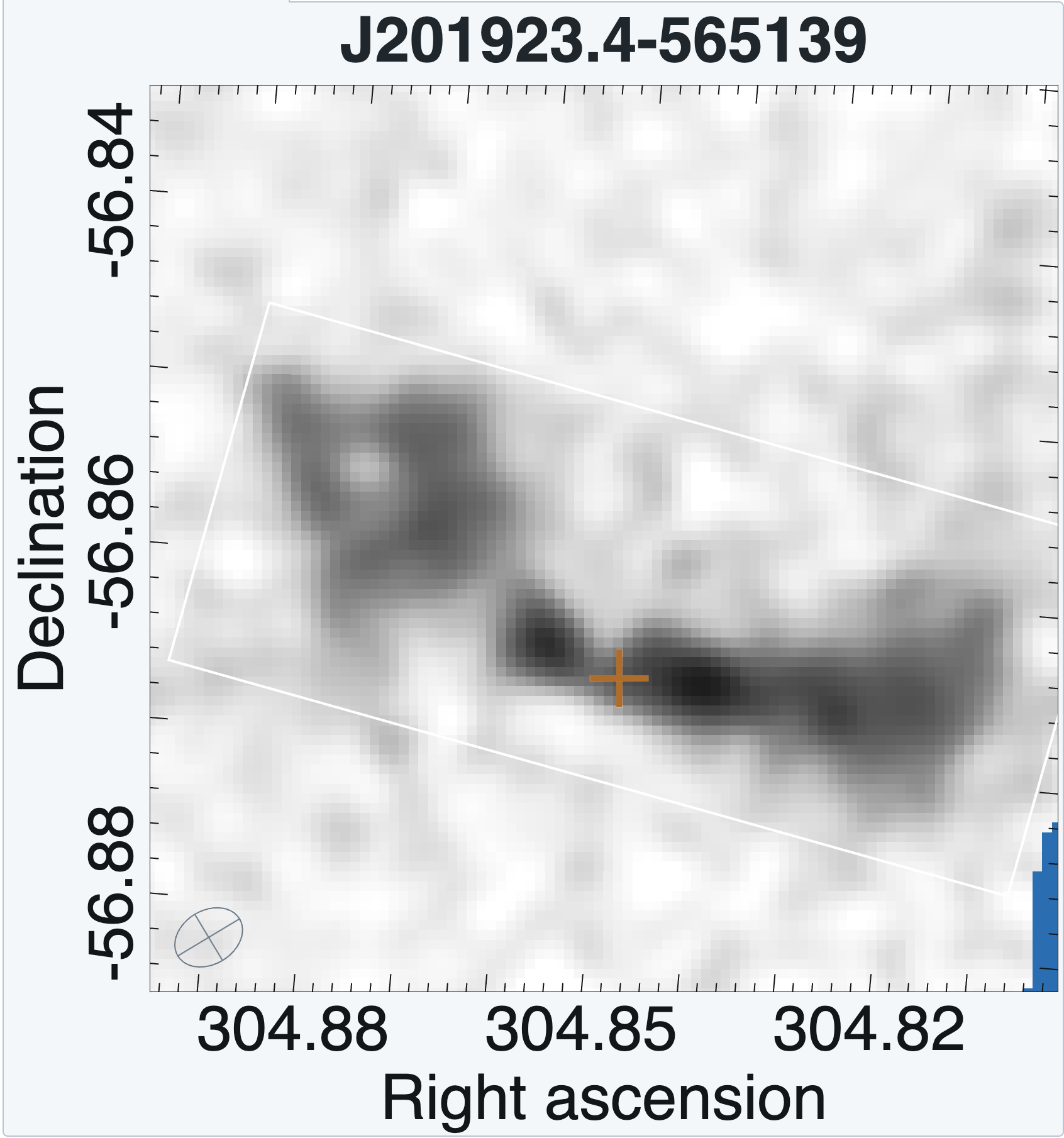}
  \includegraphics[width=4cm]{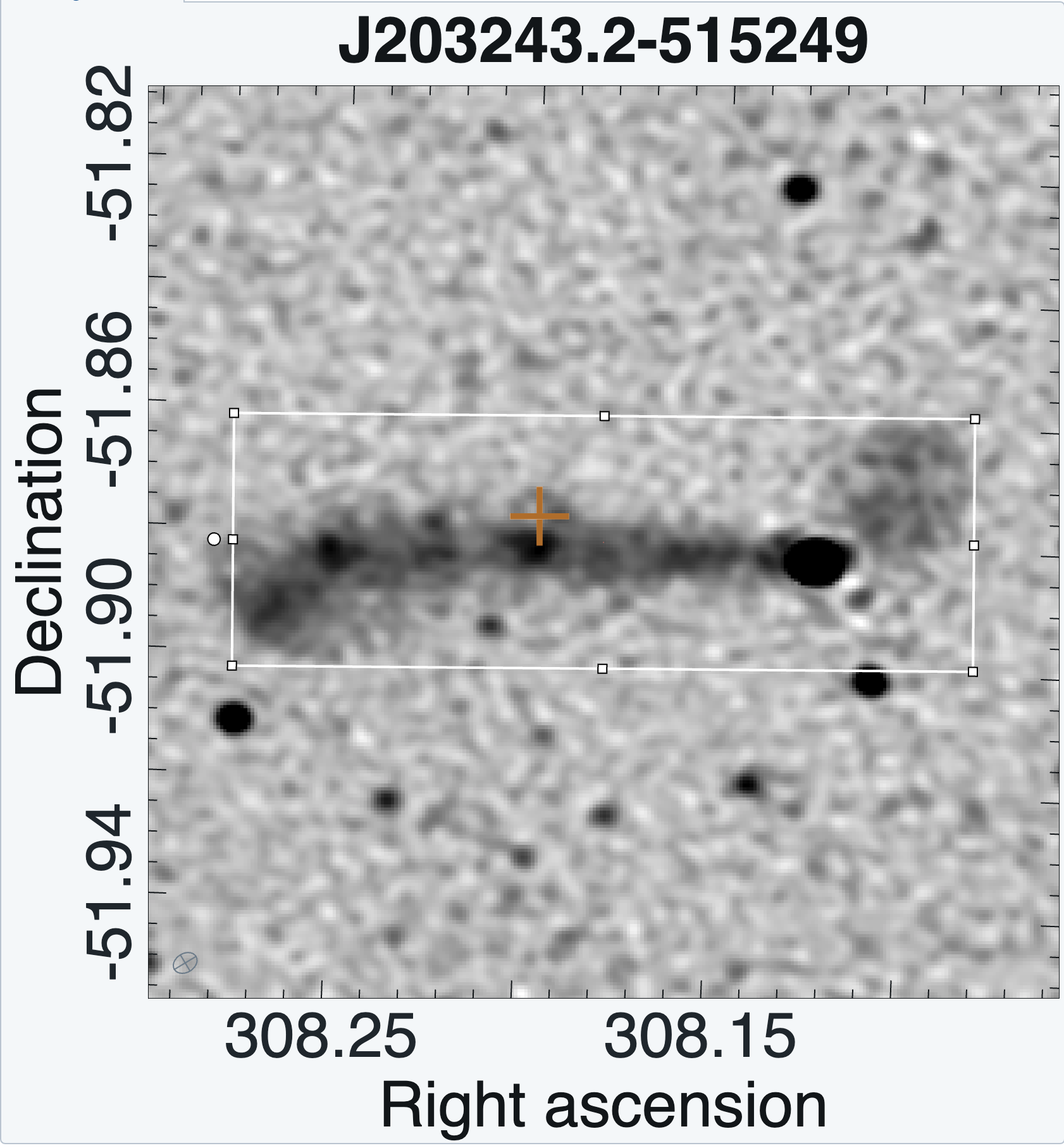}
  \includegraphics[width=4cm]{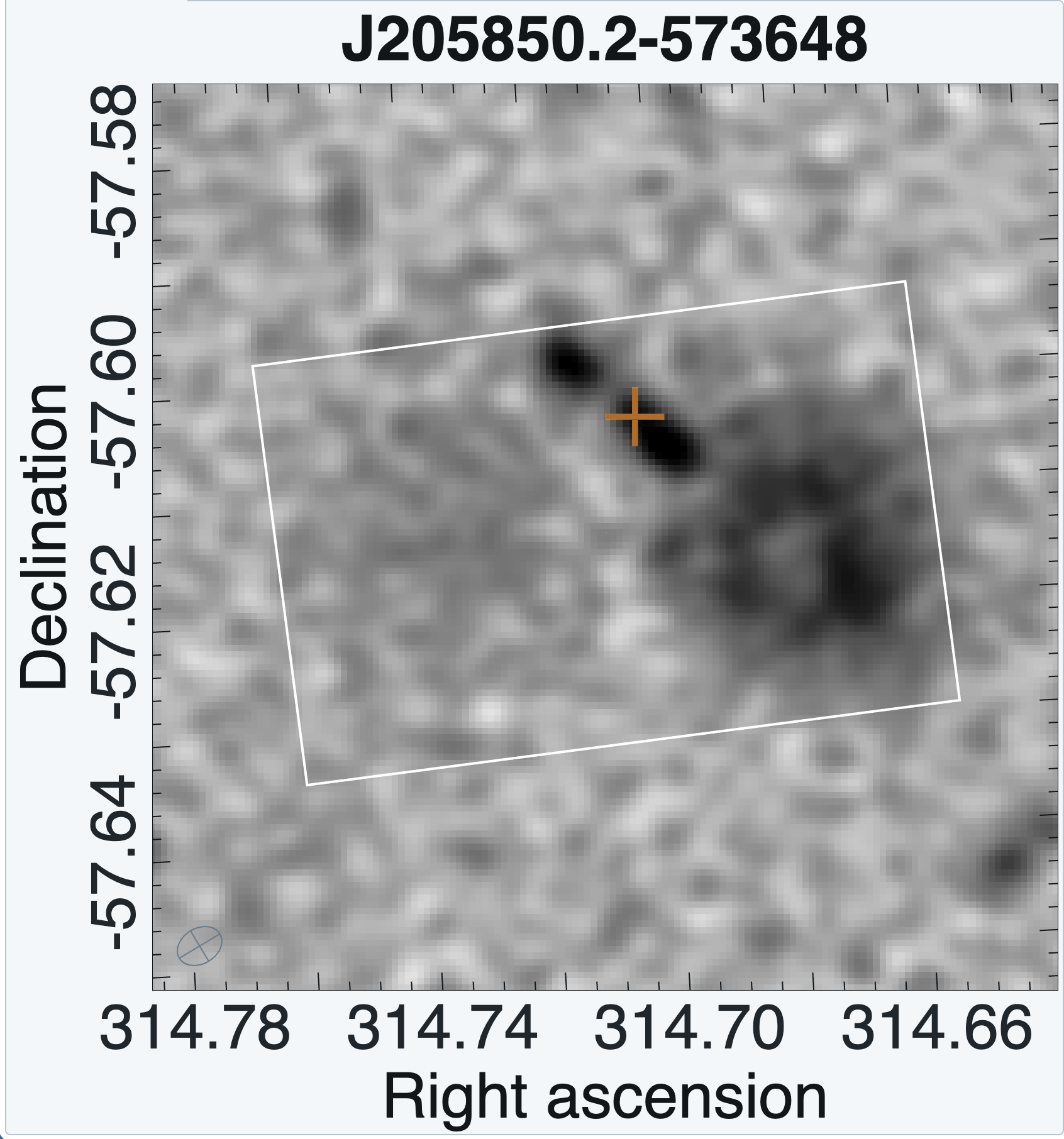} 
  \includegraphics[width=4cm]{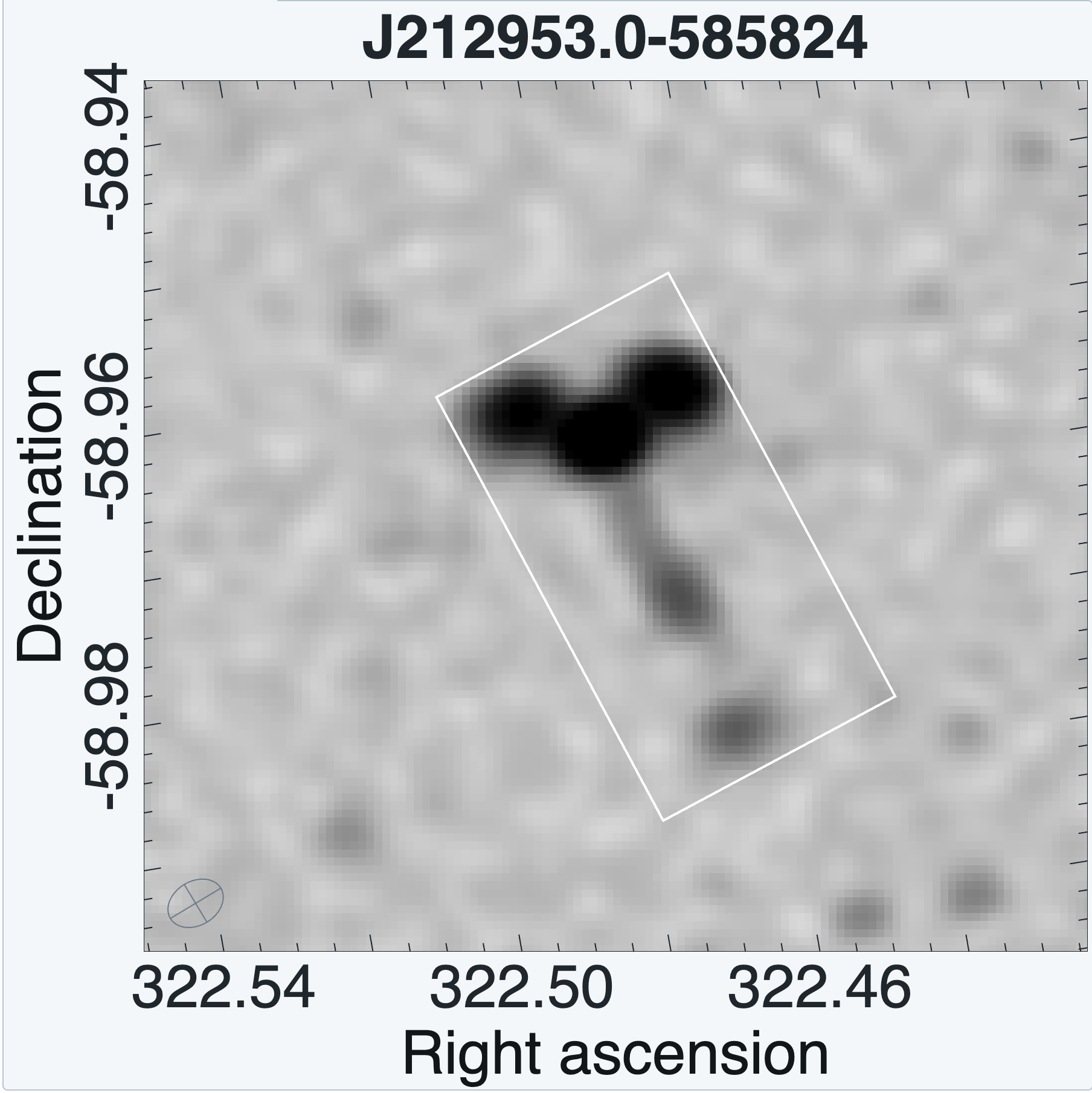}
  \includegraphics[width=4cm]{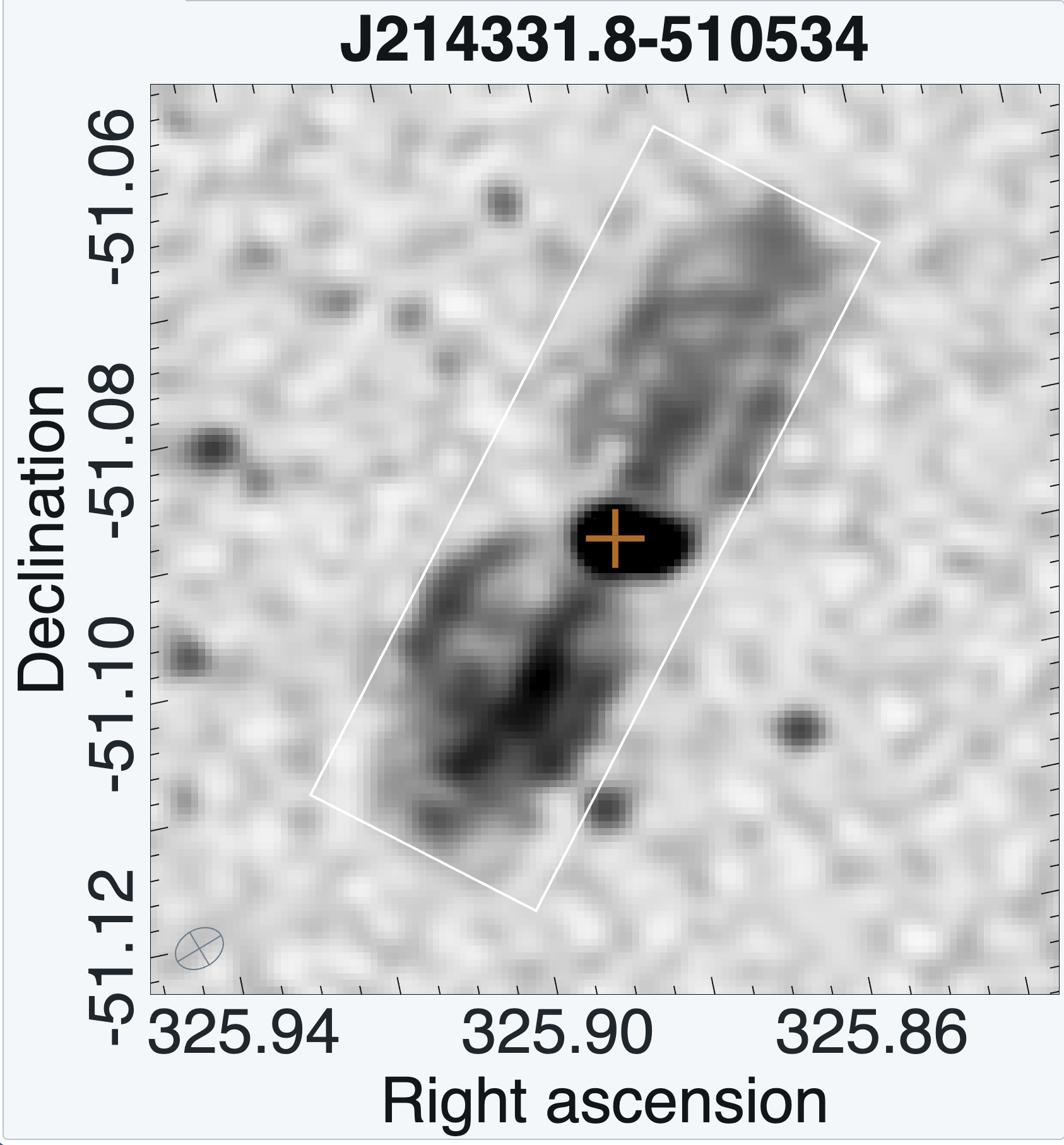}
  \includegraphics[width=4cm]{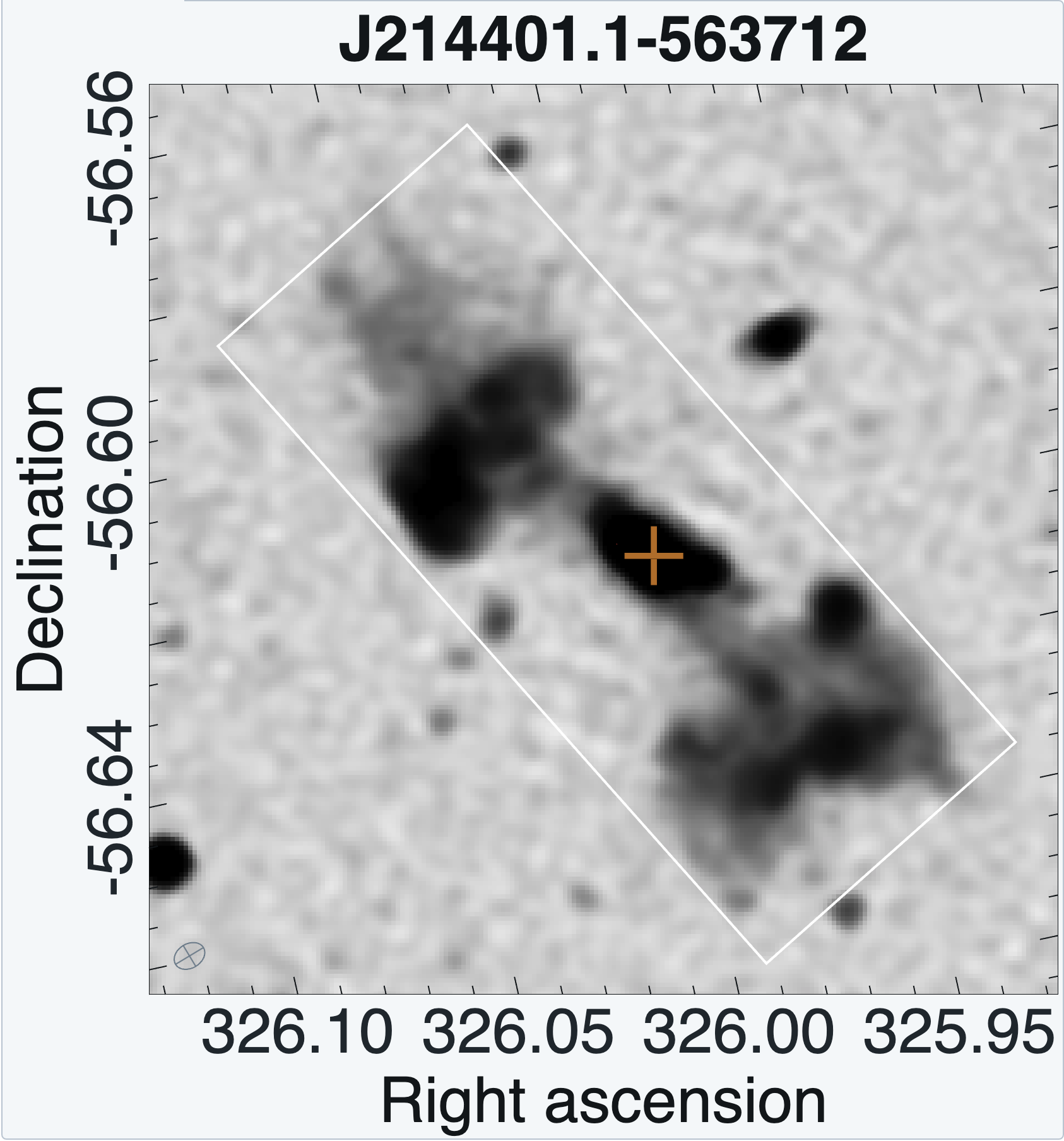}
  \includegraphics[width=4cm]{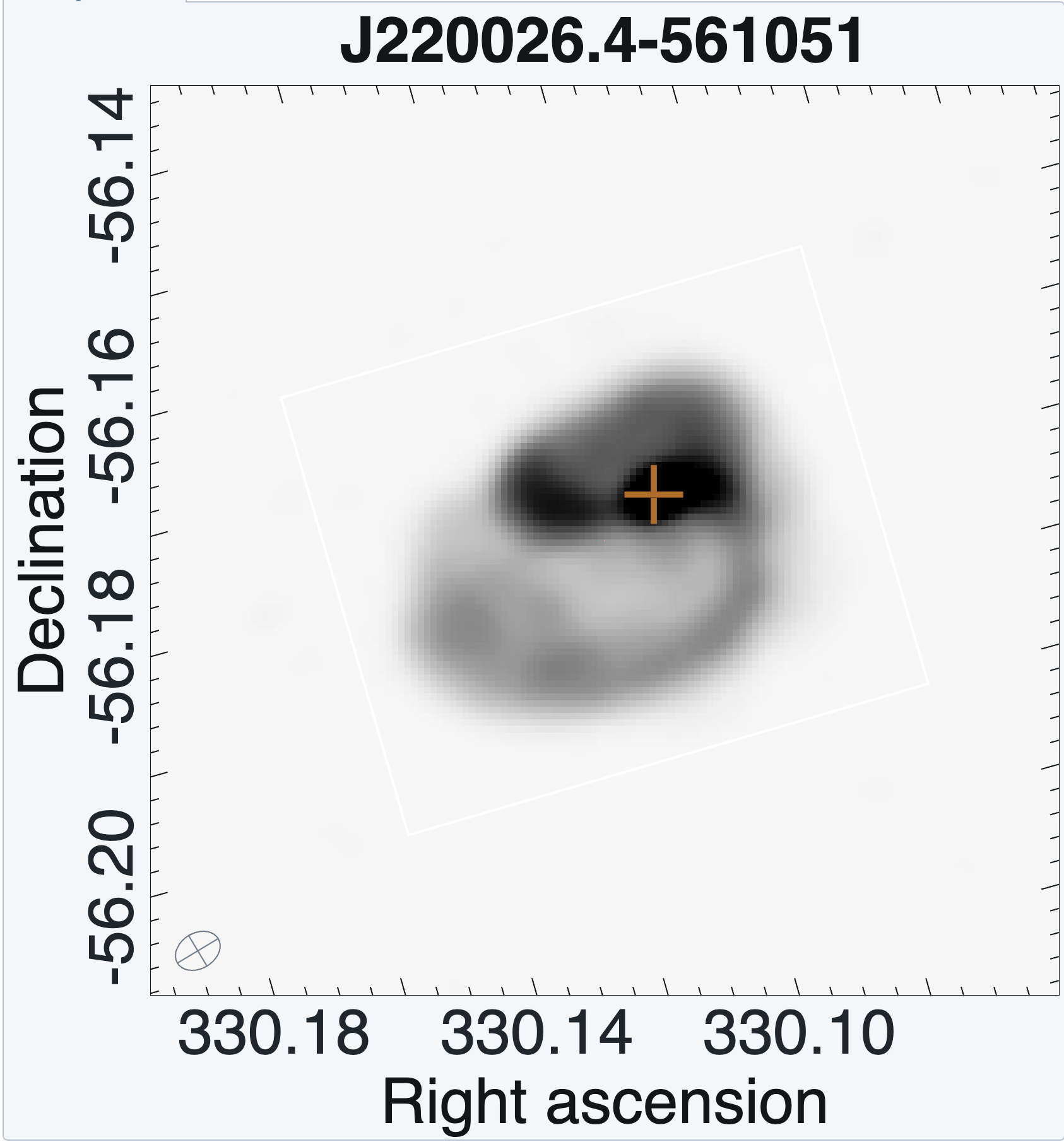}
  \includegraphics[width=4cm]{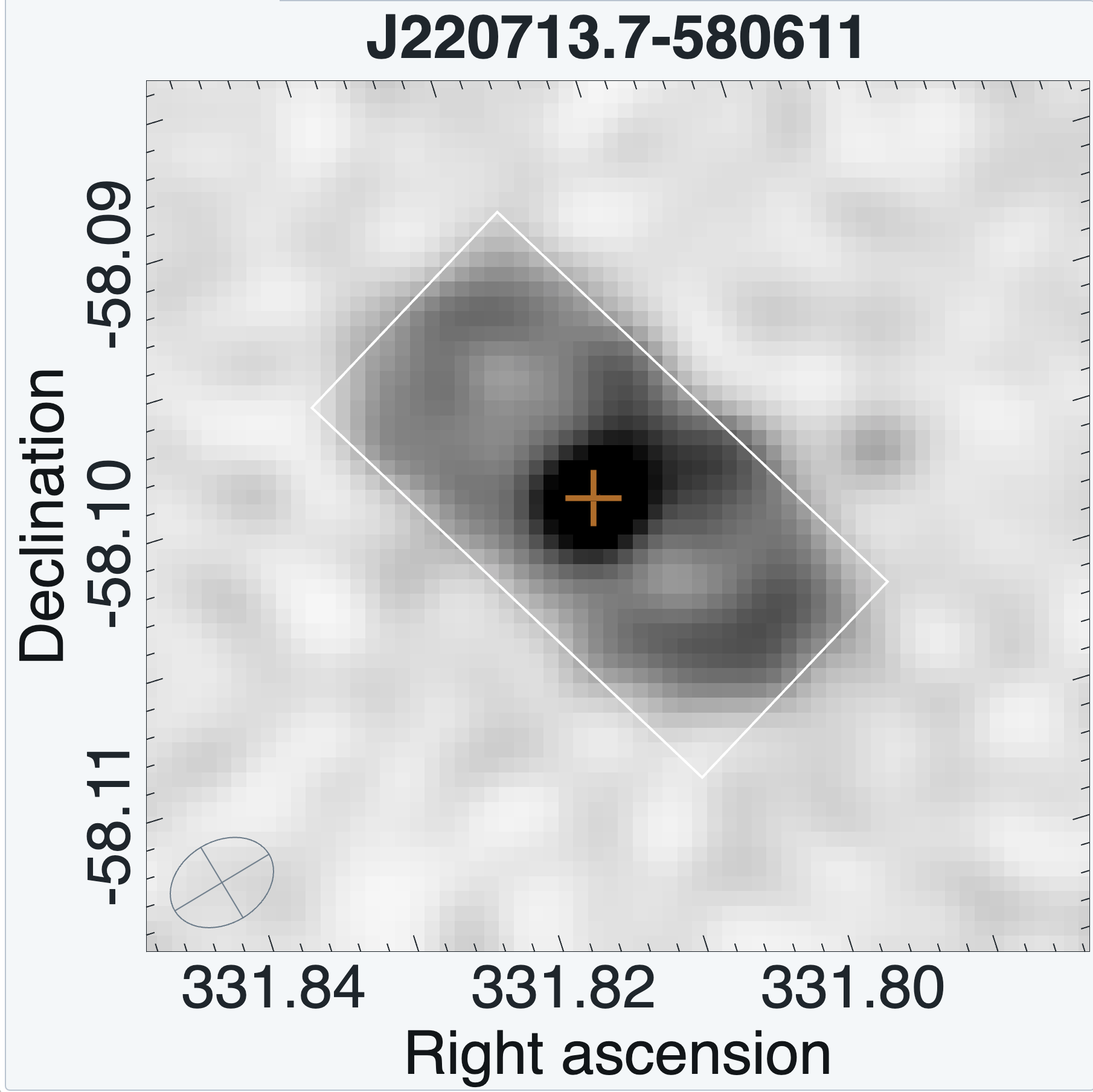}
  
  \caption{Examples of WTF\ytag\ sources, discussed in Section \ref{sec:wtf}. 
  }
  \label{fig:wtf}
\end{figure}

We find \wtf\ such WTF\ytag\ objects in our catalogue that defy description in conventional terms, and we show a sample of them in Figure \ref{fig:wtf}. It should be remembered in all cases that the two-dimensional appearance may be misleading. For example, apparently abrupt changes in jet direction may result from modest changes in a foreshortened jet seen end-on.

J214331.8-510534, which we call ``The Trident'', appears to have emission along the edge of a cavity, possibly tracing a backflow. The extension of the core to the west is due to an unresolved background source. We note that \citet{knowles22} found a very similar source in the cluster MCXC J2104.9-8243.

J214401.1-563712 shows traces of a similar morphology with a flattened triple structure at each end, but without the backflows seen in the first source. The extensions to the side of the source are perhaps related to the winged sources described in Section \ref{sec:winged}.

J205850.2-573648 is the source ORC2/3, first found by \citet{orc} and discussed in detail by \citet{macgregor25}. 

J220026.4-561051 was identified as an ORC candidate by \citet{gupta22} but may be a BT source.

J203243.2-515249 is a very asymmetric source with a long jet on one side and a diffuse blob on the other. It could perhaps be described as a HyMoRS\ytag\ (see Section \ref{sec:HyMoRS}) but that does not do justice to its extreme asymmetry. 

J201923.4-565139 is also an asymmetrical source, which could be classified as a ``double-double'' or a WAT, but the interesting feature is that one lobe has a distinct circular appearance, perhaps explained by the ORC model of \citet{shabala24}. 

J220713.7-580611 has two rings instead of lobes, also strongly suggesting the \citet{shabala24} ORC mechanism. 

We dub J212953.0-585824 the Mickey Mouse source. It may be two overlapping sources, or perhaps an extreme example of a T-shaped source. This source deserves further observations at higher resolution.

Similar collections of unusual objects have also been found in LOFAR data \citep{sasmal22,gopalkrishna24}

\subsection{Complex sources}
\label{sec:complex)}
\begin{figure}
   \includegraphics[width=4cm]{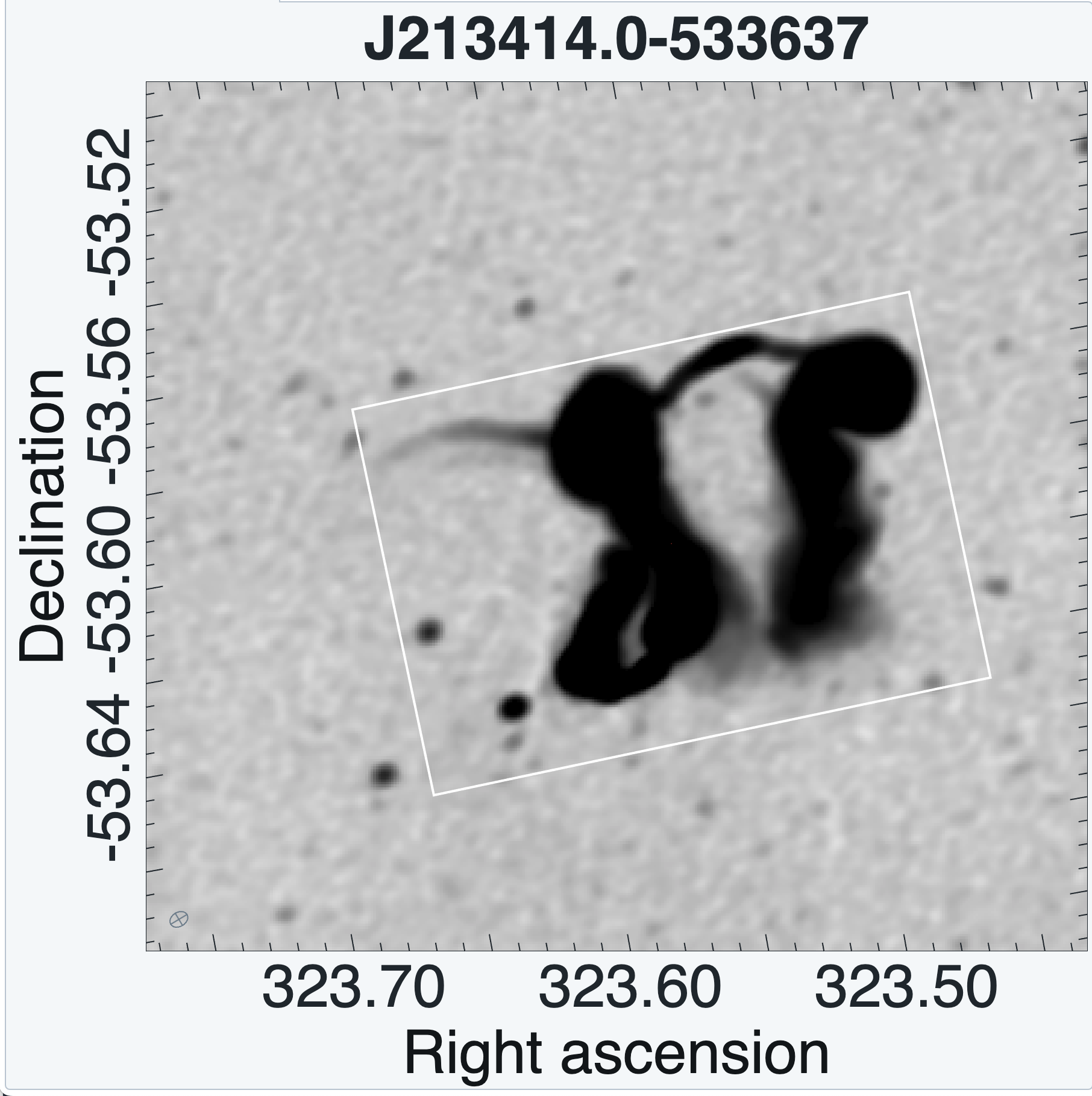}
   \includegraphics[width=4cm]{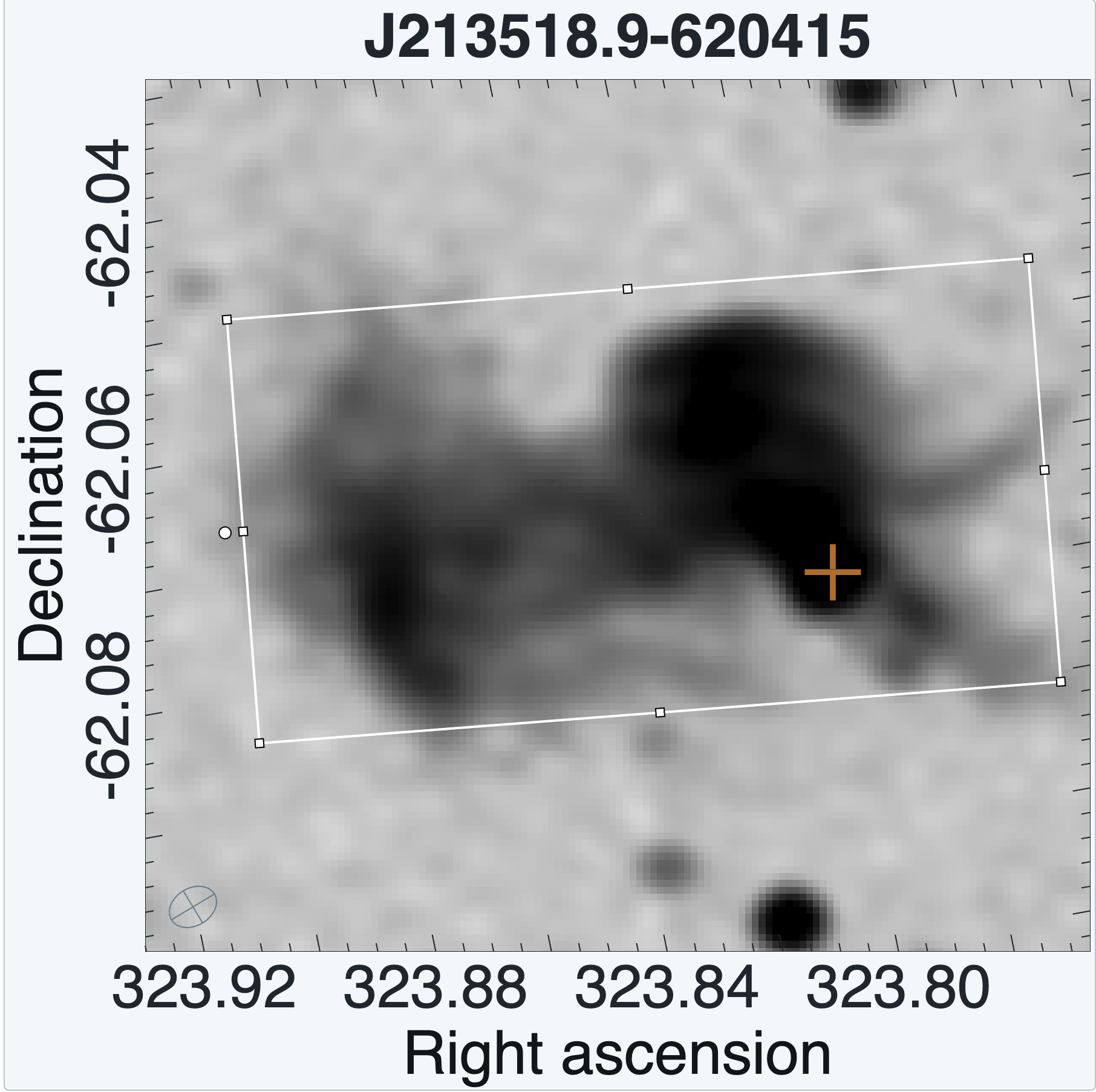}
  \caption{Two examples of CPLX\ytag\ sources that are too complex to classify as a DRAGN.}
  \label{fig:complex}
\end{figure}

In some cases, we have been unable to classify a source because of its complex structure. Two examples are shown in Figure \ref{fig:complex}. One is the well-known ``Dancing Ghosts'' source J213414.0-533637 \citep {pilot} which has been discussed in detail by \citet{velovic23}. The other, J213518.9-620415, may be a combination of cluster emission together with the strong emission from a \ac{BCG}. Both of these sources show filaments which are still poorly understood.

\subsection{Non-DRAGN sources}
\label{sec:non-AGN)}
\begin{figure}
  \includegraphics[width=4cm]{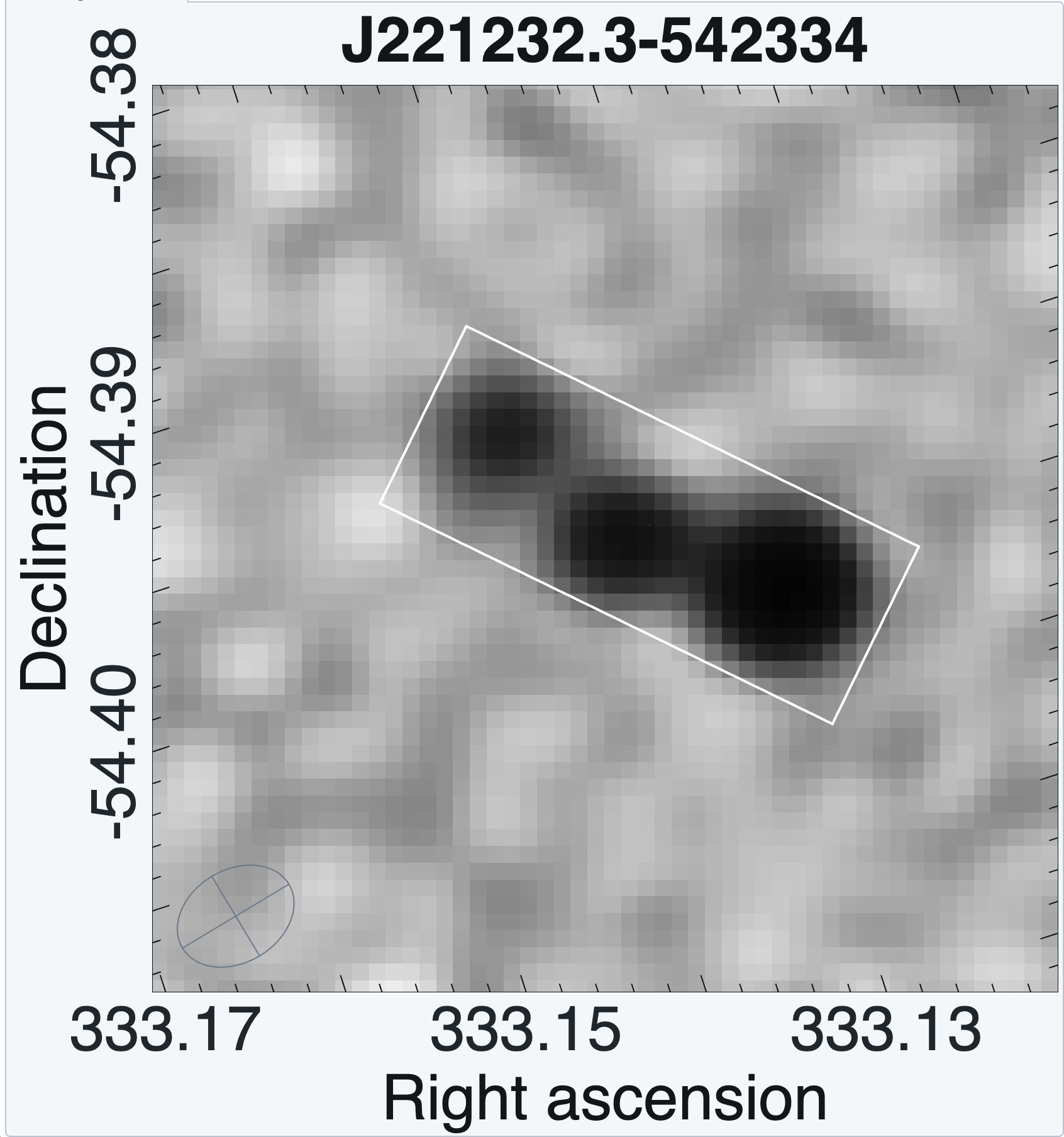}
   \includegraphics[width=4cm]{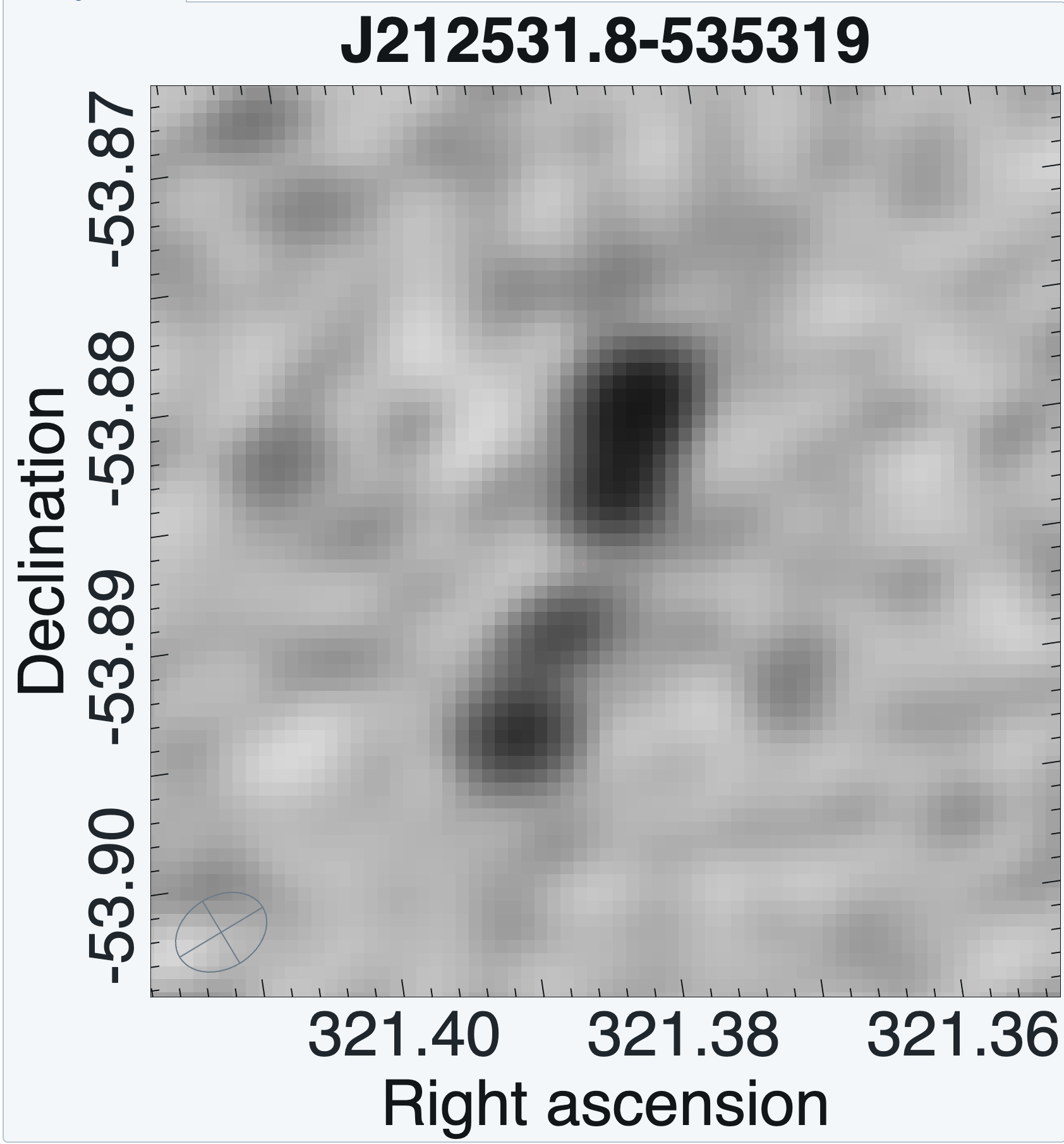}
  \caption{Two examples of non-DRAGN sources that were initially classified as a DRAGN from their radio appearance, but  the IR images show that each radio component is probably a separate galaxy.}
  \label{fig:non-AGN}
\end{figure}

\nondragn\ sources  consist of two or more nearby blobs of radio emission which resemble a DRAGN but which have no other evidence to connect them, or appear on the IR image as multiple galaxies. We list these in Table \ref{tab:non-AGN} for future reference. These sources are not included in the main catalogue or in the total number of DRAGNs. Two examples are shown in Fig. \ref{fig:non-AGN}.

\newpage
\onecolumn
\begin{table}
    \caption{Table of non-DRAGN sources that resemble DRAGNs.
    }
    \label{tab:non-AGN}
    \begin{tabular}{lllllll}
      \toprule
Name&Box centroid &Box centroid &Box major	&Box minor&Box PA &Comment	\\
& RA (\degr)& Dec (\degr)&   axis (") &  axis (") & (\degr)\\
      \midrule	
J200734.5-512314    &   301.89393   &   -51.38745   &   43.3    &   27.6    &   176.5   &   2 galaxies \\
J201148.6-591450	&	302.95285	&	-59.24724	&	68.7	&	26.1	&	30.7	&					2 galaxies	\\
J201509.4-524308	&	303.78915	&	-52.71903	&	83.5	&	53.3	&	0.0	&					spiral galaxy ESO 186-21	\\
J201926.2-582418	&	304.85936	&	-58.40510	&	136.5	&	15.9	&	100.9	&						\\
J202919.4-604521	&	307.33084	&	-60.75609	&	85.8	&	21.2	&	117.2	&						\\
J203316.6-531513	&	308.31941	&	-53.25383	&	172.9	&	37.2	&	130.5	&						\\
J204328.6-532120	&	310.86945	&	-53.35579	&	103.6	&	53.0	&	292.5	&					edge-on spiral NGC6948	\\
J204434.5-505636	&	311.14377	&	-50.94346	&	349.6	&	286.2	&	163.5	&						\\
J205051.7-594511	&	312.71540	&	-59.75320	&	148.8	&	38.7	&	207.9	&						\\
J205101.3-614302	&	312.75548	&	-61.71730	&	177.4	&	44.7	&	339.1	&					2 galaxies	\\
J205306.0-514314	&	313.27515	&	-51.72064	&	486.8	&	59.5	&	107.0	&						\\
J211200.0-570238	&	318.00036	&	-57.04392	&	155.0	&	40.1	&	324.9	&					2 galaxies	\\
J211210.6-571645	&	318.04432	&	-57.27920	&	91.2	&	37.8	&	15.0	&					edge-on spiral ESO 187-58	\\
J211214.9-511607	&	318.06225	&	-51.26866	&	66.8	&	28.2	&	345.5	&					 2 galaxies	\\
J211232.2-511511	&	318.13441	&	-51.25318	&	53.3	&	23.3	&	320.1	&					2 galaxies	\\
J211611.3-482119	&	319.04745	&	-48.35538	&	51.3	&	23.5	&	103.7	&					2 galaxies	\\
J212425.7-491640	&	321.10740	&	-49.27801	&	81.6	&	65.7	&	0.0	&					Face-on spiral ESO 236-8 with SB ring	\\
J212531.8-535319	&	321.38256	&	-53.88882	&	71.7	&	21.2	&	340.5	&					4 galaxies in a line	\\
J212910.0-524554	&	322.29176	&	-52.76517	&	278.1	&	48.6	&	92.1	&					 edge-on spiral NGC7064	\\
J213523.1-562639	&	323.84644	&	-56.44424	&	75.4	&	34.2	&	328.4	&					2 galaxies	\\
J213558.1-520617	&	323.99230	&	-52.10484	&	82.1	&	34.1	&	277.1	&					2 galaxies	\\
J213630.0-543307	&	324.12528	&	-54.55218	&	224.6	&	102.2	&	320.2	&					edge-on spiral NGC7090. 	\\
J213801.1-602432	&	324.50460	&	-60.40904	&	49.4	&	27.3	&	353.9	&					2 galaxies	\\
J214326.2-491230	&	325.85933	&	-49.20833	&	47.0	&	23.2	&	96.6	&					 2 galaxies 	\\
J214445.6-602510	&	326.19023	&	-60.41950	&	38.3	&	17.4	&	68.8	&					2 galaxies	\\
J214806.0-503404	&	327.02507	&	-50.56798	&	120.3	&	68.4	&	342.3	&					SB ring in spiral ngc7124	\\
J215222.3-564139	&	328.09305	&	-56.69436	&	48.9	&	33.7	&	24.0	&					2 galaxies	\\
J215504.6-503930	&	328.76928	&	-50.65847	&	88.8	&	37.5	&	277.2	&					2 galaxies + extended IR emission	\\
J215554.4-532801	&	328.97675	&	-53.46703	&	63.6	&	45.5	&	355.7	&					2 galaxies	\\
J215910.9-532538	&	329.79580	&	-53.42738	&	44.2	&	20.1	&	7.8	&					2 galaxies	\\
J220001.4-503345	&	330.00607	&	-50.56253	&	92.1	&	50.5	&	58.9	&					4 galaxies in a line	\\
J220239.4-560627	&	330.66441	&	-56.10774	&	69.1	&	22.5	&	138.9	&						\\
J220714.0-505919	&	331.80850	&	-50.98870	&	53.0	&	20.3	&	305.1	&					2 galaxies	\\
J220833.4-572634	&	332.13919	&	-57.44299	&	205.2	&	134.4	&	83.6	&					rings in spiral NGC7205	\\
J221205.8-484324 & 333.02444 & -48.72351 &58.1 &28.3 &313.6 &2 galaxies\\
J221232.3-542334 &333.13477 &-54.39298 & 59.9 &23.5 &64.0\\
    \end{tabular}
\end{table}

\twocolumn

\section{Comparison with other catalogues}
\label{sec:comparison}

Comparison with other catalogues is difficult because (a)~each survey has a different resolution, observing frequency, and sensitivity, (b)~terminology and criteria vary from survey to survey, and (c)~many authors combine two or more tags in their table of results. For example, hotspots in lobes tend to have a flatter spectral index than the diffuse emission \citep[e.g.][]{treichel01}, so, in a low-resolution observation, the position of the lobe peak, and potentially the FR1/FR2 classification, may change with observing frequency, as discussed in Section \ref{LoTSS}. Nevertheless, it would be remiss not to attempt to compare our results with other catalogues. 

\begin{figure}
    \centering
    \includegraphics[width=\columnwidth]{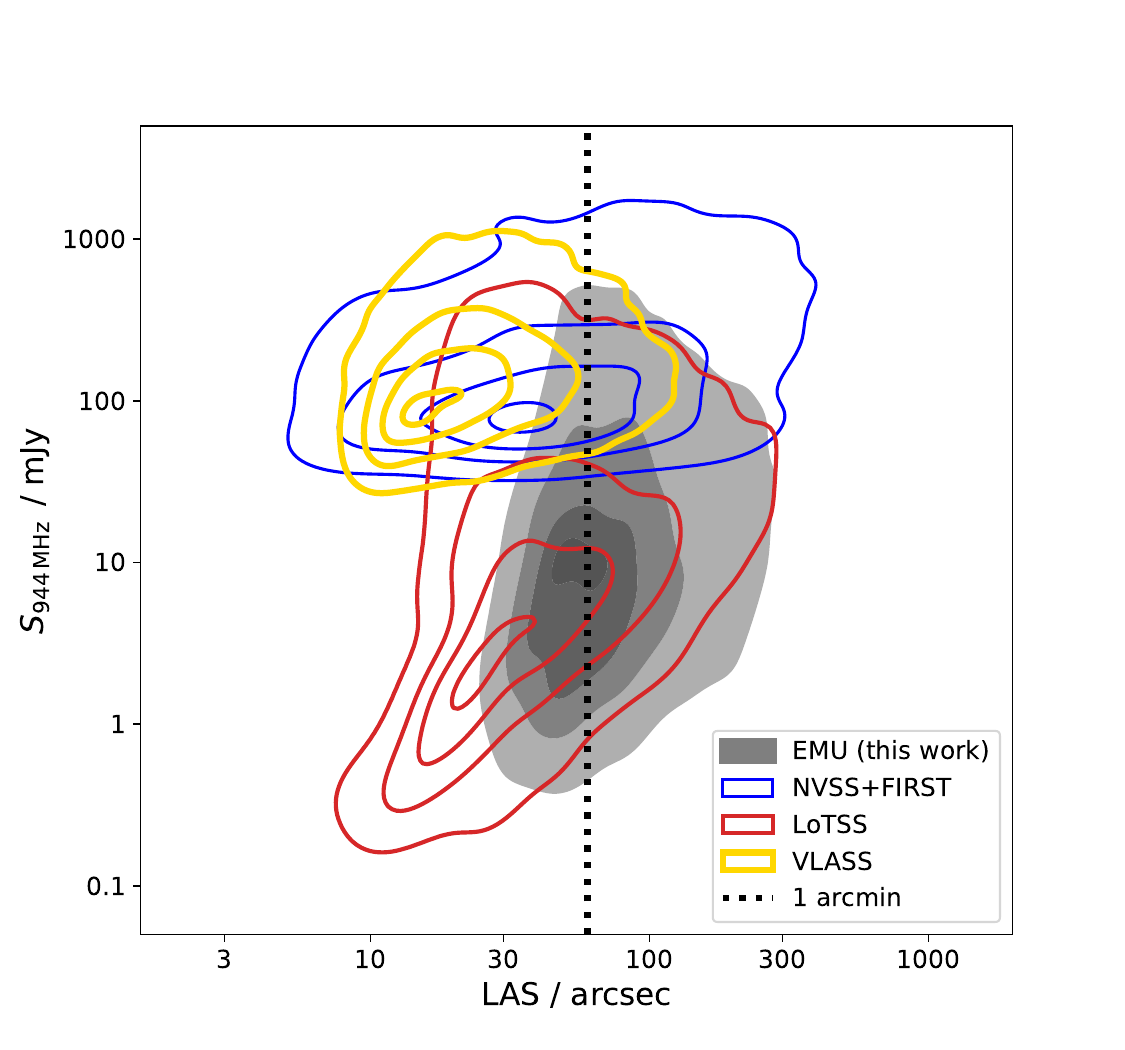}
    \caption{
    A comparison of the size-flux space occupied by DRAGNs identified by EMU in this work (solid grey), by NVSS and FIRST \citep[][blue]{mirabest17}, LoTSS {  DR1} \citep[][red]{mingo19}, and VLASS \citep[][yellow]{gordon23}.
    The contour levels contain $95\,$\%, $67\,$\%, $33\,$\%, and $5\,$\% of the distributions.
    The flux densities from other surveys are extrapolated to the EMU frequency of $944\,$MHz assuming a spectral index of $\alpha=-0.7$.
    Each of the surveys is at a different frequency, so comparing them tells us about the frequency dependence of the DRAGN population.}
    \label{fig:sizefluxcomp}
\end{figure}

In Figure \ref{fig:sizefluxcomp} we show the distribution of the EMU DRAGNs identified in this work compared to three other recent catalogues of DRAGNs, with the flux densities scaled to $944\,$MHz assuming a typical spectral index of $\alpha = -0.7$.
Figure \ref{fig:sizefluxcomp} highlights how the different source finding methodologies employed and observational data sets used can identify vastly different populations of DRAGNs.
Our EMU-identified DRAGNs probe substantially fainter radio galaxies than those found in \ac{NVSS}, \ac{FIRST} or \ac{VLASS}.
LoTSS identifies fainter galaxies at smaller sizes than in our catalogue but, notably, the EMU DRAGNs extend to larger angular sizes than the \citet{mingo19} catalogue at $S_{944\,\text{MHz}} \lesssim 10\,$mJy.
We discuss comparisons with DRAGN catalogues in more detail below.

\subsection{3CRR}
{  The 3CRR sample \citep{laing83} is a reliable catalogue of strong radio sources revised from the earlier 3CR survey \citep{bennett62}.} Table 1 shows that only 76\% of the sources in our DRAGN catalogue are FR sources (i.e. FR1\ytag, FR2\ytag, or FRX\ytag). The remaining sources have morphologies such as LTS\ytag, OSS\ytag, and HT\ytag\ which cannot be classified as FR sources. By comparison, at least 89\% of the 181 {  3CRR}
strong radio sources listed in table 3 of \citet{laing83} are FR sources (the actual breakdown is 132 are FR2, 29 are FR1 and 20 are core-dominated sources). In this and other early catalogues focusing on strong sources, the other morphologies discussed in this paper (e.g. LTS, OSS) are virtually absent. 

This illustrates a major difference between the early catalogues of strong sources and the current generation of large catalogues of much fainter sources: fainter sources show a much wider range of morphologies, and much more morphological disturbance, than the strong sources.

\subsection{NVSS and FIRST}
\label{ssec:nvssfirst}
\citet{gendre10} compiled and classified a subset of sources from NVSS, which they call the CONFIG sample, and which should constitute a valuable comparison with our catalogue. Unfortunately, the value to us is reduced because their classification process is rather different from ours. For example, Gendre et al. included core-jet sources in the FR2 class, and HyMoRS sources were included in either FR1 or FR2 categories, depending on their most prominent features. CONFIG consists of four sub-samples with different areas and sensitivities, none of which are comparable with the survey here. Nevertheless, we summarise the morphological classifications of the total sample in Table \ref{tab:mirabest}.

The MiraBest sample \citep{porter23, mirabest17} is a labelled sample of 1256 low-redshift (0.03 < z < 0.1) radio-loud AGN from NVSS and FIRST. which are limited to FR1, FR2, and Hybrid (corresponding to our \ac{HyMoRS}\ytag) morphologies. Here we reject those that are classified as ``uncertain'', leaving 852 sources that they label as ``confident''. We rearrange Table 2 of \citet{porter23} and compare their results with ours in Table \ref{tab:mirabest}.

\begin{table}
    \caption{Comparison of our tags with the MiraBest and CONFIG samples. In each case, a source may be assigned more than one tag} 
    \label{tab:mirabest}
    \begin{tabular}{llll}
      \toprule
      Tag	&	\# Sources & \# Sources & \# Sources  \\
  & in our catalogue	    & in MiraBest & in CONFIG\\
      \midrule
FR1 &238 (6.7\%)&397 (46.6\%)&68 (6.9\%) \\
FR2 &1410 (39.6\%)&436 (51.2\%)&350 (35.5\%)\\
BT &244 (6.9\%)&52 (6.1\%)&\\
HT &16 (0.5\%)&9 (1.1\%)& \\
DD &21 (0.6\%)&4 (0.5\%)& \\
HyMoRS &43 (1.2\%)&19 (2.2\%)& \\
      \bottomrule
    \end{tabular}
    \medskip\\
\end{table}

The sensitivities, resolutions, and areas covered by the surveys are too different to draw any physical significance from Table \ref{tab:mirabest} . 
For example,  Figure \ref{fig:sizefluxcomp} shows the \citet{mirabest17} catalogue only covers the upper tip of the EMU-PS1 flux range because of the deeper EMU-PS1 data,  but finds  smaller DRAGNs because of the higher resolution of FIRST.
However, two things are notable. First, the FR1/FR2 ratio of our catalogue and CONFIG are similar, but both are much higher than the MiraBest catalogue. It is unclear why this should be, given that CONFIG and MiraBest are drawn from the same data, but is a useful illustration of the hazards of comparing catalogues made with different selection criteria. Second, the fractions of BT, HT, DD, and HyMoRS in our catalogue and MiraBest are comparable (i.e. within a factor of about two).

\subsection{VLASS}
\label{sec:vlass}

The Karl G. Jansky Very Large Array (VLA) Sky Survey \citep[VLASS;][]{lacy20} is a multi-epoch survey of  the entire sky north of -40\degr\ at $\sim$ 3 GHz at high resolution. The first epoch of VLASS was completed in 2019 \citep{gordon2021}, and Quick Look images covering $\sim$ 34,000\sqdeg\ are publicly available.

\citet{gordon23} searched the VLASS Quick-look images for DRAGNs using an algorithm they called \textsc{DRAGNhunter}, and identified 17724 DRAGNs. 
A detailed comparison of their data with ours is complex because the two surveys differ so greatly in observing frequency, resolution, and sensitivity.
At a higher frequency than EMU, VLASS {  at $3\,$GHz} may be more biased towards identifying DRAGNs with bright hotspots or younger radio lobes.
VLASS was observed using the VLA in BnA configuration, a setup that lacks the short baselines needed to detect radio structures larger than a few tens of arcseconds across.
A consequence of this, for example, is that while two hotspots of a large FR II may be identifiable in VLASS, large and diffuse FR Is are likely missing from the \citet{gordon23} catalogue.
The point source sensitivity of VLASS is substantially lower than that of EMU, with a typical rms noise in VLASS of $130\,\mu\text{Jy}\,\text{beam}^{-1}$.
For objects with a typical spectral index of $\alpha=-0.7$, the faintest point sources in VLASS ($S_{3\,\text{GHz}}\sim 0.65\,$mJy) would appear to be $\sim 1.5\,$mJy in EMU, approximately $15$ times brighter than the faintest objects in EMU.
This results in a much lower source density for VLASS (0.51  deg$^{-2}$) than for our survey (13.11 deg$^{-2}$).
The effects of these differences are seen clearly in Figure \ref{fig:sizefluxcomp}, where the DRAGNs identified in this work (grey shading) occupy an almost totally different region of the angular size / flux density plane to the DRAGNs identified by \citet[yellow contours]{gordon23}.

\subsection{LoTSS}
\label{LoTSS}
The LoTSS survey \citep{shimwell17, shimwell19} at 144 MHz has a median sensitivity of 71 \ujybm\ and a resolution of 6\arcsec. Assuming a spectral index $\alpha = -0.7$, this is equivalent to a sensitivity of 19 \ujybm\ at 944 MHz, which is comparable to that of EMU-PS1. A catalogue of double sources, based on LoTSS, similar to that presented here, has been published by \citet{mingo19}, with the primary goal of studying the FR dichotomy. Of all the catalogues discussed here, this is the most similar to ours, and so comparisons might in principle yield information about the differences between sources selected at 944 MHz and those selected at 144 MHz.

Figure \ref{fig:sizefluxcomp} shows that the DRAGNs in LoTSS occupy a similar location in size/flux space but differs in two respects: 1) LoTSS finds a population of smaller DRAGNs that our survey misses, because of their higher resolution, 2) our survey finds larger DRAGNs at low flux densities than LoTSS, possibly because of better sensitivity to low surface brightness emission by EMU-PS1, or differences in methodology. 
{  We note that the newer LOTSS DR2 \citep{shimwell22} has greater low surface brightness sensitivity than the LOTSS DR1 survey used by \citet{mingo19}.} 

\citet{mingo19} used the sample of 23344 radio-loud sources with optical/IR identifications and redshifts listed by \citet{hardcastle19}. They further restricted the sample to those with a size > 12\arcsec and S$_{150}$>2 mJy, and further reliability filtering, to obtain a catalogue of 5805 radio sources. Their sample is therefore a factor of 5805/\total = 1.6 larger than ours.

They then applied a combination of automated classification and visual checking, to obtain samples of FR1, FR2, and other classifications. Here we refer to their classifications using the tags FR1\mtag\ and FR2\mtag.

Comparing the results from \citet{mingo19} with those in this paper suffers from a number of challenges, which are illustrated by Table \ref{tab:mingo}, in which we show a comparison of the numbers of FR1/FR2 sources found by morphological type. Here we have used the numbers from Table 3 of Mingo et al., and restricted it to their class S2 (size $> 60\arcsec$) and F3 (total flux  $> 50\,$mJy). We apply the same criteria to our data, scaling the total flux limit to S$_{944MHz}$>13.4 mJy, assuming a spectral index of $\alpha = -0.7$. The two samples should thus be similar, other than the observing frequency.

It is clear there is a major difference, with  Mingo et al seeing far more FR1\mtag\ than FR2\mtag, while we see far more FR2\ytag\ than FR1\ytag. In addition to any astrophysical reason for this, we can identify six main effects that would lead to this difference.

{  First, we note that, as discussed in Section \ref{sec:fr12}, Mingo et al. adopted a definition of an FR1\mtag\ which includes the radio emission associated with the host galaxy, whereas our definition of an FR1\ytag\ depends only on the maxima of the extended radio jets/lobes, and ignores any central radio emission associated with the host galaxy.
For example, we would classify as an FR2\ytag\ the example of an FR1\mtag\ source given in Figure 2(a) of their paper.}

Second, we recognise a class of sources (FRX\ytag) for which we would say the classification is uncertain, whereas all sources in Mingo et al. are classified  as FR1\mtag\ or FR2\mtag. Their confidence in their classification was helped by the higher resolution of the LoTSS data compared to EMU data, but some marginal classifications are presumably included in their counts of FR1\mtag\ and FR2\mtag.

Third, we recognise a category of source (LTS\ytag, or Linear triple source, discussed further in Section \ref{sec:LTS}) in which two jets are seen, but either one or both do not have maxima that would enable them to be classified as FR sources. These would also be classified by Mingo et al. as FR1\mtag.

\begin{figure}
  \includegraphics[width=8cm]{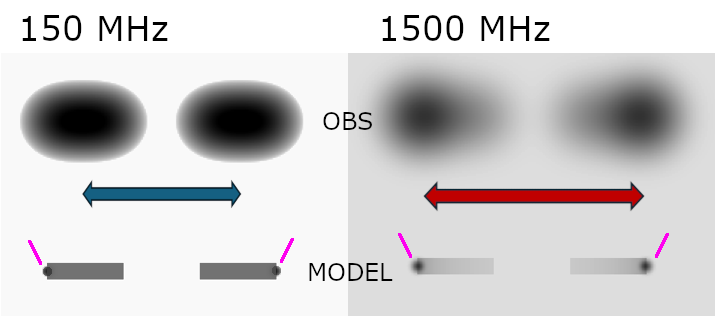}
  \caption{A cartoon simulation of an FR2 source consisting of two rectangular ``jets'' or ``lobes",  $10''\times 50''$, separated by $50''$,  with spectral index $\alpha = -1.5$. At the outer end of each jet/lobe is a hotspot consisting of a $5''$ circular Gaussian with spectral index $\alpha = -0.7$. The model at each observing frequency (150~MHz and 1500~MHz) is shown at the bottom, with the flatter spectrum hot spots indicated by magenta lines.  Their contrast against the steep spectrum jets/lobes is much lower at 150 MHz.  At the top, the observations at a resolution of $30\arcsec$, (5 beams across the source), are shown. The double-sided arrows indicate the observed separation between peaks, as used in the FR1/II definition. In this example, the peaks are $150\arcsec$ apart in the model, but observed to be $143\arcsec$, and $108\arcsec$ apart in the 1500~MHz and 150~MHz observations, respectively.}
  \label{fig:cartoon}
\end{figure}

Fourth, hotspots tend to be flatter spectrum than the diffuse jet or lobe emission of an FR2 \citep[e.g.][]{treichel01, mahatma20}, so they are less prominent at low frequencies. Thus, even with the same number of beams across the source, at low frequencies, the separation of the emission peaks will be closer together than at high frequencies. This becomes even more important when coupled with the increase in beam size at low frequency; it can cause a source that is clearly FR2 at higher frequencies to appear as an FR1 at low frequencies, as shown in Figure \ref{fig:cartoon}.

Fifth,  Mingo et al. used an automated approach which is likely to produce a higher error rate for individual objects and may misclassify the most unusual objects, although it does have the advantage of being immediately scalable to larger survey areas. However, we note that for large strong sources used here (categories S2 and F3), their reliability appears to be high (84\% for FR1\mtag\ and 94\% for FR2\mtag) and so this should not be a major factor. 

Sixth, the  Mingo et al. catalogue is restricted to sources with an optical/IR identification, which will tend to bias it to lower redshift and lower excitation sources compared to ours. However, since Mingo et al. include sources even if they only have a WISE identification, and since 90\% of our sources do have a WISE identification (see Table 1), this will probably only have a small effect on the comparison.

These differences may help explain why Mingo et al. see many more FR1's than we do, and why we find many more FR2's than they do. 
These differences will be explored further in Paper IV.

\begin{table}
  \begin{center}
    \caption{Comparison of FR1/FR2 sources from our paper and \citet{mingo19}, limited to FR/LTS sources with size $> 60\arcsec$ (i.e. S2 class in \citet{mingo19}) and $S_{150}>50\,$mJy, $S_{944}>13.4\,$mJy (i.e. F3 class in \citet{mingo19}).
    }
    \label{tab:mingo}
    \scriptsize
    \begin{tabular}{llllll}
      \toprule
       		& Total & FR1 & FR2 & FRX & LTS\\
      \midrule
Mingo et al.&629&430 (68\%)&199 (32\%)\\
This paper&900&136 (15\%)&495 (55\%)&223 (25\%)&46 (5\%)\\
      \bottomrule
    \end{tabular}
    \medskip\\
  \end{center}
\end{table}

\section{Conclusion}
This paper presents one of the largest morphological catalogues of radio sources to date, which is distinguished by the fact that all classifications have been carried out by human eye, without relying on automated classification techniques.

It is primarily motivated by the need to understand the physical processes that drive the formation of radio jets and other structures around supermassive black holes, which we observe as DRAGNs. In this paper, we provide data to test our models of these processes, by
\begin{itemize}
    \item Finding hitherto unknown classes of source -- the 
    ``unknown unknowns'', or WTFs. 
    \item Finding features and unusual morphologies of DRAGNs that were not apparent in earlier observations. 
    \item Examining how the properties of \acp{DRAGN} change as we probe to fainter populations of sources.
\end{itemize}

It is also a reliable ``ground truth'' dataset for training and labelling automated algorithms, and for validating citizen science projects. 

Our catalogue is based on the EMU-PS1  Survey, which observed an area of about 270 square degrees, or about 1\% of the total area of EMU, and catalogued about 200,000 radio sources in that area. Among those sources, we find  \dragn\ DRAGN sources larger than about 1 arcmin, which constitute about 1.8\% of all sources. We estimate from these numbers that the entire EMU survey will detect about 20 million radio sources, of which about one quarter will be AGN, and the rest star-forming sources. Of the AGN, about 350,000 will be DRAGNs larger than about 1 arcmin.

Earlier low-sensitivity surveys suggested that most DRAGNs are relatively simple structures consisting of a core with two jets and two lobes. This catalogue differs from those earlier large surveys in having a greater sensitivity, due to the higher survey speed of the ASKAP telescope, and a better sensitivity to low surface brightness emission, due to the large number of short spacings in the antenna configuration of ASKAP. As a result, we see more diffuse structure and a plethora of more complex structures, ranging from wings of radio emission on the side of the jets, to types of object which have not been seen in earlier observations.
These new types of object in this survey include Odd Radio Circles, and the ``Trident'' sources. Possibly related to the trident sources are many otherwise unremarkable FR1/FR2 sources that are winged, such as X-shaped, Z-shaped, and T-shaped radio sources.

We also find a large number of Linear Triple Sources, which appear to be double-jetted or double-lobed AGN but without the maxima on either side that would enable them to be classified as FR1\ytag\ or FR2\ytag.

We also find significant numbers of other rare types of radio source such as HyMoRS and one-sided jets, as well as a large range of bent-tail and head-tail sources.

In this paper, we have presented the catalogue and classified the different types of morphology. In subsequent papers, we will add optical identifications and redshifts, and examine the astrophysical implications of this sample.

\section{Data Availability}
The radio images of the EMU Pilot Survey 1 are available on https://doi.org/10.25919/exq5-t894.
In the Supplementary Material, we include
\begin{itemize}
    \item The full table of results (Table \ref{catalogue}) in FITs format (Table7.fits) and csv format (Table7.csv).
    \item A catalogue file of DRAGNs suitable for CARTA and other packages (CARTA\_catalog.fits)
    \item A catalogue file of CATWISE hosts suitable for CARTA and other packages (CATWISE\_catalog.fits)
    \item A region file in DS9 format with xy position angles suitable for CARTA and DS9 (regions.ds9)
    \item A region file in DS9 format with WCS position angles suitable for Aladin (regions.wcs)
    \item a pdf file with cutout images of every source in this paper (cutouts.pdf)
\end{itemize}

\section*{Acknowledgements}
The original inspiration for this project was the PhD project of co-author Yew \citep{yew22}, but the results in this paper are largely independent of the data in that thesis.
 
This scientific work uses data obtained from Inyarrimanha Ilgari Bundara / the Murchison Radio-astronomy Observatory. We acknowledge the Wajarri Yamaji People as the Traditional Owners and native title holders of the Observatory site. The Australian SKA Pathfinder is part of the Australia Telescope National Facility (https://ror.org/05qajvd42) which is managed by CSIRO. Operation of ASKAP is funded by the Australian Government with support from the National Collaborative Research Infrastructure Strategy. ASKAP uses the resources of the Pawsey Supercomputing Centre. Establishment of ASKAP, the Murchison Radio-astronomy Observatory and the Pawsey Supercomputing Centre are initiatives of the Australian Government, with support from the Government of Western Australia and the Science and Industry Endowment Fund.

This paper includes archived data obtained through the CSIRO ASKAP Science Data Archive, CASDA (\url{http://data.csiro.au}).

We thank Lachlan Barnes, Baerbel Koribalski, Sarah White, and Ivy Wong for helpful comments on a draft of this paper.

\clearpage
\newpage\onecolumn
\begin{landscape}
\tiny
    \begin{longtable}{lllllllllllll}
    \caption{The source catalogue. The column descriptions are as follows: Columns 1 and 2 contain an index number and a J2000 name, which is derived from Columns 3 and 4. Columns 3 and 4 give the position of the centre of a rectangular region that just contains the source. Columns 5 and 6 give the major and minor axis of the region (parameters $b$ and $c$ in Section \ref{extraction}).
    Column 7 gives the position angle of this box (north through east) relative to the pixel grid of the image, as used by CARTA and DS9. Column 8 gives the true position angle of this box (north through east) relative to the WCS declination axis.  Column 9 gives the separation between the two peaks of the source (parameter $a$ in Section \ref{extraction}). Column 10 gives the total flux density integrated over the region, and Column  11 gives the CATWISE20 ID. Column 12 gives the tags  as described in Section \ref{sec:morphology}. The final column contains informal notes on the source.}
    \label{catalogue}\\
    \toprule
	& 		            & Box centroid  & Box centroid	& Box major	& Box minor	& Box	xy    &
    Box WCS &
    Peak	 & Total Flux& CATWISE & Tags	& 	Comments 	 \\
Index &  	Name	        &  	RA (\degr)	&  Dec (\degr)	&  axis ('')& axis ('')	& PA (\degr) & PA (\degr)&  	Sepn.('')& Dens (mJy)& ID & 	&  		 \\
1	&	J200613.7-523237	&	301.55734	&	-52.54366	&	96.8	&	53.9	&	78.3	&	85.7	&	66.1	&	101.9	&	J200611.73-523233.6	&	FR2\ytag,BT\ytag	&		\\
2	&	J200626.4-614932	&	301.61005	&	-61.82561	&	114.4	&	36.2	&	11.1	&	19.3	&	85.0	&	19.2	&	J200627.21-614924.9	&	FR2\ytag	&	W cpt is not part of source	\\
3	&	J200633.1-512447	&	301.63819	&	-51.41324	&	57.1	&	25.7	&	46.7	&	54.0	&	17.0	&	22.7	&	J200633.12-512447.0	&	FR1\ytag	&		\\
4	&	J200652.7-533819	&	301.71996	&	-53.63875	&	33.5	&	20.2	&	4.6	&	12.0	&	16.4	&	6.9	&	J200652.99-533814.4	&	FRX\ytag	&		\\
5	&	J200701.6-493302	&	301.75691	&	-49.55056	&	43.0	&	21.3	&	33.2	&	40.2	&	14.9	&	2.0	&	J200701.54-493301.7	&	FRX\ytag	&		\\
6	&	J200713.2-532345	&	301.80502	&	-53.39608	&	67.2	&	37.7	&	131.7	&	139.0	&	31.2	&	101.7	&	J200713.36-532345.1	&	FRX\ytag	&		\\
7	&	J200719.2-515252	&	301.83011	&	-51.88123	&	46.1	&	14.6	&	34.4	&	41.5	&	27.2	&	2.7	&	J200719.45-515251.6	&	FRX\ytag	&		\\
8	&	J200723.5-532428	&	301.84803	&	-53.40794	&	51.8	&	28.0	&	62.5	&	69.7	&	17.8	&	63.7	&	J200723.80-532427.0	&	FR1\ytag	&		\\
9	&	J200728.1-521307	&	301.86721	&	-52.21878	&	182.5	&	29.3	&	60.8	&	67.9	&	124.8	&	14.0	&	J200729.43-521307.9	&	FR2\ytag	&		\\
10	&	J200729.1-530909	&	301.87154	&	-53.15252	&	542.0	&	199.5	&	175.7	&	2.9	&	258.0	&	87.7	&	J200731.72-530832.8	&	FR1\ytag,BT\ytag	&		\\
11	&	J200732.0-503623	&	301.88250	&	-50.60646	&	98.2	&	32.2	&	89.3	&	96.3	&	139.8	&	7.8	&	J200732.63-503624.3	&	HyMoRS\ytag	&		\\
12	&	J200751.6-624026	&	301.96504	&	-62.67416	&	76.0	&	29.7	&	86.0	&	94.1	&	38.1	&	81.1	&	J200750.99-624026.7	&	FRX\ytag	&		\\
13	&	J200754.8-502329	&	301.97843	&	-50.39156	&	109.6	&	30.5	&	109.1	&	115.9	&	86.2	&	4.9	&	J200753.73-502329.5	&	FR2\ytag	&		\\
14	&	J200800.5-512318	&	302.00210	&	-51.38849	&	68.8	&	27.5	&	5.9	&	12.8	&	44.3	&	2.3	&		&	FR2\ytag	&		\\
15	&	J200804.4-623744	&	302.01872	&	-62.62895	&	41.0	&	25.7	&	36.0	&	44.0	&	19.4	&	3.0	&	J200804.30-623743.4	&	FRX\ytag	&		\\
16	&	J200814.6-523048	&	302.06099	&	-52.51340	&	95.4	&	29.7	&	143.2	&	150.2	&	55.0	&	4.1	&	J200814.43-523051.2	&	FR2\ytag	&		\\
17	&	J200816.0-512716	&	302.06693	&	-51.45457	&	66.3	&	32.7	&	32.1	&	39.0	&	0.0	&	3.2	&	J200815.28-512724.6	&	CPLX\ytag	&	blobby diffuse object	\\
18	&	J200825.8-525522	&	302.10787	&	-52.92288	&	153.6	&	23.0	&	102.0	&	109.0	&	126.0	&	6.8	&	J200825.17-525523.9	&	FR2\ytag	&		\\
19	&	J200826.7-513605	&	302.11140	&	-51.60150	&	59.4	&	23.5	&	25.8	&	32.6	&	39.1	&	3.2	&	J200826.79-513606.9	&	FR2\ytag	&		\\
20	&	J200828.0-530013	&	302.11683	&	-53.00368	&	104.8	&	37.8	&	68.9	&	75.9	&	57.0	&	77.8	&	J200828.58-530012.9	&	FRX\ytag	&	mildly twisted FR2	\\
21	&	J200837.2-515537	&	302.15521	&	-51.92719	&	71.8	&	36.9	&	124.2	&	131.1	&	33.5	&	30.5	&		&	FRX\ytag	&		\\
22	&	J200839.6-511046	&	302.16505	&	-51.17960	&	83.1	&	31.4	&	98.7	&	105.5	&	48.7	&	17.6	&	J200839.70-511043.2	&	FR2\ytag	&		\\
23	&	J200847.6-523010	&	302.19849	&	-52.50280	&	78.8	&	24.7	&	44.0	&	50.9	&	0.0	&	2.4	&	J200847.51-523009.1	&	LTS\ytag	&		\\
24	&	J200849.4-512756	&	302.20615	&	-51.46568	&	42.2	&	27.3	&	38.6	&	45.3	&	0.0	&	1.6	&	J200849.33-512803.7	&	TRG\ytag	&		\\
25	&	J200909.8-490005	&	302.29118	&	-49.00161	&	88.4	&	36.9	&	103.0	&	109.5	&	65.4	&	16.7	&		&	FR2\ytag	&	one lobe is double	\\
26	&	J200915.5-583606	&	302.31461	&	-58.60192	&	41.7	&	27.0	&	126.6	&	134.0	&	17.2	&	17.5	&	J200915.54-583606.7	&	FRX\ytag	&		\\
27	&	J200919.8-494023	&	302.33284	&	-49.67329	&	107.2	&	32.2	&	66.6	&	73.1	&	70.7	&	100.5	&	J200920.82-494020.6	&	FR2\ytag	&		\\
28	&	J200920.3-522359	&	302.33476	&	-52.39976	&	78.8	&	25.6	&	128.4	&	135.2	&	64.8	&	16.8	&	J200921.14-522405.7	&	FR2\ytag	&		\\
29	&	J200921.4-513808	&	302.33921	&	-51.63581	&	45.0	&	27.7	&	153.9	&	160.6	&	0.0	&	1.9	&	J200921.30-513806.2	&	LTS\ytag,BT\ytag	&		\\
30	&	J200932.4-531016	&	302.38518	&	-53.17119	&	53.6	&	36.6	&	117.5	&	124.2	&	24.6	&	9.2	&	J200933.79-531012.5	&	FRX\ytag,TRG\ytag	&	See Fig. 13	\\
      \bottomrule
    \end{longtable}
\end{landscape}
\onecolumn

\twocolumn
\bibliography{main} 

\begin{thebibliography}{}
\expandafter\ifx\csname natexlab\endcsname\relax\def\natexlab#1{#1}\fi

\bibitem[{{Banfield} {et~al.}(2015){Banfield}, {Wong}, {Willett}, {Norris}, {Rudnick}, {Shabala}, {Simmons}, {Snyder}, {Garon}, {Seymour}, {Middelberg}, {Andernach}, {Lintott}, {Jacob}, {Kapi{\'n}ska}, {Mao}, {Masters}, {Jarvis}, {Schawinski}, {Paget}, {Simpson}, {Kl{\"o}ckner}, {Bamford}, {Burchell}, {Chow}, {Cotter}, {Fortson}, {Heywood}, {Jones}, {Kaviraj}, {L{\'o}pez-S{\'a}nchez}, {Maksym}, {Polsterer}, {Borden}, {Hollow}, \& {Whyte}}]{banfield15}
{Banfield}, J.~K., {Wong}, O.~I., {Willett}, K.~W., {et~al.} 2015, \mnras, 453, 2326

\bibitem[{{Barkus} {et~al.}(2022){Barkus}, {Croston}, {Piotrowska}, {Mingo}, {Best}, {Hardcastle}, {Mostert}, {R{\"o}ttgering}, {Sabater}, {Webster}, \& {Williams}}]{barkus22}
{Barkus}, B., {Croston}, J.~H., {Piotrowska}, J., {et~al.} 2022, \mnras, 509, 1

\bibitem[{{Barnes}(2025)}]{barnes25}
{Barnes}, L.~e. 2025, \pasa, submitted

\bibitem[{{Baum} {et~al.}(1995){Baum}, {Zirbel}, \& {O'Dea}}]{baum95}
{Baum}, S.~A., {Zirbel}, E.~L., \& {O'Dea}, C.~P. 1995, \apj, 451, 88

\bibitem[{{Becker} {et~al.}(1995){Becker}, {White}, \& {Helfand}}]{Becker1995}
{Becker}, R.~H., {White}, R.~L., \& {Helfand}, D.~J. 1995, \apj, 450, 559

\bibitem[{{Bennett}(1962)}]{bennett62}
{Bennett}, A.~S. 1962, \memras, 68, 163

\bibitem[{{Bera} {et~al.}(2020){Bera}, {Pal}, {Sasmal}, \& {Mondal}}]{bera20}
{Bera}, S., {Pal}, S., {Sasmal}, T.~K., \& {Mondal}, S. 2020, \apjs, 251, 9

\bibitem[{{Best}(2009)}]{best09}
{Best}, P.~N. 2009, Astronomische Nachrichten, 330, 184

\bibitem[{{Bicknell}(1986)}]{bicknell86}
{Bicknell}, G.~V. 1986, \apj, 300, 591

\bibitem[{{Bicknell}(1995)}]{bicknell95}
---. 1995, \apjs, 101, 29

\bibitem[{{Blandford} \& {Icke}(1978)}]{blandford78}
{Blandford}, R.~D., \& {Icke}, V. 1978, \mnras, 185, 527

\bibitem[{{Bonaldi} {et~al.}(2019){Bonaldi}, {Bonato}, {Galluzzi}, {Harrison}, {Massardi}, {Kay}, {De Zotti}, \& {Brown}}]{bonaldi19}
{Bonaldi}, A., {Bonato}, M., {Galluzzi}, V., {et~al.} 2019, \mnras, 482, 2

\bibitem[{{Bowles} {et~al.}(2023){Bowles}, {Tang}, {Vardoulaki}, {Alexander}, {Luo}, {Rudnick}, {Walmsley}, {Porter}, {Scaife}, {Slijepcevic}, {Adams}, {Drabent}, {Dugdale}, {G{\"u}rkan}, {Hopkins}, {Jimenez-Andrade}, {Leahy}, {Norris}, {Rahman}, {Ouyang}, {Segal}, {Shabala}, \& {Wong}}]{bowles23}
{Bowles}, M., {Tang}, H., {Vardoulaki}, E., {et~al.} 2023, \mnras, 522, 2584

\bibitem[{{Boyce} {et~al.}(2023){Boyce}, {Hopkins}, {Riggi}, {Rudnick}, {Ramsay}, {Hale}, {Marvil}, {Whiting}, {Venkataraman}, {O'Dea}, {Baum}, {Gordon}, {Vantyghem}, {Dionyssiou}, {Andernach}, {Collier}, {English}, {Koribalski}, {Leahy}, {Micha{\l}owski}, {Safi-Harb}, {Vaccari}, {Alexander}, {Cowley}, {Kapinska}, {Robotham}, \& {Tang}}]{boyce23}
{Boyce}, M.~M., {Hopkins}, A.~M., {Riggi}, S., {et~al.} 2023, \pasa, 40, e027

\bibitem[{{Brienza} {et~al.}(2017){Brienza}, {Godfrey}, {Morganti}, {Prandoni}, {Harwood}, {Mahony}, {Hardcastle}, {Murgia}, {R{\"o}ttgering}, {Shimwell}, \& {Shulevski}}]{brienza17}
{Brienza}, M., {Godfrey}, L., {Morganti}, R., {et~al.} 2017, \aap, 606, A98

\bibitem[{{Brocksopp} {et~al.}(2007){Brocksopp}, {Kaiser}, {Schoenmakers}, \& {de Bruyn}}]{brocksopp07}
{Brocksopp}, C., {Kaiser}, C.~R., {Schoenmakers}, A.~P., \& {de Bruyn}, A.~G. 2007, \mnras, 382, 1019

\bibitem[{{Cheung}(2007)}]{cheung07}
{Cheung}, C.~C. 2007, \aj, 133, 2097

\bibitem[{Comrie {et~al.}(2018)Comrie, Wang, Hwang, Moraghan, Harris, Pińska, Raul-Omar, Chiang, Chang, Hsu, Pang, Simmonds, Lin, \& Jan}]{carta}
Comrie, A., Wang, K.-S., Hwang, Y.-H., {et~al.} 2018, {CARTA: The Cube Analysis and Rendering Tool for Astronomy}, doi:\url{10.5281/zenodo.3377984}

\bibitem[{{Condon} {et~al.}(1998){Condon}, {Cotton}, {Greisen}, {Yin}, {Perley}, {Taylor}, \& {Broderick}}]{condon98}
{Condon}, J.~J., {Cotton}, W.~D., {Greisen}, E.~W., {et~al.} 1998, \aj, 115, 1693

\bibitem[{{Contigiani} {et~al.}(2017){Contigiani}, {de Gasperin}, {Miley}, {Rudnick}, {Andernach}, {Banfield}, {Kapi{\'n}ska}, {Shabala}, \& {Wong}}]{contigiani17}
{Contigiani}, O., {de Gasperin}, F., {Miley}, G.~K., {et~al.} 2017, \mnras, 472, 636

\bibitem[{{Cotton} {et~al.}(2020){Cotton}, {Thorat}, {Condon}, {Frank}, {J{\'o}zsa}, {White}, {Deane}, {Oozeer}, {Atemkeng}, {Bester}, {Fanaroff}, {Kupa}, {Smirnov}, {Mauch}, {Krishnan}, \& {Camilo}}]{cotton20}
{Cotton}, W.~D., {Thorat}, K., {Condon}, J.~J., {et~al.} 2020, \mnras, 495, 1271

\bibitem[{{Courvoisier}(1998)}]{courvoisier98}
{Courvoisier}, T. J.~L. 1998, \aapr, 9, 1

\bibitem[{{Dabhade} {et~al.}(2024){Dabhade}, {Chavan}, {Saikia}, {Oei}, \& {Rottgering}}]{dabhade24}
{Dabhade}, P., {Chavan}, K., {Saikia}, D.~J., {Oei}, M. S.~S.~L., \& {Rottgering}, H. J.~A. 2024, arXiv e-prints, arXiv:2408.13607

\bibitem[{{de la Rosa Vald{\'e}s} \& {Andernach}(2019)}]{rosa19}
{de la Rosa Vald{\'e}s}, P.~A., \& {Andernach}, H. 2019, arXiv e-prints, arXiv:1908.09988

\bibitem[{{de Ruiter} {et~al.}(1990){de Ruiter}, {Parma}, {Fanti}, \& {Fanti}}]{ruiter90}
{de Ruiter}, H.~R., {Parma}, P., {Fanti}, C., \& {Fanti}, R. 1990, \aap, 227, 351

\bibitem[{{Dekel} \& {Birnboim}(2006)}]{dekel06}
{Dekel}, A., \& {Birnboim}, Y. 2006, \mnras, 368, 2

\bibitem[{{Duchesne} {et~al.}(2024){Duchesne}, {Botteon}, {Koribalski}, {Loi}, {Rajpurohit}, {Riseley}, {Rudnick}, {Vernstrom}, {Andernach}, {Hopkins}, {Kapinska}, {Norris}, \& {Zafar}}]{duchesne24}
{Duchesne}, S.~W., {Botteon}, A., {Koribalski}, B.~S., {et~al.} 2024, \pasa, 41, e026

\bibitem[{{Ekers} {et~al.}(1978){Ekers}, {Fanti}, {Lari}, \& {Parma}}]{ekers78}
{Ekers}, R.~D., {Fanti}, R., {Lari}, C., \& {Parma}, P. 1978, \nat, 276, 588

\bibitem[{{Fanaroff} \& {Riley}(1974)}]{FR}
{Fanaroff}, B.~L., \& {Riley}, J.~M. 1974, \mnras, 167, 31P

\bibitem[{{Galvin} {et~al.}(2020){Galvin}, {Huynh}, {Norris}, {Wang}, {Hopkins}, {Polsterer}, {Ralph}, {O'Brien}, \& {Heald}}]{galvin20}
{Galvin}, T.~J., {Huynh}, M.~T., {Norris}, R.~P., {et~al.} 2020, \mnras, 497, 2730

\bibitem[{{Gawro{\'n}ski} {et~al.}(2006){Gawro{\'n}ski}, {Marecki}, {Kunert-Bajraszewska}, \& {Kus}}]{gawronski06}
{Gawro{\'n}ski}, M.~P., {Marecki}, A., {Kunert-Bajraszewska}, M., \& {Kus}, A.~J. 2006, \aap, 447, 63

\bibitem[{{Gendre} {et~al.}(2010){Gendre}, {Best}, \& {Wall}}]{gendre10}
{Gendre}, M.~A., {Best}, P.~N., \& {Wall}, J.~V. 2010, \mnras, 404, 1719

\bibitem[{{Giacintucci} \& {Venturi}(2009)}]{giacintucci09}
{Giacintucci}, S., \& {Venturi}, T. 2009, \aap, 505, 55

\bibitem[{{Gopal-Krishna} {et~al.}(2025){Gopal-Krishna}, {Patra}, \& {Joshi}}]{gopalkrishna24}
{Gopal-Krishna}, {Patra}, D., \& {Joshi}, R. 2025, Journal of Astrophysics and Astronomy, 46, 36

\bibitem[{{Gopal-Krishna} \& {Wiita}(2000)}]{gopalkrishna00}
{Gopal-Krishna}, \& {Wiita}, P.~J. 2000, \aap, 363, 507

\bibitem[{{Gopal-Krishna} \& {Wiita}(2001)}]{gopalkrishna01}
---. 2001, \aap, 373, 100

\bibitem[{{Gopal-Krishna} \& {Wiita}(2002)}]{gopalkrishna02}
---. 2002, \nar, 46, 357

\bibitem[{{Gordon} {et~al.}(2021){Gordon}, {Boyce}, {O'Dea}, {Rudnick}, {Andernach}, {Vantyghem}, {Baum}, {Bui}, {Dionyssiou}, \& {Sander}}]{gordon2021}
{Gordon}, Y.~A., {Boyce}, M.~M., {O'Dea}, C.~P., {et~al.} 2021, arXiv e-prints, arXiv:2102.11753

\bibitem[{{Gordon} {et~al.}(2023){Gordon}, {Rudnick}, {Andernach}, {Morabito}, {O'Dea}, {Achong}, {Baum}, {Bayona-Figueroa}, {Hooper}, {Mingo}, {Morris}, \& {Vantyghem}}]{gordon23}
{Gordon}, Y.~A., {Rudnick}, L., {Andernach}, H., {et~al.} 2023, \apjs, 267, 37

\bibitem[{{Gupta} {et~al.}(2024{\natexlab{a}}){Gupta}, {Hayder}, {Norris}, {Huynh}, \& {Petersson}}]{gupta24a}
{Gupta}, N., {Hayder}, Z., {Norris}, R.~P., {Huynh}, M., \& {Petersson}, L. 2024{\natexlab{a}}, \pasa, 41, e001

\bibitem[{{Gupta} {et~al.}(2023{\natexlab{a}}){Gupta}, {Hayder}, {Norris}, {Hyunh}, \& {Petersson}}]{gupta23b}
{Gupta}, N., {Hayder}, Z., {Norris}, R.~P., {Hyunh}, M., \& {Petersson}, L. 2023{\natexlab{a}}, NeurIPS ML4PS 2023, arXiv:2312.06728

\bibitem[{{Gupta} {et~al.}(2022){Gupta}, {Huynh}, {Norris}, {Wang}, {Hopkins}, {Andernach}, {Koribalski}, \& {Galvin}}]{gupta22}
{Gupta}, N., {Huynh}, M., {Norris}, R.~P., {et~al.} 2022, \pasa, 39, e051

\bibitem[{{Gupta} {et~al.}(2023{\natexlab{b}}){Gupta}, {Hayder}, {Norris}, {Huynh}, {Petersson}, {Wang}, {Andernach}, {Koribalski}, {Yew}, \& {Crawford}}]{gupta23}
{Gupta}, N., {Hayder}, Z., {Norris}, R.~P., {et~al.} 2023{\natexlab{b}}, \pasa, 40, e044

\bibitem[{{Gupta} {et~al.}(2024{\natexlab{b}}){Gupta}, {Norris}, {Hayder}, {Huynh}, {Petersson}, {Rosalind Wang}, {Hopkins}, {Andernach}, {Gordon}, {Riggi}, {Yew}, {Crawford}, {Koribalski}, {Filipovi{\'c}}, {Kapi{\'n}ska}, {Shabala}, {Vernstrom}, \& {Marvil}}]{gupta24b}
{Gupta}, N., {Norris}, R.~P., {Hayder}, Z., {et~al.} 2024{\natexlab{b}}, \pasa, 41, e027

\bibitem[{{Gupta} {et~al.}(2025{\natexlab{a}}){Gupta}, {Norris}, {Hayder}, {Huynh}, {Andernach}, {Hopkins}, {Shabala}, {Rudnick}, {Filipovi{\'c}}, {Koribalski}, {Petersson}, \& {Wang}}]{gupta25submit}
---. 2025{\natexlab{a}}, arXiv e-prints, arXiv:2506.08439

\bibitem[{{Gupta} {et~al.}(2025{\natexlab{b}}){Gupta}, {Hayder}, {Huynh}, {Norris}, {Petersson}, {Hopkins}, {Riggi}, {Koribalski}, \& {Filipovi{\'c}}}]{gupta25emuse}
{Gupta}, N., {Hayder}, Z., {Huynh}, M., {et~al.} 2025{\natexlab{b}}, arXiv e-prints, arXiv:2506.15090

\bibitem[{{Hale} {et~al.}(2023){Hale}, {Whittam}, {Jarvis}, {Best}, {Thomas}, {Heywood}, {Prescott}, {Adams}, {Afonso}, {An}, {Bowler}, {Collier}, {Cook}, {Dav{\'e}}, {Frank}, {Glowacki}, {Hatfield}, {Kolwa}, {Lovell}, {Maddox}, {Marchetti}, {Morabito}, {Murphy}, {Prandoni}, {Randriamanakoto}, \& {Taylor}}]{mightee}
{Hale}, C.~L., {Whittam}, I.~H., {Jarvis}, M.~J., {et~al.} 2023, \mnras, 520, 2668

\bibitem[{{Hardcastle} {et~al.}(2019){Hardcastle}, {Williams}, {Best}, {Croston}, {Duncan}, {R{\"o}ttgering}, {Sabater}, {Shimwell}, {Tasse}, {Callingham}, {Cochrane}, {de Gasperin}, {G{\"u}rkan}, {Jarvis}, {Mahatma}, {Miley}, {Mingo}, {Mooney}, {Morabito}, {O'Sullivan}, {Prandoni}, {Shulevski}, \& {Smith}}]{hardcastle19}
{Hardcastle}, M.~J., {Williams}, W.~L., {Best}, P.~N., {et~al.} 2019, \aap, 622, A12

\bibitem[{{Hardcastle} {et~al.}(2023){Hardcastle}, {Horton}, {Williams}, {Duncan}, {Alegre}, {Barkus}, {Croston}, {Dickinson}, {Osinga}, {R{\"o}ttgering}, {Sabater}, {Shimwell}, {Smith}, {Best}, {Botteon}, {Br{\"u}ggen}, {Drabent}, {de Gasperin}, {G{\"u}rkan}, {Hajduk}, {Hale}, {Hoeft}, {Jamrozy}, {Kunert-Bajraszewska}, {Kondapally}, {Magliocchetti}, {Mahatma}, {Mostert}, {O'Sullivan}, {Pajdosz-{\'S}mierciak}, {Petley}, {Pierce}, {Prandoni}, {Schwarz}, {Shulewski}, {Siewert}, {Stott}, {Tang}, {Vaccari}, {Zheng}, {Bailey}, {Desbled}, {Goyal}, {Gonano}, {Hanset}, {Kurtz}, {Lim}, {Mielle}, {Molloy}, {Roth}, {Terentev}, \& {Torres}}]{hardcastle23}
{Hardcastle}, M.~J., {Horton}, M.~A., {Williams}, W.~L., {et~al.} 2023, \aap, 678, A151

\bibitem[{{Hardcastle} {et~al.}(2025){Hardcastle}, {Pierce}, {Duncan}, {G{\"u}rkan}, {Gong}, {Horton}, {Mingo}, {R{\"o}ttgering}, \& {Smith}}]{hardcastle25}
{Hardcastle}, M.~J., {Pierce}, J.~C.~S., {Duncan}, K.~J., {et~al.} 2025, \mnras, 539, 1856

\bibitem[{{Harwood} {et~al.}(2020){Harwood}, {Vernstrom}, \& {Stroe}}]{harwood20}
{Harwood}, J.~J., {Vernstrom}, T., \& {Stroe}, A. 2020, \mnras, 491, 803

\bibitem[{{Hopkins} {et~al.}(2025){Hopkins}, {Kapinska}, {Marvil}, {Vernstrom}, {Collier}, {Norris}, {Gordon}, {Duchesne}, {Rudnick}, {Gupta}, {Carretti}, {Anderson}, {Dai}, {G{\"u}rkan}, {Parkinson}, {Prandoni}, {Riggi}, {Shekhar Saraf}, {Ma}, {Filipovi{\'c}}, {Umana}, {Bahr-Kalus}, {Koribalski}, {Lenc}, {Ingallinera}, {Afonso}, {Ahmad}, {Ahmed}, {Alexander}, {Andernach}, {Asorey}, {Battisti}, {Bilicki}, {Botteon}, {Brown}, {Br{\"u}ggen}, {Cowley}, {Dage}, {Hale}, {Hardcastle}, {Kothes}, {Lazarevi{\'c}}, {Lin}, {Luken}, {Moss}, {Prathap}, {ur Rahman}, {Reiprich}, {Riseley}, {Salvato}, {Seymour}, {Shabala}, {Smith}, {Vaccari}, {van Loon}, {Wong}, {Zainal Alsaberi}, {Asher}, {Ball}, {Barbosa}, {Biava}, {Bradley}, {Carvajal}, {Crawford}, {Galvin}, {Huynh}, {Leahy}, {Matute}, {Moss}, {Pappalardo}, {Smeaton}, {Velovi{\'c}}, \& {Zafar}}]{hopkins25}
{Hopkins}, A., {Kapinska}, A., {Marvil}, J., {et~al.} 2025, \pasa, 42, e071

\bibitem[{{Hotan} {et~al.}(2021){Hotan}, {Bunton}, {Chippendale}, {Whiting}, {Tuthill}, {Moss}, {McConnell}, {Amy}, {Huynh}, {Allison}, {Anderson}, {Bannister}, {Bastholm}, {Beresford}, {Bock}, {Bolton}, {Chapman}, {Chow}, {Collier}, {Cooray}, {Cornwell}, {Diamond}, {Edwards}, {Feain}, {Franzen}, {George}, {Gupta}, {Hampson}, {Harvey-Smith}, {Hayman}, {Heywood}, {Jacka}, {Jackson}, {Jackson}, {Jeganathan}, {Johnston}, {Kesteven}, {Kleiner}, {Koribalski}, {Lee-Waddell}, {Lenc}, {Lensson}, {Mackay}, {Mahony}, {McClure-Griffiths}, {McConigley}, {Mirtschin}, {Ng}, {Norris}, {Pearce}, {Phillips}, {Pilawa}, {Raja}, {Reynolds}, {Roberts}, {Roxby}, {Sadler}, {Shields}, {Schinckel}, {Serra}, {Shaw}, {Sweetnam}, {Troup}, {Tzioumis}, {Voronkov}, \& {Westmeier}}]{hotan21}
{Hotan}, A.~W., {Bunton}, J.~D., {Chippendale}, A.~P., {et~al.} 2021, \pasa, 38, e009

\bibitem[{{Johnston} {et~al.}(2007){Johnston}, {Bailes}, {Bartel}, {Baugh}, {Bietenholz}, {Blake}, {Braun}, {Brown}, {Chatterjee}, {Darling}, {Deller}, {Dodson}, {Edwards}, {Ekers}, {Ellingsen}, {Feain}, {Gaensler}, {Haverkorn}, {Hobbs}, {Hopkins}, {Jackson}, {James}, {Joncas}, {Kaspi}, {Kilborn}, {Koribalski}, {Kothes}, {Landecker}, {Lenc}, {Lovell}, {Macquart}, {Manchester}, {Matthews}, {McClure-Griffiths}, {Norris}, {Pen}, {Phillips}, {Power}, {Protheroe}, {Sadler}, {Schmidt}, {Stairs}, {Staveley-Smith}, {Stil}, {Taylor}, {Tingay}, {Tzioumis}, {Walker}, {Wall}, \& {Wolleben}}]{johnston07}
{Johnston}, S., {Bailes}, M., {Bartel}, N., {et~al.} 2007, \pasa, 24, 174

\bibitem[{{Johnston} {et~al.}(2008){Johnston}, {Taylor}, {Bailes}, {Bartel}, {Baugh}, {Bietenholz}, {Blake}, {Braun}, {Brown}, {Chatterjee}, {Darling}, {Deller}, {Dodson}, {Edwards}, {Ekers}, {Ellingsen}, {Feain}, {Gaensler}, {Haverkorn}, {Hobbs}, {Hopkins}, {Jackson}, {James}, {Joncas}, {Kaspi}, {Kilborn}, {Koribalski}, {Kothes}, {Landecker}, {Lenc}, {Lovell}, {Macquart}, {Manchester}, {Matthews}, {McClure-Griffiths}, {Norris}, {Pen}, {Phillips}, {Power}, {Protheroe}, {Sadler}, {Schmidt}, {Stairs}, {Staveley-Smith}, {Stil}, {Tingay}, {Tzioumis}, {Walker}, {Wall}, \& {Wolleben}}]{johnston08}
{Johnston}, S., {Taylor}, R., {Bailes}, M., {et~al.} 2008, Experimental Astronomy, 22, 151

\bibitem[{{Jonas} \& {MeerKAT Team}(2016)}]{jonas16}
{Jonas}, J., \& {MeerKAT Team}. 2016, in MeerKAT Science: On the Pathway to the SKA, 1

\bibitem[{{Joshi} {et~al.}(2019){Joshi}, {Krishna}, {Yang}, {Shi}, {Yu}, {Wiita}, {Ho}, {Wu}, {An}, {Wang}, {Subramanian}, \& {Yesuf}}]{joshi19}
{Joshi}, R., {Krishna}, G., {Yang}, X., {et~al.} 2019, \apj, 887, 266

\bibitem[{{Kaiser} \& {Best}(2007)}]{kaiser07}
{Kaiser}, C.~R., \& {Best}, P.~N. 2007, \mnras, 381, 1548

\bibitem[{{Kapahi}(1981)}]{kapahi81}
{Kapahi}, V.~K. 1981, Journal of Astrophysics and Astronomy, 2, 43

\bibitem[{{Kapi{\'n}ska} {et~al.}(2017){Kapi{\'n}ska}, {Terentev}, {Wong}, {Shabala}, {Andernach}, {Rudnick}, {Storer}, {Banfield}, {Willett}, {de Gasperin}, {Lintott}, {L{\'o}pez-S{\'a}nchez}, {Middelberg}, {Norris}, {Schawinski}, {Seymour}, \& {Simmons}}]{kapinska17}
{Kapi{\'n}ska}, A.~D., {Terentev}, I., {Wong}, O.~I., {et~al.} 2017, \aj, 154, 253

\bibitem[{{Keel} {et~al.}(2022){Keel}, {Tate}, {Wong}, {Banfield}, {Lintott}, {Masters}, {Simmons}, {Scarlata}, {Cardamone}, {Smethurst}, {Fortson}, {Shanahan}, {Kruk}, {Garland}, {Hancock}, \& {O'Ryan}}]{keel22}
{Keel}, W.~C., {Tate}, J., {Wong}, O.~I., {et~al.} 2022, \aj, 163, 150

\bibitem[{{Kellermann} \& {Owen}(1988)}]{kellermann88}
{Kellermann}, K.~I., \& {Owen}, F.~N. 1988, in Galactic and Extragalactic Radio Astronomy, ed. K.~I. {Kellermann} \& G.~L. {Verschuur} (Springer-Verlag), 563--602, {Reproduced in ned.ipac.caltech.edu/level5/Sept04/Kellermann2/Kellermann\_contents.html}

\bibitem[{{Knowles} {et~al.}(2022){Knowles}, {Cotton}, {Rudnick}, {Camilo}, {Goedhart}, {Deane}, {Ramatsoku}, {Bietenholz}, {Br{\"u}ggen}, {Button}, {Chen}, {Chibueze}, {Clarke}, {de Gasperin}, {Ianjamasimanana}, {J{\'o}zsa}, {Hilton}, {Kesebonye}, {Kolokythas}, {Kraan-Korteweg}, {Lawrie}, {Lochner}, {Loubser}, {Marchegiani}, {Mhlahlo}, {Moodley}, {Murphy}, {Namumba}, {Oozeer}, {Parekh}, {Pillay}, {Passmoor}, {Ramaila}, {Ranchod}, {Retana-Montenegro}, {Sebokolodi}, {Sikhosana}, {Smirnov}, {Thorat}, {Venturi}, {Abbott}, {Adam}, {Adams}, {Aldera}, {Bauermeister}, {Bennett}, {Bode}, {Botha}, {Botha}, {Brederode}, {Buchner}, {Burger}, {Cheetham}, {de Villiers}, {Dikgale-Mahlakoana}, {du Toit}, {Esterhuyse}, {Fadana}, {Fanaroff}, {Fataar}, {Foley}, {Fourie}, {Frank}, {Gamatham}, {Gatsi}, {Geyer}, {Gouws}, {Gumede}, {Heywood}, {Hlakola}, {Hokwana}, {Hoosen}, {Horn}, {Horrell}, {Hugo}, {Isaacson}, {Jonas}, {Jordaan}, {Joubert}, {Julie}, {Kapp}, {Kasper}, {Kenyon}, {Kotz{\'e}}, {Kotze}, {Kriek}, {Kriel}, {Krishnan},
  {Kusel}, {Legodi}, {Lehmensiek}, {Liebenberg}, {Lord}, {Lunsky}, {Madisa}, {Magnus}, {Main}, {Makhaba}, {Makhathini}, {Malan}, {Manley}, {Marais}, {Maree}, {Martens}, {Mauch}, {McAlpine}, {Merry}, {Millenaar}, {Mokone}, {Monama}, {Mphego}, {New}, {Ngcebetsha}, {Ngoasheng}, {Ockards}, {Otto}, {Patel}, {Peens-Hough}, {Perkins}, {Ramanujam}, {Ramudzuli}, {Ratcliffe}, {Renil}, {Robyntjies}, {Rust}, {Salie}, {Sambu}, {Schollar}, {Schwardt}, {Schwartz}, {Serylak}, {Siebrits}, {Sirothia}, {Slabber}, {Sofeya}, {Taljaard}, {Tasse}, {Tiplady}, {Toruvanda}, {Twum}, {van Balla}, {van der Byl}, {van der Merwe}, {van Dyk}, {Van Tonder}, {Van Wyk}, {Venter}, {Venter}, {Welz}, {Williams}, \& {Xaia}}]{knowles22}
{Knowles}, K., {Cotton}, W.~D., {Rudnick}, L., {et~al.} 2022, \aap, 657, A56

\bibitem[{{Koribalski} \& {et al.}(2025)}]{koribalski25}
{Koribalski}, B., \& {et al.} 2025, \pasa, in preparation

\bibitem[{{Koribalski} {et~al.}(2020){Koribalski}, {Staveley-Smith}, {Westmeier}, {Serra}, {Spekkens}, {Wong}, {Lee-Waddell}, {Lagos}, {Obreschkow}, {Ryan-Weber}, {Zwaan}, {Kilborn}, {Bekiaris}, {Bekki}, {Bigiel}, {Boselli}, {Bosma}, {Catinella}, {Chauhan}, {Cluver}, {Colless}, {Courtois}, {Crain}, {de Blok}, {D{\'e}nes}, {Duffy}, {Elagali}, {Fluke}, {For}, {Heald}, {Henning}, {Hess}, {Holwerda}, {Howlett}, {Jarrett}, {Jones}, {Jones}, {J{\'o}zsa}, {Jurek}, {J{\"u}tte}, {Kamphuis}, {Karachentsev}, {Kerp}, {Kleiner}, {Kraan-Korteweg}, {L{\'o}pez-S{\'a}nchez}, {Madrid}, {Meyer}, {Mould}, {Murugeshan}, {Norris}, {Oh}, {Oosterloo}, {Popping}, {Putman}, {Reynolds}, {Rhee}, {Robotham}, {Ryder}, {Schr{\"o}der}, {Shao}, {Stevens}, {Taylor}, {van{\^A} der Hulst}, {Verdes-Montenegro}, {Wakker}, {Wang}, {Whiting}, {Winkel}, \& {Wolf}}]{wallaby}
{Koribalski}, B.~S., {Staveley-Smith}, L., {Westmeier}, T., {et~al.} 2020, \apss, 365, 118

\bibitem[{{Krause} {et~al.}(2012){Krause}, {Alexander}, {Riley}, \& {Hopton}}]{krause12}
{Krause}, M., {Alexander}, P., {Riley}, J., \& {Hopton}, D. 2012, \mnras, 427, 3196

\bibitem[{{Kumari} \& {Pal}(2022)}]{kumari22}
{Kumari}, S., \& {Pal}, S. 2022, \mnras, 514, 4290

\bibitem[{{Lacy} {et~al.}(2020){Lacy}, {Baum}, {Chandler}, {Chatterjee}, {Clarke}, {Deustua}, {English}, {Farnes}, {Gaensler}, {Gugliucci}, {Hallinan}, {Kent}, {Kimball}, {Law}, {Lazio}, {Marvil}, {Mao}, {Medlin}, {Mooley}, {Murphy}, {Myers}, {Osten}, {Richards}, {Rosolowsky}, {Rudnick}, {Schinzel}, {Sivakoff}, {Sjouwerman}, {Taylor}, {White}, {Wrobel}, {Andernach}, {Beasley}, {Berger}, {Bhatnager}, {Birkinshaw}, {Bower}, {Brandt}, {Brown}, {Burke-Spolaor}, {Butler}, {Comerford}, {Demorest}, {Fu}, {Giacintucci}, {Golap}, {G{\"u}th}, {Hales}, {Hiriart}, {Hodge}, {Horesh}, {Ivezi{\'c}}, {Jarvis}, {Kamble}, {Kassim}, {Liu}, {Loinard}, {Lyons}, {Masters}, {Mezcua}, {Moellenbrock}, {Mroczkowski}, {Nyland}, {O'Dea}, {O'Sullivan}, {Peters}, {Radford}, {Rao}, {Robnett}, {Salcido}, {Shen}, {Sobotka}, {Witz}, {Vaccari}, {van Weeren}, {Vargas}, {Williams}, \& {Yoon}}]{lacy20}
{Lacy}, M., {Baum}, S.~A., {Chandler}, C.~J., {et~al.} 2020, \pasp, 132, 035001

\bibitem[{{Laing} \& {Bridle}(2002)}]{laing02}
{Laing}, R.~A., \& {Bridle}, A.~H. 2002, \mnras, 336, 1161

\bibitem[{{Laing} {et~al.}(1999){Laing}, {Parma}, {de Ruiter}, \& {Fanti}}]{laing99}
{Laing}, R.~A., {Parma}, P., {de Ruiter}, H.~R., \& {Fanti}, R. 1999, \mnras, 306, 513

\bibitem[{{Laing} {et~al.}(1983){Laing}, {Riley}, \& {Longair}}]{laing83}
{Laing}, R.~A., {Riley}, J.~M., \& {Longair}, M.~S. 1983, \mnras, 204, 151

\bibitem[{{Lal} \& {Rao}(2005)}]{lal05}
{Lal}, D.~V., \& {Rao}, A.~P. 2005, \mnras, 356, 232

\bibitem[{{Lao} {et~al.}(2025){Lao}, {Andernach}, {Yang}, {Zhang}, {Zhao}, {Zhao}, {Yu}, {Sun}, \& {Qin}}]{lao25}
{Lao}, B., {Andernach}, H., {Yang}, X., {et~al.} 2025, \apjs, 276, 46

\bibitem[{{Leahy}(1993)}]{leahy93}
{Leahy}, J.~P. 1993, in Jets in Extragalactic Radio Sources, ed. H.-J. {R{\"o}ser} \& K.~{Meisenheimer}, Vol. 421, 1

\bibitem[{{Leahy} \& {Williams}(1984)}]{leahy84}
{Leahy}, J.~P., \& {Williams}, A.~G. 1984, \mnras, 210, 929

\bibitem[{{Ledlow} \& {Owen}(1996)}]{ledlow96}
{Ledlow}, M.~J., \& {Owen}, F.~N. 1996, \aj, 112, 9

\bibitem[{{Macgregor} \& {et al.}(2025)}]{macgregor25}
{Macgregor}, P., \& {et al.} 2025, \pasa, in preparation

\bibitem[{{Mackay}(1971)}]{mackay71}
{Mackay}, C.~D. 1971, \mnras, 154, 209

\bibitem[{{Mahatma} {et~al.}(2020){Mahatma}, {Hardcastle}, {Croston}, {Harwood}, {Ineson}, \& {Moldon}}]{mahatma20}
{Mahatma}, V.~H., {Hardcastle}, M.~J., {Croston}, J.~H., {et~al.} 2020, \mnras, 491, 5015

\bibitem[{{Mahatma} {et~al.}(2019){Mahatma}, {Hardcastle}, {Williams}, {Best}, {Croston}, {Duncan}, {Mingo}, {Morganti}, {Brienza}, {Cochrane}, {G{\"u}rkan}, {Harwood}, {Jarvis}, {Jamrozy}, {Jurlin}, {Morabito}, {R{\"o}ttgering}, {Sabater}, {Shimwell}, {Smith}, {Shulevski}, \& {Tasse}}]{mahatma19}
{Mahatma}, V.~H., {Hardcastle}, M.~J., {Williams}, W.~L., {et~al.} 2019, \aap, 622, A13

\bibitem[{{Mao} {et~al.}(2009){Mao}, {Johnston-Hollitt}, {Stevens}, \& {Wotherspoon}}]{mao09}
{Mao}, M.~Y., {Johnston-Hollitt}, M., {Stevens}, J.~B., \& {Wotherspoon}, S.~J. 2009, \mnras, 392, 1070

\bibitem[{{Marocco} {et~al.}(2020){Marocco}, {Eisenhardt}, {Fowler}, {Kirkpatrick}, {Meisner}, {Schlafly}, {Stanford}, {Garcia}, {Caselden}, {Cushing}, {Cutri}, {Faherty}, {Gelino}, {Gonzalez}, {Jarrett}, {Koontz}, {Mainzer}, {Marchese}, {Mobasher}, {Schlegel}, {Stern}, {Teplitz}, \& {Wright}}]{catwise}
{Marocco}, F., {Eisenhardt}, P. R.~M., {Fowler}, J.~W., {et~al.} 2020, arXiv e-prints, arXiv:2012.13084

\bibitem[{{Meier}(1999)}]{meier99}
{Meier}, D.~L. 1999, \apj, 522, 753

\bibitem[{{Meisner} {et~al.}(2019){Meisner}, {Lang}, {Schlafly}, \& {Schlegel}}]{meisner19}
{Meisner}, A.~M., {Lang}, D., {Schlafly}, E.~F., \& {Schlegel}, D.~J. 2019, \pasp, 131, 124504

\bibitem[{{Meisner} {et~al.}(2017){Meisner}, {Lang}, \& {Schlegel}}]{meisner17}
{Meisner}, A.~M., {Lang}, D., \& {Schlegel}, D.~J. 2017, \aj, 154, 161

\bibitem[{{Miley}(1980)}]{miley80}
{Miley}, G. 1980, \araa, 18, 165

\bibitem[{{Miley} {et~al.}(1972){Miley}, {Perola}, {van der Kruit}, \& {van der Laan}}]{miley72}
{Miley}, G.~K., {Perola}, G.~C., {van der Kruit}, P.~C., \& {van der Laan}, H. 1972, \nat, 237, 269

\bibitem[{{Mingo} {et~al.}(2019){Mingo}, {Croston}, {Hardcastle}, {Best}, {Duncan}, {Morganti}, {Rottgering}, {Sabater}, {Shimwell}, {Williams}, {Brienza}, {Gurkan}, {Mahatma}, {Morabito}, {Prandoni}, {Bondi}, {Ineson}, \& {Mooney}}]{mingo19}
{Mingo}, B., {Croston}, J.~H., {Hardcastle}, M.~J., {et~al.} 2019, \mnras, 488, 2701

\bibitem[{{Miraghaei} \& {Best}(2017)}]{mirabest17}
{Miraghaei}, H., \& {Best}, P.~N. 2017, \mnras, 466, 4346

\bibitem[{{Missaglia} {et~al.}(2019){Missaglia}, {Massaro}, {Capetti}, {Paolillo}, {Kraft}, {Baldi}, \& {Paggi}}]{missaglia19}
{Missaglia}, V., {Massaro}, F., {Capetti}, A., {et~al.} 2019, \aap, 626, A8

\bibitem[{{Mohan} \& {Rafferty}(2015)}]{mohan15}
{Mohan}, N., \& {Rafferty}, D. 2015, {PyBDSF: Python Blob Detection and Source Finder}, Astrophysics Source Code Library, record ascl:1502.007

\bibitem[{{Mostert} {et~al.}(2021){Mostert}, {Duncan}, {R{\"o}ttgering}, {Polsterer}, {Best}, {Brienza}, {Br{\"u}ggen}, {Hardcastle}, {Jurlin}, {Mingo}, {Morganti}, {Shimwell}, {Smith}, \& {Williams}}]{mostert21}
{Mostert}, R. I.~J., {Duncan}, K.~J., {R{\"o}ttgering}, H. J.~A., {et~al.} 2021, \aap, 645, A89

\bibitem[{{Norris}(2017{\natexlab{a}})}]{norris17b}
{Norris}, R.~P. 2017{\natexlab{a}}, \pasa, 34, e007

\bibitem[{{Norris}(2017{\natexlab{b}})}]{norris17a}
---. 2017{\natexlab{b}}, Nature Astronomy, 1, 671

\bibitem[{{Norris} {et~al.}(2006){Norris}, {Afonso}, {Appleton}, {Boyle}, {Ciliegi}, {Croom}, {Huynh}, {Jackson}, {Koekemoer}, {Lonsdale}, {Middelberg}, {Mobasher}, {Oliver}, {Polletta}, {Siana}, {Smail}, \& {Voronkov}}]{norris06}
{Norris}, R.~P., {Afonso}, J., {Appleton}, P.~N., {et~al.} 2006, \aj, 132, 2409

\bibitem[{{Norris} {et~al.}(2011){Norris}, {Hopkins}, {Afonso}, {Brown}, {Condon}, {Dunne}, {Feain}, {Hollow}, {Jarvis}, {Johnston-Hollitt}, {Lenc}, {Middelberg}, {Padovani}, {Prandoni}, {Rudnick}, {Seymour}, {Umana}, {Andernach}, {Alexander}, {Appleton}, {Bacon}, {Banfield}, {Becker}, {Brown}, {Ciliegi}, {Jackson}, {Eales}, {Edge}, {Gaensler}, {Giovannini}, {Hales}, {Hancock}, {Huynh}, {Ibar}, {Ivison}, {Kennicutt}, {Kimball}, {Koekemoer}, {Koribalski}, {L{\'o}pez-S{\'a}nchez}, {Mao}, {Murphy}, {Messias}, {Pimbblet}, {Raccanelli}, {Randall}, {Reiprich}, {Roseboom}, {R{\"o}ttgering}, {Saikia}, {Sharp}, {Slee}, {Smail}, {Thompson}, {Urquhart}, {Wall}, \& {Zhao}}]{emu}
{Norris}, R.~P., {Hopkins}, A.~M., {Afonso}, J., {et~al.} 2011, \pasa, 28, 215

\bibitem[{{Norris} {et~al.}(2021{\natexlab{a}}){Norris}, {Marvil}, {Collier}, {Kapi{\'n}ska}, {O'Brien}, {Rudnick}, {Andernach}, {Asorey}, {Brown}, {Br{\"u}ggen}, {Crawford}, {English}, {Rahman}, {Filipovi{\'c}}, {Gordon}, {G{\"u}rkan}, {Hale}, {Hopkins}, {Huynh}, {HyeongHan}, {James Jee}, {Koribalski}, {Lenc}, {Luken}, {Parkinson}, {Prandoni}, {Raja}, {Reiprich}, {Riseley}, {Shabala}, {Sheil}, {Vernstrom}, {Whiting}, {Allison}, {Anderson}, {Ball}, {Bell}, {Bunton}, {Galvin}, {Gupta}, {Hotan}, {Jacka}, {Macgregor}, {Mahony}, {Maio}, {Moss}, {Pandey-Pommier}, \& {Voronkov}}]{pilot}
{Norris}, R.~P., {Marvil}, J., {Collier}, J.~D., {et~al.} 2021{\natexlab{a}}, \pasa, 38, e046

\bibitem[{{Norris} {et~al.}(2021{\natexlab{b}}){Norris}, {Intema}, {Kapi{\'n}ska}, {Koribalski}, {Lenc}, {Rudnick}, {Alsaberi}, {Anderson}, {Anderson}, {Crawford}, {Crocker}, {English}, {Filipovi{\'c}}, {Galvin}, {Hopkins}, {Hurley-Walker}, {Inoue}, {Luken}, {Macgregor}, {Manojlovi{\'c}}, {Marvil}, {O'Brien}, {Park}, {Raja}, {Shobhana}, {Venturi}, {Collier}, {Hale}, {Hotan}, {Moss}, \& {Whiting}}]{orc}
{Norris}, R.~P., {Intema}, H.~T., {Kapi{\'n}ska}, A.~D., {et~al.} 2021{\natexlab{b}}, \pasa, 38, e003

\bibitem[{{Osinga} {et~al.}(2020){Osinga}, {Miley}, {van Weeren}, {Shimwell}, {Duncan}, {Hardcastle}, {Mechev}, {R{\"o}ttgering}, {Tasse}, \& {Williams}}]{osinga20}
{Osinga}, E., {Miley}, G.~K., {van Weeren}, R.~J., {et~al.} 2020, \aap, 642, A70

\bibitem[{{Owen} \& {Rudnick}(1976)}]{owen76}
{Owen}, F.~N., \& {Rudnick}, L. 1976, \apjl, 205, L1

\bibitem[{{Pal} \& {Kumari}(2023)}]{pal23}
{Pal}, S., \& {Kumari}, S. 2023, Journal of Astrophysics and Astronomy, 44, 17

\bibitem[{{Parma} {et~al.}(1987){Parma}, {Fanti}, {Fanti}, {Morganti}, \& {de Ruiter}}]{parma87}
{Parma}, P., {Fanti}, C., {Fanti}, R., {Morganti}, R., \& {de Ruiter}, H.~R. 1987, \aap, 181, 244

\bibitem[{{Perucho} \& {Mart{\'\i}}(2007)}]{perucho07}
{Perucho}, M., \& {Mart{\'\i}}, J.~M. 2007, \mnras, 382, 526

\bibitem[{{Perucho} {et~al.}(2014){Perucho}, {Mart{\'\i}}, {Laing}, \& {Hardee}}]{perucho14}
{Perucho}, M., {Mart{\'\i}}, J.~M., {Laing}, R.~A., \& {Hardee}, P.~E. 2014, \mnras, 441, 1488

\bibitem[{{Porter} \& {Scaife}(2023)}]{porter23}
{Porter}, F. A.~M., \& {Scaife}, A. M.~M. 2023, RAS Techniques and Instruments, 2, 293

\bibitem[{{Proctor}(2011)}]{proctor11}
{Proctor}, D.~D. 2011, \apjs, 194, 31

\bibitem[{{Reynolds} {et~al.}(1996){Reynolds}, {Fabian}, {Celotti}, \& {Rees}}]{reynolds96}
{Reynolds}, C.~S., {Fabian}, A.~C., {Celotti}, A., \& {Rees}, M.~J. 1996, \mnras, 283, 873

\bibitem[{{Riley}(1972)}]{riley72}
{Riley}, J.~M. 1972, \mnras, 157, 349

\bibitem[{{Rubinur} {et~al.}(2017){Rubinur}, {Das}, {Kharb}, \& {Honey}}]{rubinur17}
{Rubinur}, K., {Das}, M., {Kharb}, P., \& {Honey}, M. 2017, \mnras, 465, 4772

\bibitem[{{Rudnick}(2002)}]{rudnick02}
{Rudnick}, L. 2002, \pasp, 114, 427

\bibitem[{{Rudnick}(2021)}]{rudnick21}
---. 2021, Galaxies, 9, 85

\bibitem[{{Rudnick} \& {Edgar}(1984)}]{rudnick84}
{Rudnick}, L., \& {Edgar}, B.~K. 1984, \apj, 279, 74

\bibitem[{{Rudnick} \& {Owen}(1976)}]{rudnick76}
{Rudnick}, L., \& {Owen}, F.~N. 1976, \apjl, 203, L107

\bibitem[{{Ryle} \& {Windram}(1968)}]{ryle68}
{Ryle}, M., \& {Windram}, M.~D. 1968, \mnras, 138, 1

\bibitem[{{Saikia} {et~al.}(1990){Saikia}, {Junor}, {Cornwell}, {Muxlow}, \& {Shastri}}]{saikia90}
{Saikia}, D.~J., {Junor}, W., {Cornwell}, T.~J., {Muxlow}, T.~W.~B., \& {Shastri}, P. 1990, \mnras, 245, 408

\bibitem[{{Saikia} {et~al.}(1989){Saikia}, {Junor}, {Muxlow}, \& {Tzioumis}}]{saikia89}
{Saikia}, D.~J., {Junor}, W., {Muxlow}, T.~W.~B., \& {Tzioumis}, A.~K. 1989, \nat, 339, 286

\bibitem[{{Saikia} {et~al.}(2006){Saikia}, {Konar}, \& {Kulkarni}}]{saikia06}
{Saikia}, D.~J., {Konar}, C., \& {Kulkarni}, V.~K. 2006, \mnras, 366, 1391

\bibitem[{{Saikia} {et~al.}(1986){Saikia}, {Shastri}, {Cornwell}, \& {Salter}}]{saikia86}
{Saikia}, D.~J., {Shastri}, P., {Cornwell}, T.~J., \& {Salter}, C.~J. 1986, in IAU Symposium, Vol. 119, Quasars, ed. G.~{Swarup} \& V.~K. {Kapahi}, 219

\bibitem[{{Saikia} {et~al.}(1996){Saikia}, {Thomasson}, {Jackson}, {Salter}, \& {Junor}}]{saikia96}
{Saikia}, D.~J., {Thomasson}, P., {Jackson}, N., {Salter}, C.~J., \& {Junor}, W. 1996, \mnras, 282, 837

\bibitem[{{Saripalli} \& {Subrahmanyan}(2009)}]{saripalli09}
{Saripalli}, L., \& {Subrahmanyan}, R. 2009, \apj, 695, 156

\bibitem[{{Sasmal} {et~al.}(2022{\natexlab{a}}){Sasmal}, {Bera}, \& {Mondal}}]{sasmal22}
{Sasmal}, T.~K., {Bera}, S., \& {Mondal}, S. 2022{\natexlab{a}}, Astronomische Nachrichten, 343, e20210083

\bibitem[{{Sasmal} {et~al.}(2022{\natexlab{b}}){Sasmal}, {Bera}, {Pal}, \& {Mondal}}]{sasma22}
{Sasmal}, T.~K., {Bera}, S., {Pal}, S., \& {Mondal}, S. 2022{\natexlab{b}}, \apjs, 259, 31

\bibitem[{{Schlafly} {et~al.}(2019){Schlafly}, {Meisner}, \& {Green}}]{schlafly19}
{Schlafly}, E.~F., {Meisner}, A.~M., \& {Green}, G.~M. 2019, \apjs, 240, 30

\bibitem[{{Schoenmakers} {et~al.}(2000){Schoenmakers}, {de Bruyn}, {R{\"o}ttgering}, {van der Laan}, \& {Kaiser}}]{schoenmakers00}
{Schoenmakers}, A.~P., {de Bruyn}, A.~G., {R{\"o}ttgering}, H.~J.~A., {van der Laan}, H., \& {Kaiser}, C.~R. 2000, \mnras, 315, 371

\bibitem[{{Segal} {et~al.}(2023){Segal}, {Parkinson}, {Norris}, {Hopkins}, {Andernach}, {Alexander}, {Carretti}, {Koribalski}, {Legodi}, {Leslie}, {Luo}, {Pierce}, {Tang}, {Vardoulaki}, \& {Vernstrom}}]{segal23}
{Segal}, G., {Parkinson}, D., {Norris}, R., {et~al.} 2023, \mnras, 521, 1429

\bibitem[{{Shabala}(2018)}]{shabala18}
{Shabala}, S.~S. 2018, \mnras, 478, 5074

\bibitem[{{Shabala} {et~al.}(2024){Shabala}, {Yates-Jones}, {Jerrim}, {Turner}, {Krause}, {Norris}, {Koribalski}, {Filipovi{\'c}}, {Rudnick}, {Power}, \& {Crocker}}]{shabala24}
{Shabala}, S.~S., {Yates-Jones}, P.~M., {Jerrim}, L.~A., {et~al.} 2024, \pasa, 41, e024

\bibitem[{{Shimwell} {et~al.}(2017){Shimwell}, {R{\"o}ttgering}, {Best}, {Williams}, {Dijkema}, {de Gasperin}, {Hardcastle}, {Heald}, {Hoang}, {Horneffer}, {Intema}, {Mahony}, {Mandal}, {Mechev}, {Morabito}, {Oonk}, {Rafferty}, {Retana-Montenegro}, {Sabater}, {Tasse}, {van Weeren}, {Br{\"u}ggen}, {Brunetti}, {Chy{\.z}y}, {Conway}, {Haverkorn}, {Jackson}, {Jarvis}, {McKean}, {Miley}, {Morganti}, {White}, {Wise}, {van Bemmel}, {Beck}, {Brienza}, {Bonafede}, {Calistro Rivera}, {Cassano}, {Clarke}, {Cseh}, {Deller}, {Drabent}, {van Driel}, {Engels}, {Falcke}, {Ferrari}, {Fr{\"o}hlich}, {Garrett}, {Harwood}, {Heesen}, {Hoeft}, {Horellou}, {Israel}, {Kapi{\'n}ska}, {Kunert-Bajraszewska}, {McKay}, {Mohan}, {Orr{\'u}}, {Pizzo}, {Prandoni}, {Schwarz}, {Shulevski}, {Sipior}, {Smith}, {Sridhar}, {Steinmetz}, {Stroe}, {Varenius}, {van der Werf}, {Zensus}, \& {Zwart}}]{shimwell17}
{Shimwell}, T.~W., {R{\"o}ttgering}, H.~J.~A., {Best}, P.~N., {et~al.} 2017, \aap, 598, A104

\bibitem[{{Shimwell} {et~al.}(2019){Shimwell}, {Tasse}, {Hardcastle}, {Mechev}, {Williams}, {Best}, {R{\"o}ttgering}, {Callingham}, {Dijkema}, {de Gasperin}, {Hoang}, {Hugo}, {Mirmont}, {Oonk}, {Prandoni}, {Rafferty}, {Sabater}, {Smirnov}, {van Weeren}, {White}, {Atemkeng}, {Bester}, {Bonnassieux}, {Br{\"u}ggen}, {Brunetti}, {Chy{\.z}y}, {Cochrane}, {Conway}, {Croston}, {Danezi}, {Duncan}, {Haverkorn}, {Heald}, {Iacobelli}, {Intema}, {Jackson}, {Jamrozy}, {Jarvis}, {Lakhoo}, {Mevius}, {Miley}, {Morabito}, {Morganti}, {Nisbet}, {Orr{\'u}}, {Perkins}, {Pizzo}, {Schrijvers}, {Smith}, {Vermeulen}, {Wise}, {Alegre}, {Bacon}, {van Bemmel}, {Beswick}, {Bonafede}, {Botteon}, {Bourke}, {Brienza}, {Calistro Rivera}, {Cassano}, {Clarke}, {Conselice}, {Dettmar}, {Drabent}, {Dumba}, {Emig}, {En{\ss}lin}, {Ferrari}, {Garrett}, {G{\'e}nova-Santos}, {Goyal}, {G{\"u}rkan}, {Hale}, {Harwood}, {Heesen}, {Hoeft}, {Horellou}, {Jackson}, {Kokotanekov}, {Kondapally}, {Kunert-Bajraszewska}, {Mahatma}, {Mahony}, {Mandal}, {McKean},
  {Merloni}, {Mingo}, {Miskolczi}, {Mooney}, {Nikiel-Wroczy{\'n}ski}, {O'Sullivan}, {Quinn}, {Reich}, {Roskowi{\'n}ski}, {Rowlinson}, {Savini}, {Saxena}, {Schwarz}, {Shulevski}, {Sridhar}, {Stacey}, {Urquhart}, {van der Wiel}, {Varenius}, {Webster}, \& {Wilber}}]{shimwell19}
{Shimwell}, T.~W., {Tasse}, C., {Hardcastle}, M.~J., {et~al.} 2019, \aap, 622, A1

\bibitem[{{Shimwell} {et~al.}(2022){Shimwell}, {Hardcastle}, {Tasse}, {Best}, {R{\"o}ttgering}, {Williams}, {Botteon}, {Drabent}, {Mechev}, {Shulevski}, {van Weeren}, {Bester}, {Br{\"u}ggen}, {Brunetti}, {Callingham}, {Chy{\.z}y}, {Conway}, {Dijkema}, {Duncan}, {de Gasperin}, {Hale}, {Haverkorn}, {Hugo}, {Jackson}, {Mevius}, {Miley}, {Morabito}, {Morganti}, {Offringa}, {Oonk}, {Rafferty}, {Sabater}, {Smith}, {Schwarz}, {Smirnov}, {O'Sullivan}, {Vedantham}, {White}, {Albert}, {Alegre}, {Asabere}, {Bacon}, {Bonafede}, {Bonnassieux}, {Brienza}, {Bilicki}, {Bonato}, {Calistro Rivera}, {Cassano}, {Cochrane}, {Croston}, {Cuciti}, {Dallacasa}, {Danezi}, {Dettmar}, {Di Gennaro}, {Edler}, {En{\ss}lin}, {Emig}, {Franzen}, {Garc{\'\i}a-Vergara}, {Grange}, {G{\"u}rkan}, {Hajduk}, {Heald}, {Heesen}, {Hoang}, {Hoeft}, {Horellou}, {Iacobelli}, {Jamrozy}, {Jeli{\'c}}, {Kondapally}, {Kukreti}, {Kunert-Bajraszewska}, {Magliocchetti}, {Mahatma}, {Ma{\l}ek}, {Mandal}, {Massaro}, {Meyer-Zhao}, {Mingo}, {Mostert}, {Nair},
  {Nakoneczny}, {Nikiel-Wroczy{\'n}ski}, {Orr{\'u}}, {Pajdosz-{\'S}mierciak}, {Pasini}, {Prandoni}, {van Piggelen}, {Rajpurohit}, {Retana-Montenegro}, {Riseley}, {Rowlinson}, {Saxena}, {Schrijvers}, {Sweijen}, {Siewert}, {Timmerman}, {Vaccari}, {Vink}, {West}, {Wo{\l}owska}, {Zhang}, \& {Zheng}}]{shimwell22}
{Shimwell}, T.~W., {Hardcastle}, M.~J., {Tasse}, C., {et~al.} 2022, \aap, 659, A1

\bibitem[{{Simonte} {et~al.}(2023){Simonte}, {Andernach}, {Br{\"u}ggen}, {Best}, \& {Osinga}}]{simonte23}
{Simonte}, M., {Andernach}, H., {Br{\"u}ggen}, M., {Best}, P.~N., \& {Osinga}, E. 2023, \aap, 672, A178

\bibitem[{{Stroe} {et~al.}(2022){Stroe}, {Catlett}, {Harwood}, {Vernstrom}, \& {Mingo}}]{stroe22}
{Stroe}, A., {Catlett}, V., {Harwood}, J.~J., {Vernstrom}, T., \& {Mingo}, B. 2022, \apj, 941, 136

\bibitem[{{Taylor} \& {Jagannathan}(2016)}]{taylor16}
{Taylor}, A.~R., \& {Jagannathan}, P. 2016, \mnras, 459, L36

\bibitem[{{Taylor} {et~al.}(1990){Taylor}, {Perley}, {Inoue}, {Kato}, {Tabara}, \& {Aizu}}]{taylor98}
{Taylor}, G.~B., {Perley}, R.~A., {Inoue}, M., {et~al.} 1990, \apj, 360, 41

\bibitem[{{Taylor}(2005)}]{topcat}
{Taylor}, M.~B. 2005, in Astronomical Society of the Pacific Conference Series, Vol. 347, Astronomical Data Analysis Software and Systems XIV, ed. P.~{Shopbell}, M.~{Britton}, \& R.~{Ebert}, 29

\bibitem[{{Tchekhovskoy} \& {Bromberg}(2016)}]{tchekhovskoy16}
{Tchekhovskoy}, A., \& {Bromberg}, O. 2016, \mnras, 461, L46

\bibitem[{{Treichel} {et~al.}(2001){Treichel}, {Rudnick}, {Hardcastle}, \& {Leahy}}]{treichel01}
{Treichel}, K., {Rudnick}, L., {Hardcastle}, M.~J., \& {Leahy}, J.~P. 2001, \apj, 561, 691

\bibitem[{{Turner} \& {Shabala}(2020)}]{turner20}
{Turner}, R.~J., \& {Shabala}, S.~S. 2020, \mnras, 493, 5181

\bibitem[{{van Groningen} {et~al.}(1980){van Groningen}, {Miley}, \& {Norman}}]{groningen80}
{van Groningen}, E., {Miley}, G.~K., \& {Norman}, C.~A. 1980, \aap, 90, L7

\bibitem[{{van Haarlem} {et~al.}(2013){van Haarlem}, {Wise}, {Gunst}, {Heald}, {McKean}, {Hessels}, {de Bruyn}, {Nijboer}, {Swinbank}, {Fallows}, {Brentjens}, {Nelles}, {Beck}, {Falcke}, {Fender}, {H{\"o}randel}, {Koopmans}, {Mann}, {Miley}, {R{\"o}ttgering}, {Stappers}, {Wijers}, {Zaroubi}, {van den Akker}, {Alexov}, {Anderson}, {Anderson}, {van Ardenne}, {Arts}, {Asgekar}, {Avruch}, {Batejat}, {B{\"a}hren}, {Bell}, {Bell}, {van Bemmel}, {Bennema}, {Bentum}, {Bernardi}, {Best}, {B{\^\i}rzan}, {Bonafede}, {Boonstra}, {Braun}, {Bregman}, {Breitling}, {van de Brink}, {Broderick}, {Broekema}, {Brouw}, {Br{\"u}ggen}, {Butcher}, {van Cappellen}, {Ciardi}, {Coenen}, {Conway}, {Coolen}, {Corstanje}, {Damstra}, {Davies}, {Deller}, {Dettmar}, {van Diepen}, {Dijkstra}, {Donker}, {Doorduin}, {Dromer}, {Drost}, {van Duin}, {Eisl{\"o}ffel}, {van Enst}, {Ferrari}, {Frieswijk}, {Gankema}, {Garrett}, {de Gasperin}, {Gerbers}, {de Geus}, {Grie{\ss}meier}, {Grit}, {Gruppen}, {Hamaker}, {Hassall}, {Hoeft}, {Holties},
  {Horneffer}, {van der Horst}, {van Houwelingen}, {Huijgen}, {Iacobelli}, {Intema}, {Jackson}, {Jelic}, {de Jong}, {Juette}, {Kant}, {Karastergiou}, {Koers}, {Kollen}, {Kondratiev}, {Kooistra}, {Koopman}, {Koster}, {Kuniyoshi}, {Kramer}, {Kuper}, {Lambropoulos}, {Law}, {van Leeuwen}, {Lemaitre}, {Loose}, {Maat}, {Macario}, {Markoff}, {Masters}, {McFadden}, {McKay-Bukowski}, {Meijering}, {Meulman}, {Mevius}, {Middelberg}, {Millenaar}, {Miller-Jones}, {Mohan}, {Mol}, {Morawietz}, {Morganti}, {Mulcahy}, {Mulder}, {Munk}, {Nieuwenhuis}, {van Nieuwpoort}, {Noordam}, {Norden}, {Noutsos}, {Offringa}, {Olofsson}, {Omar}, {Orr{\'u}}, {Overeem}, {Paas}, {Pandey-Pommier}, {Pandey}, {Pizzo}, {Polatidis}, {Rafferty}, {Rawlings}, {Reich}, {de Reijer}, {Reitsma}, {Renting}, {Riemers}, {Rol}, {Romein}, {Roosjen}, {Ruiter}, {Scaife}, {van der Schaaf}, {Scheers}, {Schellart}, {Schoenmakers}, {Schoonderbeek}, {Serylak}, {Shulevski}, {Sluman}, {Smirnov}, {Sobey}, {Spreeuw}, {Steinmetz}, {Sterks}, {Stiepel}, {Stuurwold},
  {Tagger}, {Tang}, {Tasse}, {Thomas}, {Thoudam}, {Toribio}, {van der Tol}, {Usov}, {van Veelen}, {van der Veen}, {ter Veen}, {Verbiest}, {Vermeulen}, {Vermaas}, {Vocks}, {Vogt}, {de Vos}, {van der Wal}, {van Weeren}, {Weggemans}, {Weltevrede}, {White}, {Wijnholds}, {Wilhelmsson}, {Wucknitz}, {Yatawatta}, {Zarka}, \& {Zensus}}]{lofar}
{van Haarlem}, M.~P., {Wise}, M.~W., {Gunst}, A.~W., {et~al.} 2013, \aap, 556, A2

\bibitem[{{Vantyghem} {et~al.}(2024){Vantyghem}, {Galvin}, {Sebastian}, {O'Dea}, {Gordon}, {Boyce}, {Rudnick}, {Polsterer}, {Andernach}, {Dionyssiou}, {Venkataraman}, {Norris}, {Baum}, {Wang}, \& {Huynh}}]{vantyghem24}
{Vantyghem}, A.~N., {Galvin}, T.~J., {Sebastian}, B., {et~al.} 2024, Astronomy and Computing, 47, 100824

\bibitem[{{Velovi{\'c}} {et~al.}(2023){Velovi{\'c}}, {Cotton}, {Filipovi{\'c}}, {Norris}, {Barnes}, \& {Condon}}]{velovic23}
{Velovi{\'c}}, V., {Cotton}, W.~D., {Filipovi{\'c}}, M.~D., {et~al.} 2023, \mnras, 523, 1933

\bibitem[{{Whiting} \& {Humphreys}(2012)}]{whiting12}
{Whiting}, M., \& {Humphreys}, B. 2012, \pasa, 29, 371

\bibitem[{{Williams} {et~al.}(2019){Williams}, {Hardcastle}, {Best}, {Sabater}, {Croston}, {Duncan}, {Shimwell}, {R{\"o}ttgering}, {Nisbet}, {G{\"u}rkan}, {Alegre}, {Cochrane}, {Goyal}, {Hale}, {Jackson}, {Jamrozy}, {Kondapally}, {Kunert-Bajraszewska}, {Mahatma}, {Mingo}, {Morabito}, {Prandoni}, {Roskowinski}, {Shulevski}, {Smith}, {Tasse}, {Urquhart}, {Webster}, {White}, {Beswick}, {Callingham}, {Chy{\.z}y}, {de Gasperin}, {Harwood}, {Hoeft}, {Iacobelli}, {McKean}, {Mechev}, {Miley}, {Schwarz}, \& {van Weeren}}]{williams19}
{Williams}, W.~L., {Hardcastle}, M.~J., {Best}, P.~N., {et~al.} 2019, \aap, 622, A2

\bibitem[{{Wing} \& {Blanton}(2011)}]{wing11}
{Wing}, J.~D., \& {Blanton}, E.~L. 2011, \aj, 141, 88

\bibitem[{{Wong} {et~al.}(2025){Wong}, {Garon}, {Alger}, {Rudnick}, {Shabala}, {Willett}, {Banfield}, {Andernach}, {Norris}, {Swan}, {Hardcastle}, {Lintott}, {White}, {Seymour}, {Kapi{\'n}ska}, {Tang}, {Simmons}, \& {Schawinski}}]{wong25}
{Wong}, O.~I., {Garon}, A.~F., {Alger}, M.~J., {et~al.} 2025, \mnras, 536, 3488

\bibitem[{{Wright} {et~al.}(2010){Wright}, {Eisenhardt}, {Mainzer}, {Ressler}, {Cutri}, {Jarrett}, {Kirkpatrick}, {Padgett}, {McMillan}, {Skrutskie}, {Stanford}, {Cohen}, {Walker}, {Mather}, {Leisawitz}, {Gautier}, {McLean}, {Benford}, {Lonsdale}, {Blain}, {Mendez}, {Irace}, {Duval}, {Liu}, {Royer}, {Heinrichsen}, {Howard}, {Shannon}, {Kendall}, {Walsh}, {Larsen}, {Cardon}, {Schick}, {Schwalm}, {Abid}, {Fabinsky}, {Naes}, \& {Tsai}}]{wright10}
{Wright}, E.~L., {Eisenhardt}, P. R.~M., {Mainzer}, A.~K., {et~al.} 2010, \aj, 140, 1868

\bibitem[{{Wu} {et~al.}(2019){Wu}, {Wong}, {Rudnick}, {Shabala}, {Alger}, {Banfield}, {Ong}, {White}, {Garon}, {Norris}, {Andernach}, {Tate}, {Lukic}, {Tang}, {Schawinski}, \& {Diakogiannis}}]{wu19}
{Wu}, C., {Wong}, O.~I., {Rudnick}, L., {et~al.} 2019, \mnras, 482, 1211

\bibitem[{{Yang} {et~al.}(2021){Yang}, {Paragi}, {Beswick}, {Chen}, {van Bemmel}, {Wu}, {An}, {Wu}, {Fan}, {Oonk}, {Liu}, \& {Wang}}]{yang21}
{Yang}, J., {Paragi}, Z., {Beswick}, R.~J., {et~al.} 2021, \mnras, 503, 3886

\bibitem[{Yang {et~al.}(2019)Yang, Joshi, Gopal-Krishna, An, Ho, Wiita, Liu, Yang, Wang, Wu, \& Yang}]{yang19}
Yang, X., Joshi, R., Gopal-Krishna, {et~al.} 2019, The Astrophysical Journal Supplement Series, 245, 17

\bibitem[{{Yew}(2022)}]{yew22}
{Yew}, M. 2022, PhD thesis, {School of Computing, Engineering, and Maths, Western Sydney University}

\bibitem[{{Young} {et~al.}(2024){Young}, {Turner}, {Shabala}, {Stewart}, \& {Yates-Jones}}]{young25}
{Young}, S.~A., {Turner}, R.~J., {Shabala}, S.~S., {Stewart}, G. S.~C., \& {Yates-Jones}, P.~M. 2024, arXiv e-prints, arXiv:2412.14433

\end{thebibliography}
\newpage 
\appendix

\end{document}